\documentclass[final,5p,times,twocolumn]{elsarticle}

\usepackage{graphicx}
\usepackage{subfigure}
\usepackage{float}

\usepackage{amssymb}

\biboptions{sort&compress}
\journal{Computers $\&$ Fluids}

\begin{document}

\begin{frontmatter}



\title{High-order methods for decaying two-dimensional homogeneous isotropic turbulence}


\author{Omer San\corref{cor1}}
\ead{omersan@vt.edu}
\cortext[cor1]{Corresponding author.}
\author{and Anne E. Staples}

\address{Department of Engineering Science and Mechanics\\ Virginia Tech, Blacksburg, VA 24061, USA}

\begin{abstract}
Numerical schemes used for the integration of complex flow simulations should provide accurate solutions for the long time integrations these flows require. To this end, the performance of various high-order accurate numerical schemes is investigated for direct numerical simulations (DNS) of homogeneous isotropic two-dimensional decaying turbulent flows. The numerical accuracy of compact difference, explicit central difference, Arakawa, and dispersion-relation-preserving schemes are analyzed and compared with the Fourier-Galerkin pseudospectral scheme. In addition, several explicit Runge-Kutta schemes for time integration are investigated. We demonstrate that the centered schemes suffer from spurious Nyquist signals that are generated almost instantaneously and propagate into much of the field when the numerical resolution is insufficient. We further show that the order of the scheme becomes increasingly important for increasing cell Reynolds number. Surprisingly, the sixth-order schemes are found to be in perfect agreement with the pseudospectral method. Considerable reduction in computational time compared to the pseudospectral method is also reported in favor of the finite difference schemes. Among the fourth-order schemes, the compact scheme provides better accuracy than the others for fully resolved computations. The fourth-order Arakawa scheme provides more accurate results for under-resolved computations, however, due to its conservation properties. Our results show that, contrary to conventional wisdom, difference methods demonstrate superior performance in terms of accuracy and efficiency for fully resolved DNS computations of the complex flows considered here. For under-resolved simulations, however, the choice of difference method should be made with care.
\end{abstract}


\end{frontmatter}


\section{Introduction}
\label{sec:intro}

The physics of two-dimensional turbulence have been elucidated substantially during the past decades by theoretical models, intensive numerical investigations, and dedicated soap film experiments \cite{bruneau2009influence}. Two-dimensional turbulence research efforts have applicability in geophysics, astronomy and plasma physics, in which numerical experiments play a large role. One of the most important reasons for studying two-dimensional turbulence is to improve our understanding of geophysical flows in the atmosphere and ocean \cite{lilly1971numerical,herring1974decay,mcwilliams1984emergence,maltrud1991energy,lindborg1999can,ferziger2002numerical,san2011approximate}. We may also find two-dimensional flows in a wide variety of situations such as flows in rapidly rotating systems and flows in a fluid film on top of the surface of another fluid or a rigid object \cite{kolvin2009energy}.

Two-dimensional turbulence behaves in a profoundly different way from three-dimensional turbulence due to different energy cascade behavior, and follows the Kraichnan-Batchelor-Leith (KBL) theory  \citep{kraichnan1967inertial,batchelor1969computation,leith1971atmospheric}. In three-dimensional turbulence, energy is transferred forward, from large scales to smaller scales, via vortex stretching. In two dimensions that mechanism is absent, and it turns out that under most forcing and dissipation conditions energy will be transferred from smaller scales to larger scales. This is largely because of another quadratic invariant, the potential enstrophy, defined as the integral of the square of the potential vorticity. Despite the apparent simplicity in dealing with two rather than three spatial dimensions, two-dimensional turbulence is possibly richer in its dynamics than three-dimensional turbulence due to its conservation properties, such as its inverse energy and forward enstrophy cascading mechanisms. Danilov and Gurarie \cite{danilov2000quasi} and Tabeling \cite{tabeling2002two} reviewed both theoretical and experimental two-dimensional turbulence studies along with extensions into geophysical flow settings. More recent reviews on two-dimensional turbulence are also provided by Clercx and van Heijst \cite{clercx2009two} and Boffetta and Ecke \cite{boffetta2012two}. Recent studies in two-dimensional turbulence, both forced (stationary) turbulence \citep{danilov2001forced,boffetta2010evidence,bracco2010reynolds,vallgren2011enstrophy} and unforced (decaying) turbulence \citep{lindborg2010testing,kuznetsov2010sharp,fox2010freely} provide high resolution computational confirmation of the KBL theory.

Simulation of turbulent and other convection-dominated unsteady flows using direct numerical simulation (DNS) requires a numerical method that properly resolves all the multiscale flow structures \citep{orlandi2000fluid}. Since high accuracy is crucial in numerical simulation of complex flows with multiscale structures, such as the unsteady evolution of a turbulent flow field, most two-dimensional turbulence studies have been performed using pseudospectral methods based on fast Fourier transform (FFT) algorithms \citep{lindborg2010testing,boffetta2012two}. Simulations performed by the lattice Boltzmann method (LBM) have been also presented for two-dimensional decaying turbulence \cite{házi2006simulation}. Pseudospectral methods are highly accurate but mostly limited to ideal geometries such as rectangular or circular domains. Discretization methods such as finite difference, finite element, or finite volume methods are often preferred in more realistic problems. Finite difference methods offer an attractive alternative to spectral methods in the direct and large eddy simulations (LES) of turbulence providing reasonable accuracy coupled with relative ease of implementation in simple and complex flow geometries \cite{tafti1996comparison,jordan1996efficient,ekaterinaris2005high,merle2010finite}. Computational algorithms developed in the past were mainly designed for solving large-scale fluid dynamics problems using second-order spatial accuracy \cite{harlow1965numerical,lilly1965computational,mavriplis1997unstructured}. These algorithms usually have rather significant dispersion errors and if they are not centered schemes they also have large dissipation errors, making it hard to accurately compute fine structures in the flow field using them \cite{mattsson2007high}. There are two ways to improve the resolution of these methods; one is to refine the grid and the other is to construct a high-order accurate scheme. Our approach here is to test and evaluate different high-order formulations for instantaneous and statistical properties of two-dimensional turbulence and compare their accuracy and efficiency with those of the pseudospectral and the second-order schemes. Furthermore, it has been shown by Kravchenko and Moin \cite{kravchenko1997effect} that the subgrid-scale models in LES are effective only if central discretization of order higher than two is employed. With this in mind, we will investigate the behavior of four different families of high-order accurate finite difference methods in the decay of two-dimensional isotropic turbulence.

High-order finite difference schemes can be formulated to reduce the truncation errors associated with the difference approximations. A straightforward Taylor series expansion of a pointwise discretization under certain assumptions results in a family of the explicit difference (ED) schemes. The compact difference (CD) schemes feature high-order accuracy with smaller stencils and smaller truncation errors than the ED schemes, and have been employed as an alternative to spectral methods in simulations of turbulence with great flexibility \citep{lele1992compact}. On the other hand, increasing the stencil size allows us to optimize the weight coefficients in the difference equation. This strategy leads to the dispersion-relation-preserving (DRP) schemes \citep{tam1993dispersion}, which have been used mostly in acoustics. Another strategy to construct a numerical scheme is based on the conservation properties of the discrete form of the equations. Arakawa \cite{arakawa1966computational} suggested that the conservation of energy, enstrophy, and skew-symmetry is sufficient to avoid computational instabilities stemming from nonlinear interactions. The conservation and stability properties of the Arakawa scheme were investigated by Lilly \cite{lilly1965computational} by means of spectral analysis along with several first and second-order time integration methods. In the present work, we test several Runge-Kutta methods for time integration, although the primary goal here is to analyze the accuracy of these high-order accurate spatial differencing methods for the long-term evolution of complex two-dimensional turbulent flows. For finite difference schemes, the combination of differentiation errors and nonlinear truncation and aliasing errors, which usually manifest themselves in the high wavenumbers of the resolved scales, determines the overall error at the small scales. Looking at the accuracy of the whole solution procedure we also investigate the resolution requirements for these finite difference families, the effects of the order of the schemes, and the importance of the global conservation properties.

The paper is organized as follows: the mathematical formulation of the problem is given in Section~\ref{sec:mathmod}. The numerical methods are presented in Section~\ref{sec:NumS} with descriptions of high-order accurate spatial discretization schemes, temporal discretization algorithms, and an efficient fast Poisson solver algorithm. These schemes are validated in Section~\ref{sec:vTGV} by simulating the Taylor-Green decaying vortex benchmark problem for the unsteady incompressible Navier-Stokes equations. The effective accuracies of these methods are also provided in this section, and are confirmed to be the theoretical accuracies of the schemes. Section~\ref{sec:dsl} presents a careful numerical investigation of their performance for a challenging benchmark problem which consists of strong shear layers. The results for two-dimensional isotropic homogeneous decaying turbulence are provided in Section~\ref{sec:decayR}. The behavior of these nine different spatial schemes are tested in terms of accuracy and efficiency. The effects of several explicit Runge-Kutta time advancement techniques on the whole solution procedure are also analyzed. In addition, the Reynolds number ($Re$) dependency of the turbulence statistics is illustrated in this section. Final conclusions and some comments on the performance of these schemes are drawn in section~\ref{sec:decayC}.

\section{Mathematical model}
\label{sec:mathmod}
The governing equations for two-dimensional incompressible flows can be written in a dimensionless form of the vorticity-stream function formulation as
\begin{equation}
\frac{\partial \omega}{\partial t} + \frac{\partial \psi}{\partial y}\frac{\partial \omega}{\partial x} - \frac{\partial \psi}{\partial x}\frac{\partial \omega}{\partial y} = \frac{1}{Re}(\frac{\partial^2 \omega}{\partial x^2} + \frac{\partial^2 \omega}{\partial y^2})
\label{eq:ge}
\end{equation}
along with the kinematic relationship between vorticity and stream function according to a Poisson equation, which is given as
\begin{equation}
\frac{\partial^2 \psi}{\partial x^2} + \frac{\partial^2 \psi}{\partial y^2} = -\omega.
\label{eq:ke}
\end{equation}

From a computational point of view, this formulation has several advantages over the primitive variable formulation. It eliminates pressure from the Navier-Stokes equations and hence has no corresponding odd-even decoupling between the pressure and velocity components, as well as projection inaccuracies usually observed in fractional step approaches \cite{brown2001accurate}. Therefore, the usage of a collocated grid does not produce any spurious modes in the vorticity-stream function formulation. The vorticity-stream function formulation automatically satisfies the divergence-free condition and allows one to reduce the number of equations to be solved.

The main objective of our work is to test and evaluate different frameworks for high-order accurate finite difference schemes and compare them with a spectrally accurate pseudospectral method for two-dimensional isotropic turbulent flows. In fact, to be able to compare the numerical schemes more precisely we restricted ourselves to periodic boundary conditions and a uniform Cartesian grid. Consequently, we eliminated errors coming from the mesh non-uniformities and inconsistent boundary schemes. It should also be noted that using the vorticity-stream function formulation on a collocated grid provides us with an ideal computational setting in which to test the characteristics of the numerical schemes by eliminating any possible errors coming from projection inaccuracies.

\section{Numerical methods}
\label{sec:NumS}
The objective of the present work is to test and evaluate different frameworks for high-order accurate finite difference schemes and compare them with a spectrally accurate pseudospectral method for two-dimensional isotropic turbulence flows. We will also compare them with their classical second-order accurate formulations in order to analyze the effects of order of accuracy. In the text that follows, the high-order finite difference approximations for spatial discretization, the pseudospectral method, an efficient procedure for the Poisson equation, and the time integration methods that we utilize for time advancement are given briefly for completeness.
\subsection{Explicit difference (ED) scheme}
\label{sec:ed}
For any scalar pointwise value of $f$ the classical centered finite difference schemes for the first derivative up the the sixth order accuracy are given by \cite{strikwerda2004finite}
\begin{eqnarray}
f'_i&=&\frac{1}{2h}(-f_{i-1} + f_{i+1}) + O(h^2) \\
f'_i&=&\frac{1}{12h}(f_{i-2} - 8f_{i-1} + 8f_{i+1} - f_{i+2}) + O(h^4) \\
f'_i&=&\frac{1}{60h}(-f_{i-3} +9f_{i-2} - 45f_{i-1} \nonumber \\
    & & \quad \quad +45f_{i+1} - 9f_{i+2} + f_{i+3}) + O(h^6)
\label{eq:ex2}
\end{eqnarray}
Similarly, the second derivative approximations become
\begin{eqnarray}
f''_i&=&\frac{1}{h^2}(f_{i-1} - 2f_{i} + f_{i+1}) + O(h^2) \\
f''_i&=&\frac{1}{12h^2}(-f_{i-2} +16f_{i-1} -30f_{i} \nonumber\\
     & &\quad \quad + 16f_{i+1} - f_{i+2}) + O(h^4) \\
f''_i&=&\frac{1}{180h^2}(2f_{i-3} -27f_{i-2} +270f_{i-1} -490f_{i} \nonumber\\
     & &\quad \quad+ 270f_{i+1} - 27f_{i+2} + 2f_{i+3}) + O(h^6)
\label{eq:ex2}
\end{eqnarray}
where $h$ is the step size in the derivative direction.

\subsection{Compact difference (CD) scheme}
\label{sec:cd}
Apart from the explicit centered schemes, with the cost of a tridiagonal system solution (i.e., by using the well-known Thomas algorithm), the first derivative can be computed accordingly \cite{lele1992compact}
\begin{equation}
\alpha f'_{i-1} + f'_{i} +  \alpha f'_{i+1} = a\frac{f_{i+1}-f_{i-1}}{2h} + b\frac{f_{i+2}-f_{i-2}}{4h}
\label{eq:com1}
\end{equation}
which gives rise to an $\alpha$-family of tridiagonal schemes with $a=\frac{2}{3}(\alpha+2)$, and $b=\frac{1}{3}(4\alpha-1)$. Here, $\alpha=0$ leads to the explicit non-compact fourth-order scheme for the first derivative. A classical compact fourth-order scheme, which is also known as the Pad\'{e} scheme, is obtained by setting $\alpha=1/4$. The truncation error in the Eq.~(\ref{eq:com1}) is $\frac{4}{5!}(3\alpha-1)h^4 f^{(5)}$. Therefore, a sixth-order compact scheme is obtained by choosing $\alpha=1/3$.

The second derivative compact centered scheme is given by
\begin{eqnarray}
\alpha f''_{i-1} + f''_{i} +  \alpha f''_{i+1} &=& a\frac{f_{i+1}-2f_{i} + f_{i-1}}{h^2} \nonumber\\
                                               &+& b\frac{f_{i+2}-2f_{i} +f_{i-2}}{4h^2}
\label{eq:com2}
\end{eqnarray}
where $a=\frac{4}{3}(1-\alpha)$, and $b=\frac{1}{3}(10\alpha-1)$. For $\alpha=1/10$ the classical fourth-order Pad\'{e} scheme is obtained. The truncation error in the Eq.~(\ref{eq:com2}) is $-\frac{4}{6!}(11\alpha-2)h^4 f^{(6)}$. Therefore, a sixth-order compact scheme is also obtained by choosing $\alpha=2/11$.

These compact schemes for the first and second derivatives show better spectral accuracy then their explicit counterparts. However, they involve solving tridiagonal matrix equations and are therefore effectively non-local. This requires special care for designing parallel computations. However, in this study, we focus on a single domain without domain decomposition, and so the compact schemes can be adopted easily.

\subsection{Dispersion-relation-preserving (DRP) scheme}
\label{sec:drp}
An optimized dispersion-preserving scheme was proposed in \cite{tam1993dispersion}. The scheme increases the stencil size, but improves the resolution in wavespace. The finite difference approximation for the first derivative is
\begin{equation}
f'_{i} = \frac{1}{h}\sum_{j=-N}^{M}a_{j}f_{i+j}
\label{eq:DRP}
\end{equation}
Fourth-order accuracy is obtained for $N=M=3$ (centered DRP scheme) and the optimized coefficients are $a_0=0$, $a_1=-a_{-1}=0.79926643$, $a_2=-a_{-2}=-0.18941314$, and $a_3=-a_{-3}=0.02651995$. The viscous terms can be computed by using the explicit difference schemes.

\subsection{Arakawa scheme}
\label{sec:arakawa}
Arakawa \cite{arakawa1966computational} suggested that the conservation of energy, enstrophy and skew symmetry is sufficient to avoid computational instabilities arising from nonlinear interactions. The nonlinear terms in Eq.~(\ref{eq:ge}) were defined as the Jacobian
\begin{equation}
J(\omega,\psi) = \frac{\partial \psi}{\partial y}\frac{\partial \omega}{\partial x} - \frac{\partial \psi}{\partial x}\frac{\partial \omega}{\partial y}
\label{eq:jac}
\end{equation}
The second-order Arakawa scheme for the Jacobian is
\begin{equation}
J_{I}(\omega,\psi) =\frac{1}{3}(J_{1}(\omega,\psi)+J_{2}(\omega,\psi)+J_{3}(\omega,\psi))
\label{eq:ja1}
\end{equation}
where the discrete parts of the Jacobian
\begin{eqnarray}
J_{1}(\omega,\psi) &=& \frac{1}{4h_xh_y}[(\omega_{i+1,j}-\omega_{i-1,j})(\psi_{i,j+1}-\psi_{i,j-1}) \nonumber \\
&-&(\omega_{i,j+1}-\omega_{i,j-1})(\psi_{i+1,j}-\psi_{i-1,j})]
\label{eq:j1}
\end{eqnarray}
\begin{eqnarray}
J_{2}(\omega,\psi) &=& \frac{1}{4h_xh_y}[\omega_{i+1,j}(\psi_{i+1,j+1}-\psi_{i+1,j-1}) \nonumber \\ &-&\omega_{i-1,j}(\psi_{i-1,j+1}-\psi_{i-1,j-1}) \nonumber \\
&-&\omega_{i,j+1}(\psi_{i+1,j+1}-\psi_{i-1,j+1}) \nonumber \\
&+&\omega_{i,j-1}(\psi_{i+1,j-1}-\psi_{i-1,j-1}) ]
\label{eq:j2}
\end{eqnarray}
\begin{eqnarray}
J_{3}(\omega,\psi) &=& \frac{1}{4h_xh_y}[\omega_{i+1,j+1}(\psi_{i,j+1}-\psi_{i+1,j}) \nonumber\\
&-&\omega_{i-1,j-1}(\psi_{i-1,j}-\psi_{i,j-1}) \nonumber \\
&-&\omega_{i-1,j+1}(\psi_{i,j+1}-\psi_{i-1,j}) \nonumber \\
&+&\omega_{i+1,j-1}(\psi_{i+1,j}-\psi_{i,j-1}) ]
\label{eq:j3}
\end{eqnarray}
The fourth-order accurate Arakawa discretization of the Jacobian becomes
\begin{equation}
J_{II}(\omega,\psi)  =\frac{1}{3}(J_{4}(\omega,\psi)+J_{5}(\omega,\psi)+J_{6}(\omega,\psi))
\label{eq:ja2}
\end{equation}
where
\begin{eqnarray}
J_{4}(\omega,\psi) &=& \frac{1}{8h_xh_y}[(\omega_{i+1,j+1}-\omega_{i-1,j-1})(\psi_{i-1,j+1}-\psi_{i+1,j-1}) \nonumber \\
&-&(\omega_{i-1,j+1}-\omega_{i+1,j-1})(\psi_{i+1,j+1}-\psi_{i-1,j-1})]
\label{eq:j4}
\end{eqnarray}
\begin{eqnarray}
J_{5}(\omega,\psi) &=& \frac{1}{8h_xh_y}[\omega_{i+1,j+1}(\psi_{i,j+2}-\psi_{i+2,j}) \nonumber\\ &-&\omega_{i-1,j-1}(\psi_{i-2,j}-\psi_{i,j-2}) \nonumber \\
&-&\omega_{i-1,j+1}(\psi_{i,j+2}-\psi_{i-2,j}) \nonumber\\
&+&\omega_{i+1,j-1}(\psi_{i+2,j}-\psi_{i,j-2}) ]
\label{eq:j5}
\end{eqnarray}
\begin{eqnarray}
J_{6}(\omega,\psi) &=& \frac{1}{8h_xh_y}[\omega_{i+2,j}(\psi_{i+1,j+1}-\psi_{i+1,j-1}) \nonumber\\ &-&\omega_{i-2,j}(\psi_{i-1,j+1}-\psi_{i-1,j-1}) \nonumber \\
&-&\omega_{i,j+2}(\psi_{i+1,j+1}-\psi_{i-1,j+1}) \nonumber\\
&+&\omega_{i,j-2}(\psi_{i+1,j-1}-\psi_{i-1,j-1}) ]
\label{eq:j6}
\end{eqnarray}
Arakawa showed that $J_{II}$ conserves enstrophy and energy and the following Jacobian
\begin{equation}
J(\omega,\psi) = 2J_{I}(\omega,\psi)  - J_{II}(\omega,\psi)  + O(h^{4})
\label{eq:ja2}
\end{equation}
has fourth-order accuracy. The viscous terms can be discretized with the 4th-order explicit difference scheme given in the previous section. This scheme was used to compute two-dimensional isotropic turbulence by Orlandi \cite{orlandi2000fluid}.

\subsection{Fourier-Galerkin pseudospectral method}
\label{sec:spectral}
Fourier series expansion based methods are often used for solving problems with periodic boundary conditions. One of the most accurate methods for solving the Navier-Stokes equations in periodic domains is the pseudospectral method, which exploits fast Fourier transform (FFT) algorithms, resulting in spectral accuracy \cite{canuto2006spectral}. It should be noted that the discrete Fourier transform is $O(N_x^2N_y^2)$, which becomes very expensive for larger resolutions. On the other hand, the discrete Fourier transform can, in fact, be computed in $O(N_xN_y log(N_xN_y))$ operations with well-known FFT algorithms \cite{press1992numerical}. Given a set of $N_x$ and $N_y$ uniformly distributed points on the interval [0, $L_x$]$\times$ [0, $L_y$], the forward Fourier transformation of any discrete function $u$ is
\begin{equation}
u_{i,j} = \sum_{m=-\frac{N_x}{2}}^{\frac{N_x}{2}-1}\sum_{n=-\frac{N_y}{2}}^{\frac{N_y}{2}-1}\tilde{u}_{m,n}e^{\textbf{\emph{i}}(\frac{2 \pi m}{L_x}x_i +\frac{2 \pi n}{L_y}y_j)}
\label{eq:fft1}
\end{equation}
and its inverse transform to find the Fourier coefficients is
\begin{equation}
\tilde{u}_{m,n} = \frac{1}{N_x\times N_y}\sum_{i=0}^{N_x-1}\sum_{j=0}^{N_y-1}u_{i,j}e^{-\textbf{\emph{i}}(\frac{2 \pi m}{L_x}x_i +\frac{2 \pi n}{L_y}y_j)}
\label{eq:fft2}
\end{equation}
where $\textbf{\emph{i}}=\sqrt{-1}$ and $i$, and $j$ represent indices for physical space, and $m$, and $n$ are the Fourier space indices. The discrete grid coordinates for $i=0,1,..N_x$ and $j=0,1,..N_y$ are given by
\begin{equation}
x_i = \frac{i L_x}{N_x},  \quad  y_j = \frac{j L_y}{N_y}.
\label{eq:xy}
\end{equation}
Since the domain is periodic, $x_0=x_{N_x}$ and $y_0=y_{N_y}$. Defining the wave numbers
\begin{equation}
k_x = \frac{2 \pi m}{L_x}, \quad k_y = \frac{2 \pi n}{L_y}.
\label{eq:kx}
\end{equation}
Now, we can easily perform differentiation in both directions. The first and second order differentiation of any function $u$ in discrete domain becomes
\begin{equation}
\frac{\partial u_{i,j}}{\partial x} = \sum_{m=-\frac{N_x}{2}}^{\frac{N_x}{2}-1}\sum_{n=-\frac{N_y}{2}}^{\frac{N_y}{2}-1}\tilde{u}_{m,n}(\textbf{\emph{i}}k_x) e^{\textbf{\emph{i}}(k_x x_i + k_y y_j)}
\label{eq:d1}
\end{equation}
\begin{equation}
\frac{\partial^2 u_{i,j}}{\partial x^2} = \sum_{m=-\frac{N_x}{2}}^{\frac{N_x}{2}-1}\sum_{n=-\frac{N_y}{2}}^{\frac{N_y}{2}-1}\tilde{u}_{m,n}(-k_x^2) e^{\textbf{\emph{i}}(k_x x_i + k_y y_j)}
\label{eq:d2}
\end{equation}
and similarly for the $y$ direction derivatives. By transforming Eq.~(\ref{eq:ge}) and Eq.~(\ref{eq:ke}) to Fourier space, the governing equations become
\begin{equation}
\frac{\partial \tilde{\omega}_{m,n}}{\partial t} + \tilde{N} = \frac{1}{Re}[(-k_x^2 - k_y^2) \tilde{\omega}_{m,n}]
\label{eq:gef}
\end{equation}
\begin{equation}
(-k_x^2 - k_y^2) \tilde{\psi}_{m,n} = -\tilde{\omega}_{m,n}
\label{eq:kef}
\end{equation}
where $\tilde{N}$ represents the nonlinear terms (Fourier transform of the Jacobian) and needs to be computed according to the following convolution
\begin{equation}
\tilde{N} = (\textbf{\emph{i}} k_y  \tilde{\psi}_{m,n}) \circ (\textbf{\emph{i}} k_x  \tilde{\omega}_{m,n}) - (\textbf{\emph{i}} k_x  \tilde{\psi}_{m,n}) \circ (\textbf{\emph{i}} k_y  \tilde{\omega}_{m,n})
\label{eq:convl}
\end{equation}
The convolution sum for these nonlinear terms is actually computed in the physical domain, with the help of the convolution theorem. One of the basic techniques for removing the aliasing error for the nonlinear terms is the 3/2-rule which is known as the padding or truncation algorithm. The key to this dealiasing algorithm is the use of the discrete transform of $M_x$ and $M_y$ rather than $N_x$ and $N_y$ points, so that $M_x\geq 3N_x/2$ and $M_y\geq 3N_y/2$. For example, let's denote $\tilde{a}_{m,n}$ and $\tilde{b}_{m,n}$, the Fourier coefficients of two multiplicand terms, as being in $\tilde{N}$, and we would like to compute the Fourier coefficients of the multiplication of these two terms, $\tilde{c}_{m,n} = \tilde{a}_{m,n} \circ \tilde{b}_{m,n}$. In the pseudospectral (PS) method, we can compute this term as $\tilde{c}_{m,n} = F^{-1}(F(\tilde{a}_{m,n})F(\tilde{b}_{m,n}))$, where $F$ and $F^{-1}$ represent the forward and inverse transforms, respectively. First we need to use the forward transformation using $M_x$ and $M_y$ wavenumbers such that
\begin{equation}
A_{i,j} = \sum_{m=-\frac{M_x}{2}}^{\frac{M_x}{2}-1}\sum_{n=-\frac{M_y}{2}}^{\frac{M_y}{2}-1}\tilde{a}_{m,n}e^{\textbf{\emph{i}}(k_X X_i + k_Y Y_j)}
\label{eq:fftn1}
\end{equation}
\begin{equation}
B_{i,j} = \sum_{m=-\frac{M_x}{2}}^{\frac{M_x}{2}-1}\sum_{n=-\frac{M_y}{2}}^{\frac{M_y}{2}-1}\tilde{b}_{m,n}e^{\textbf{\emph{i}}(k_X X_i + k_Y Y_j)}
\label{eq:fftn2}
\end{equation}
where $X_i = \frac{i L_x}{M_x}$, $Y_j = \frac{j L_y}{M_y}$. The Fourier coefficients are padded with zeros for the additional wavenumbers such as
\begin{equation}
\tilde{a}_{m,n} =
\bigg\{ \begin{array}{c} \tilde{a}_{m,n} \\ 0 \end{array} \quad \begin{array}{c} |m|\leq N_x, \quad |n|\leq N_y   \\ \mbox{otherwise} \end{array}
\label{eq:pad1}
\end{equation}
\begin{equation}
\tilde{b}_{m,n} =
\bigg\{ \begin{array}{c} \tilde{b}_{m,n} \\ 0 \end{array} \quad \begin{array}{c} |m|\leq N_x, \quad |n|\leq N_y   \\ \mbox{otherwise} \end{array}
\label{eq:pad2}
\end{equation}
Multiplication is performed in physical space but with padded higher wavenumbers in the following way
\begin{equation}
C_{i,j}=A_{i,j}B_{i,j}.
\label{eq:mul}
\end{equation}
Then, the inverse transform gives the dealiased Fourier coefficients as
\begin{equation}
\tilde{C}_{m,n} = \frac{1}{N_x\times N_y}\sum_{i=0}^{M_x-1}\sum_{j=0}^{M_y-1}C_{i,j}e^{-\textbf{\emph{i}}(k_X X_i +k_Y Y_j)}
\label{eq:fftw}
\end{equation}
in which we are only interested in $\tilde{C}_{m,n}$ for $|m|\leq N_x$, $|n|\leq N_y$ (i.e., $\tilde{c}_{m,n}=\tilde{C}_{m,n}$ for $|m|\leq N_x$, $|n|\leq N_y$). Although in practice dealiasing need only be performed with $M_x\geq3N_x/2$, we have used $M_x = 2N_x$ here since the FFT solver we are using requires resolutions that are powers of two. In order to make a fair comparison of CPU times, however, we have chosen to present the dealisased computational CPU times an appropriately interpolated factor of the actual time between $N_x$ and $M_x$. In this study, the semi-discrete vorticity transport equation in Fourier space, Eq.~(\ref{eq:gef}), is solved explicitly by third-order or forth-order accurate Runge-Kutta time marching schemes which will be given later.

\subsection{Poisson solver}
\label{sec:ps}
In order to obtain the streamfunction from the vorticity field, Eq.~(\ref{eq:ke}) needs to be solved. In general, Gauss-Seidel or successive over relaxation (SOR) types of iterative algorithms for solving the Poisson equation are of $O(N^2)$ where $N$ is the total number of grid points (i.e., $N=N_x N_y$ for two-dimensional problems). It is not feasible to use these types of iterative Poisson solvers for high resolution (and therefore high Reynolds number) computations along with long time integration. In order to accelerate these solvers, very successful multi-grid (MG) algorithms have been developed that reduce the computational effort to $O(C_{MG}N)$ where $C_{MG}$ is a proportionality constant \cite{gupta1997comparison}. On the other hand, for certain ideal problems on equally-spaced grids, fast Fourier transform (FFT)-based fast Poisson solvers (FPS) can be used to solve elliptic problems directly with an $O(C_{FPS}N log(N))$ operational cost, and they are presently the fastest available algorithms ($C_{FPS}log(N)<C_{MG}$ in relevant resolutions) for solving Poisson equations \citep{press1992numerical,moin2001fundamentals}.

\begin{figure}
\centering
\includegraphics[width=0.45\textwidth]{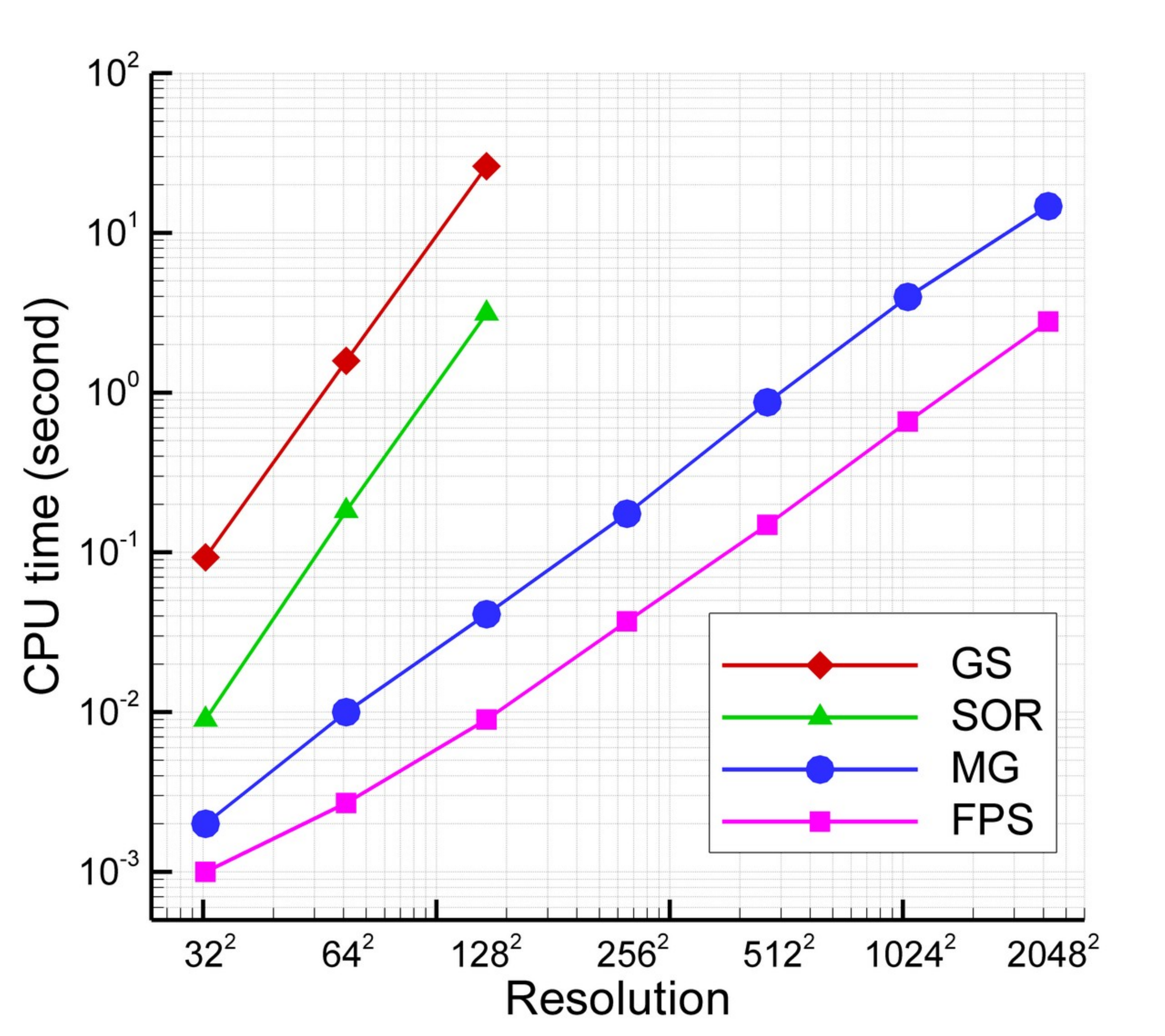}
\caption{Efficiency of Poisson solvers for a two-dimensional square domain.}
\label{fig:Poisson}
\end{figure}

A preliminary study is performed to test the computational efficiencies of different Poisson solvers for a square domain with equidistant grid spacing. The computational time (CPU time) for solving just one Poisson equation is illustrated in Fig.~\ref{fig:Poisson}. This preliminary comparison shows that the FPS is the most efficient solver for Cartesian grid problems. We take advantage of the simple rectangular shape of our domain and utilize the FPS for all the simulations in this study. The basic three-step procedure to find $\psi$ from known $\omega$ values at any time (also for the inner steps of the Runge-Kutta time stepping algorithms) has the following form.
\begin{enumerate}
  \item Use forward FFT for $\omega$ to find Fourier coefficients according to Eq.~(\ref{eq:fft1}).
  \item Compute Fourier coefficients of stream function from Eq.~(\ref{eq:kef})
\begin{equation}
\tilde{\psi}_{m,n} = \frac{\tilde{\omega}_{m,n}}{(k_x^2 + k_y^2)}.
\label{eq:kefm}
\end{equation}
  \item Use inverse FFT for $\psi$ values according to Eq.~(\ref{eq:fft2}).
\end{enumerate}

\subsection{Time integration algorithms}
\label{sec:time}
Steady calculations in fluid flows have been mostly handled using implicit time advancement algorithms and corresponding large time steps. It is tempting to adopt a similar practice in computations of turbulent flows. Unfortunately, the requirement of time accuracy over a wide range of scales does not permit very large time steps in such calculations \cite{moin1998direct}. The use of large time steps implies that the small scales can have large errors, which can corrupt the solutions. Therefore, the families of fourth-order \cite{carpenter2005fourth} and third-order \cite{williamson1980low} Runge-Kutta time integration schemes have been extensively used in turbulence simulations with their certain numerical stability requirements \cite{pruett1995spatial}. In fluid dynamics, the numerical stability of such explicit algorithms is usually addressed by using the convective Courant-Friedrichs-Lewy (CFL) or viscous Neumann stability conditions \cite{chorin1968numerical}. Usually the fourth-order Runge-Kutta schemes have a greater accuracy for a given time step and a larger allowable stability domain than the third-order Runge-Kutta schemes. To be practical, an explicit algorithm for turbulence computations must provide long-time temporal accuracy while requiring modest temporary storage resulting in the development of a low-storage Runge-Kutta family for demanding computations.

Semi-discrete ordinary differential equations (ODEs) are obtained after spatial discretization of the partial differential equations. To implement the Runge-Kutta schemes, we cast the governing vorticity transport equation in the following form
\begin{equation}
\frac{d\omega}{dt} = \pounds (\omega,\psi)
\end{equation}
where $\pounds(\omega,\psi)$ is the discrete operator of spatial derivatives including nonlinear convective terms and linear diffusive terms, and $\psi$ is obtained from the Poisson equation. A fourth-order Runge-Kutta scheme can be written in the following form \cite{pike1985accelerated}
\begin{eqnarray}
\omega^{(1)} &=& \omega^{n} +  \frac{\Delta t}{2}  \pounds(\omega^{n},\psi^{n}) \nonumber \\
\omega^{(2)} &=& \omega^{n} +  \frac{\Delta t}{2}  \pounds(\omega^{1},\psi^{1})  \nonumber \\
\omega^{(3)} &=& \omega^{n} +  \Delta t \pounds(\omega^{2},\psi^{2}) \nonumber \\
\omega^{n+1} &=& \frac{1}{3}(-\omega^{n} + \omega^{(1)} + 2\omega^{(2)} + \omega^{(3)}) + \frac{\Delta t}{6}  \pounds(\omega^{(3)},\psi^{(3)})
\label{eq:RK4}
\end{eqnarray}
The optimal third-order accurate TVD Runge-Kutta scheme is given as \cite{gottlieb1998total}
\begin{eqnarray}
\omega^{(1)} &=& \omega^{n} + \Delta t \pounds(\omega^{n},\psi^{n}) \nonumber \\
\omega^{(2)} &=& \frac{3}{4}  \omega^{n} + \frac{1}{4} \omega^{(1)} + \frac{1}{4}\Delta t \pounds (\omega^{(1)},\psi^{(1)}) \nonumber \\
\omega^{n+1} &=& \frac{1}{3}  \omega^{n} + \frac{2}{3} \omega^{(2)} + \frac{2}{3}\Delta t \pounds (\omega^{(2)},\psi^{(2)}).
\label{eq:TVDRK}
\end{eqnarray}

Alternatively, memory-effective $2N$-storage schemes that require only two sets of variables to be held in memory can be used. The iterative form of the $2N$-storage schemes can be written as
\begin{eqnarray}
g^{(i)} &=& \alpha_{i}g^{(i-1)}+\Delta t \pounds(\omega^{(i-1)},\psi^{(i-1)}) \nonumber \\
\omega^{(i)}&=&\omega^{(i-1)} + \beta_i g^{(i)}.
\label{eq:2NRK}
\end{eqnarray}
For a third-order accurate three-step scheme, we have indices $i=1,2,3$. In order to advance from $\omega^{n}$ at time $t^{n}$ to $\omega^{n+1}$ at time $t^{n+1}=t^{n}+\Delta t$ we set in Eq.~(\ref{eq:2NRK}) $\omega^{(0)}=\omega^{n}$ and, after the last step, $\omega^{n+1}=\omega^{(3)}$.
In order to be able to calculate the first step, $i=1$, for which no $g^{(0)}$ exists, we require $\alpha_1=0$. The corresponding coefficients for different types of constructed schemes are listed in Table~\ref{tab:RK}. In practice all of them are approximately equally good as advocated by Brandenburg \cite{brandenburg2003computational}. Although our main objective is to evaluate the performance of the various high-order spatial discretization schemes, we will perform a comparative study on these explicit Runge-Kutta schemes presented above to test their long-term characteristics for two-dimensional turbulence.

\begin{table}
\caption{Coefficients for various low-storage third-order Runge-Kutta methods.}
\begin{center}
\label{tab:RK}
\begin{tabular}{lcccccc}
\\
\hline
 scheme & $\alpha_2$  & $\alpha_3$  & $\beta_1$  & $\beta_2$  & $\beta_3$\\
\hline
\smallskip
symmetric & $-\frac{2}{3}$ & $-1$ & $\frac{1}{3}$ & $1$ & $\frac{1}{2}$\\
\smallskip
predictor/corrector & $-\frac{1}{4}$ & $-\frac{4}{3}$ & $\frac{1}{2}$ & $\frac{2}{3}$ & $\frac{1}{2}$\\
\smallskip
inhomogeneous & $-\frac{17}{32}$ & $-\frac{32}{27}$ & $\frac{1}{4}$ & $\frac{8}{9}$ & $\frac{3}{4}$\\
\smallskip
Williamson \cite{williamson1980low} & $-\frac{5}{9}$ & $-\frac{153}{128}$ & $\frac{1}{3}$ & $\frac{15}{16}$ & $\frac{8}{15}$\\
\hline
\end{tabular}
\end{center}
\end{table}

\section{Validation problem: Taylor-Green vortex}
\label{sec:vTGV}
In this section, the schemes under consideration are validated on a square domain by solving the Taylor-Green decaying vortex problem. The Taylor-Green vortex problem is the two-dimensional, unsteady flow of a decaying vortex, which is an exact closed form solution to the incompressible Navier-Stokes equations in Cartesian coordinates. The exact solution of this vortex flow in a $[0, 2\pi]\times[0, 2\pi]$ domain with periodic boundary conditions is given by
\begin{equation}
\omega^{e}(x,y,t)=2\kappa \mbox{cos}(\kappa x)\mbox{cos}(\kappa y)e^{-2\kappa^2t/Re}
\label{eq:exw}
\end{equation}
where $\kappa$ is an integer which represents the number of vortices in each direction.
In order to quantify the effective order of accuracy for each scheme, using the difference between exact and computed solutions, we compute the discrete $L_{2}$ norm as
\begin{equation}
\|\omega\|_{L_{2}}=\sqrt{\frac{1}{N_xN_y}\sum_{i=1}^{N_x} \sum_{j=1}^{N_y} |\omega^{e}_{i,j}-\omega_{i,j}|^{2}}.
\label{eq:err}
\end{equation}

\begin{table*}[t!]
\caption{Discrete $L_2$ error norms and corresponding convergence rates showing the effective order of the spatial difference schemes for the Taylor-Green vortex problem at $t=0.1$ for $Re=1$ and $\kappa=4$, obtained using the time step $\Delta t=10^{-4}$. The final column shows the CPU times in seconds for the $128^2$ resolution computations.}
\begin{center}
\label{tab:L2-a}
\begin{tabular}{lcccccccc}
\hline
\smallskip
Scheme &$\|\omega\|_{L_2}^{16^2}$ & $n_{32^2}^{16^2}$ & $\|\omega\|_{L_2}^{32^2}$ & $n_{64^2}^{32^2}$& $\|\omega\|_{L_2}^{64^2}$ & $n_{128^2}^{64^2}$ & $\|\omega\|_{L_2}^{128^2}$ & CPU\\
\hline
pseudospectral method (PS)        &0.76E-9&      & 6.63E-9 &     & 1.61E-8 &     & 2.36E-8  &210\\
6th-order explicit difference (ED6)&1.04E-2 &5.67 & 2.05E-4 &5.92 & 3.39E-6 & 6.00& 5.28E-8 &46 \\
4th-order explicit difference (ED4)&3.28E-2 &3.92 & 2.16E-3 &3.97 & 1.38E-4 & 4.00& 8.65E-6 &44 \\
2nd-order explicit difference (ED2)&1.44E-1 &2.29 & 2.94E-2 &2.09 & 6.91E-3 & 2.03& 1.70E-3 &42 \\
6th-order compact  difference (CD6)&2.90E-3 &6.19 & 3.98E-5 &6.06 & 5.96E-7 & 6.14& 8.47E-9 &79 \\
4th-order compact  difference (CD4)&1.58E-2 &4.17 & 8.74E-4 &4.05 & 5.28E-5 & 4.02& 3.26E-6 &76 \\
4th-order Arakawa scheme (A4)      &3.28E-2 &3.92 & 2.16E-3 &3.97 & 1.38E-4 & 4.00& 8.65E-6 &52 \\
2nd-order Arakawa scheme (A2)      &1.44E-1 &2.29 & 2.94E-2 &2.09 & 6.91E-3 & 2.03& 1.70E-3 &45 \\
4th-order DRP scheme (DRP4)        &3.28E-2 &3.92 & 2.16E-3 &3.97 & 1.38E-4 & 4.00& 8.65E-6 &45 \\
\hline
\end{tabular}
\end{center}
\end{table*}

\begin{figure}[h!]
\centering
\includegraphics[width=0.5\textwidth]{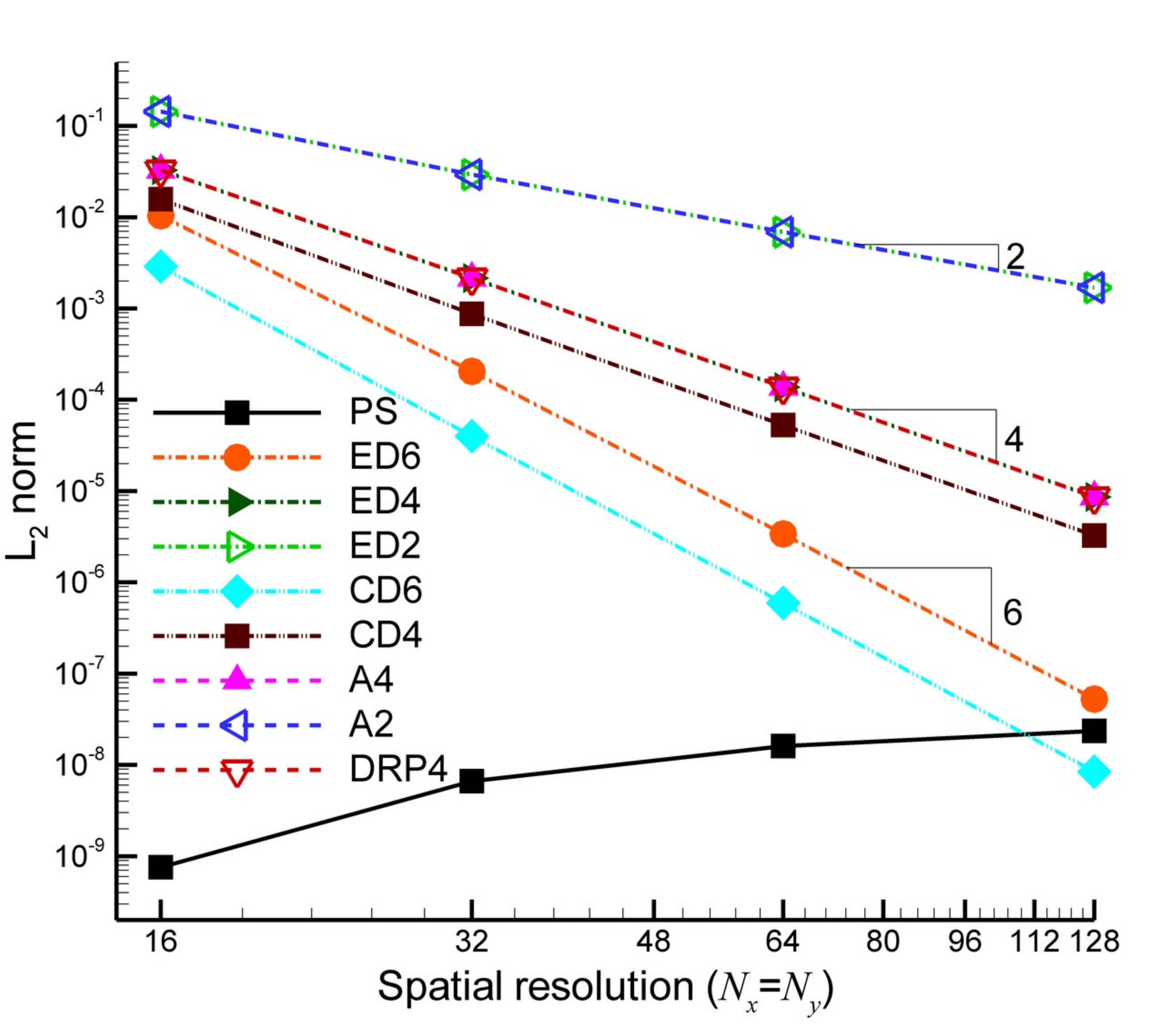}
\caption{Convergence of the discrete $L_2$ error norms at time $t=0.1$ for the Taylor-Green vortex decaying problem at $Re=1$.}
\label{fig:TGV-a}
\end{figure}

In order to test the spatial convergence rate, we first perform numerical simulations for $Re=1$ and $\kappa=4$ with all the spatial schemes introduced earlier using the third-order TVD Runge-Kutta scheme with $\Delta t=10^{-4}$. We choose this small time step (wherein the maximum CFL number is around 0.002 for the $128^2$ resolution case) to eliminate possible temporal discretization errors due to the Runge-Kutta time integration schemes. A study with $\Delta t=2\times10^{-4}$, $10^{-4}$, and $5\times10^{-5}$ shows that the predicted flow field is independent of $\Delta t$. To further verify whether the simulations are independent of the errors associated with temporal discretization, we also run the same computations using the fourth-order Runge Kutta scheme, which yields exactly the same results.

In evaluating the performance of different finite difference formulations, the computed discrete $L_{2}$ norms at time $t=0.1$ (when a 95\% decay of the initial field has happened) are tabulated in Table~\ref{tab:L2-a} and are also plotted in Fig.~\ref{fig:TGV-a}. Results show that the theoretical order of accuracies for all the schemes are obtained in practice. A linear reduction rate is obtained with the slope of all curves being their theoretical values, verifying the correct spatial convergence rate for each scheme. The most accurate difference scheme is found to be the sixth-order compact difference scheme. We also demonstrate that the errors associated with all the high-order formulations are several order magnitudes lower than the second-order formulations. It can be seen from Fig.~\ref{fig:TGV-a} that the predicted result of the 6th-order compact scheme becomes more accurate than the pseudospectral solution for the case with $128^2$ resolution. This is mainly due to the accumulation of round off errors in the FFTs of the pseudospectral method. It can also be seen that the associated error increases slightly with increasing resolution for pseudospectral computations, again, due to the round-off errors associated with the FFTs. That is actually why Lele \cite{lele1992compact} called these compact schemes ``spectral-like schemes." It can also be seen that the compact schemes give more accurate results than the explicit finite difference schemes due to their smaller truncation error. Since the viscous difference formulations are equivalent, it should be noted here that the fourth-order Arakawa, dispersion-relation-preserving, and explicit difference schemes predict almost the same answer because of the inherent dynamics of the Taylor-Green vortex validation problem. This problem does not have a convective flow field, hence, the center of the each vortex stays stationary.

\begin{table*}[t!]
\caption{Discrete $L_2$ error norms and corresponding convergence rates showing the effective order of the temporal integration schemes for the Taylor-Green vortex problem at $t=20$ for $Re=1000$ and $\kappa=4$, obtained using the pseudospectral scheme with a spatial resolution of $16^2$.}
\begin{center}
\label{tab:L2-b}
\begin{tabular}{llllllll}
\hline
\smallskip
 Scheme & $\|\omega\|_{L_2}^{\Delta t = 0.1}$ & rate & $\|\omega\|_{L_2}^{\Delta t = 0.2}$ & rate & $\|\omega\|_{L_2}^{\Delta t = 0.4}$ &rate & $\|\omega\|_{L_2}^{\Delta t = 0.8}$ \\
\hline
4th-order Runge-Kutta (RK4)   & 1.23E-12 &4.03 & 2.01E-11 & 4.01& 3.23E-10 & 4.02& 5.23E-09 \\
TVD Runge-Kutta (TVDRK3)      & 1.96E-09 &3.00 & 1.57E-08 & 3.01& 1.26E-07 & 3.01& 1.02E-06 \\
Symmetric (SYMRK3)            & 1.96E-09 &3.00 & 1.57E-08 & 3.01& 1.26E-07 & 3.01& 1.02E-06 \\
Predictor/Corrector (P/CRK3)  & 1.96E-09 &3.00 & 1.57E-08 & 3.01& 1.26E-07 & 3.01& 1.02E-06 \\
Inhomogeneous (INHRK3)        & 1.96E-09 &3.00 & 1.57E-08 & 3.01& 1.26E-07 & 3.01& 1.02E-06 \\
Williamson (WILRK3)           & 1.96E-09 &3.00 & 1.57E-08 & 3.01& 1.26E-07 & 3.01& 1.02E-06 \\
\hline
\end{tabular}
\end{center}
\end{table*}

\begin{figure}[h!]
\centering
\includegraphics[width=0.5\textwidth]{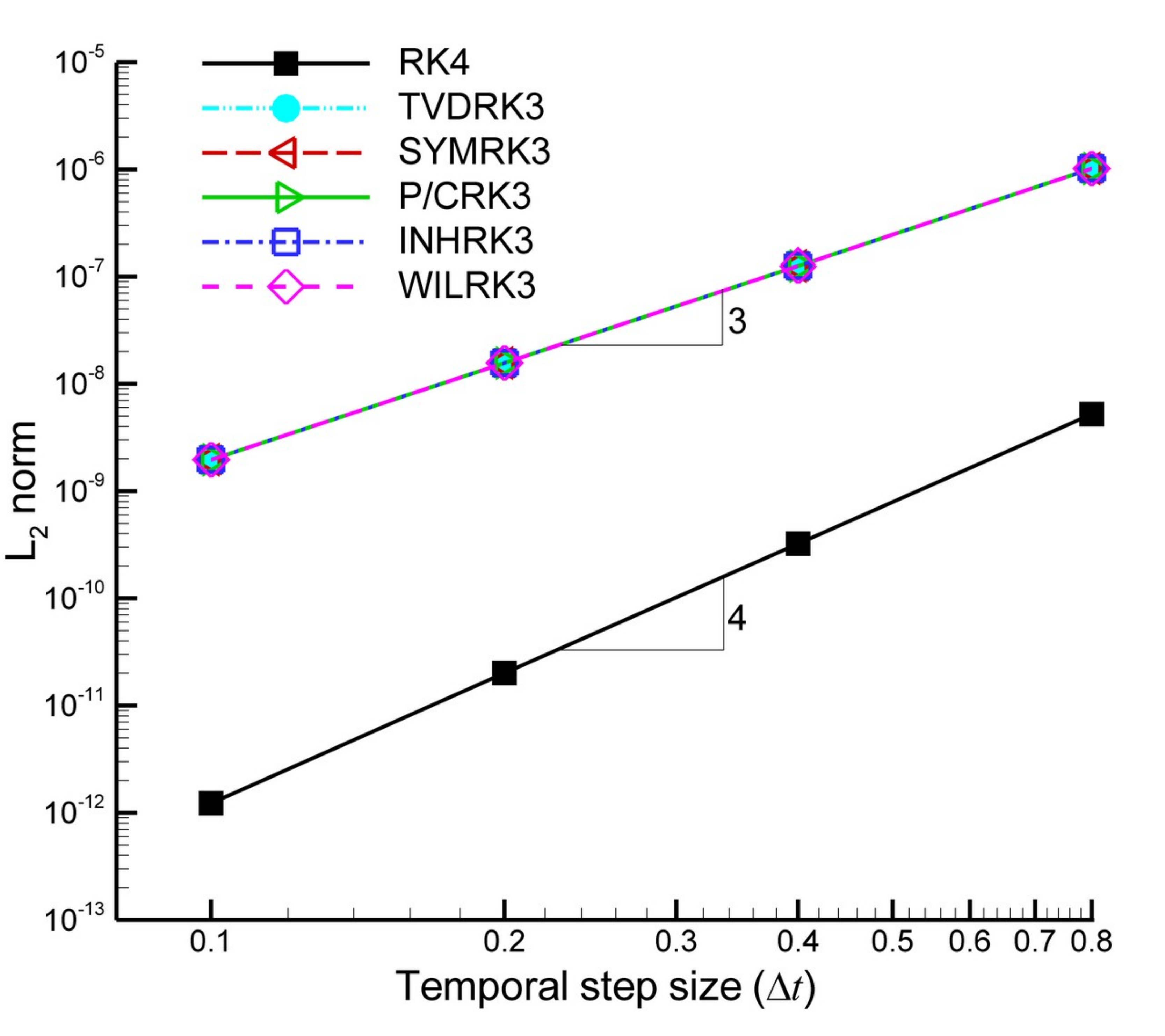}
\caption{Convergence of the discrete $L_2$ error norms for the temporal schemes at time $t=20$ for the Taylor-Green vortex decaying vortex problem at $Re=1000$.}
\label{fig:TGV-b}
\end{figure}

Next, we perform a convergence rate analysis for the temporal schemes. As we can see from the previous analysis, the pseudospectral scheme provides much more accurate results than the finite difference schemes, especially for lower resolutions. In order to be able to test the convergence rate of the temporal schemes properly, we choose our spatial discretization scheme to be the pseudospectral scheme with a  resolution of $16^2$, for which the temporal discretization error becomes greater than the spatial discretization error. The computed error norms are shown in Table~\ref{tab:L2-b} at time $t=20$ for $Re=1000$. It can be seen from Fig.~\ref{fig:TGV-b} that a linear reduction rate is obtained with the slope of all curves being 3 for all the third-order Runge-Kutta schemes, and 4 for the fourth-order Runge-Kutta scheme. We report that all the third-order Runge-Kutta schemes utilized in this study produce the same result within negligible differences due to round-off error.

The Taylor-Green vortex problem is one of the simplest incompressible flow test problems, having a smoothly decaying field. We solve it to be able to perform an exact error convergence rate analysis by comparing our results to the analytical solution of the problem. In the following sections, the effects of the order of the schemes and their global conservation properties will be analyzed for more challenging convective dominated flow problems. A truly fair comparison might require implementation of the same discretization scheme for the viscous terms. Therefore, in order to do a fair comparison among spatial difference schemes, the viscous diffusion terms are approximated using the six-order accurate compact scheme for all the finite difference schemes being considered for the rest of the paper.

\section{Double shear layer problem}
\label{sec:dsl}
The double shear layer problem was introduced by Bell, Colella and Glaz \cite{bell1989second} to test the performance of a projection method, and was studied later by Minion and Brown \citep{minion1997performance} using various schemes to determine the effects of the grid resolution on solutions of the unsteady, incompressible Navier-Stokes equations. It is a benchmark problem for testing the accuracy and resolution of a time dependent numerical method. The double shear layer problem with periodic boundary conditions in a square domain $[0,2\pi]\times[0,2\pi]$ is subjected to the following initial conditions
\begin{equation}
\omega(x,y,0) =
\bigg\{ \begin{array}{c}
\delta \ \mbox{cos}(x) -\sigma \ \mbox{cosh}^{-2}[\sigma(y-\frac{\pi}{2})] \\
\delta \ \mbox{cos}(x) +\sigma \ \mbox{cosh}^{-2}[\sigma(\frac{3\pi}{2}-y)] \end{array} \begin{array}{c} \ \mbox{if} \ y \leq \pi  \\ \  \mbox{if} \ y > \pi \end{array} \\
\label{eq:dlw}
\end{equation}
where the constants $\sigma$ and $\delta$ determine the thickness of the shear layer and the amplitude of the initial perturbation, respectively. In our computations, the thickness parameter is $\sigma = 15/\pi$, the perturbation amplitude is $\delta=0.05$, the Reynolds number is $Re=10^4$, and the time step is $\Delta t = 10^{-3}$. We choose this time step (wherein the corresponding maximum CFL number is 0.25 for the case with a resolution of $1024^2$) to concentrate on the differences among the spatial discretization schemes by eliminating the temporal integration errors in the simulations. Since the dominant error is due to the spatial difference operator rather than the temporal integration method, we employ the TVD Runge-Kutta scheme for the time advancement. We also perform, but do not report in detail here, a time step refinement study revealing that the temporal discretization error is indeed negligible. Moreover, we use the same discrete approximation, the 6th-order compact difference scheme, for the viscous terms in all the cases. In the following, we investigate the behavior of the spatial discretization schemes for convective terms by solving the double shear layer problem for various resolutions.

\begin{figure*}[t!]
\centering
\mbox{
\subfigure{\includegraphics[width=0.33\textwidth]{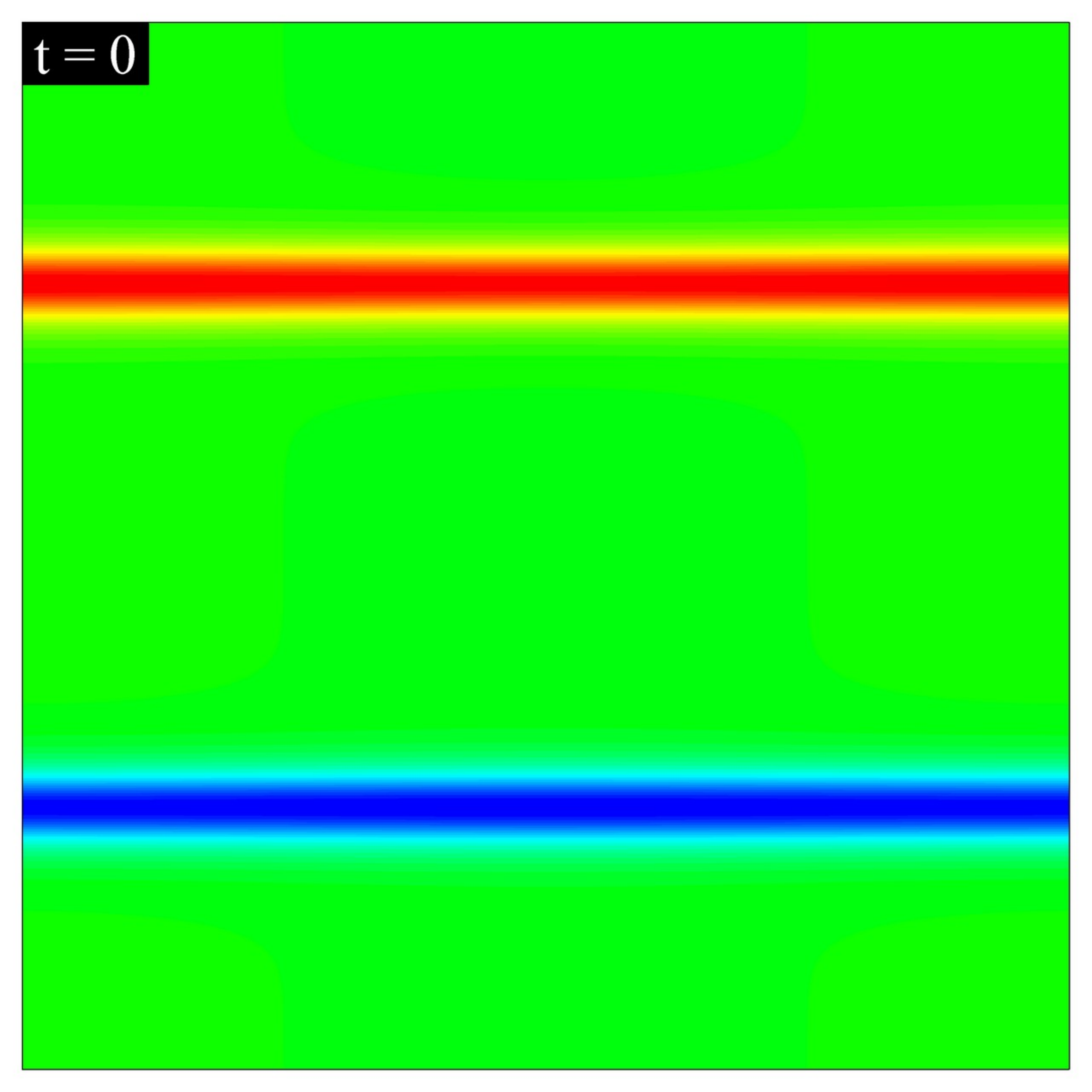}}
\subfigure{\includegraphics[width=0.33\textwidth]{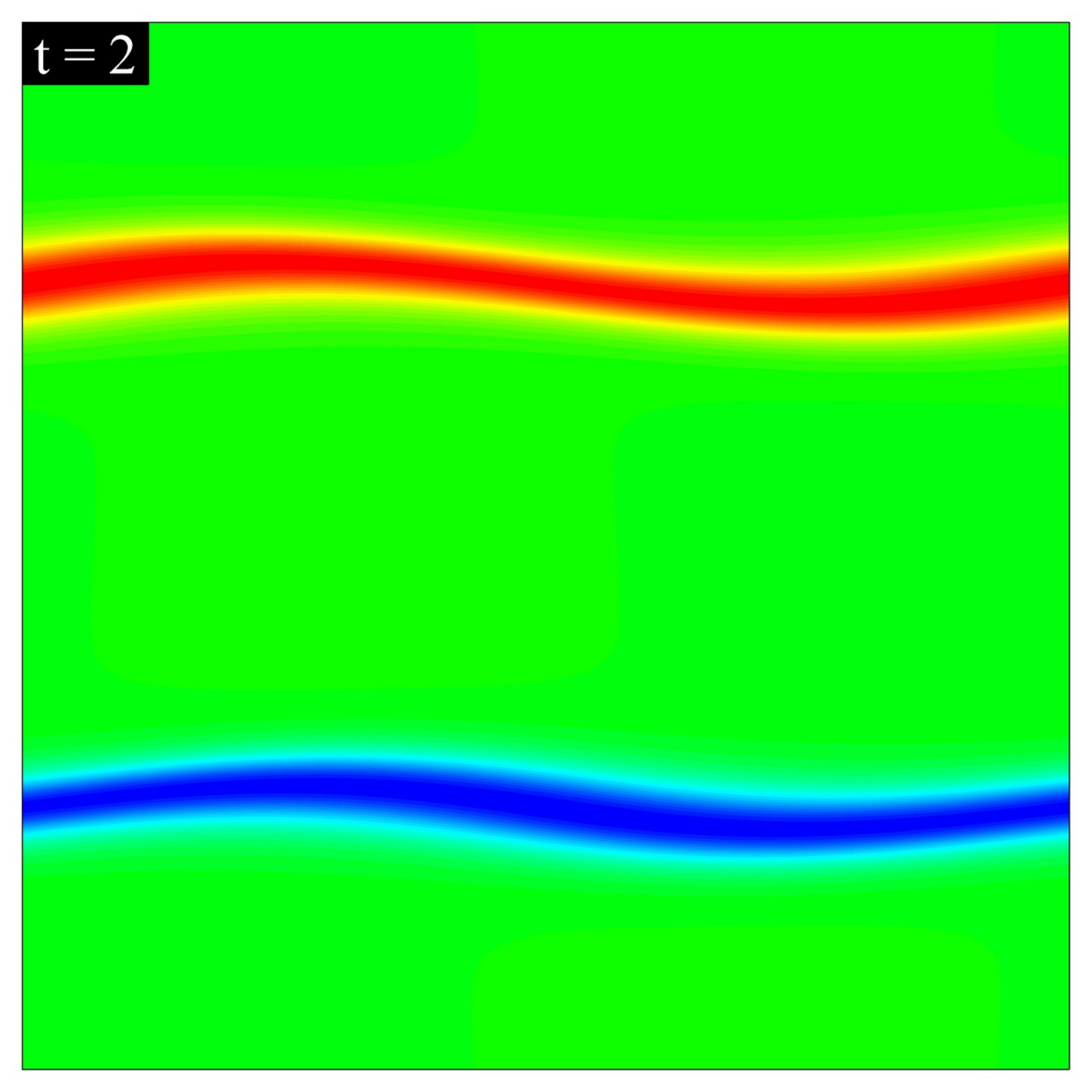}}
\subfigure{\includegraphics[width=0.33\textwidth]{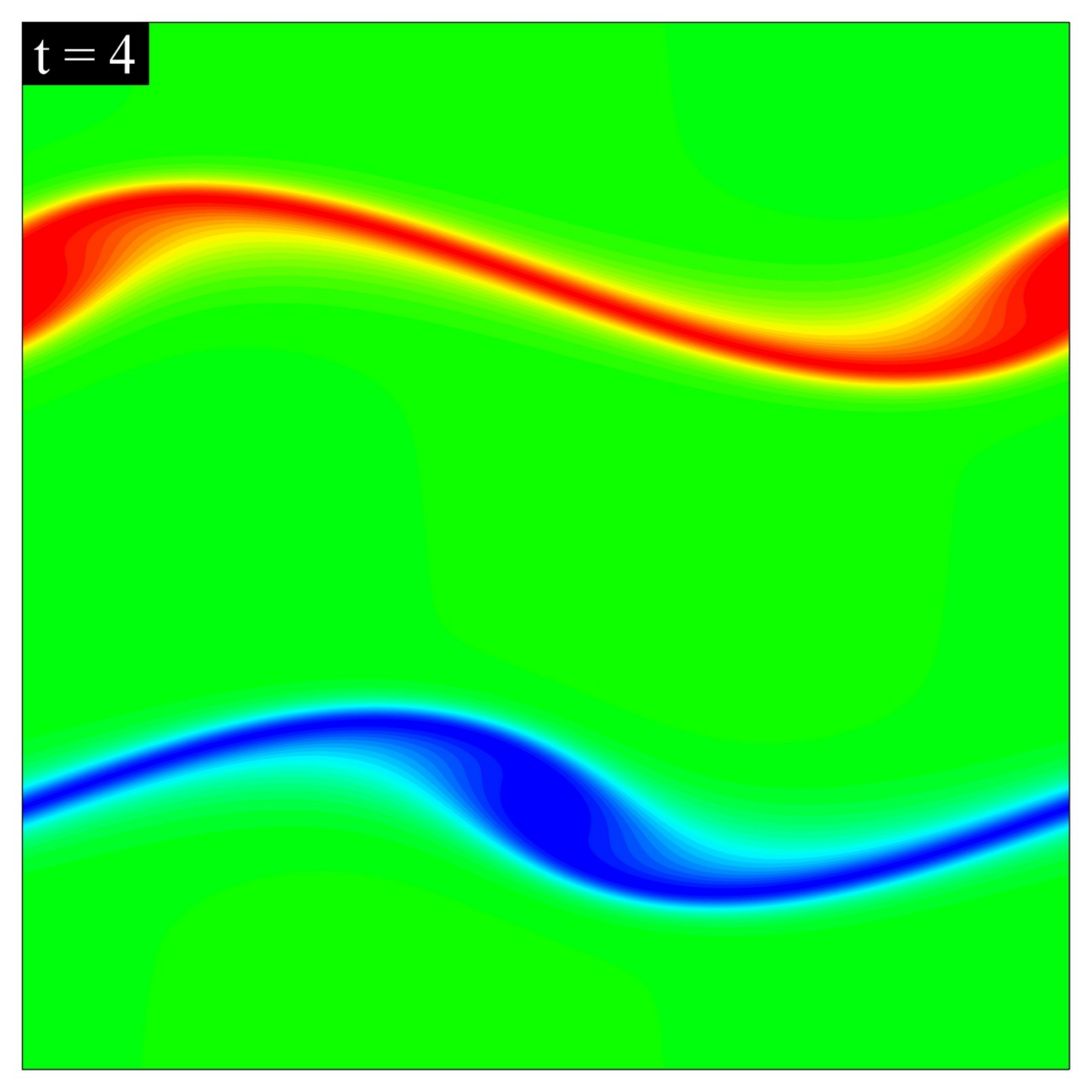}}
}
\mbox{
\subfigure{\includegraphics[width=0.33\textwidth]{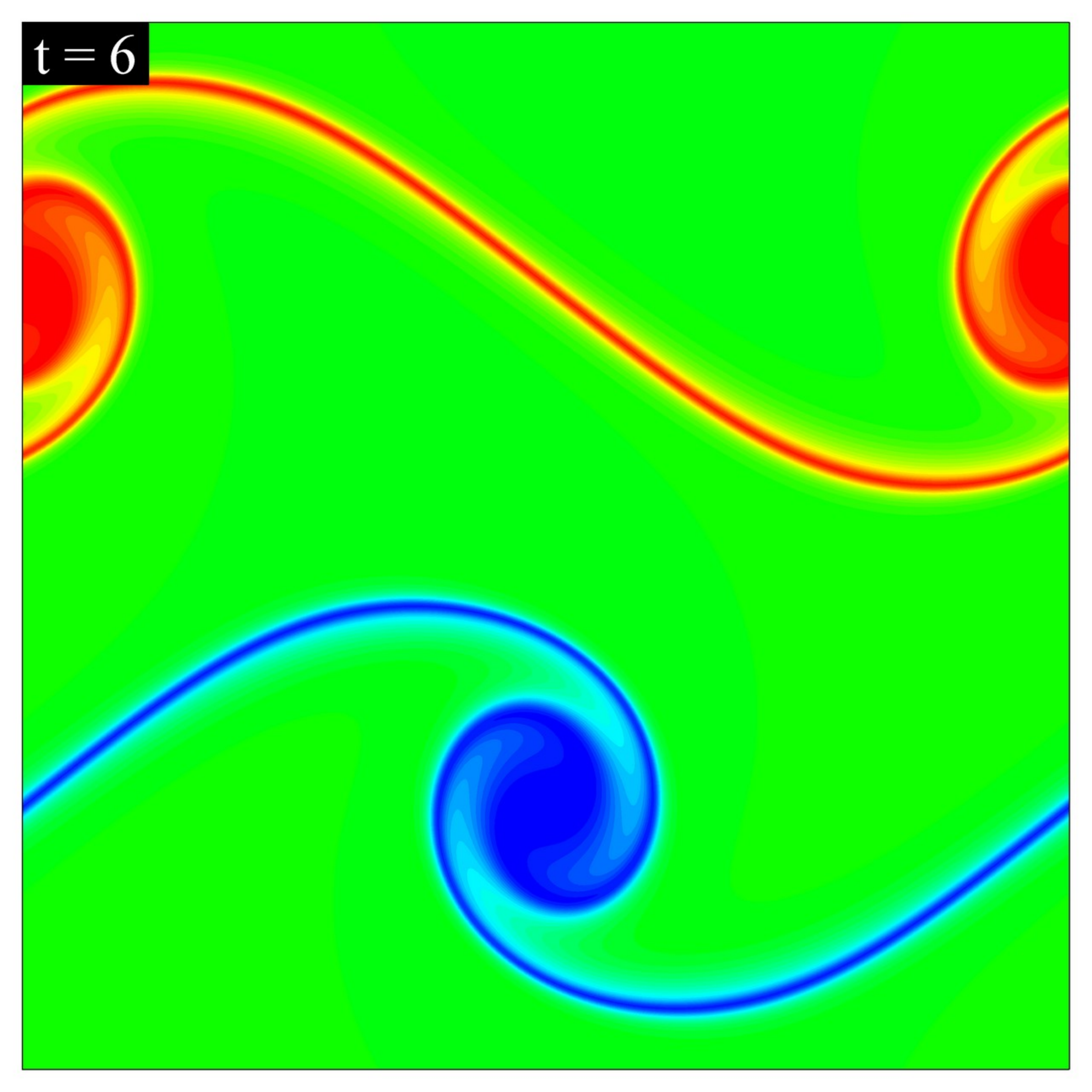}}
\subfigure{\includegraphics[width=0.33\textwidth]{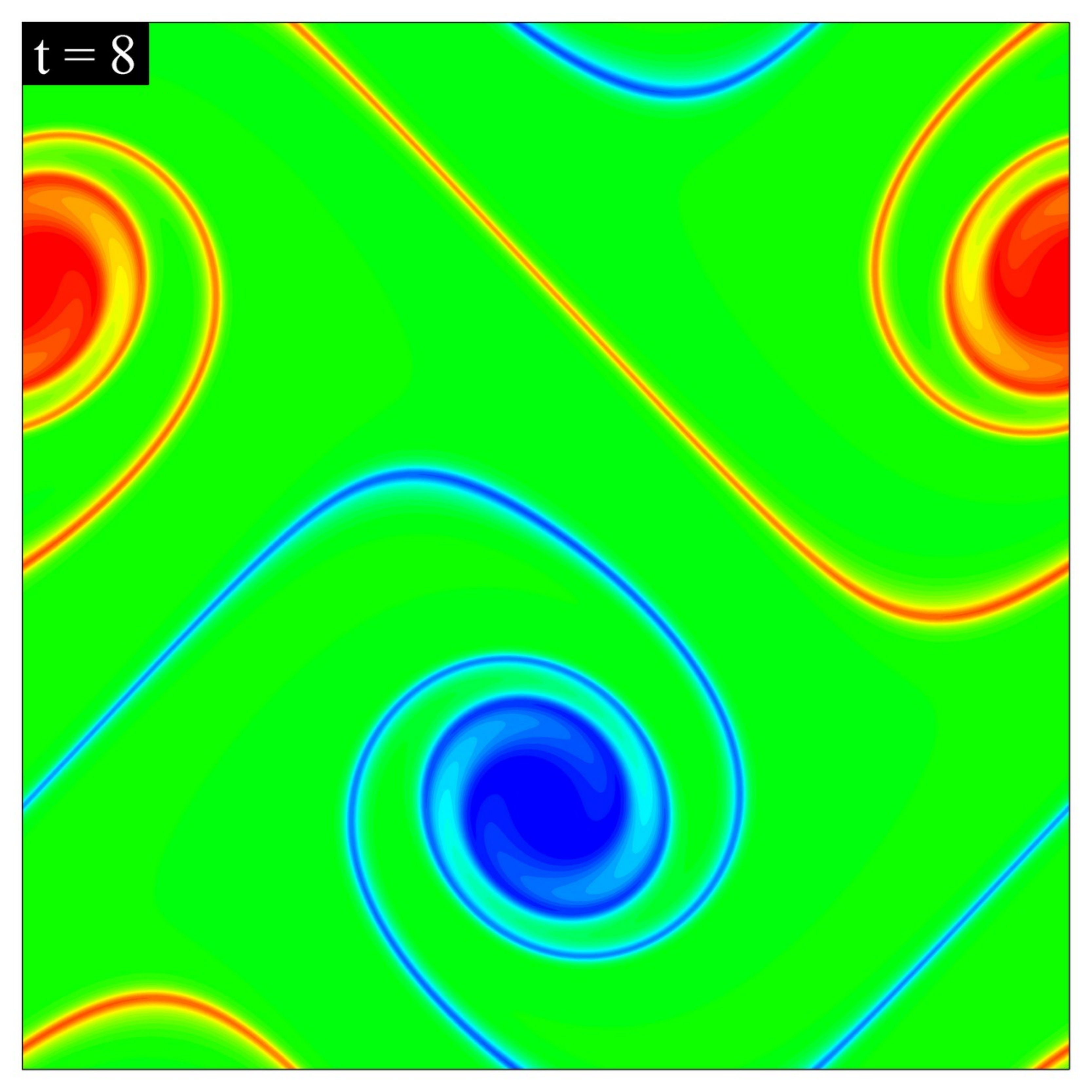}}
\subfigure{\includegraphics[width=0.33\textwidth]{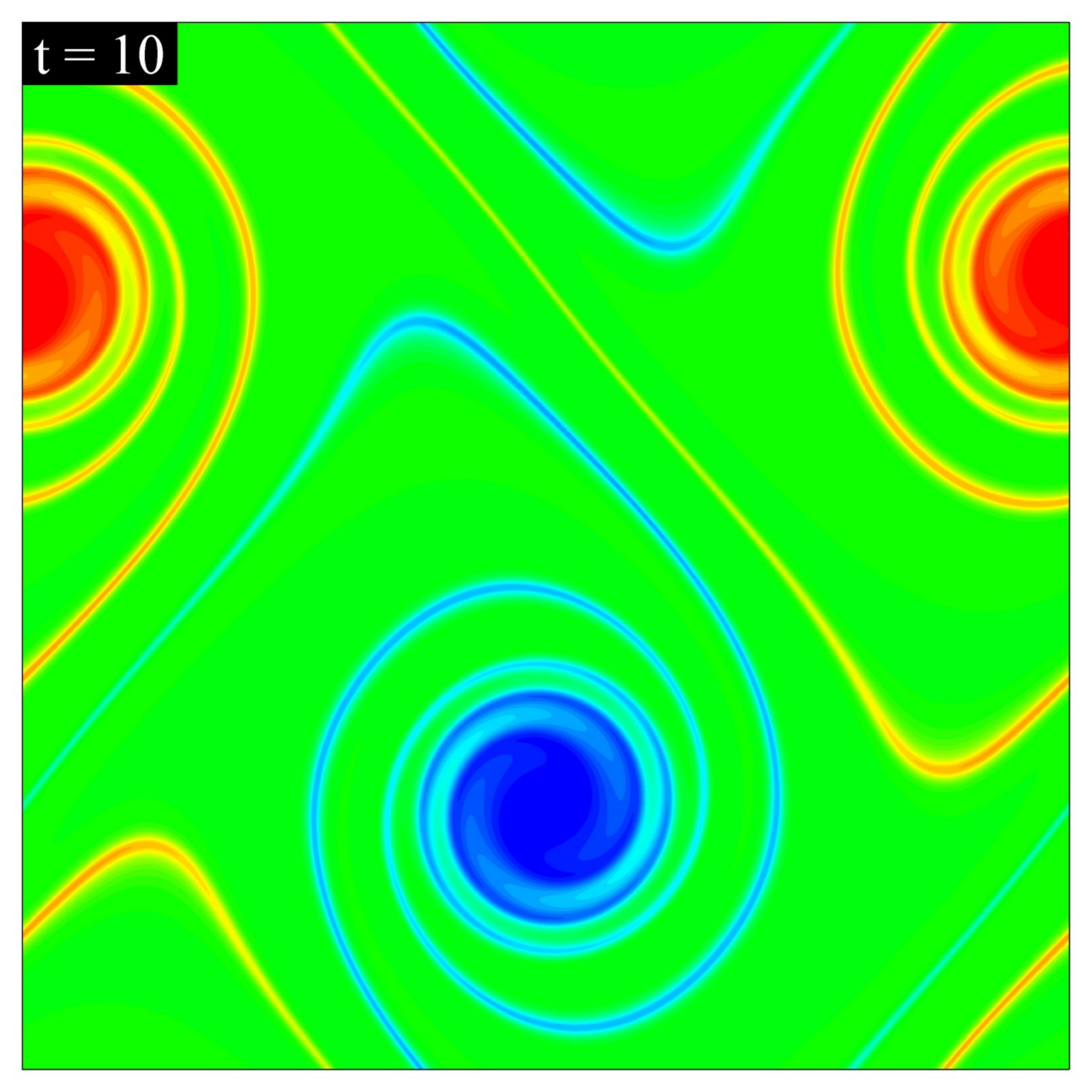}}
}
\caption{Evolution of the double shear layer problem for $Re=10^4$ with the perturbation and thickness parameters of $\delta=0.05$ and $\sigma = 15/\pi$ in a doubly periodic domain.}
\label{fig:dsl-time}
\end{figure*}

\begin{figure}[h!]
\centering
\includegraphics[width=0.45\textwidth]{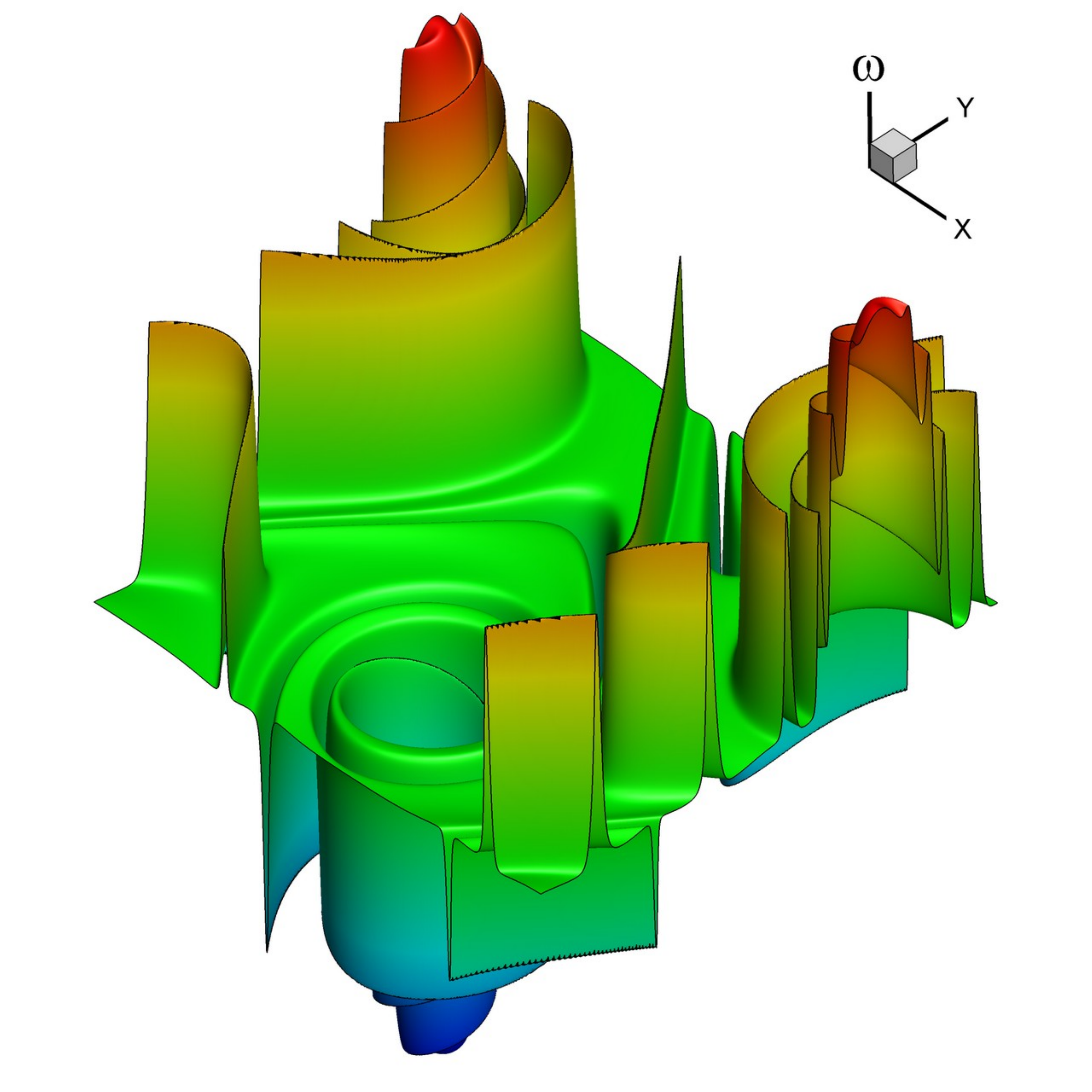}
\caption{The three-dimensional view of the shear layers at time $t=10$ by plotting the magnitude of the vorticity as the z-axis.}
\label{fig:dsl3d}
\end{figure}

\begin{table*}[t!]
\caption{Discrete $L_2$ error norms and computational efficiencies of the spatial schemes for the double shear layer problem at $t=10$ for $Re=10^{4}$. The reference solution for computing the $L_2$ norm is obtained using the pseudospectral method with a resolution of $1024^2$.}
\begin{center}
\label{tab:L2-dsl}
\begin{tabular}{llclclc}
\hline
\smallskip
Scheme &$\|\omega\|_{L_2}^{256^2}$ & CPU$_{256^2}$(hrs) & $\|\omega\|_{L_2}^{512^2}$ & CPU$_{512^2}$(hrs) & $\|\omega\|_{L_2}^{1024^2}$ & CPU$_{1024^2}$(hrs) \\
\hline
PS  &5.91E-3 &2.94 &  3.34E-6 &17.65&         & 97.03 \\
ED6 &8.10E-2 &0.67 &  6.86E-3 &3.08 & 1.87E-4 & 19.29 \\
ED4 &1.48E-1 &0.66 &  2.53E-2 &2.99 & 2.04E-3 & 17.84 \\
ED2 &3.97E-1 &0.64 &  1.81E-1 &2.95 & 5.86E-2 & 16.93 \\
CD6 &3.36E-2 &0.92 &  1.03E-3 &4.10 & 8.10E-5 & 23.60 \\
CD4 &6.93E-2 &0.90 &  6.47E-3 &4.05 & 3.71E-4 & 21.73 \\
A4  &1.45E-1 &0.76 &  2.52E-2 &3.37 & 2.04E-3 & 19.32 \\
A2  &4.45E-1 &0.69 &  2.01E-1 &3.15 & 6.46E-2 & 17.92 \\
DRP4&5.69E-2 &0.67 &  8.23E-3 &3.05 & 1.02E-3 & 18.07 \\
\hline
\end{tabular}
\end{center}
\end{table*}

The perturbed shear layers roll up into a single vortex and the shear layers become thinner and thinner as the flow evolves over time, as shown in Fig.~\ref{fig:dsl-time}. The three-dimensional view of the shear layers developed at time $t=10$ is also illustrated in Fig.~\ref{fig:dsl3d}, which shows the magnitude of the vorticity as the z-axis. This shear layer problem is particularly important in that the presence of thinner and thinner shear layers as the flow field evolves in time is not captured by low grid resolution representations. The Gibbs phenomenon, numerical oscillations occurring near sharp vorticity gradients, occurs for underresolved simulations in which the grid size is larger than the shear layer thickness. Since sharp discontinuities are observed in the vorticity field, this benchmark flow problem is very appropriate to test the overall accuracy of the methods presented in this study.

\begin{figure*}
\centering
\mbox{
\subfigure{\includegraphics[width=0.33\textwidth]{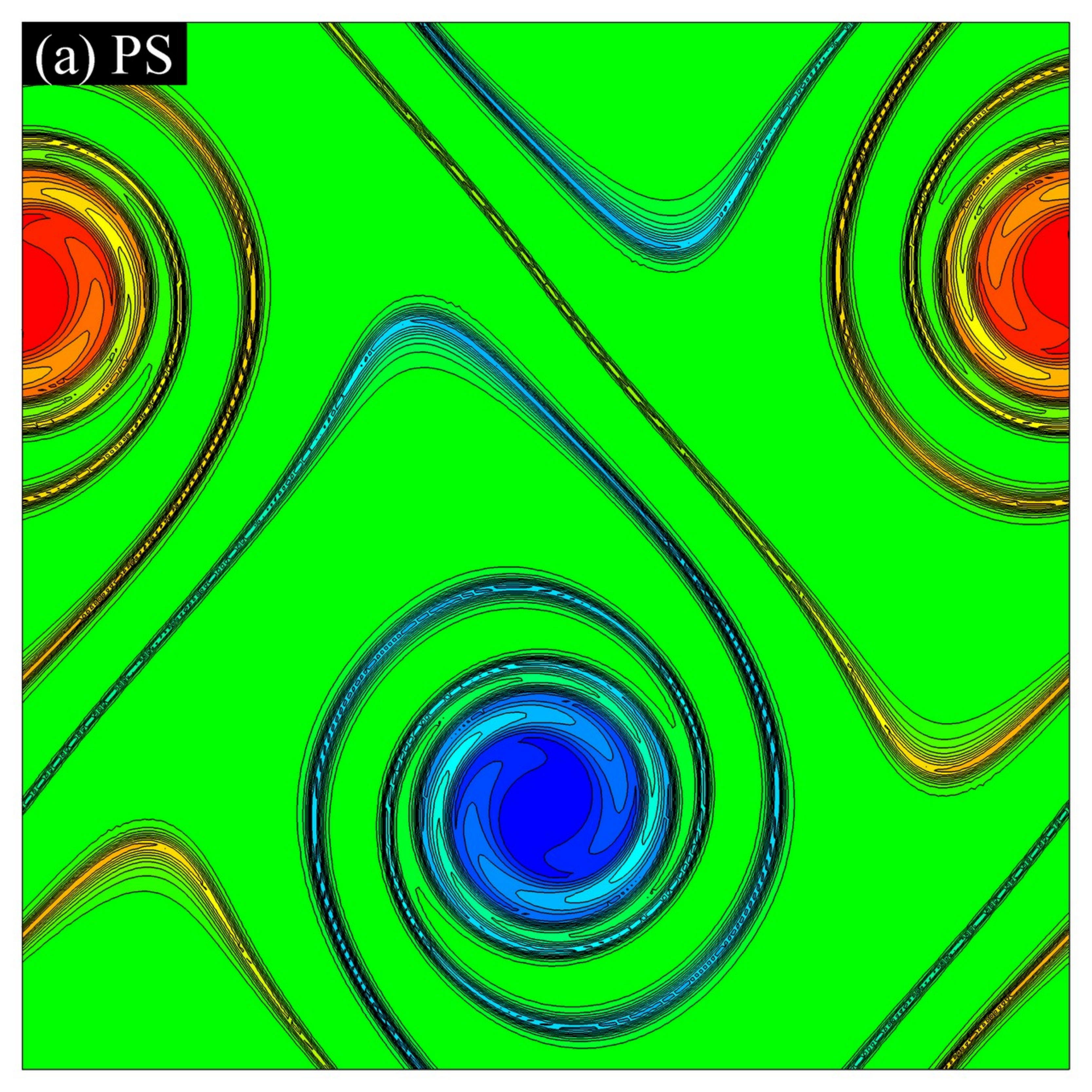}}
\subfigure{\includegraphics[width=0.33\textwidth]{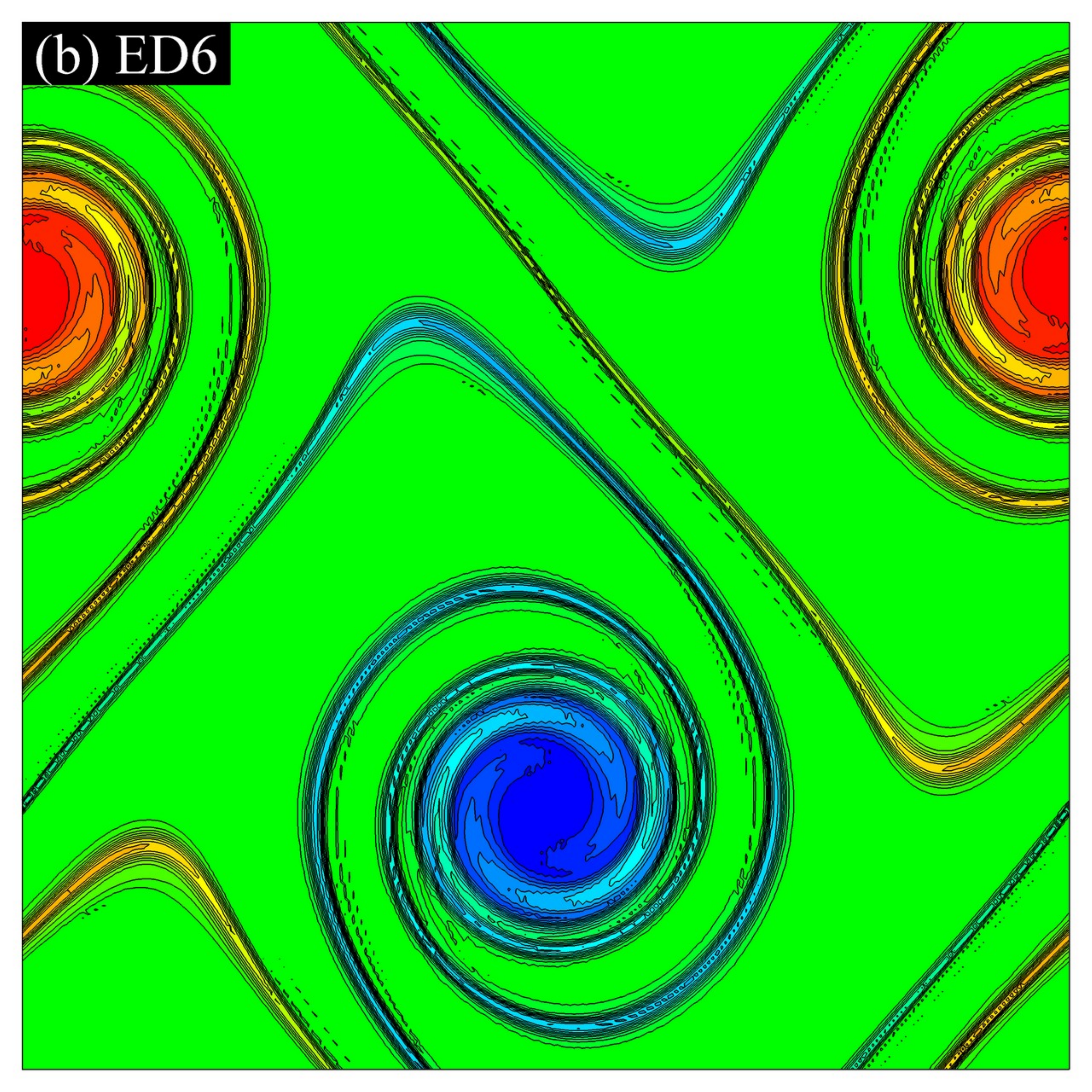}}
\subfigure{\includegraphics[width=0.33\textwidth]{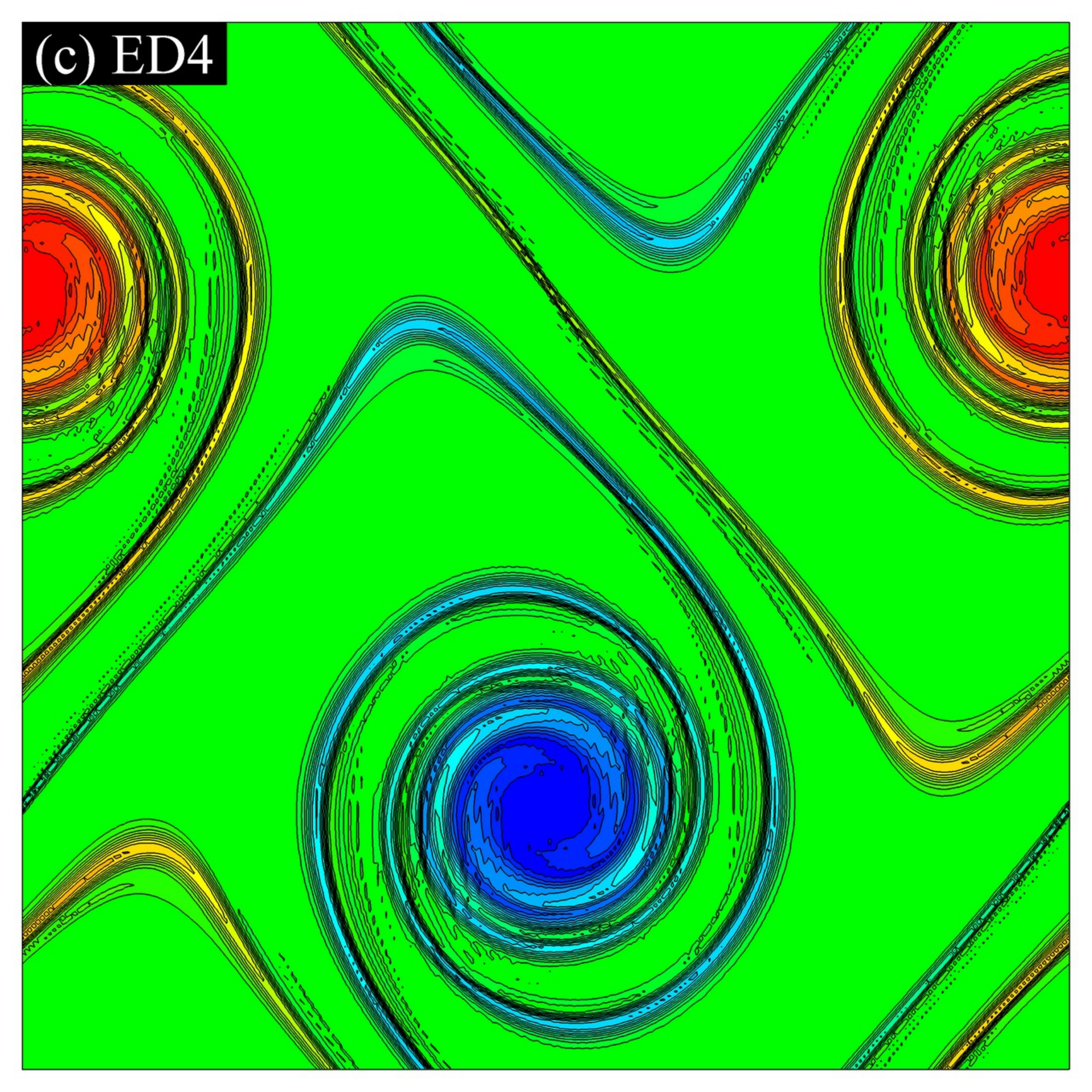}}
}
\\
\mbox{
\subfigure{\includegraphics[width=0.33\textwidth]{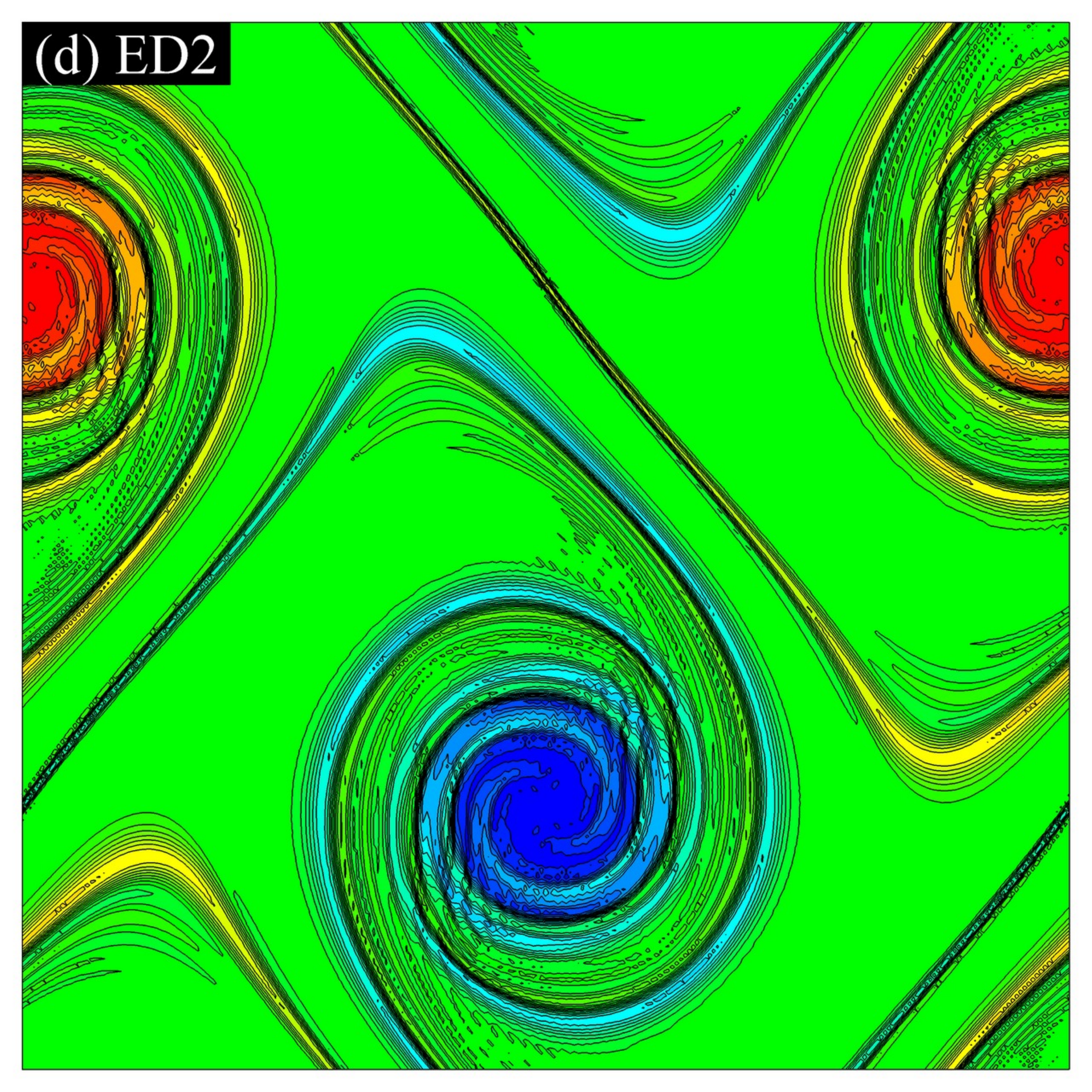}}
\subfigure{\includegraphics[width=0.33\textwidth]{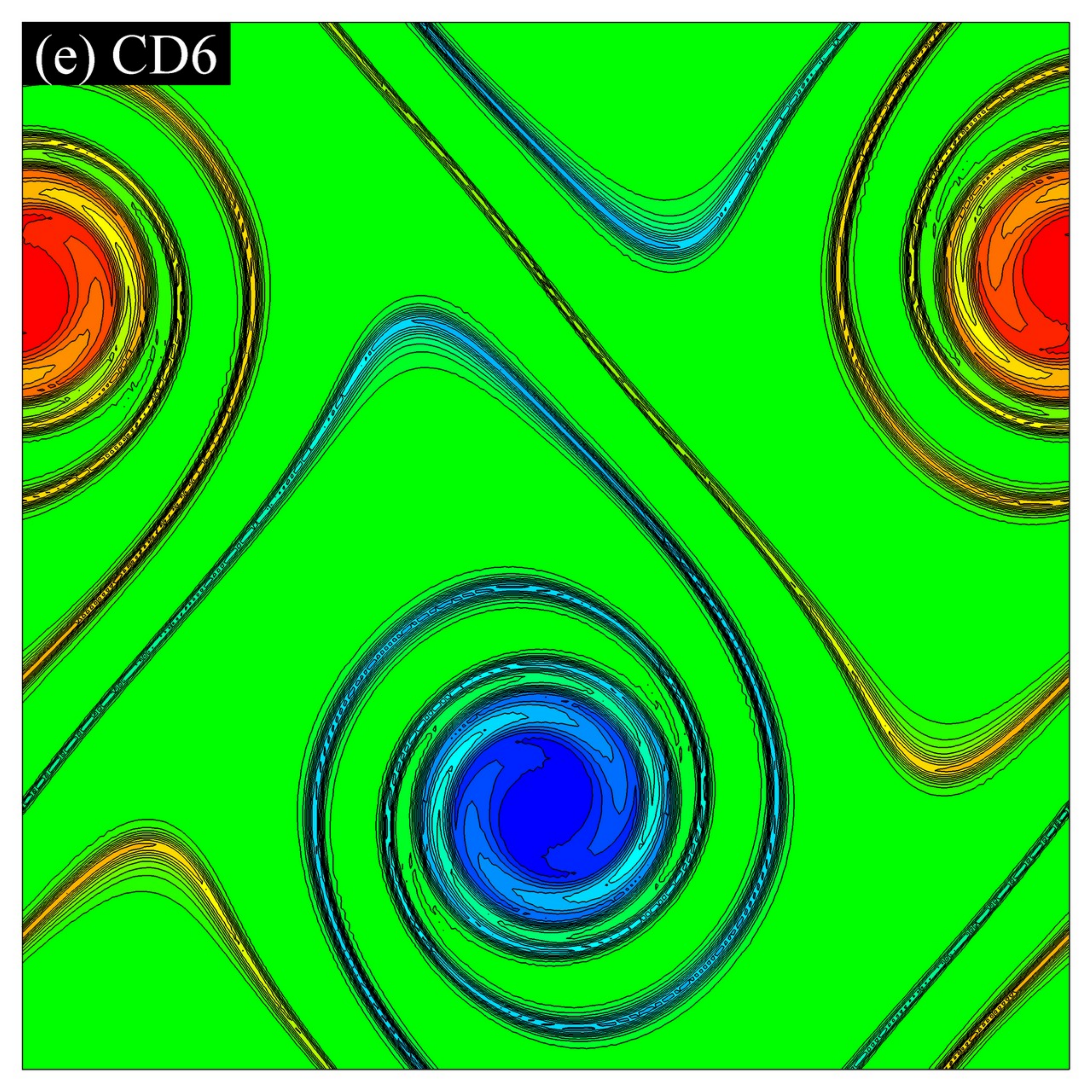}}
\subfigure{\includegraphics[width=0.33\textwidth]{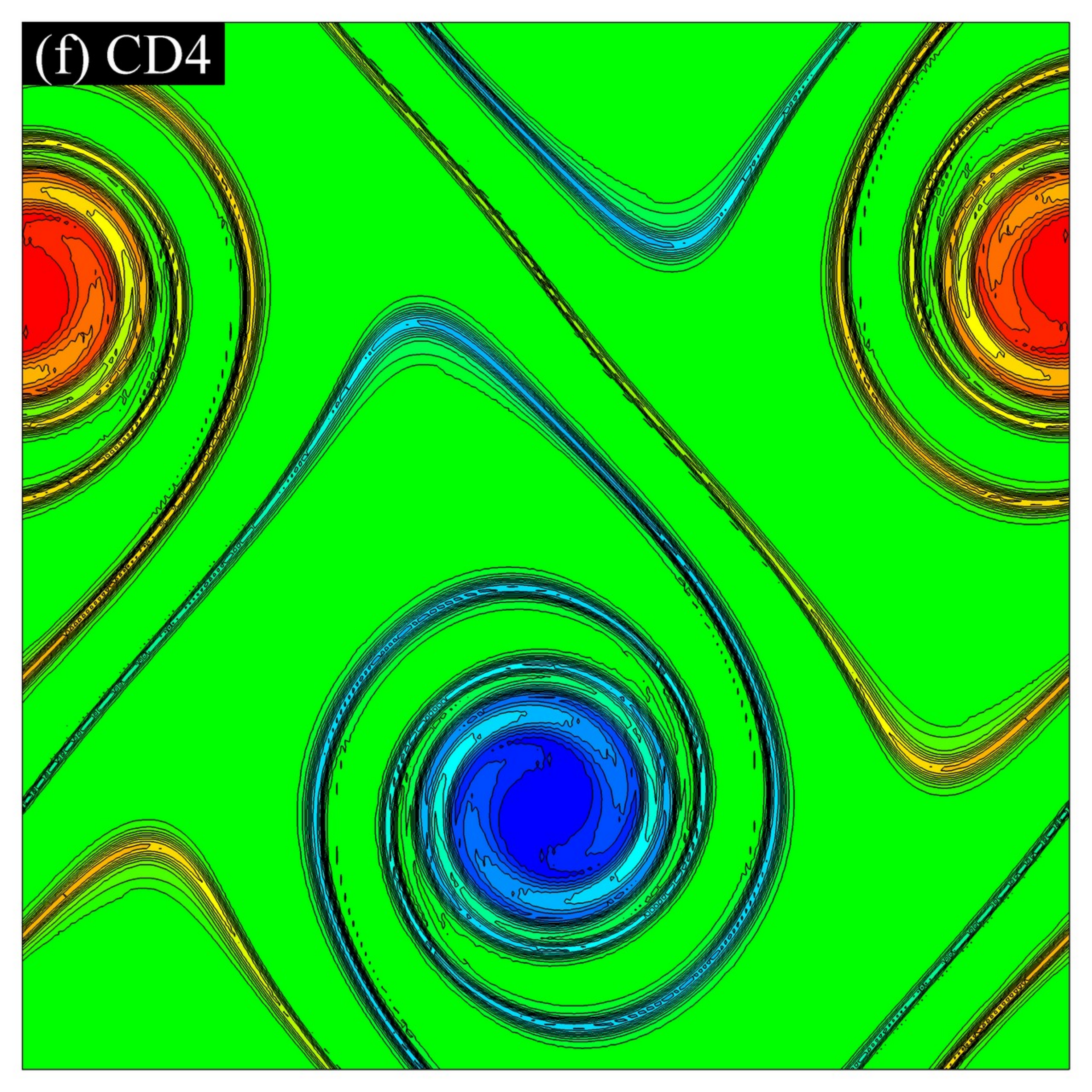}}
}
\\
\mbox{
\subfigure{\includegraphics[width=0.33\textwidth]{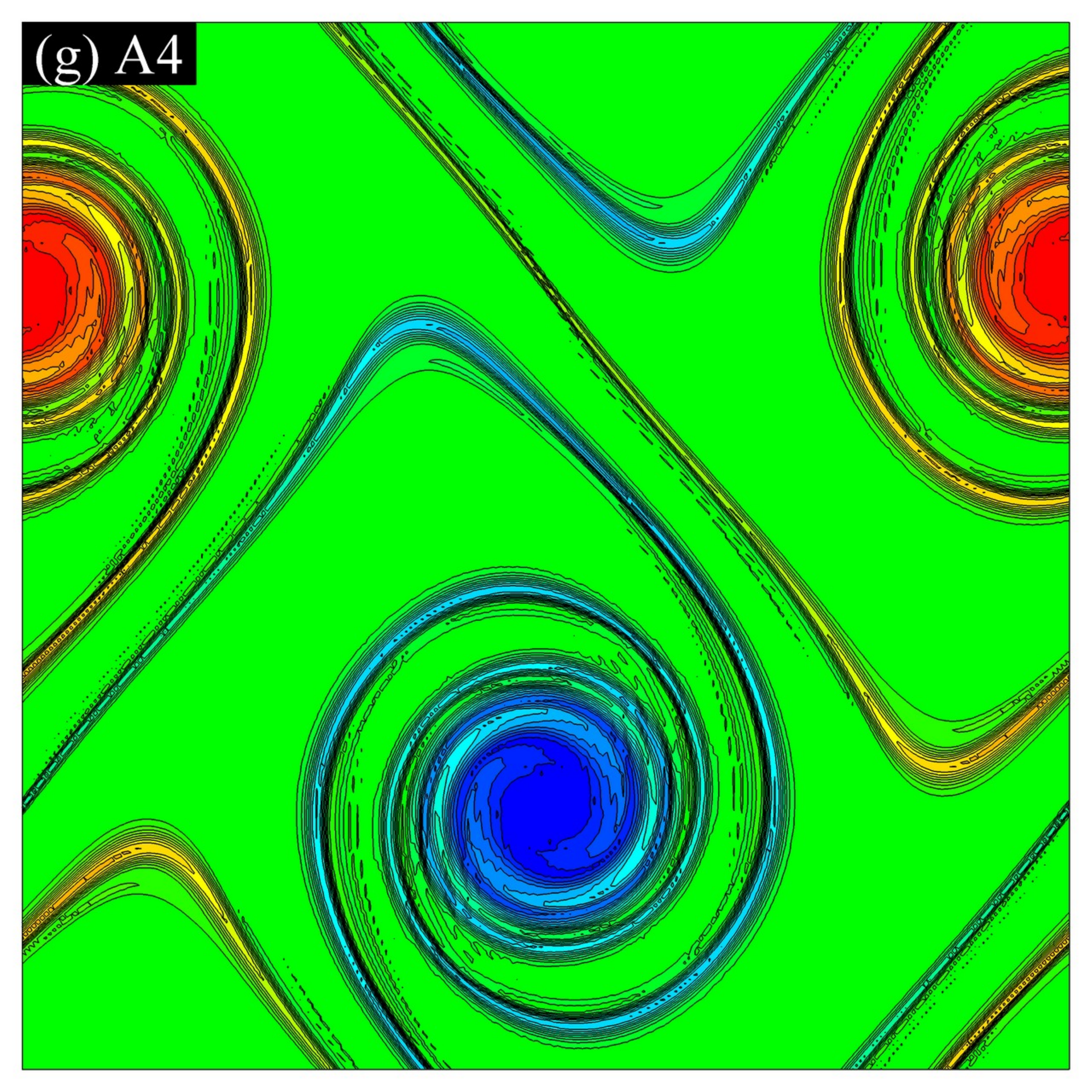}}
\subfigure{\includegraphics[width=0.33\textwidth]{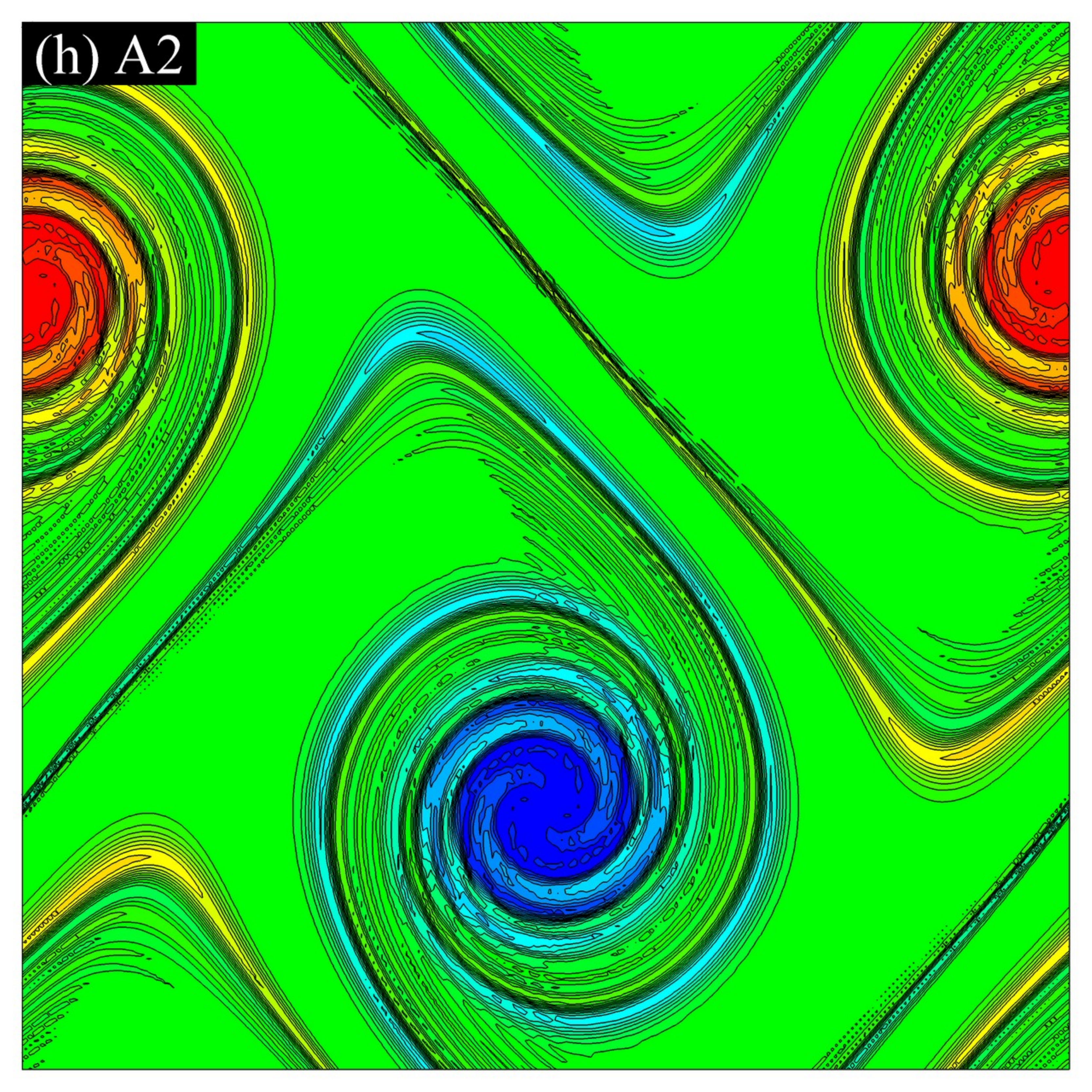}}
\subfigure{\includegraphics[width=0.33\textwidth]{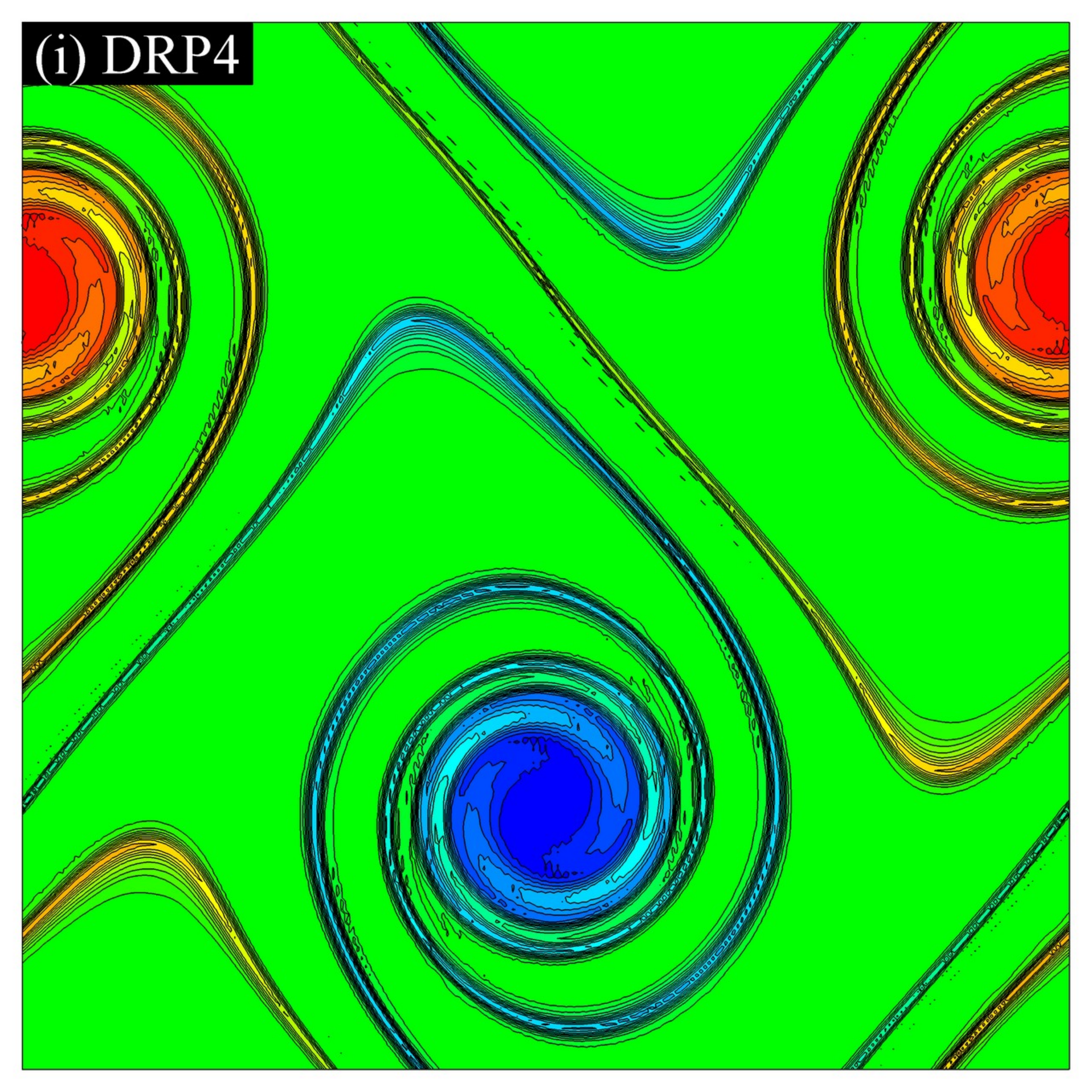}}
}
\caption{Comparison of the numerical schemes for the double shear layer problem at time $t=10$ with a resolution of $256^2$. (a) pseudospectral (PS) method, (b) sixth-order explicit difference (ED6) method, (c) fourth-order explicit difference (ED4) method, (d) second-order explicit difference (ED2) method, (e) sixth-order compact difference (CD6) method, (f) fourth-order compact difference (CD4) method, (g) fourth-order Arakawa (A4) method, (h) second-order Arakawa (A2) method, and (i) fourth-order dispersion-relation-preserving (DRP4) method. The vorticity contour layouts are identical in all nine cases illustrating 27 equidistant levels in the interval [-4.5, 4.5].}
\label{fig:dsl256}
\end{figure*}

\begin{figure*}
\centering
\mbox{
\subfigure{\includegraphics[width=0.33\textwidth]{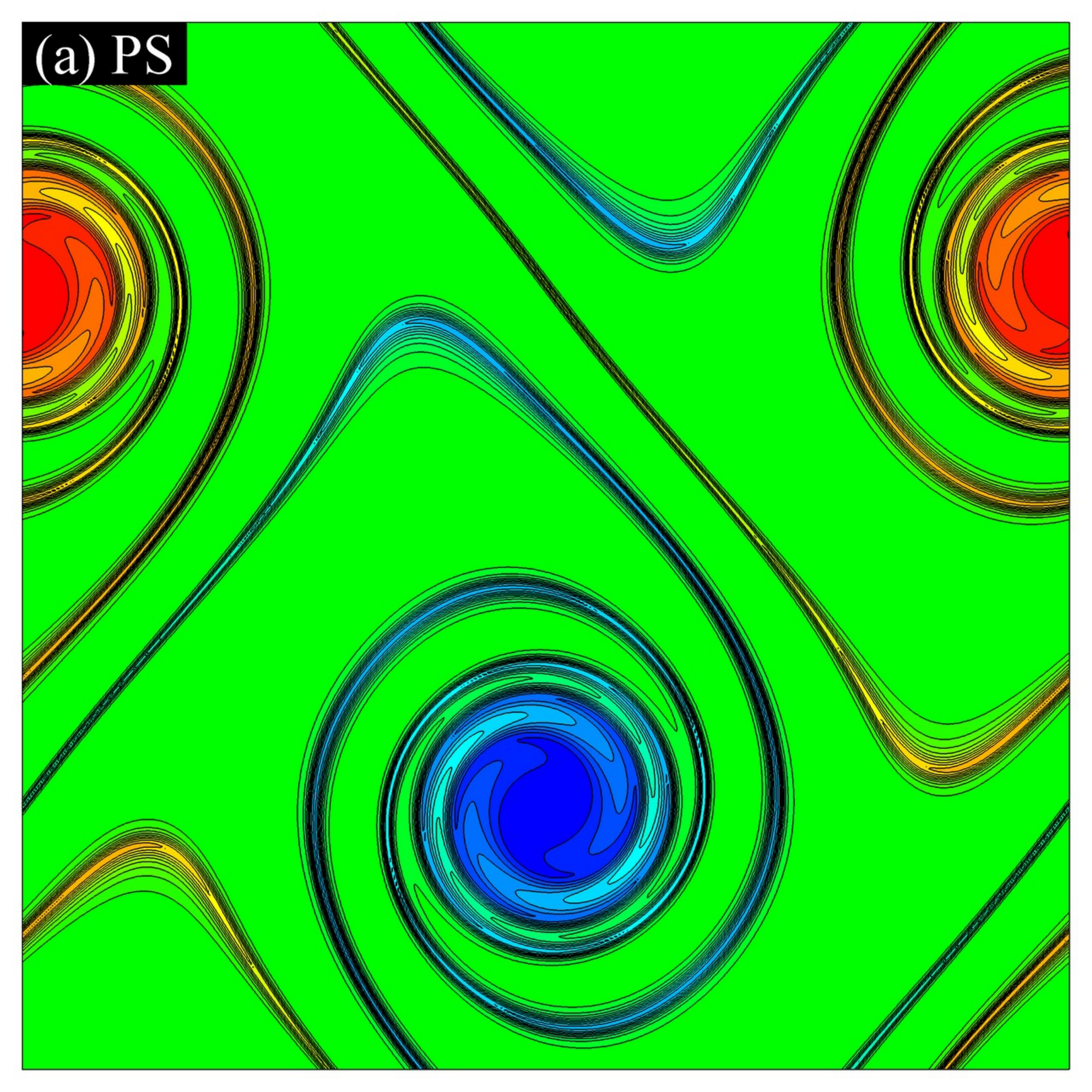}}
\subfigure{\includegraphics[width=0.33\textwidth]{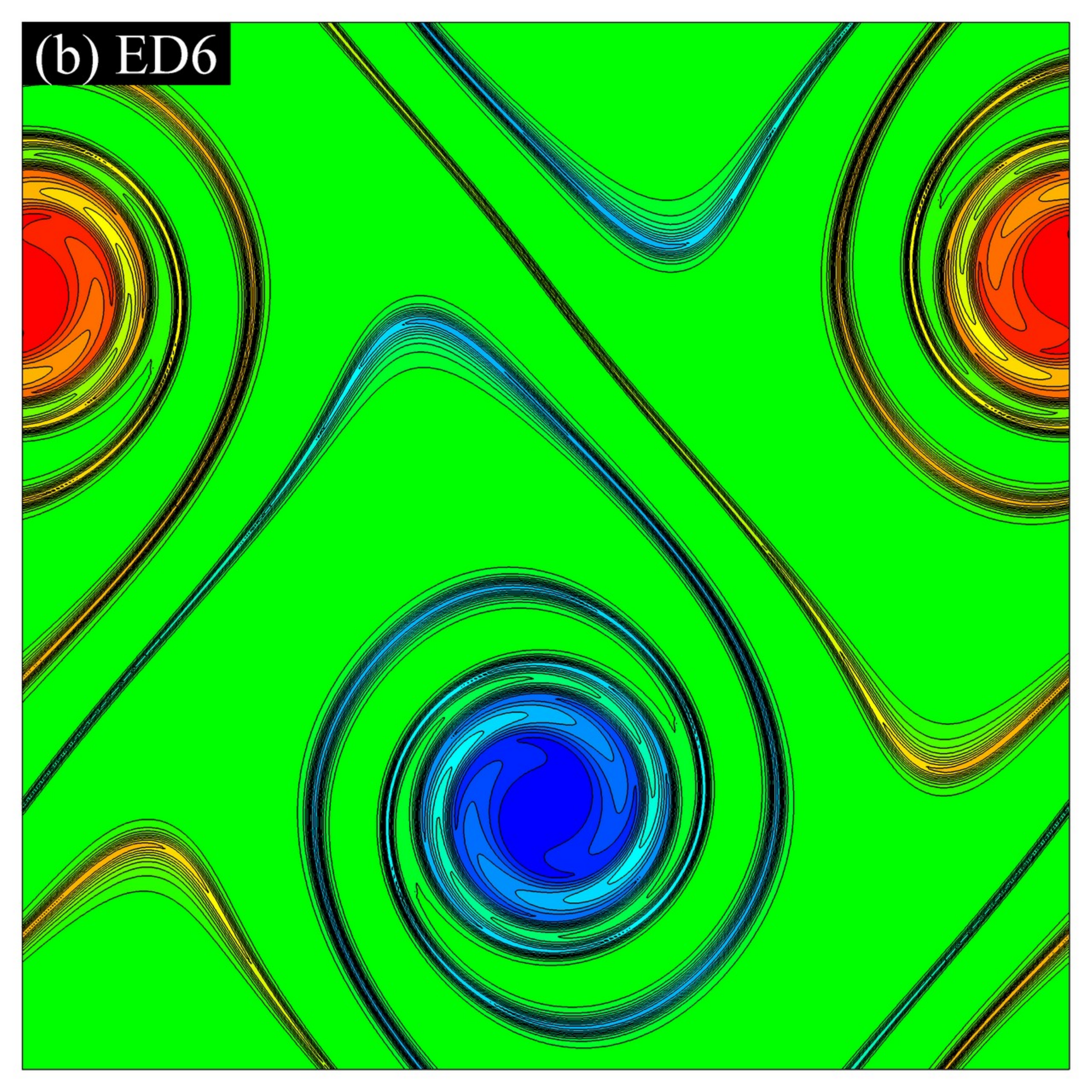}}
\subfigure{\includegraphics[width=0.33\textwidth]{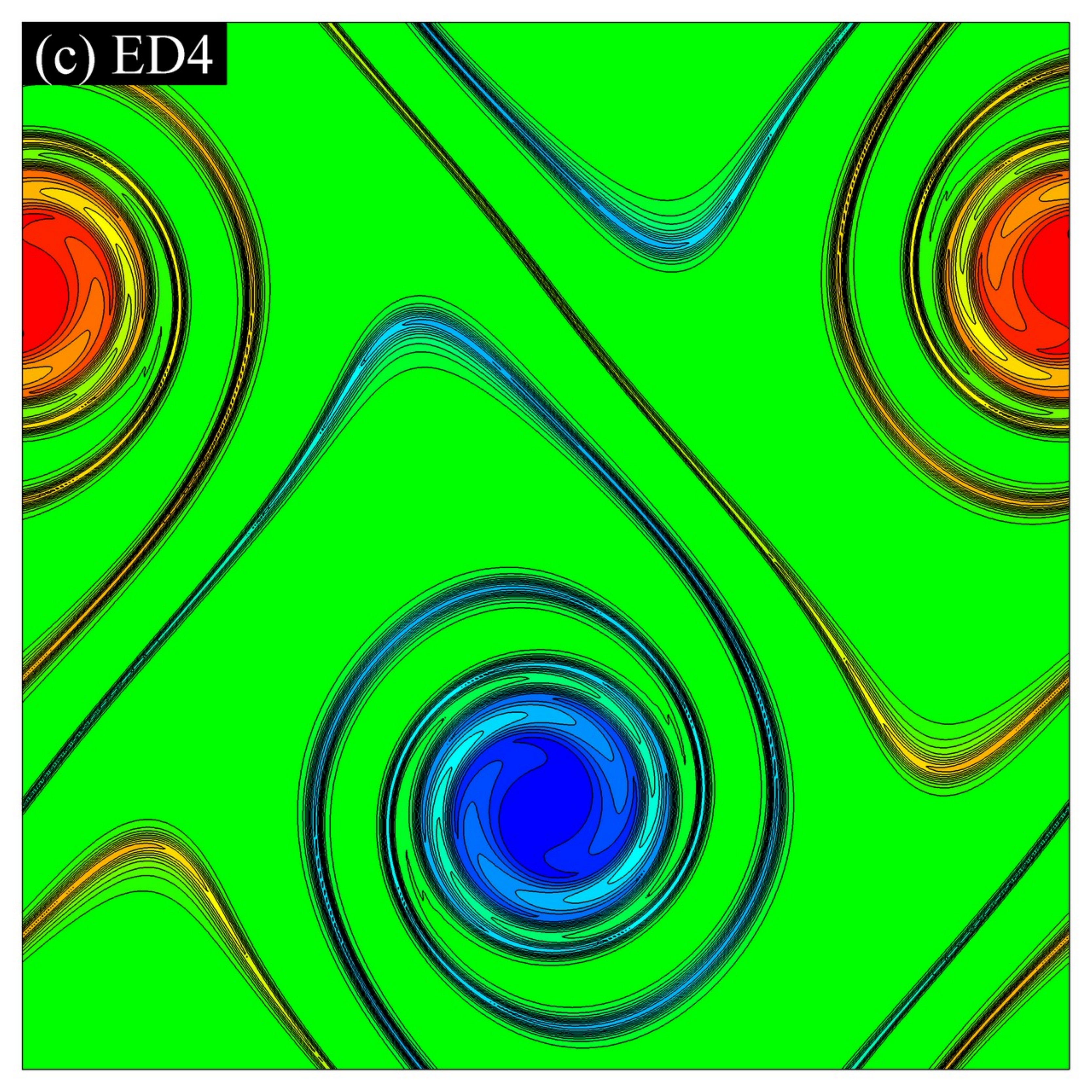}}
}
\\
\mbox{
\subfigure{\includegraphics[width=0.33\textwidth]{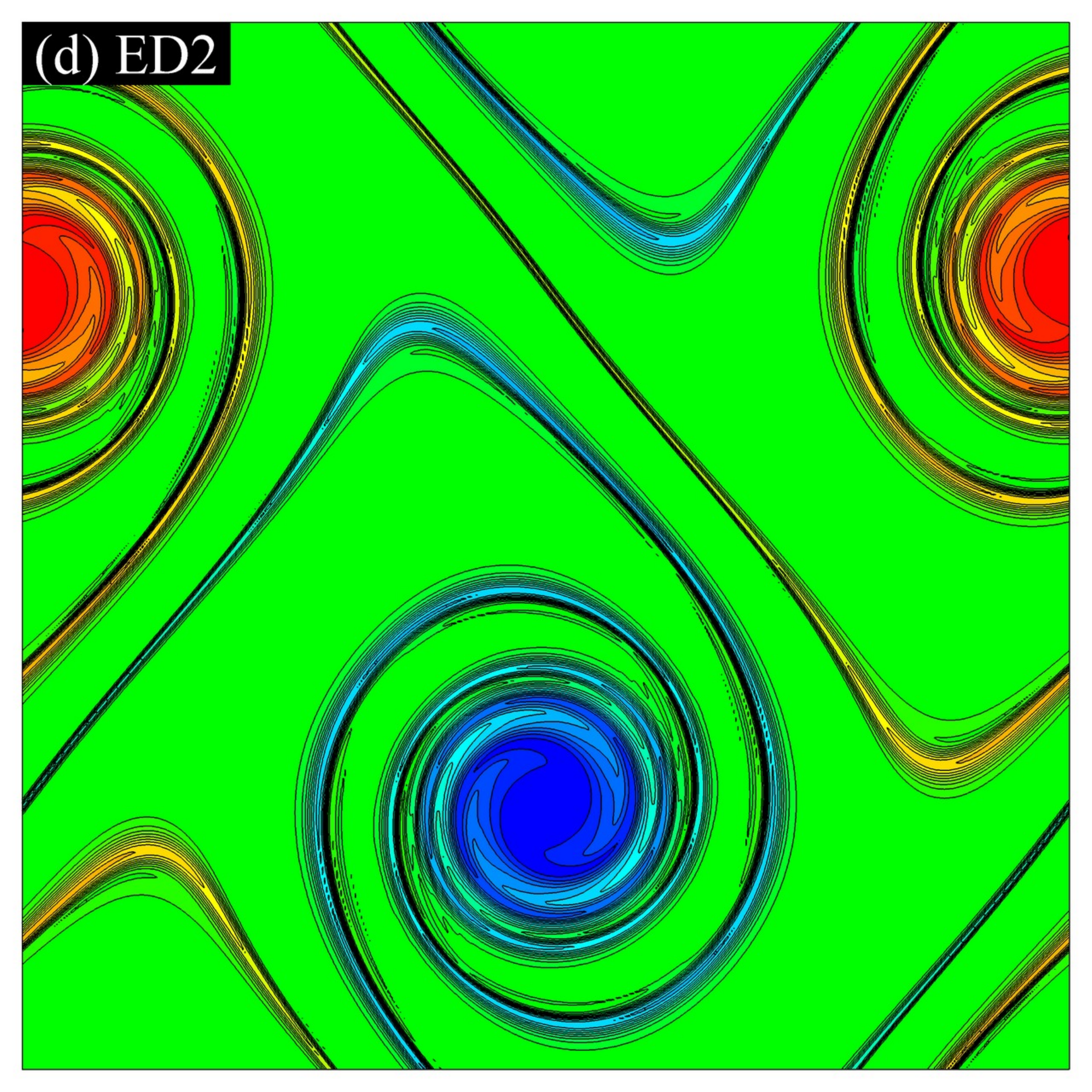}}
\subfigure{\includegraphics[width=0.33\textwidth]{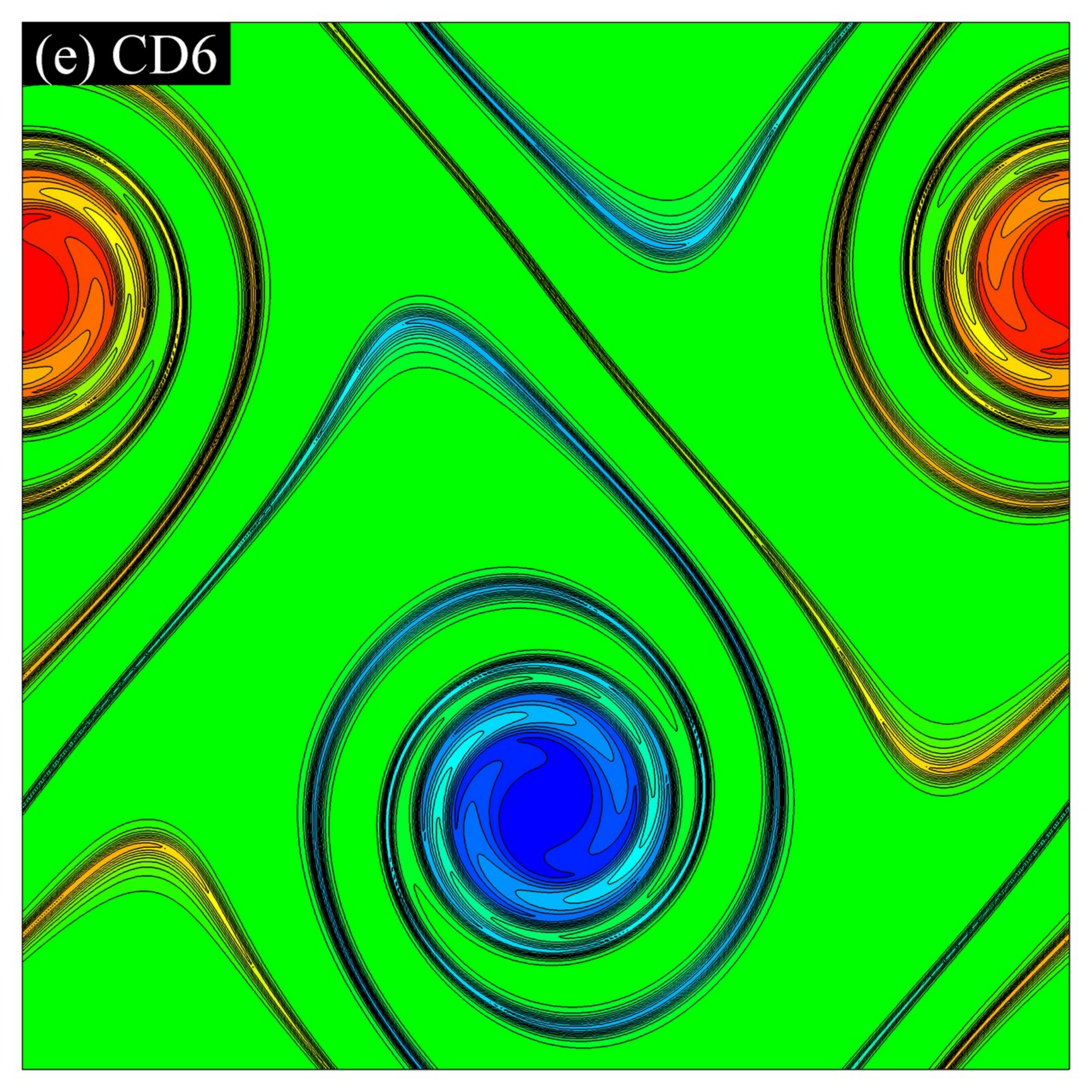}}
\subfigure{\includegraphics[width=0.33\textwidth]{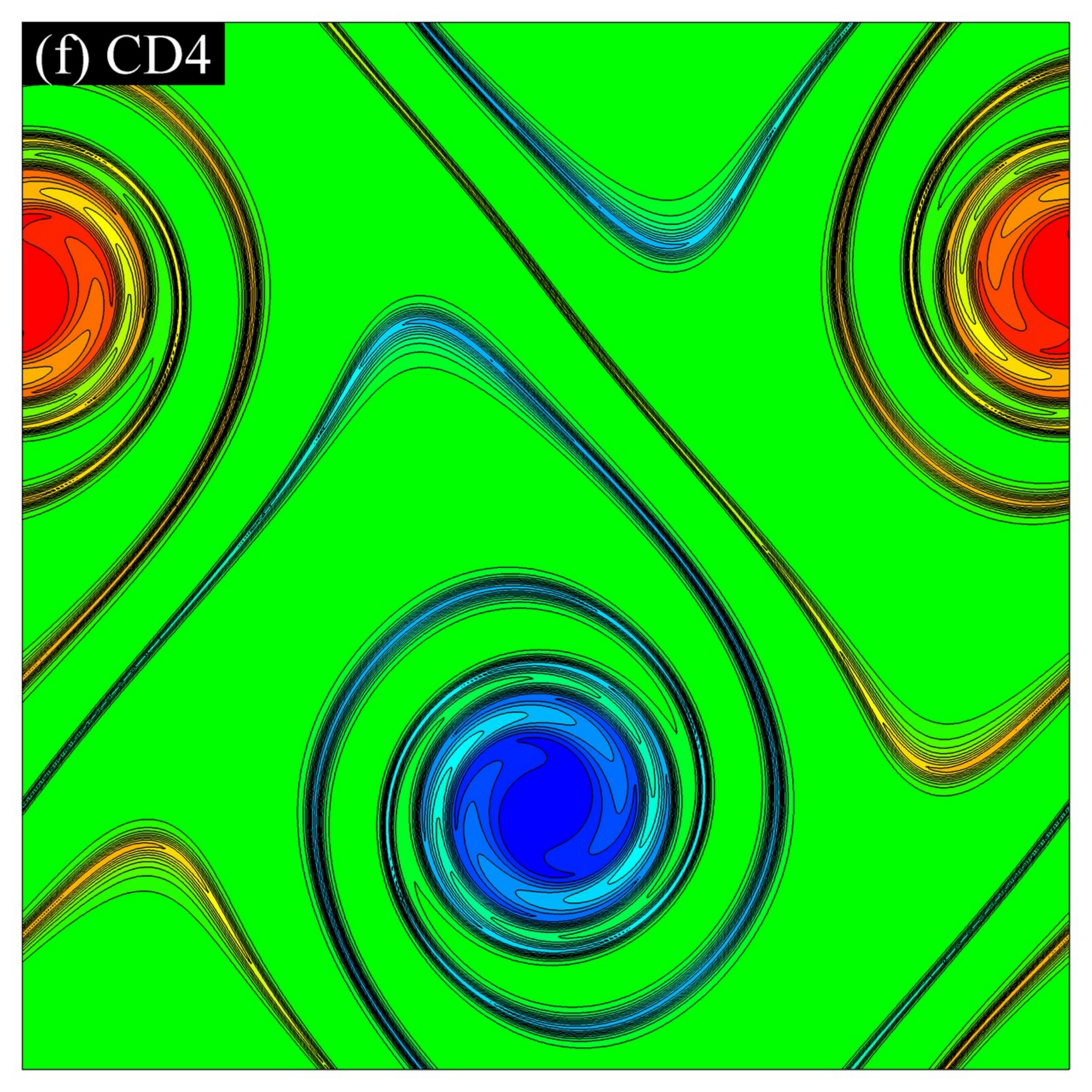}}
}
\\
\mbox{
\subfigure{\includegraphics[width=0.33\textwidth]{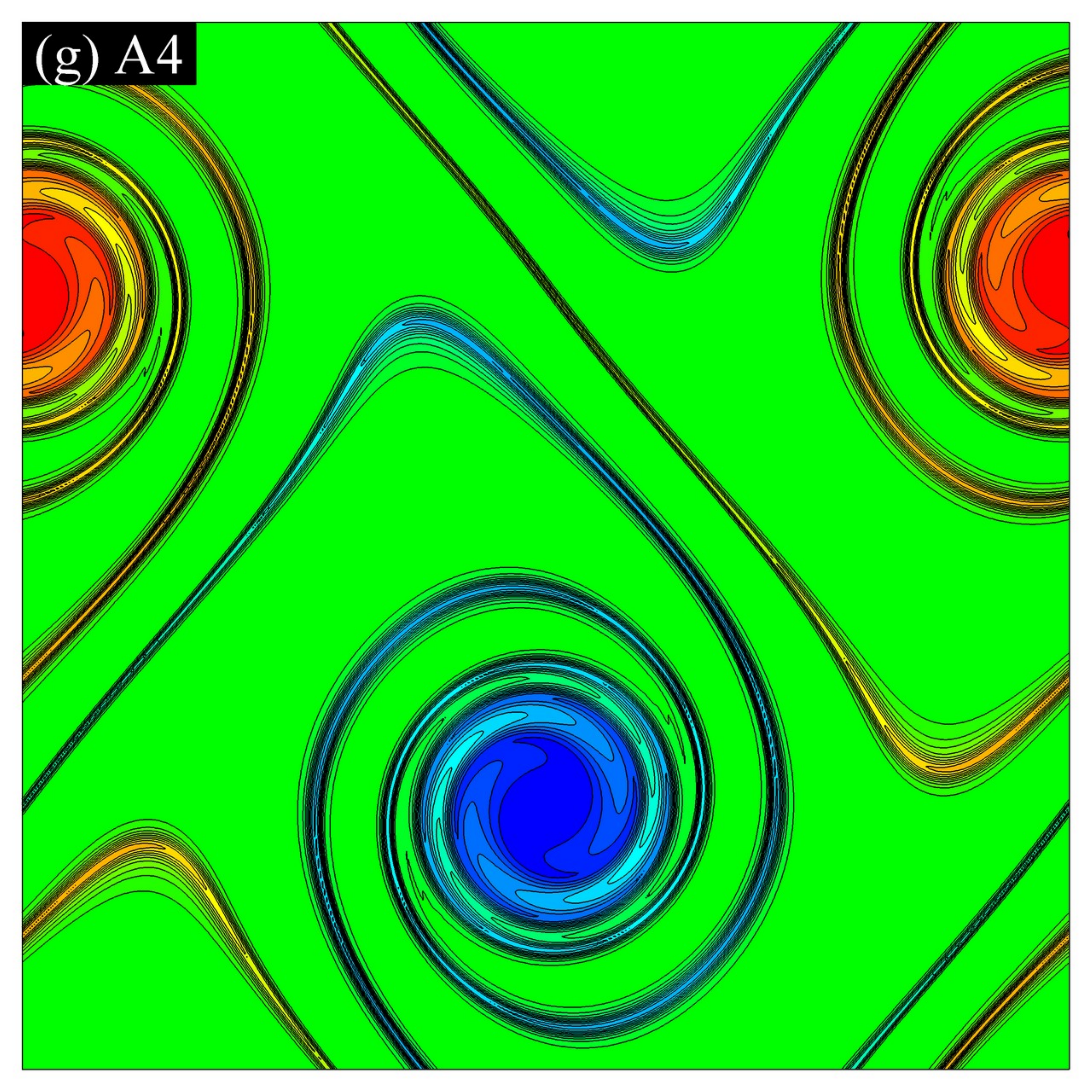}}
\subfigure{\includegraphics[width=0.33\textwidth]{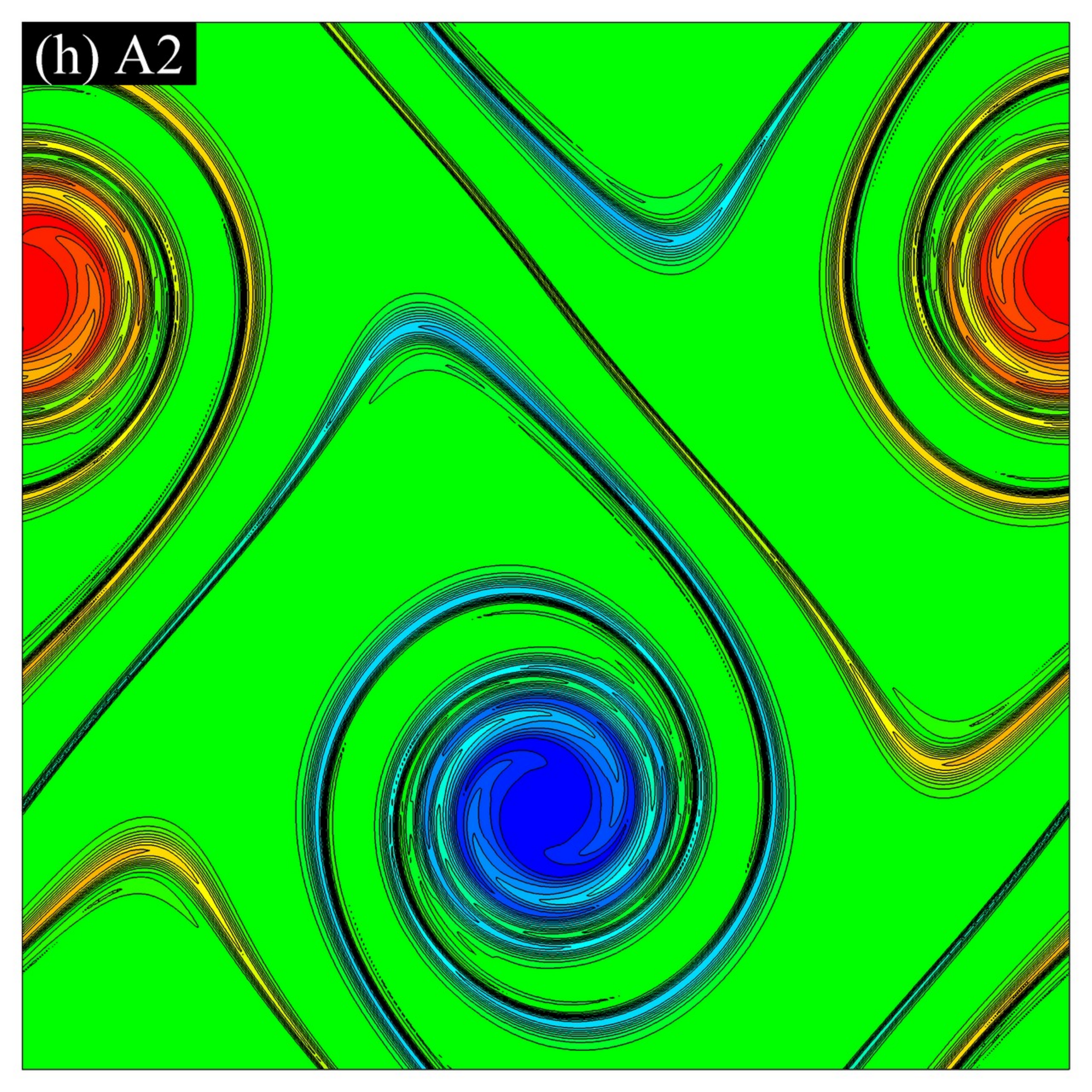}}
\subfigure{\includegraphics[width=0.33\textwidth]{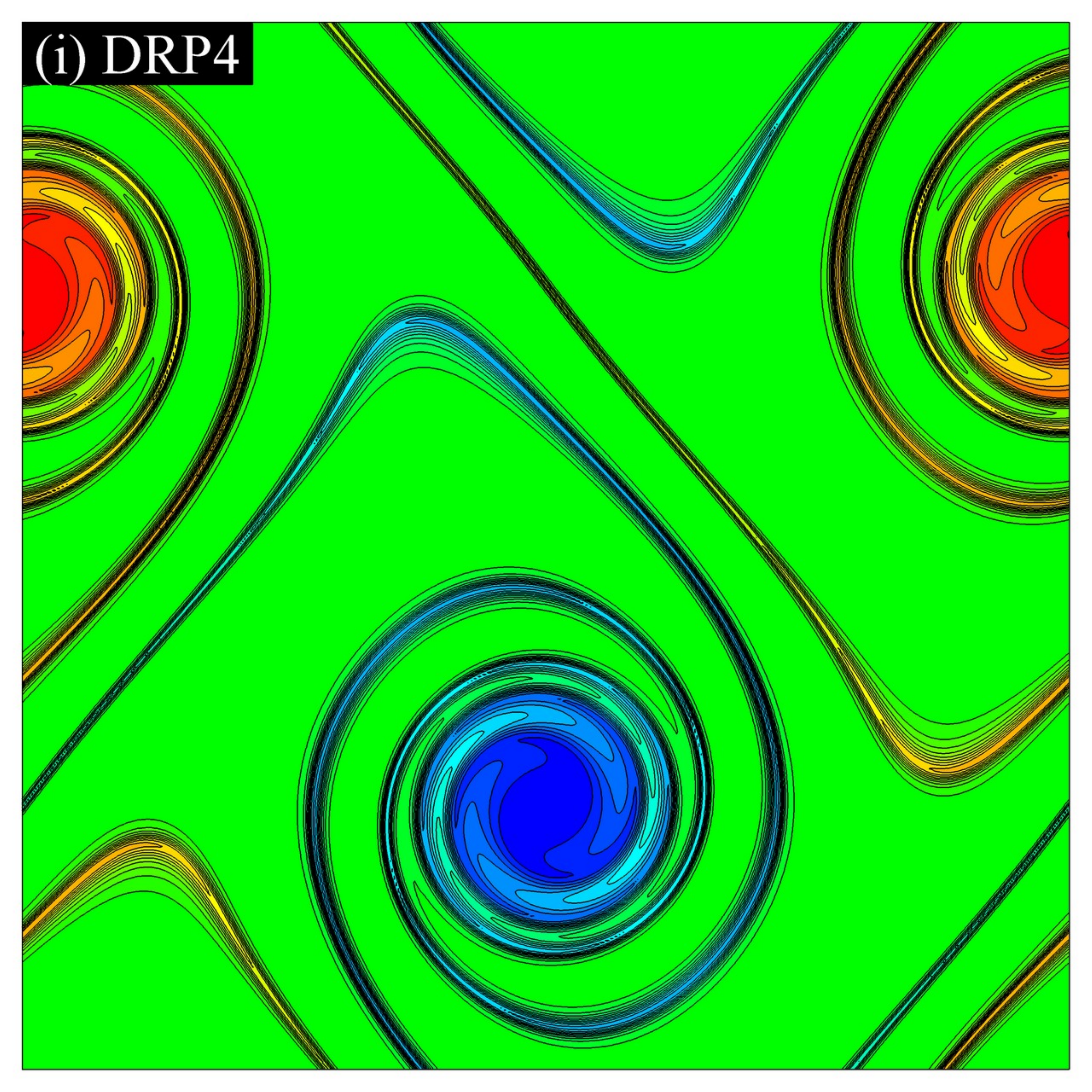}}
}
\caption{Comparison of the numerical schemes for the double shear layer problem at time $t=10$ with a resolution of $512^2$. (a) pseudospectral (PS) method, (b) sixth-order explicit difference (ED6) method, (c) fourth-order explicit difference (ED4) method, (d) second-order explicit difference (ED2) method, (e) sixth-order compact difference (CD6) method, (f) fourth-order compact difference (CD4) method, (g) fourth-order Arakawa (A4) method, (h) second-order Arakawa (A2) method, and (i) fourth-order dispersion-relation-preserving (DRP4) method. The vorticity contour layouts are identical in all nine cases illustrating 27 equidistant levels in the interval [-4.5, 4.5].}
\label{fig:dsl512}
\end{figure*}

\begin{figure*}
\centering
\mbox{
\subfigure{\includegraphics[width=0.33\textwidth]{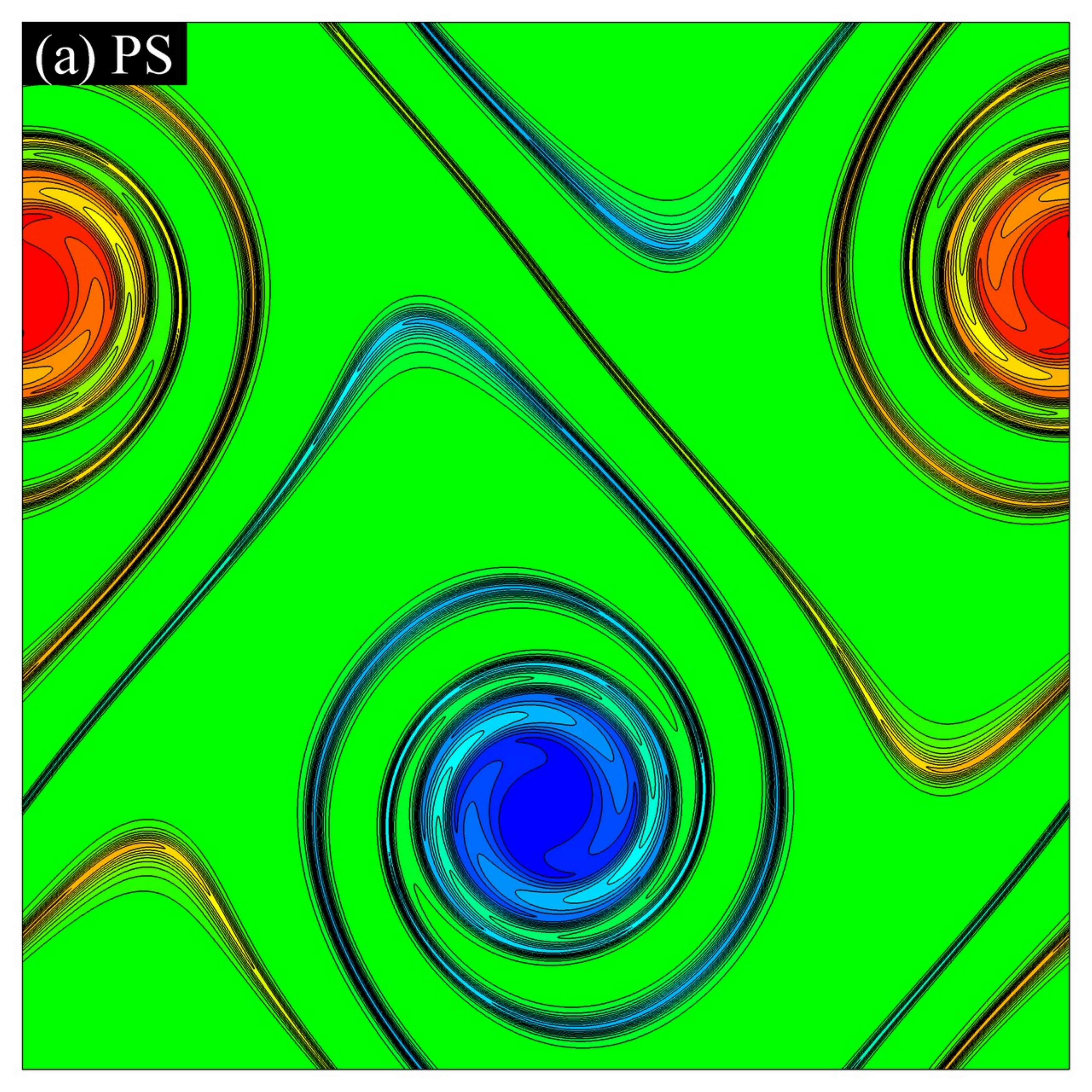}}
\subfigure{\includegraphics[width=0.33\textwidth]{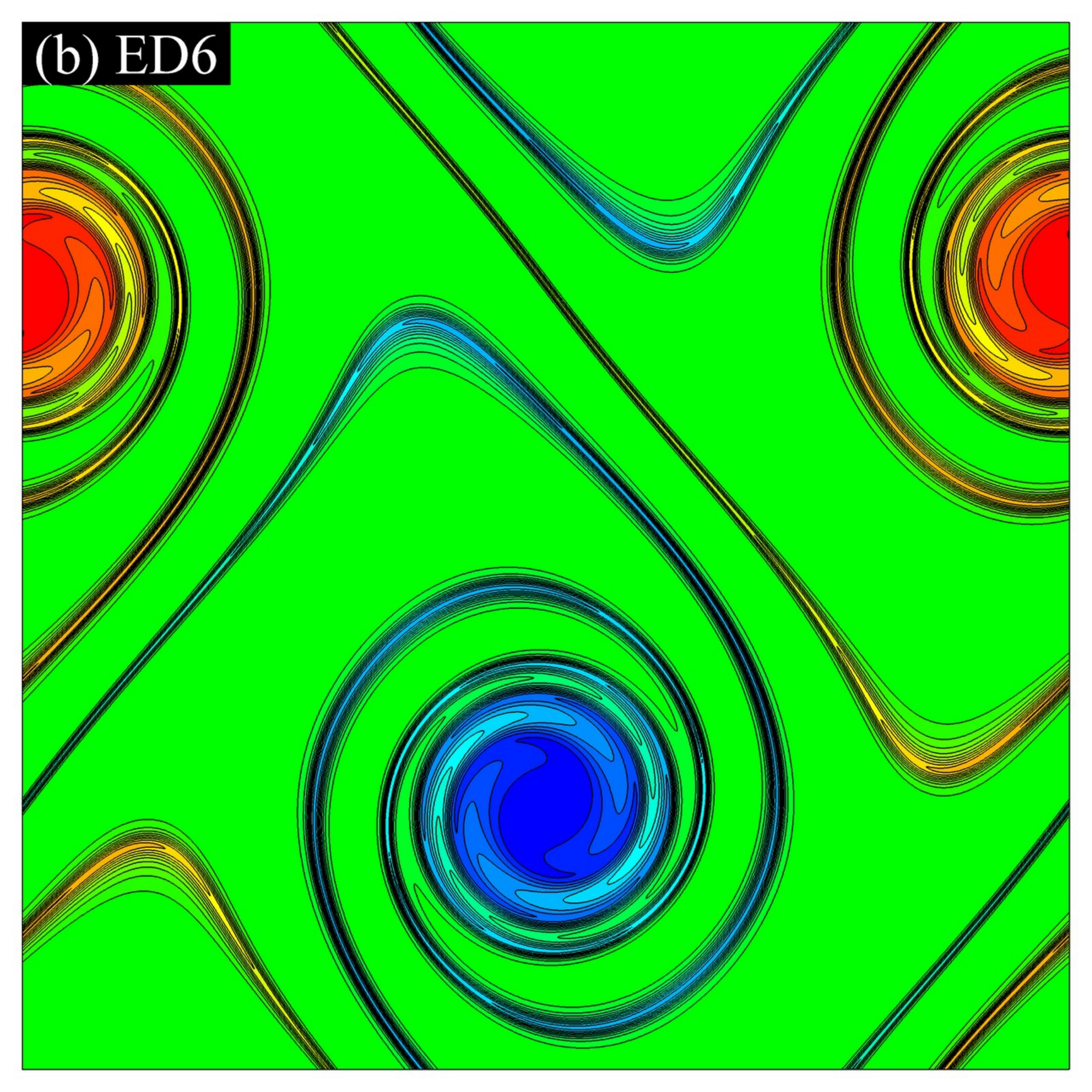}}
\subfigure{\includegraphics[width=0.33\textwidth]{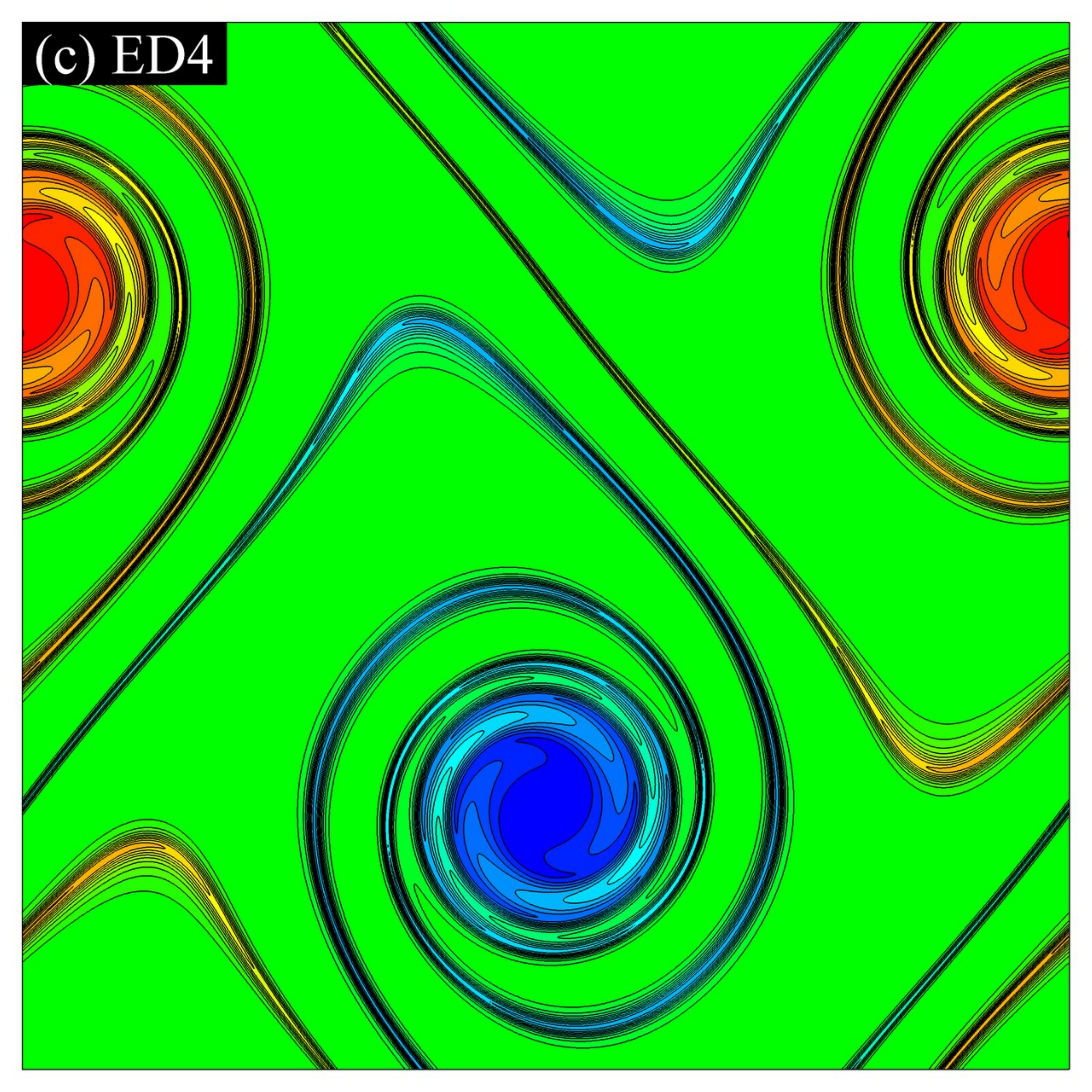}}
}
\\
\mbox{
\subfigure{\includegraphics[width=0.33\textwidth]{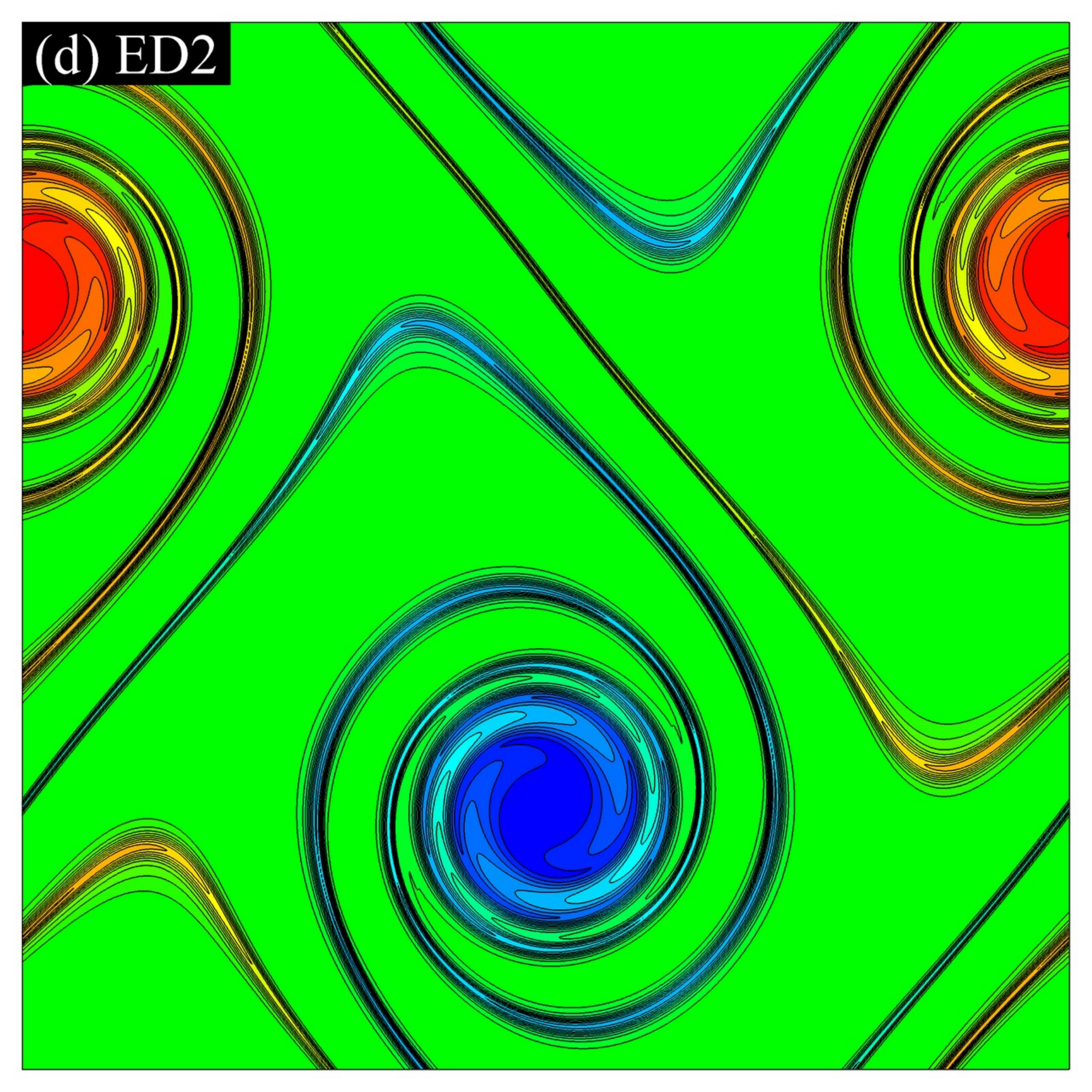}}
\subfigure{\includegraphics[width=0.33\textwidth]{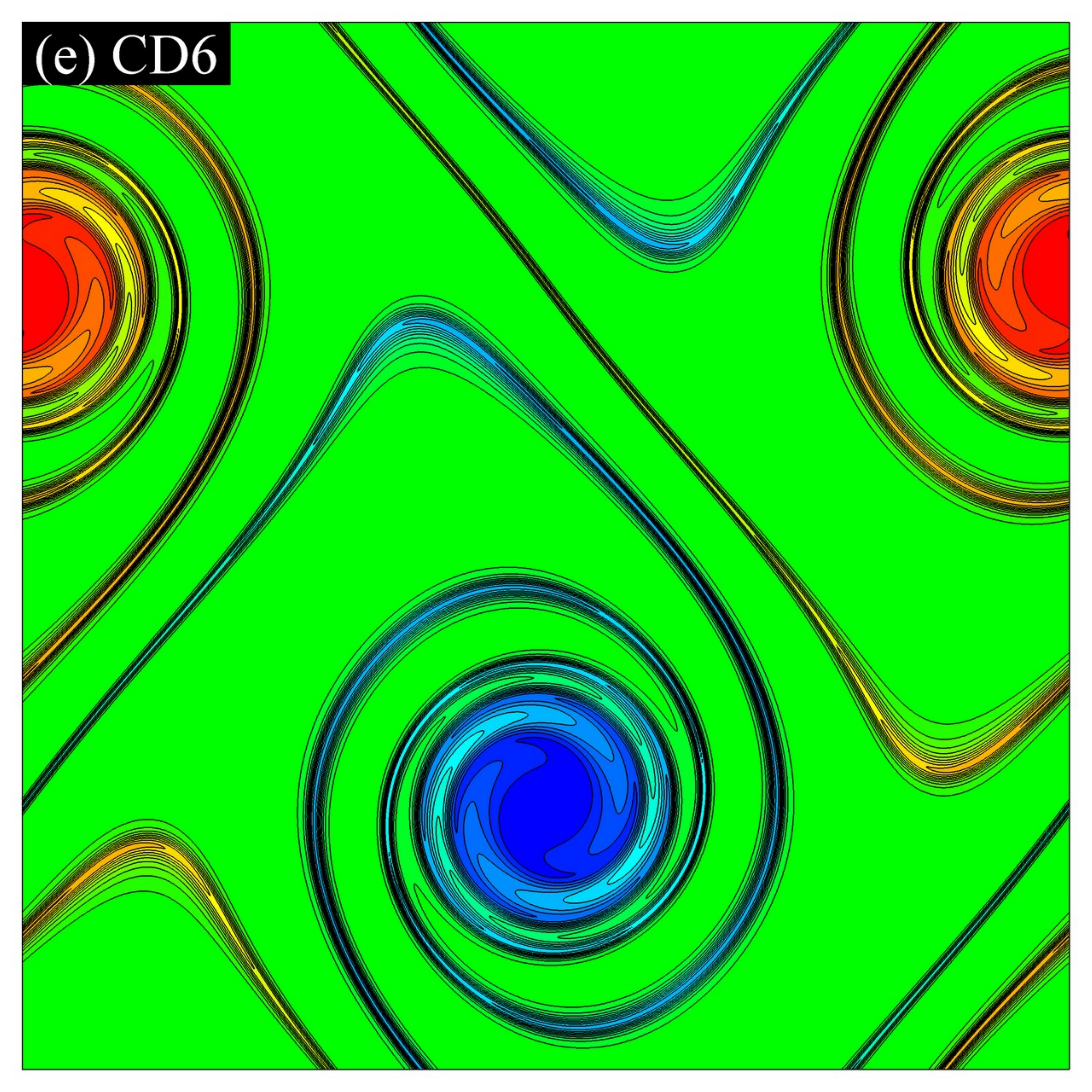}}
\subfigure{\includegraphics[width=0.33\textwidth]{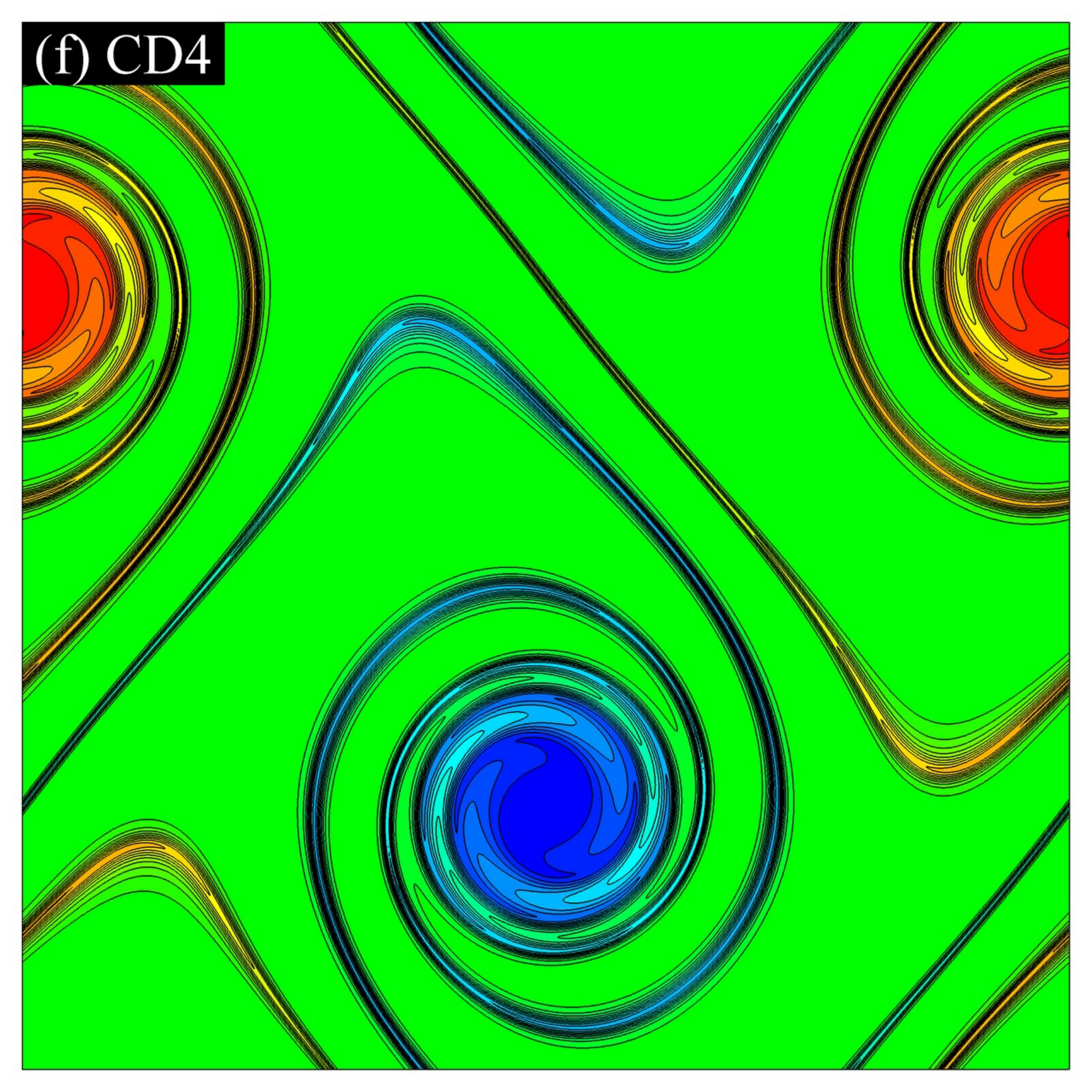}}
}
\\
\mbox{
\subfigure{\includegraphics[width=0.33\textwidth]{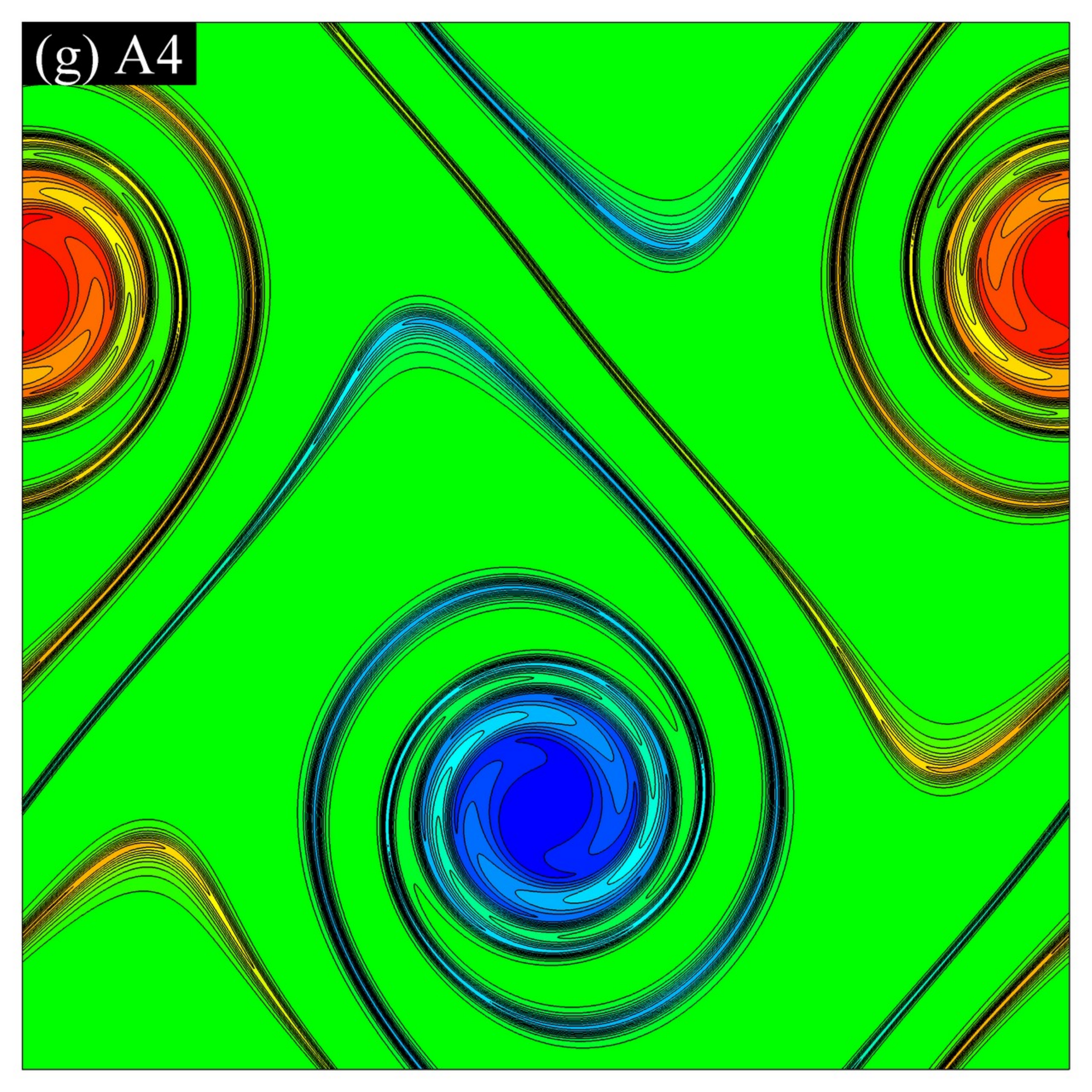}}
\subfigure{\includegraphics[width=0.33\textwidth]{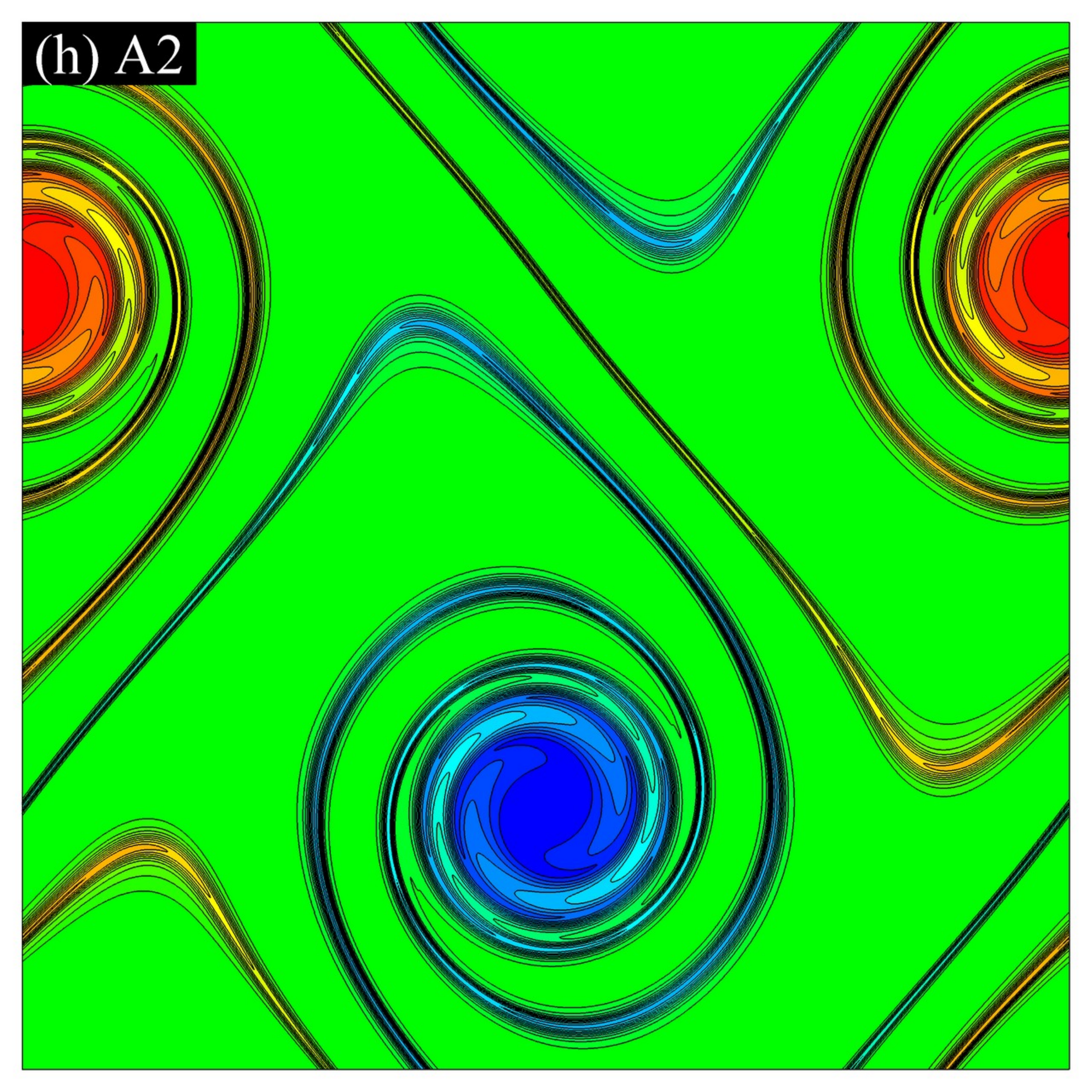}}
\subfigure{\includegraphics[width=0.33\textwidth]{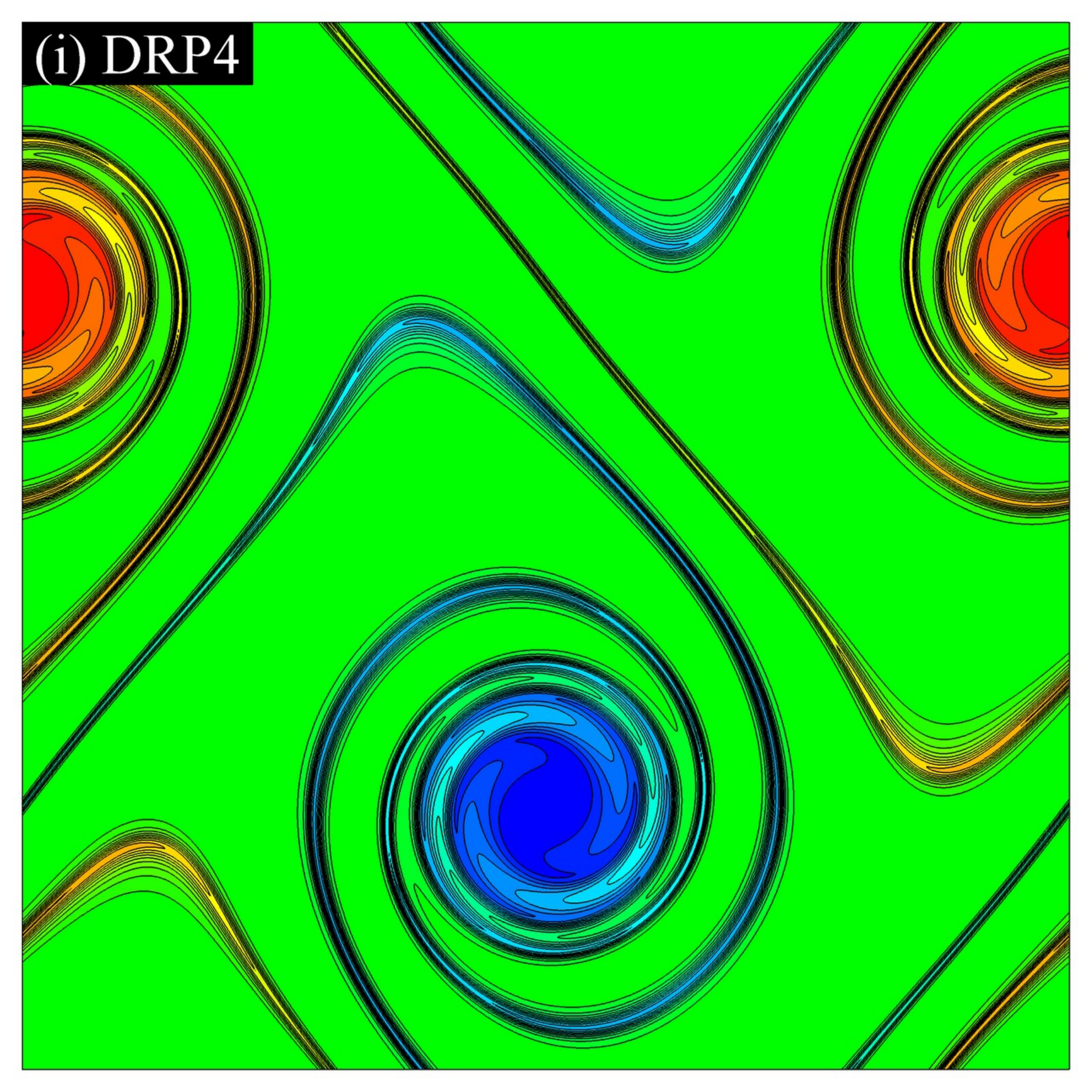}}
}
\caption{Comparison of the numerical schemes for the double shear layer problem at time $t=10$ with a resolution of $1024^2$. (a) pseudospectral (PS) method, (b) sixth-order explicit difference (ED6) method, (c) fourth-order explicit difference (ED4) method, (d) second-order explicit difference (ED2) method, (e) sixth-order compact difference (CD6) method, (f) fourth-order compact difference (CD4) method, (g) fourth-order Arakawa (A4) method, (h) second-order Arakawa (A2) method, and (i) fourth-order dispersion-relation-preserving (DRP4) method. The vorticity contour layouts are identical in all nine cases illustrating 27 equidistant levels in the interval [-4.5, 4.5].}
\label{fig:dsl1024}
\end{figure*}

\begin{figure}[h!]
\centering
\includegraphics[width=0.5\textwidth]{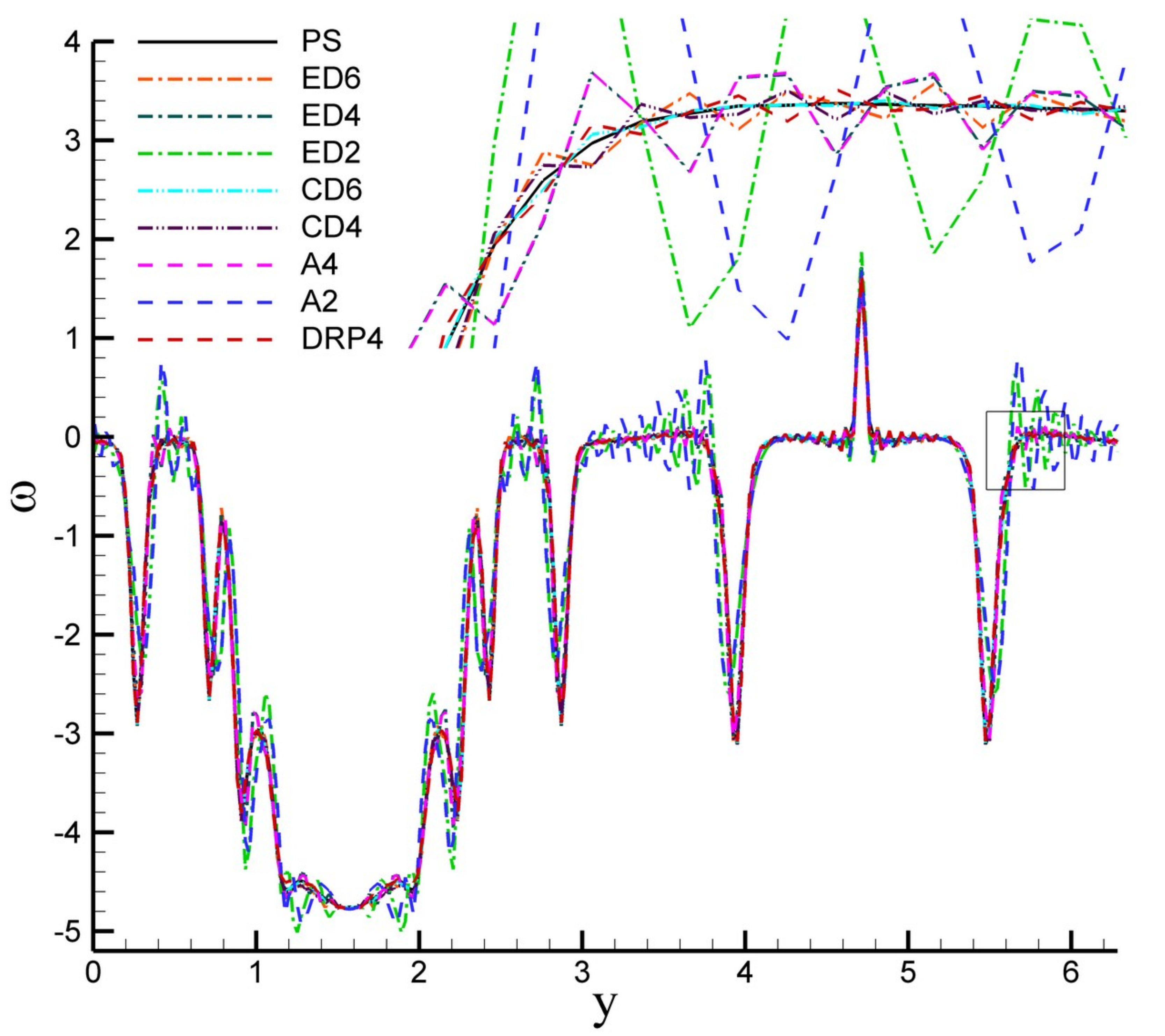}
\caption{The centerline vorticity distributions at $x=\pi$ for the double shear layer problem, plotted for a resolution of $256^2$. Comparison of the different numerical schemes at time $t=10$. Inset: close-up of boxed area.}
\label{fig:dsl-comp256}
\end{figure}

\begin{figure}[h!]
\centering
\includegraphics[width=0.5\textwidth]{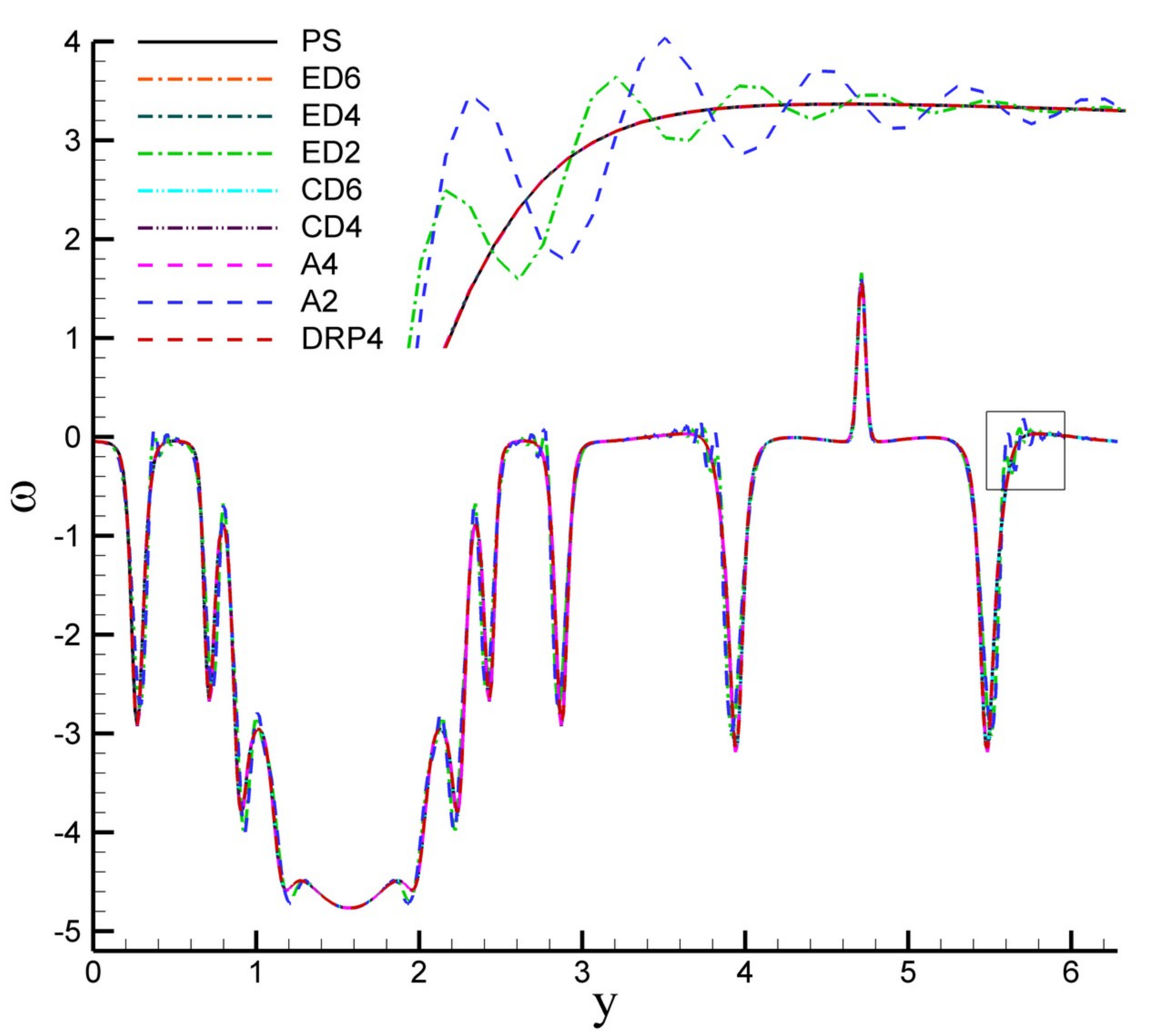}
\caption{The centerline vorticity distributions at $x=\pi$ for the double shear layer problem, plotted for a resolution of $512^2$. Comparison of the different numerical schemes at time $t=10$. Inset: close-up of boxed area.}
\label{fig:dsl-comp512}
\end{figure}

\begin{figure}[h!]
\centering
\includegraphics[width=0.5\textwidth]{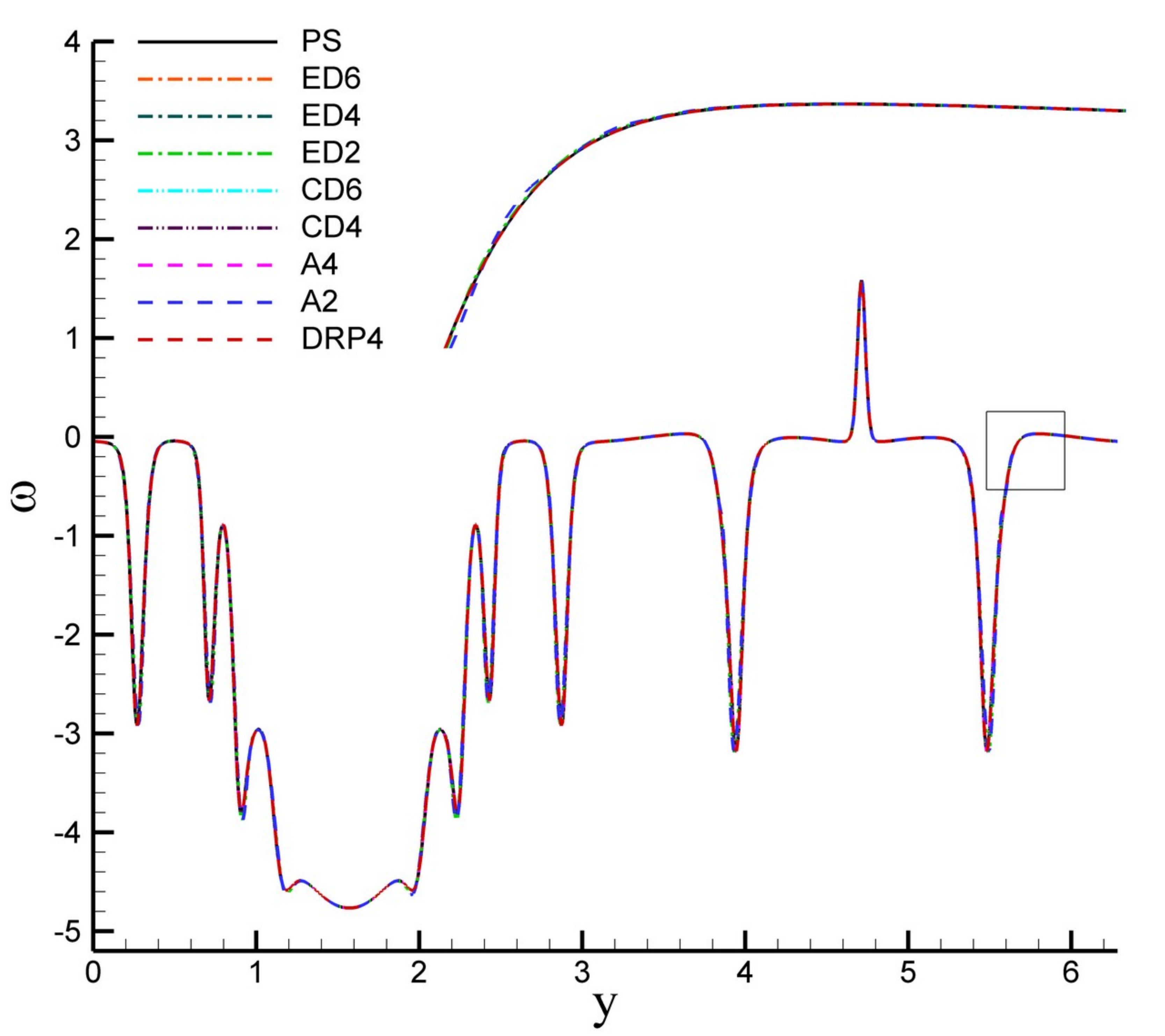}
\caption{The centerline vorticity distributions at $x=\pi$ for the double shear layer problem, plotted for a resolution of $1024^2$. Comparison of the different numerical schemes at time $t=10$. Inset: close-up of boxed area.}
\label{fig:dsl-comp1024}
\end{figure}

Vorticity contours at $t=10$ obtained using all the spatial discretization schemes introduced earlier are shown in Figs.~\ref{fig:dsl256}-\ref{fig:dsl1024} for different spatial resolutions. To make the distinctions among the schemes more clear, the centerline vorticity distributions along the $y$-axis at $t=10$ are also shown in Fig.~\ref{fig:dsl-comp256}-\ref{fig:dsl-comp1024} for different resolutions. Our comparison in these figures includes four families of fourth-order difference schemes, two second-order schemes, and two sixth-order difference schemes, as well as the pseudospectral scheme. Fig.~\ref{fig:dsl256} illustrates a comparison of vorticity fields obtained by using $256^2$ resolution for these nine different spatial discretization schemes. The corresponding comparison of centerline vorticity distributions is also provided in Fig.~\ref{fig:dsl-comp256}. As we can see from these figures, both the sixth-order compact scheme and the pseudospectral scheme yield almost the same accuracy. It can also be seen that all the high-order formulations provide much better accuracy than the second-order schemes. A truly fair comparison among all the finite difference formulations presented in this study is to look at the fourth-order discretization schemes. Results show that there is no significant difference between explicit difference and Arakawa schemes. The discrete conservation properties of the Arakawa scheme do not improve the overall accuracy of the whole solution procedure in this case. On the other hand, the compact difference and dispersion-relation-preserving schemes result in better accuracy than the Arakawa or explicit difference schemes by eliminating numerical oscillations. Increasing the resolution to $512^2$, as illustrated in Fig.~\ref{fig:dsl512} and Fig.~\ref{fig:dsl-comp512}, we show that it is possible to obtain numerical oscillation-free results only by using high-order schemes. Second-order schemes, both the explicit difference and the Arakawa scheme, still produce numerical oscillations. Finally, we perform computations by using $1024^2$ resolution and results are demonstrated in Fig.~\ref{fig:dsl1024} and Fig.~\ref{fig:dsl-comp1024} for vorticity contour fields and centerline vorticity distributions, respectively. We can see that all numerical schemes provide an accurate oscillation-free solution to the double shear layer problem with this resolution.

As a result of this study, a rule of thumb is that second-order central difference schemes, whether standard explicit difference or conservative Arakawa scheme, require about twice resolution (in each direction) to achieve the same results obtained by a high-order finite difference or pseudospectral scheme. Therefore, this study demonstrates that the numerical oscillations that occur in underresolved simulations with strong shear layers can be eliminated efficiently by using the high-order formulations with the same resolution. For example, at a relatively high Reynolds number $Re=10^4$, it can clearly be seen that a resolution of $512^2$ is not high enough to capture the correct physics without numerical oscillations using a low-order finite difference formulation. To be able to obtain all the small-scale physical layers, at least a $1024^2$ resolution needs to be used with a spatially
second-order accurate scheme for solving underlying equations. On the other hand, we demonstrated that the properly-resolved physical behavior without a Gibbs phenomenon is obtained using a resolution of $512^2$ with only a slight increase in the computational price.

In evaluating the efficiency of the different formulations, the CPU times are tabulated in Table~\ref{tab:L2-dsl}. In this table, we also present the corresponding discrete $L_2$ norms which show the average root-mean-square deviations of the vorticity field from the data obtained by well-resolved pseudospectral calculation using $1024^2$ resolution. These results demonstrate that high fidelity numerical simulations can be obtained using the high-order finite difference methods with a speed-up factor of 5 if we compare them with the pseudospectral computation. More importantly, if we compare them with their second-order counterparts, the additional cost due to high-order formulations becomes quite small. The data presented in Table~\ref{tab:L2-dsl} also demonstrates that the computational speed-up increases with increasing spatial resolution as expected. With this in mind, one of the interesting results is that the errors associated with both the second-order explicit difference (ED2) and the second-order Arakawa (A2) schemes with $1024^2$ resolution are greater than the error associated with the sixth-order compact scheme (CD6) with a much smaller $256^2$ resolution. This clearly demonstrates that the high-order difference formulations can provide an efficient way of solving flow problems even with strong shear components.

\section{Two-dimensional decaying turbulence}
\label{sec:decayR}
Two-dimensional homogenous decaying turbulence is an incompressible flow problem in which the kinetic energy decays \cite{mcwilliams1990vortices}. The problem of the free decay of two-dimensional turbulence is to determine how abundant populations of vortices freely evolve with time \cite{tabeling2002two}. In this study, we solve this problem to examine the accuracy and efficiency of the presented finite difference approximations. The results are compared with those of the pseudospectral method. According to the KBL theory for two-dimensional turbulence this system has an inertial range in the energy spectrum that is proportional to $k^{-3}$ in the inviscid limit.
\begin{figure}[h!]
\centering
\includegraphics[width=0.45\textwidth]{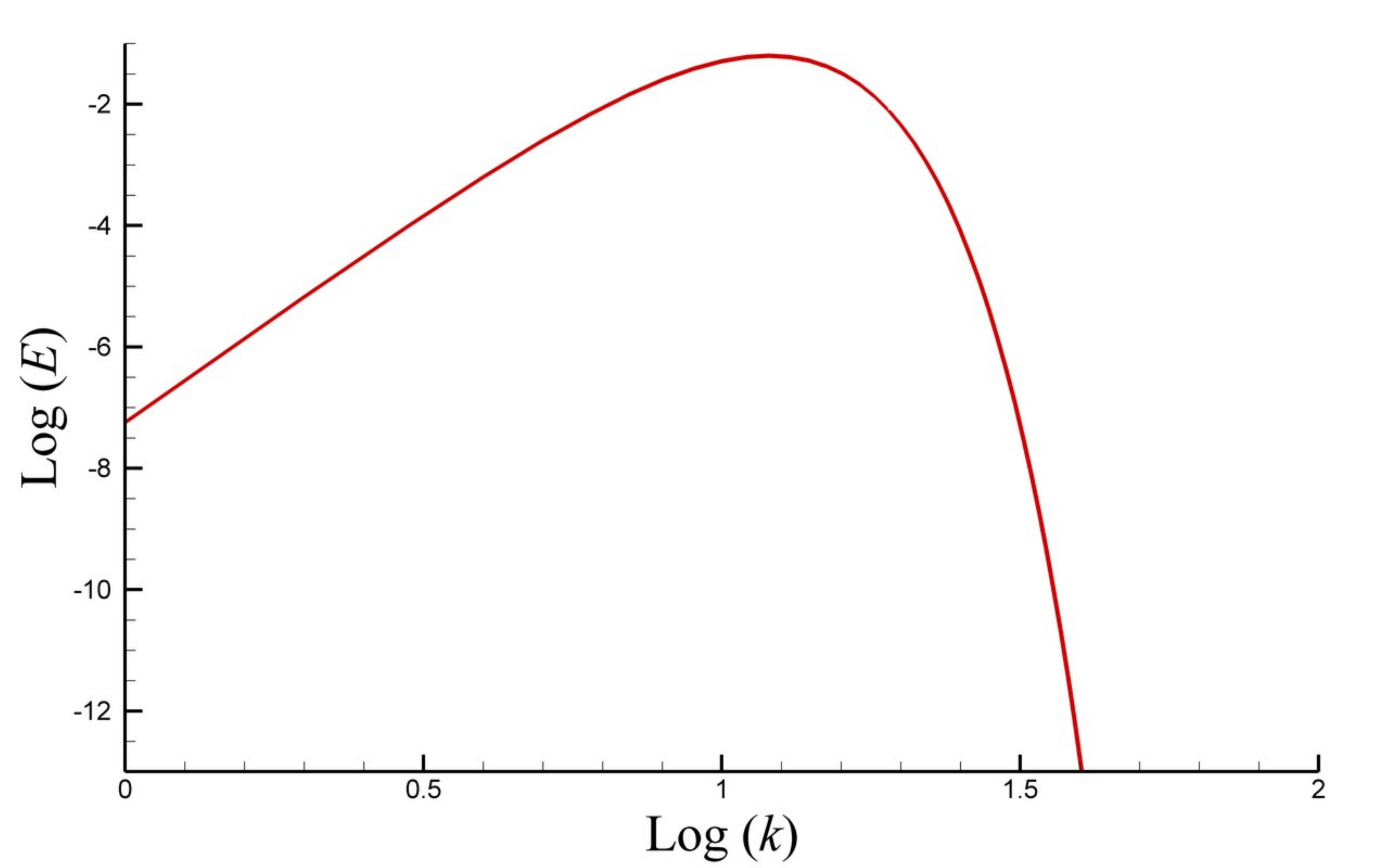}
\caption{Initial energy spectrum for the two-dimensional decaying turbulence problem.}
\label{fig:init}
\end{figure}

The computational domain is a square with sides of length $2\pi$. Periodic boundary conditions are applied. The initial energy spectrum in Fourier space is given by
\begin{equation}
E(k) = \frac{a_s}{2}\frac{1}{k_p}\left(\frac{k}{k_p}\right)^{2s+1}\mbox{exp} \bigg[-\left(s+\frac{1}{2}\right)\left(\frac{k}{k_p}\right)^{2} \bigg]
\label{eq:initE}
\end{equation}
where $k= |\textbf{k}|=\sqrt{k_x^2 + k_y^2}$. The maximum value of initial energy spectrum occurs at wavenumber $k_p$ which is assumed to be $k_p=12$ in this study. The coefficient $a_s$ normalizes the initial kinetic energy and is given by
\begin{equation}
a_s=\frac{(2s+1)^{s+1}}{2^{s}s!}
\label{eq:err}
\end{equation}
where $s$ is a shape parameter. In this study, we take $s=3$. The corresponding energy spectrum given in Eq.~(\ref{eq:initE}) is illustrated in Fig.~\ref{fig:init}. The magnitude of vorticity Fourier coefficients related to the assumed initial energy spectrum becomes
\begin{equation}
|\tilde{\omega}(\textbf{k})|=\sqrt{\frac{k}{\pi}E(k)}
\label{eq:err}
\end{equation}
The initial vorticity distribution in Fourier space is then obtained by introducing a random phase
\begin{equation}
\tilde{\omega}(\textbf{k})=\sqrt{\frac{k}{\pi}E(k)} \ e^{\textbf{\emph{i}}\zeta(\textbf{k})}
\label{eq:err}
\end{equation}
where the phase function is given by $\zeta(\textbf{k})=\xi(\textbf{k})+\eta(\textbf{k})$, where $\xi(\textbf{k})$ and $\eta(\textbf{k})$ are independent random values chosen in [$0,2\pi$] at each coordinate point in the first quadrant of the $k_x-k_y$ plane. The conjugate relations for other quadrants are
\begin{eqnarray}
\xi(-k_x,k_y)&=& -\xi(k_x,k_y) \nonumber \\
\xi(-k_x,-k_y)&=&-\xi(k_x,k_y) \nonumber \\
\xi(k_x,-k_y)&=&\xi(k_x,k_y) \nonumber \\
\eta(-k_x,k_y)&=&\eta(k_x,k_y)\nonumber \\
\eta(-k_x,-k_y)&=&-\eta(k_x,k_y)\nonumber \\
\eta(k_x,-k_y)&=&-\eta(k_x,k_y).
\label{eq:err}
\end{eqnarray}

After the randomization process described above for the phases, the initial vorticity distribution in the physical space is computed by taking the inverse FFT. However, we use a random phase generator to obtain initial flow field, it should be noted that we will use exactly same initial vorticity field when we compare the numerical schemes for the same resolution. The corresponding initial vorticity distribution is a vortex population which satisfies the divergence free condition and provides the energy spectrum given in Eq.~(\ref{eq:initE}). In the following text, we first investigate how homogeneous two-dimensional decaying turbulence varies with $Re$ when using the sixth-order compact difference scheme. Then we will focus our attention on the behavior of a selection of numerical methods for computations of two-dimensional turbulence. It should be noted here that the diffusion terms are approximated using the sixth-order compact scheme for all difference formulations in order to provide a fair comparison between them. To make the simulation independent of the errors associated with temporal discretization, a time step of $\Delta t = 2\times 10^{-4}$ is used in all simulations except when the temporal discretization schemes are evaluated by a time refinement analysis. A study with $\Delta t = 1\times 10^{-4}$, $2\times 10^{-4}$, and $4\times 10^{-4}$ also show that the predicted flow field is independent of $\Delta t$.

\subsection{Reynolds number dependence}
\label{sec:Re}
\begin{figure*}[t!]
\centering
\mbox{
\subfigure{\includegraphics[width=0.33\textwidth]{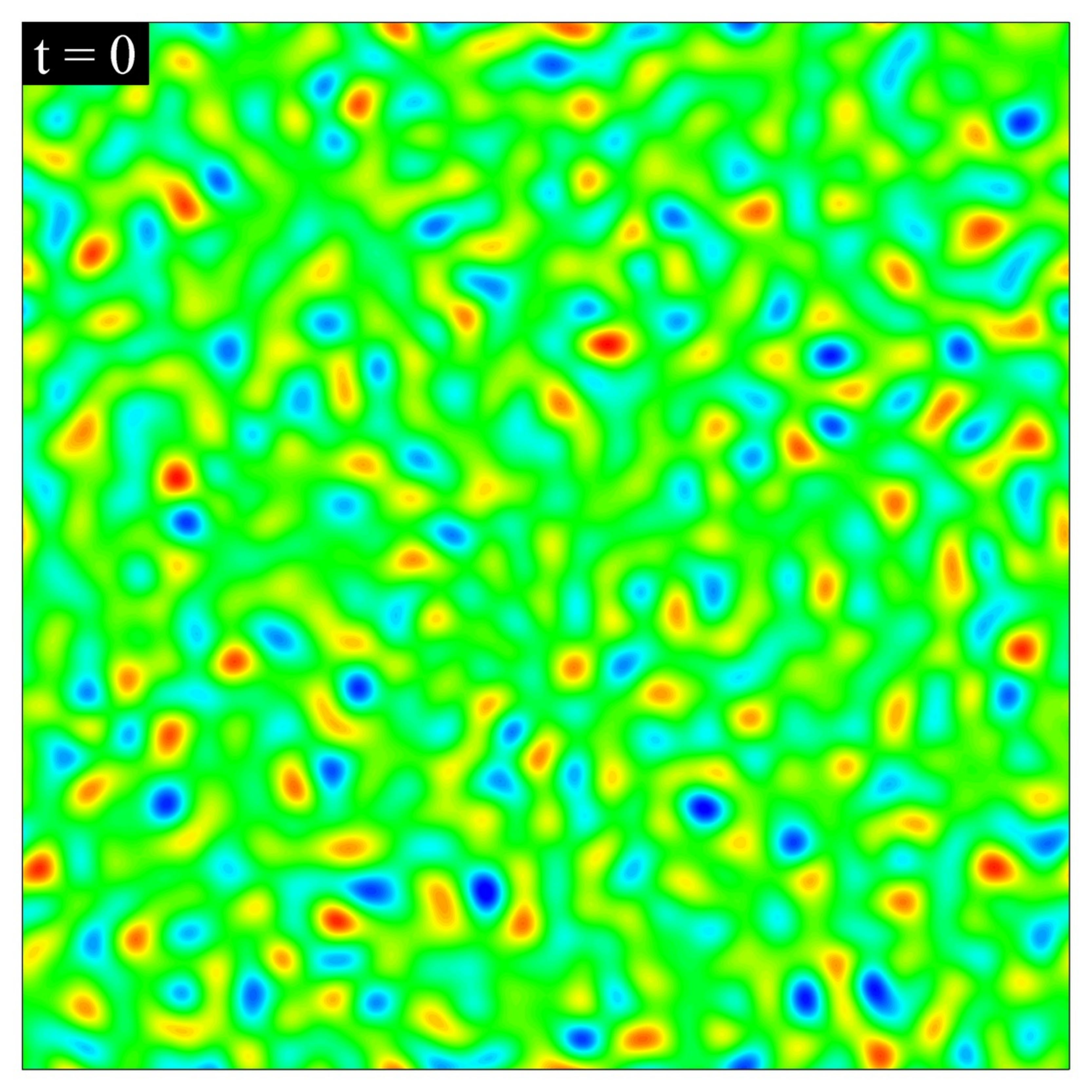}}
\subfigure{\includegraphics[width=0.33\textwidth]{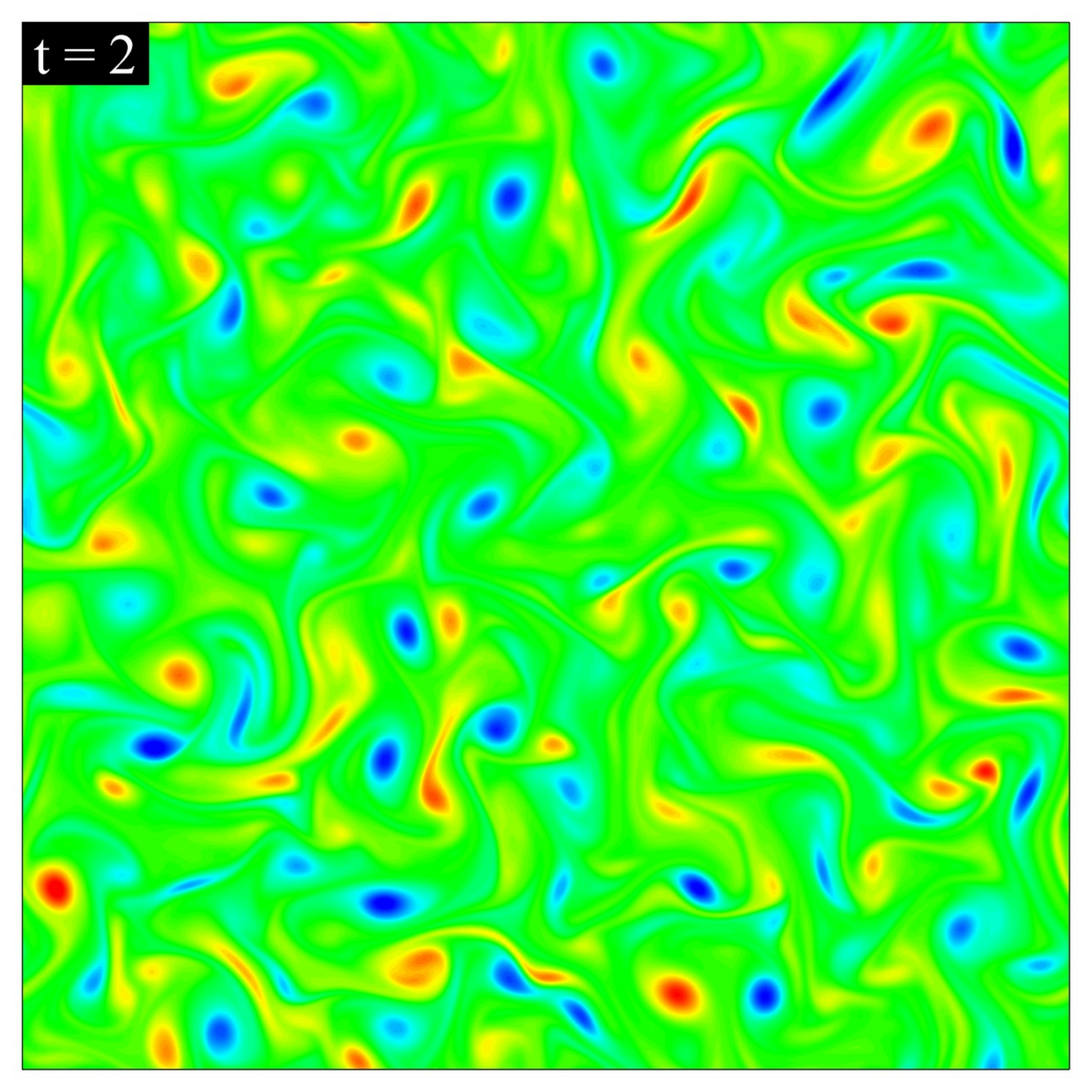}}
\subfigure{\includegraphics[width=0.33\textwidth]{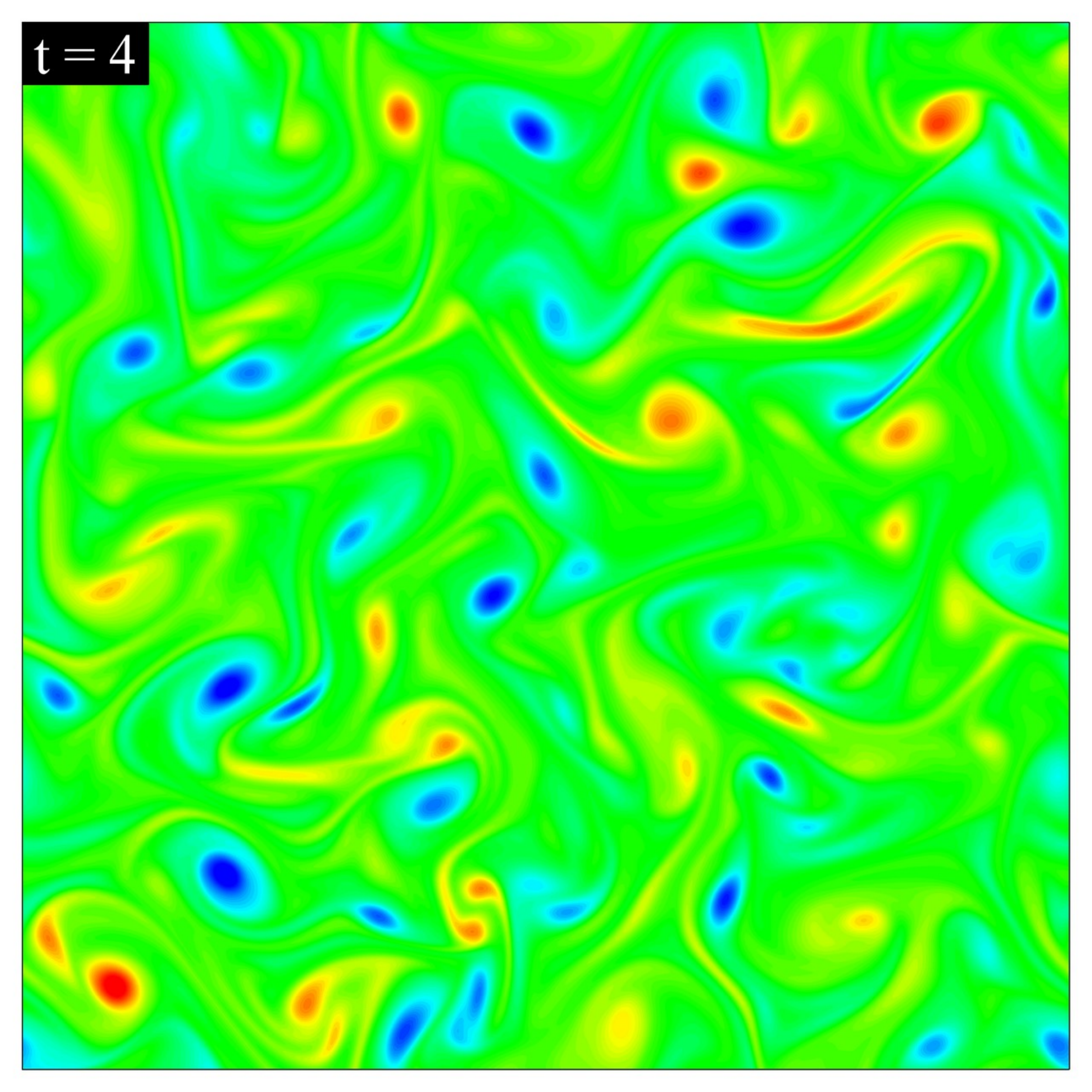}}
}
\\
\mbox{
\subfigure{\includegraphics[width=0.33\textwidth]{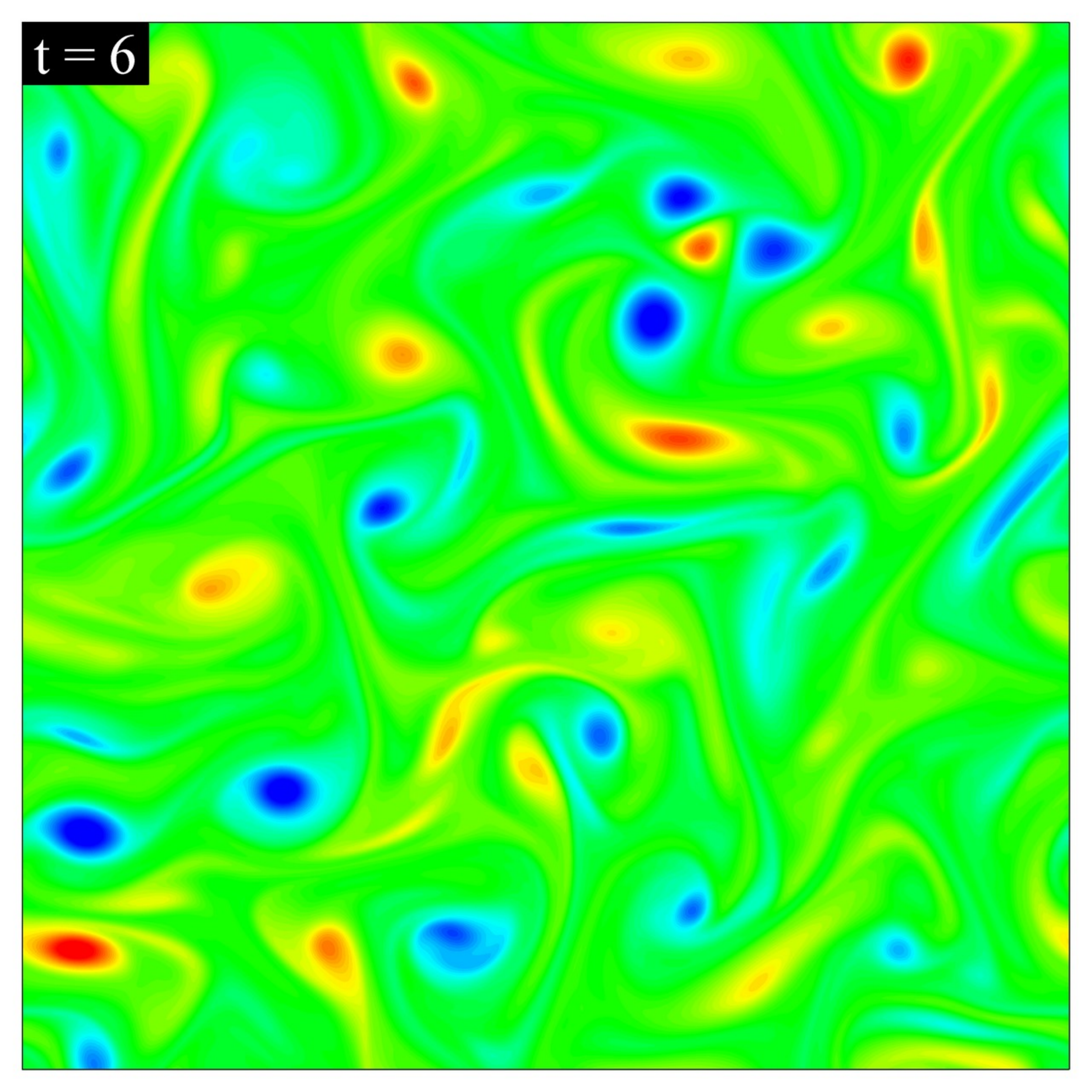}}
\subfigure{\includegraphics[width=0.33\textwidth]{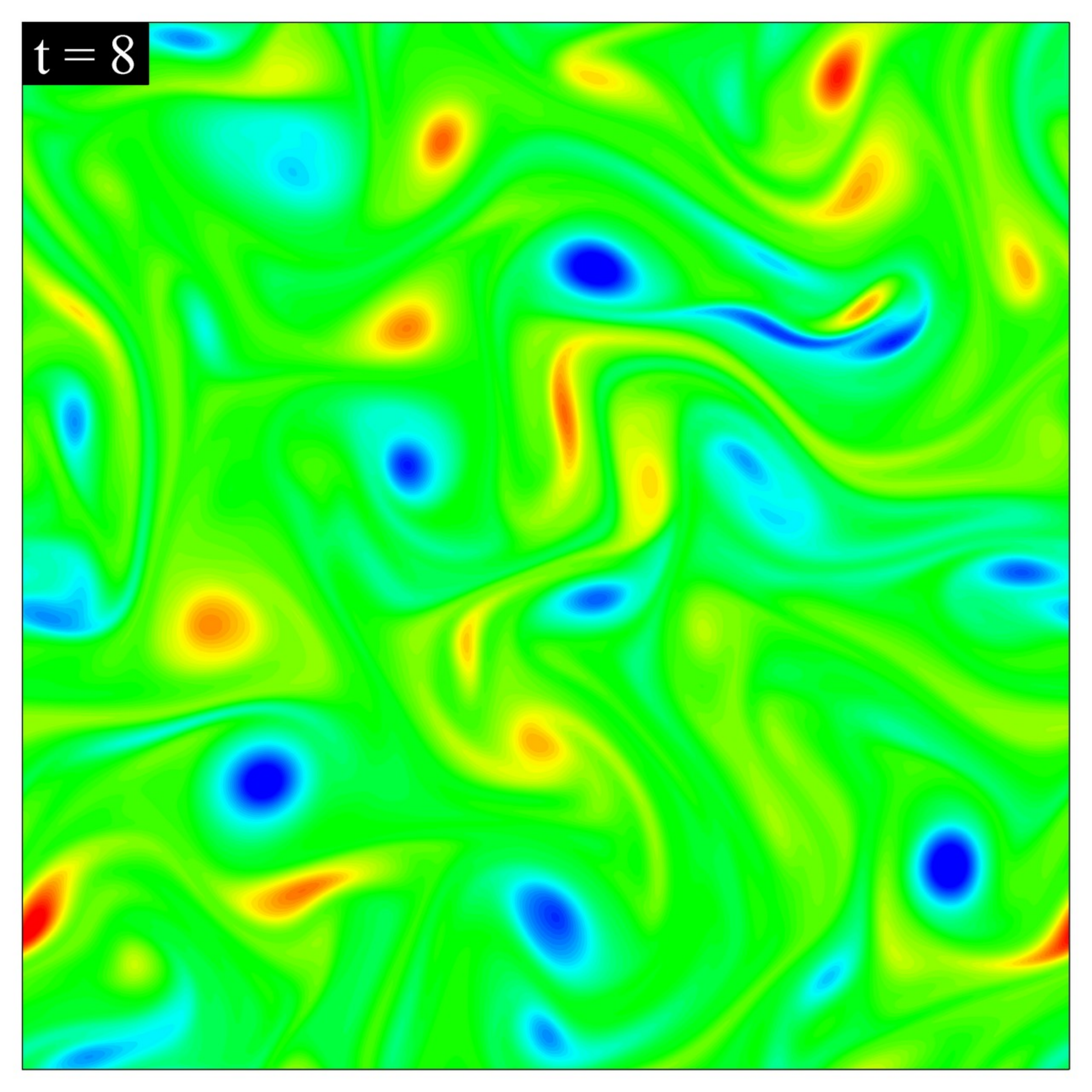}}
\subfigure{\includegraphics[width=0.33\textwidth]{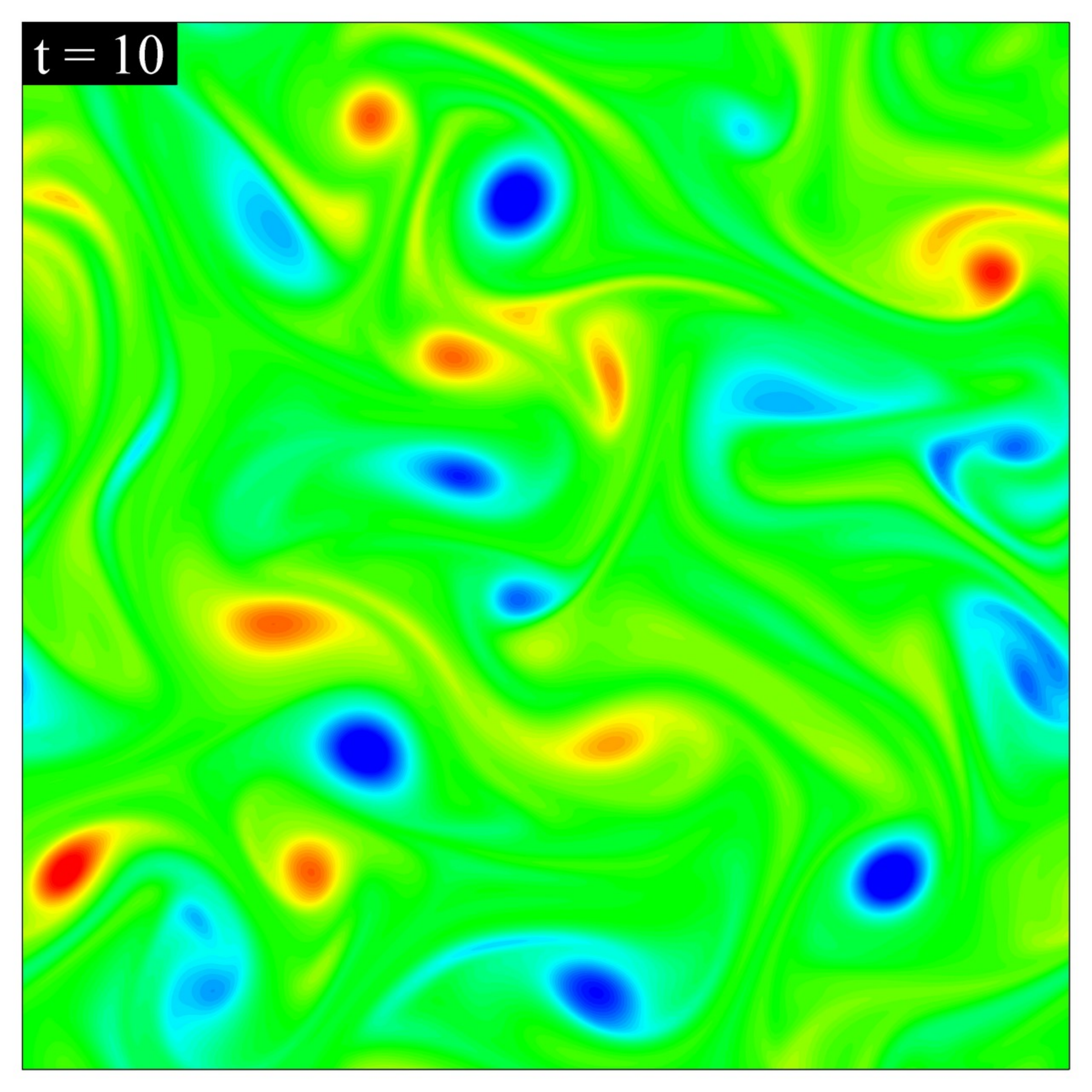}}
}
\caption{Evolution of the vorticity field in decaying turbulence for $Re=1000$. Initially randomly distributed vortices start to interact with each other and merge to form larger vortices with time.}
\label{fig:Ev1000}
\end{figure*}

Starting from a divergence free initial vorticity field having the specified energy spectrum, we solve Eq.~(\ref{eq:ge}) for various Reynolds numbers. Fig.~\ref{fig:Ev1000} shows the evolution of the vorticity field for $Re=1000$. As shown in Fig.~\ref{fig:Ev1000}, initial flow field at time $t=0$ consists of many vortices according to the spectrum given by Eq.~(\ref{eq:initE}). As we can see, a filamentation process occurs, initially randomly distributed vortices start to interact with each other and become bigger vortices with time by a vortex merging mechanism. As also illustrated in Fig.~\ref{fig:Ev1000}, as time evolves, the exterior strain field from each vortex deforms the vorticity field of the other one so that the vorticity fields wrap around each other.

In order to examine the characteristics of two-dimensional decaying turbulence simulations and their dependencies on $Re$, we first define two statistical measures. one is the energy spectrum in wave space, and the other is the structure function in physical space. The energy spectrum is defined as
\begin{equation}
\hat{E}(\textbf{k},t)=\frac{1}{2}k^2|\tilde{\psi}(\textbf{k},t)|^2
\label{eq:esp}
\end{equation}
and the angle averaged energy spectrum is
\begin{equation}
E(k,t)= \sum_{k\leq|\acute{\small\textbf{k}}|\leq k+1} \hat{E}(\acute{\textbf{k}},t).
\label{eq:Aesp}
\end{equation}
It is known from the KBL theory that the energy spectrum in the inertial range approaches the classical $k^{-3}$ scaling in the limit of infinite Reynolds number. On the other hand, the statistics of two-dimensional turbulent flow can be further investigated considering powers of vorticity differences in the physical space. A commonly used statistical quantity in two-dimensional turbulence is the second-order vorticity structure function which is defined as
\begin{equation}
\langle \delta \omega (r)^{2}\rangle = \langle |\omega(\textbf{x} + \textbf{r}) - \omega(\textbf{x})|^{2} \rangle
\label{eq:str}
\end{equation}
with $r=|\textbf{r}|$ being the spatial separation. Statistical properties of decaying two-dimensional turbulence are investigated here by numerical simulations for different Reynolds numbers.

\begin{figure*}[t!]
\centering
\mbox{
\subfigure{\includegraphics[width=0.33\textwidth]{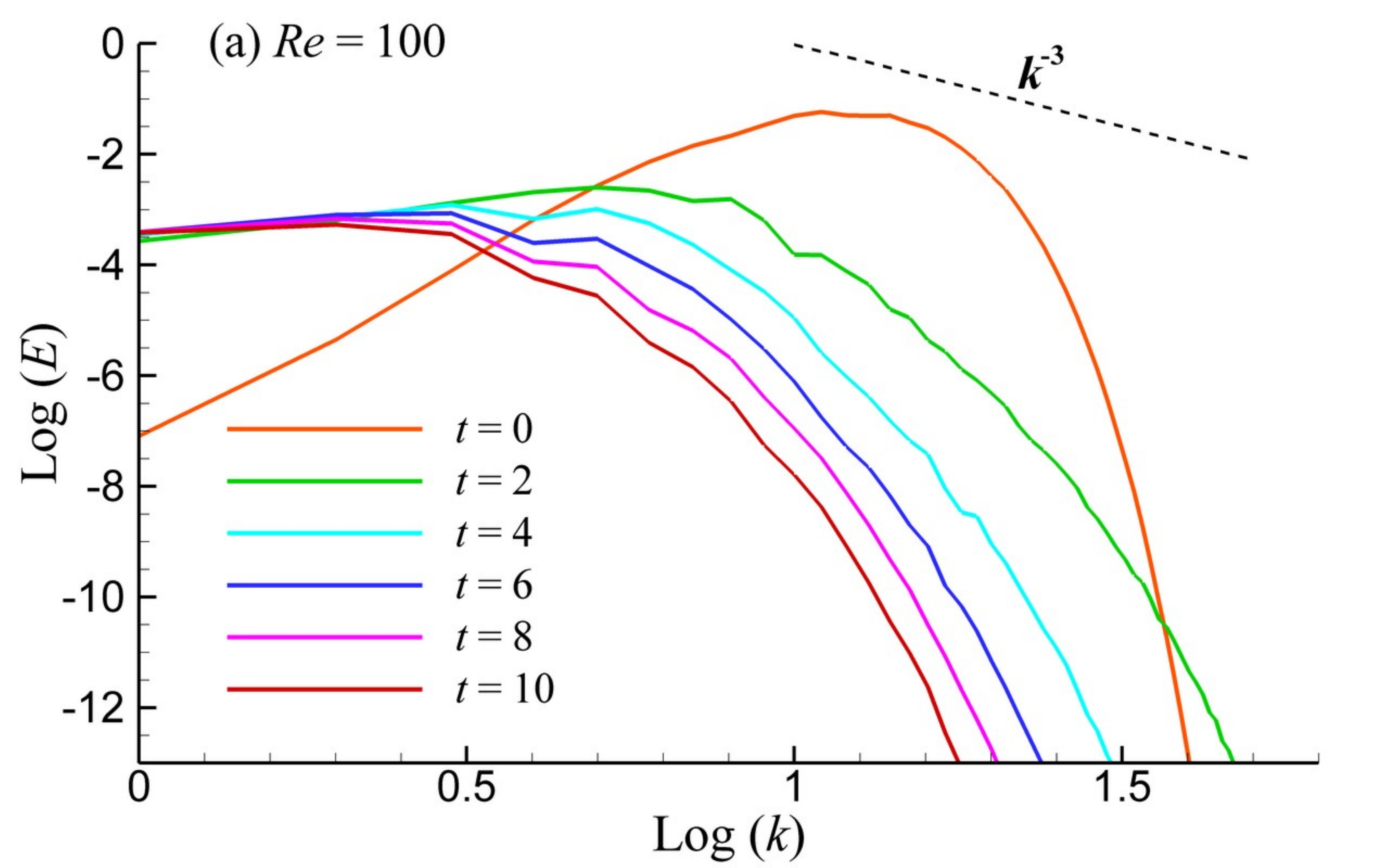}}
\subfigure{\includegraphics[width=0.33\textwidth]{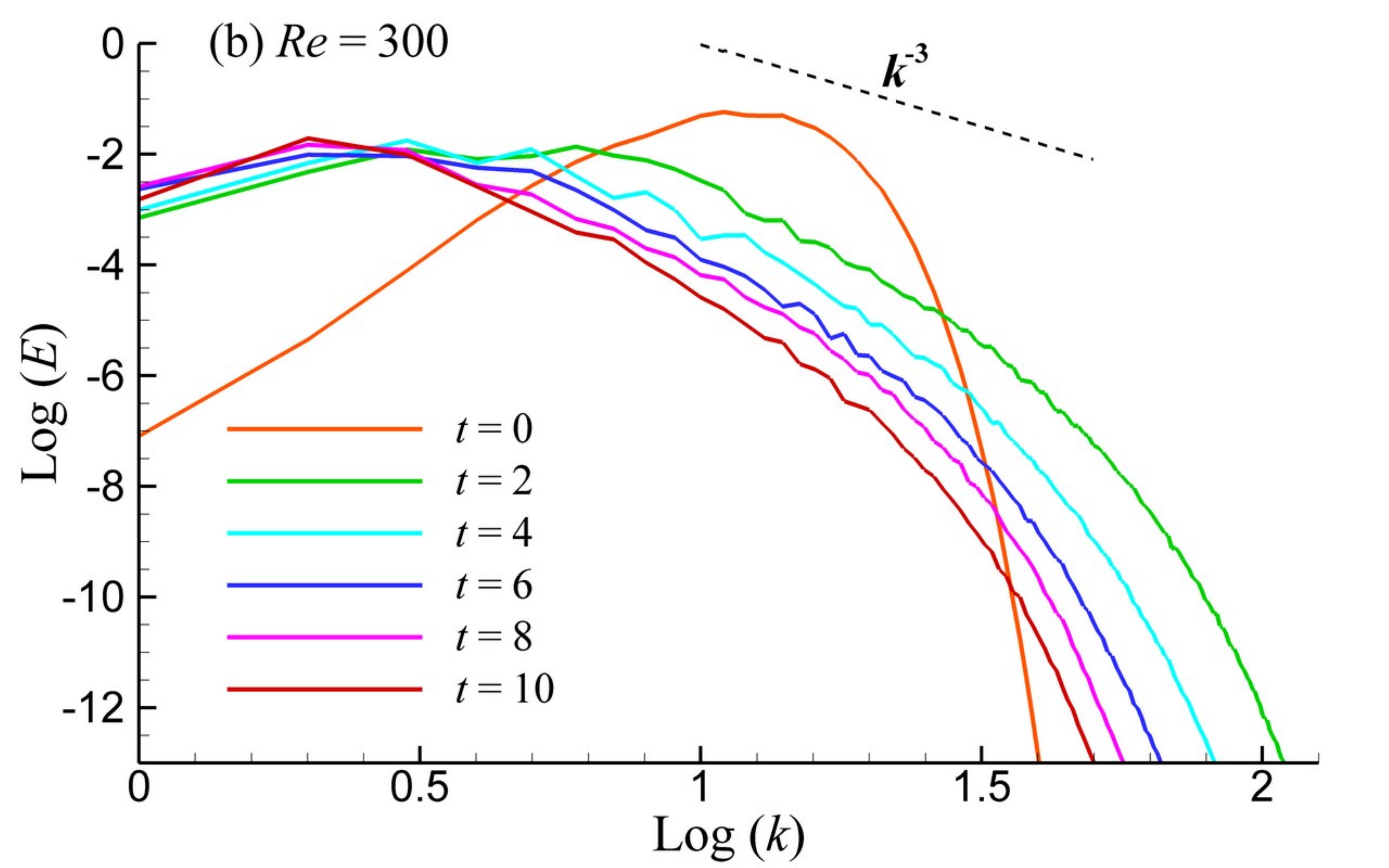}}
\subfigure{\includegraphics[width=0.33\textwidth]{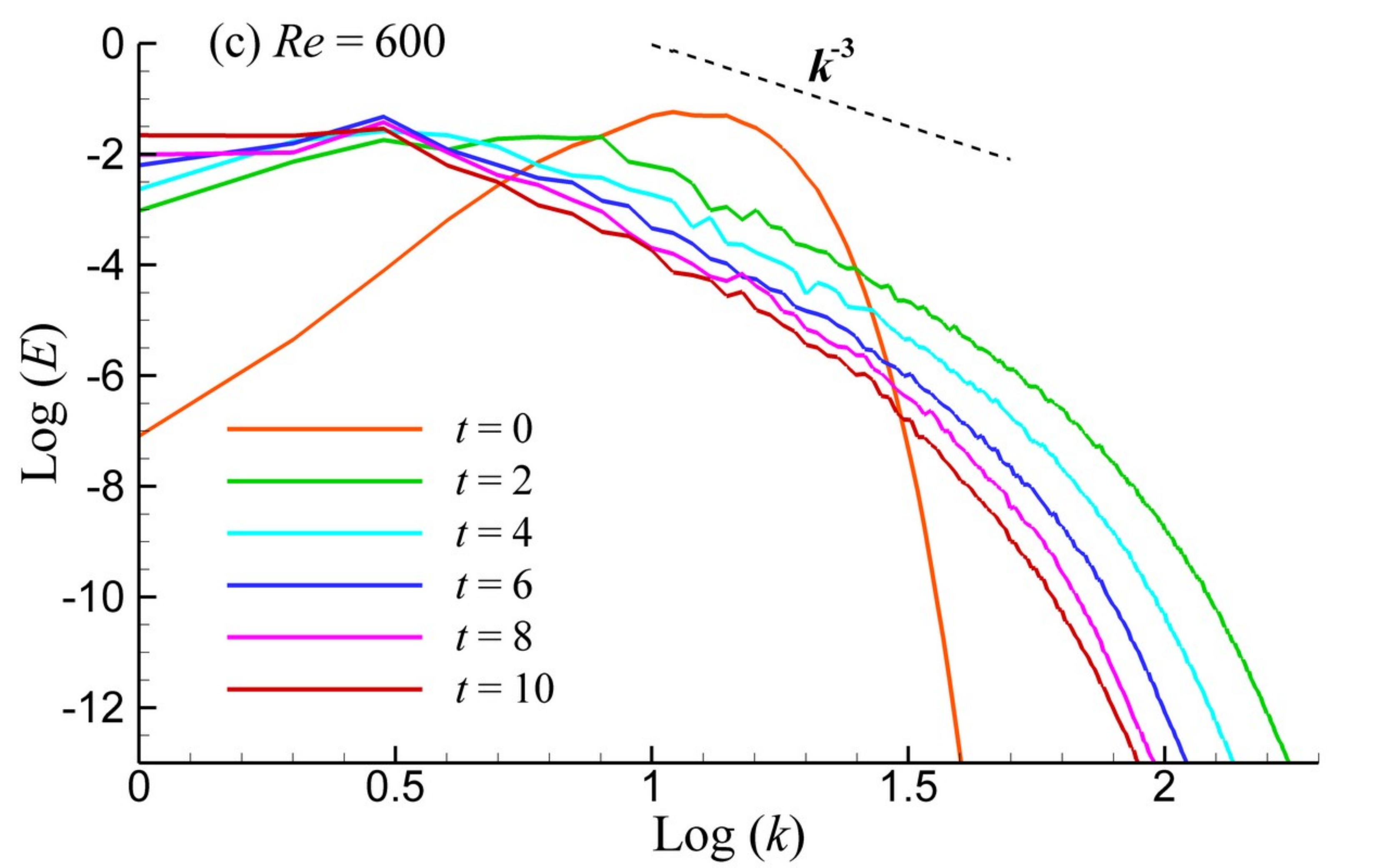}}
}
\mbox{
\subfigure{\includegraphics[width=0.33\textwidth]{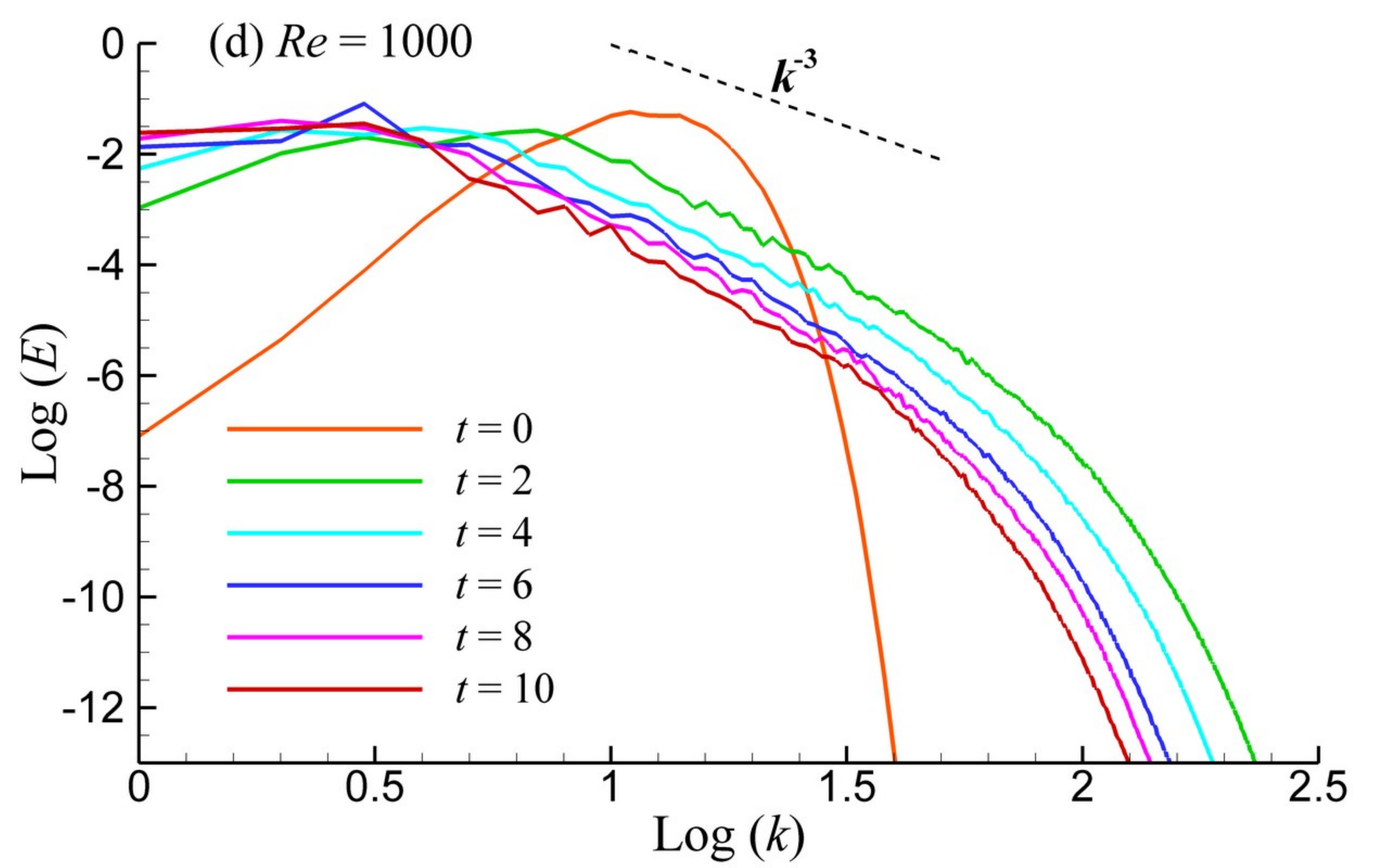}}
\subfigure{\includegraphics[width=0.33\textwidth]{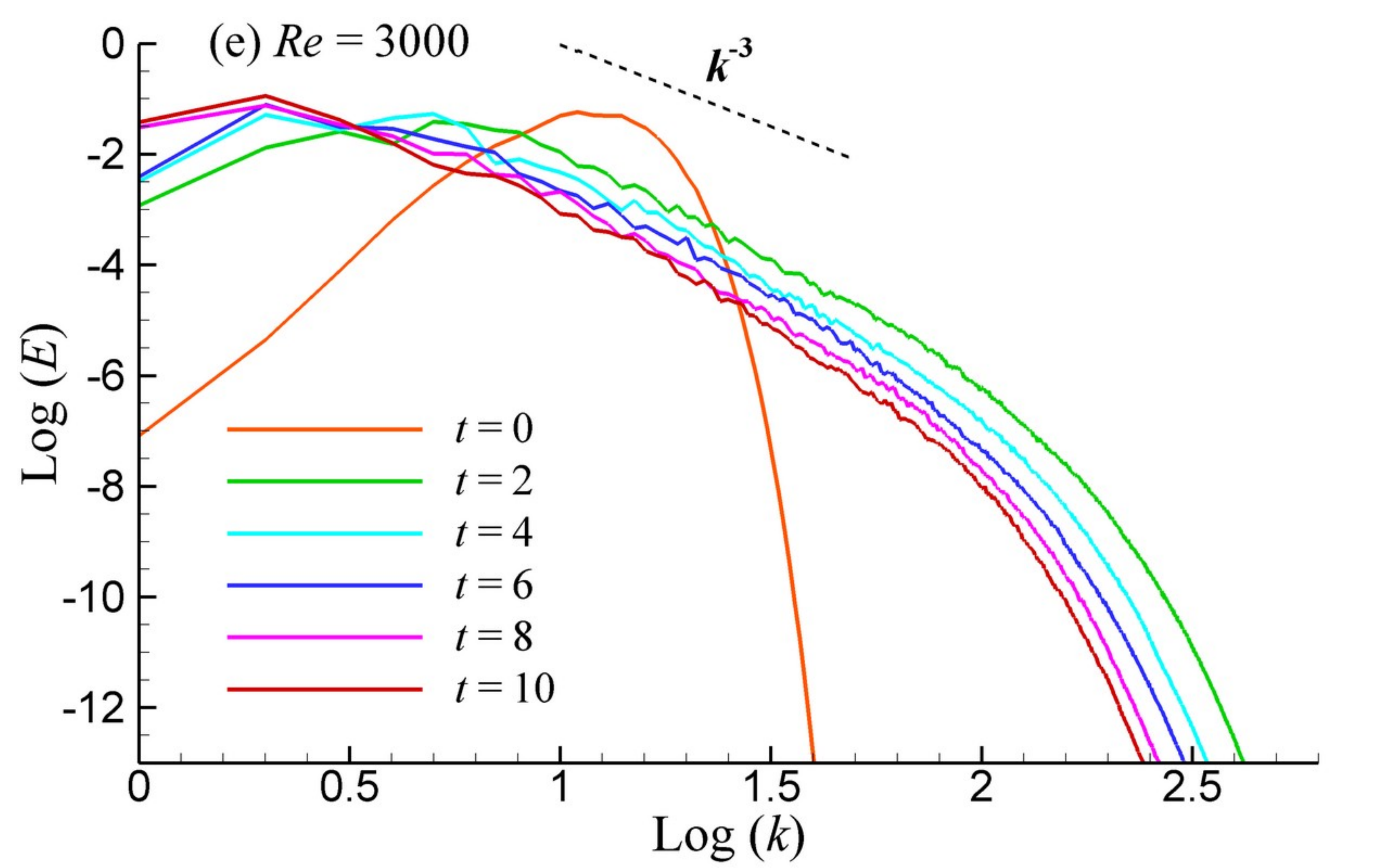}}
\subfigure{\includegraphics[width=0.33\textwidth]{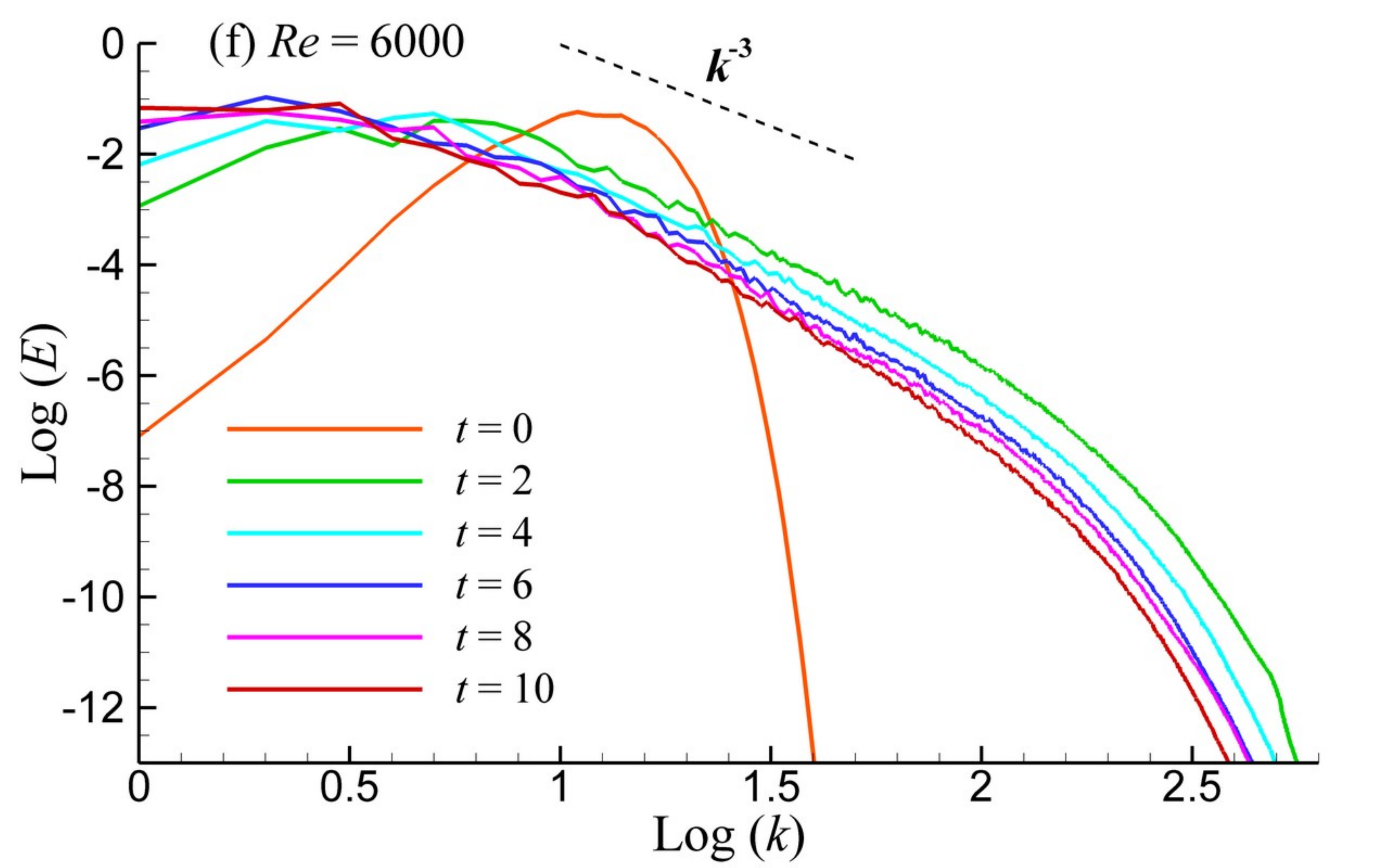}}
}
\caption{Evolution of energy spectra in decaying turbulence for (a) $Re$=100, (b) $Re$=300, (c) $Re$=600, (d) $Re$=1000, (e) $Re$=3000,  and (f) $Re$=6000. The energy spectrum in the inertial range flattens towards the classical $k^{-3}$ scaling limit as $Re$ increases, in agreement with the KBL theory of two-dimensional turbulence.}
\label{fig:ES}
\end{figure*}

\begin{figure*}[t!]
\centering
\mbox{
\subfigure{\includegraphics[width=0.33\textwidth]{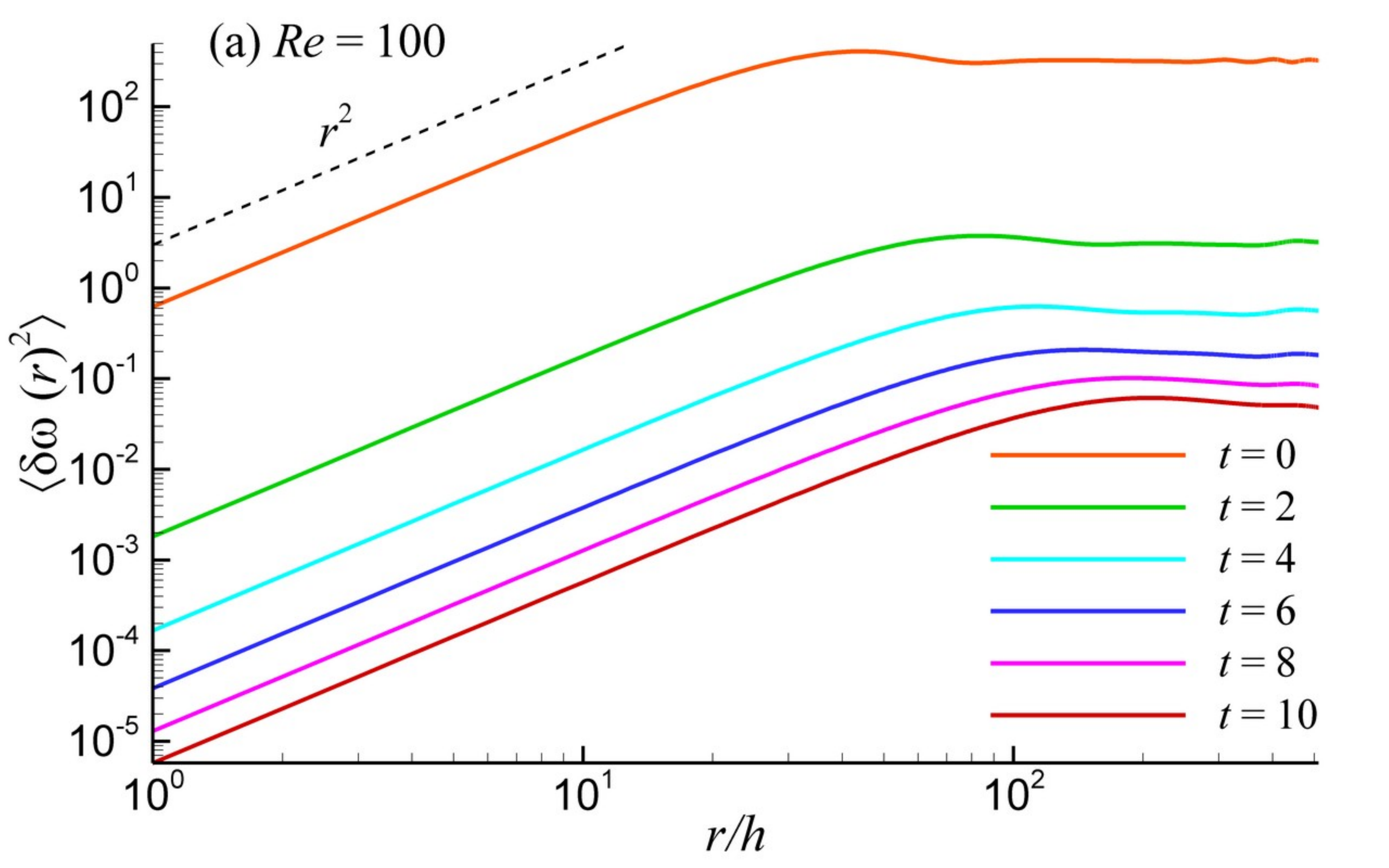}}
\subfigure{\includegraphics[width=0.33\textwidth]{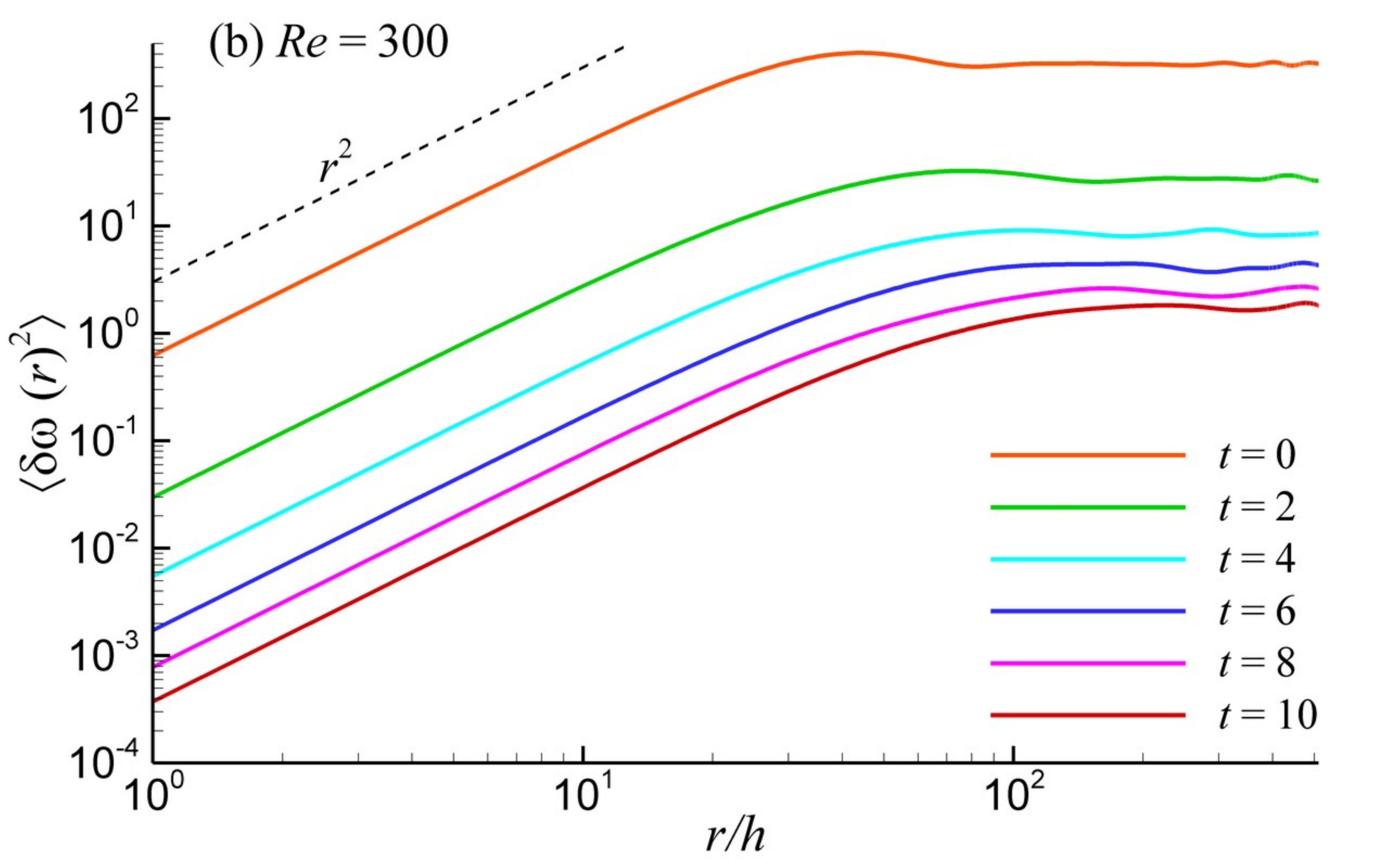}}
\subfigure{\includegraphics[width=0.33\textwidth]{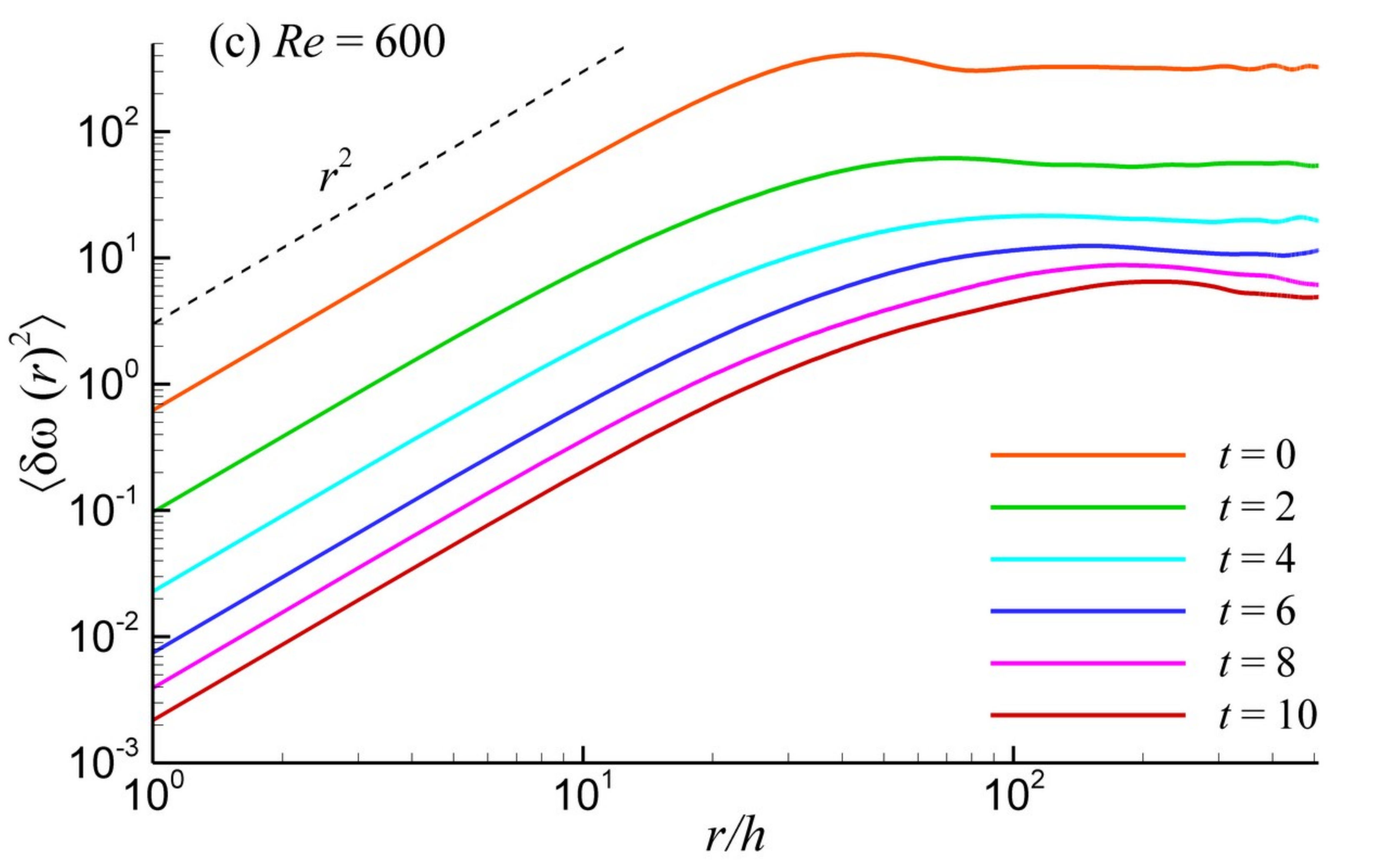}}
}
\mbox{
\subfigure{\includegraphics[width=0.33\textwidth]{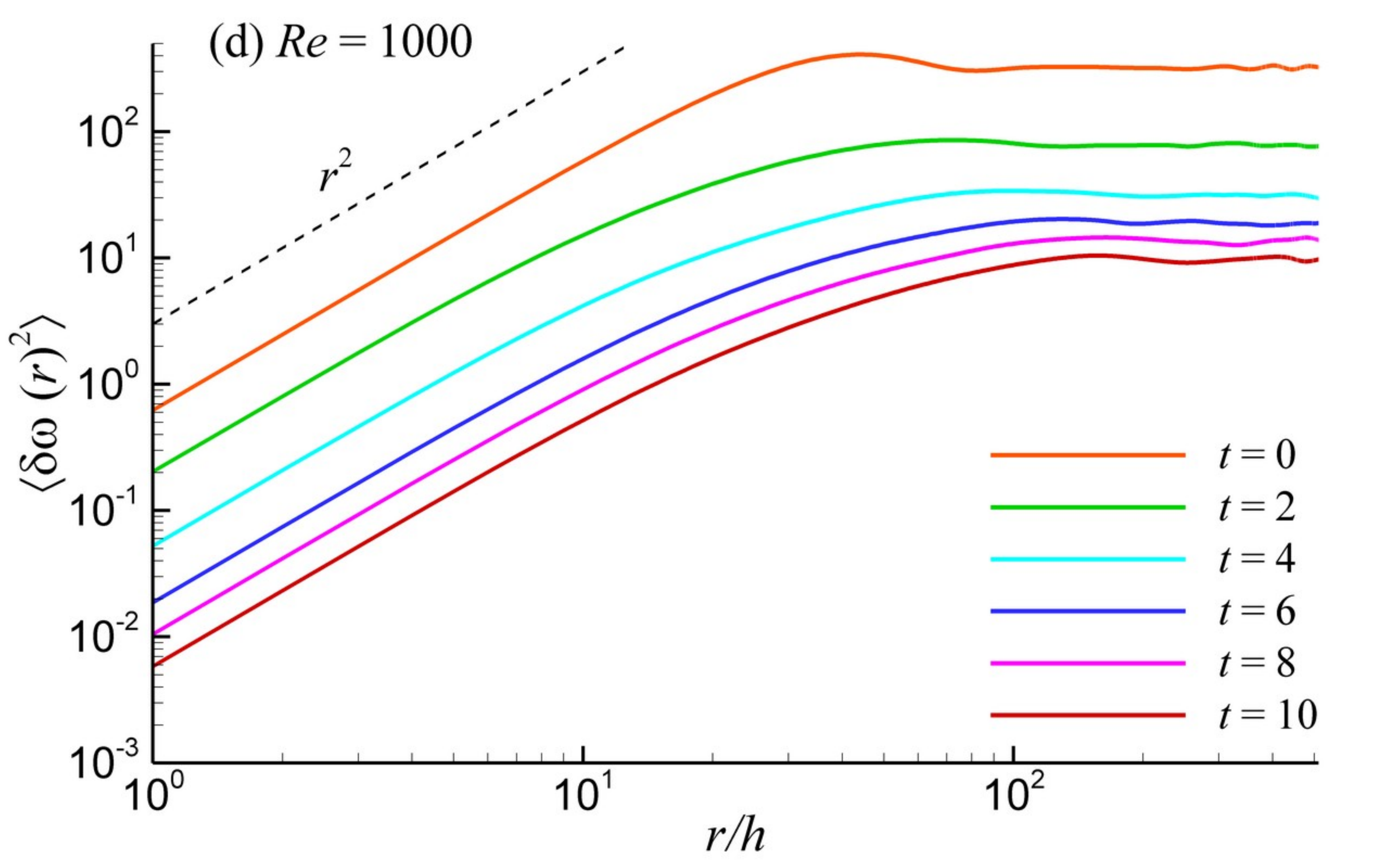}}
\subfigure{\includegraphics[width=0.33\textwidth]{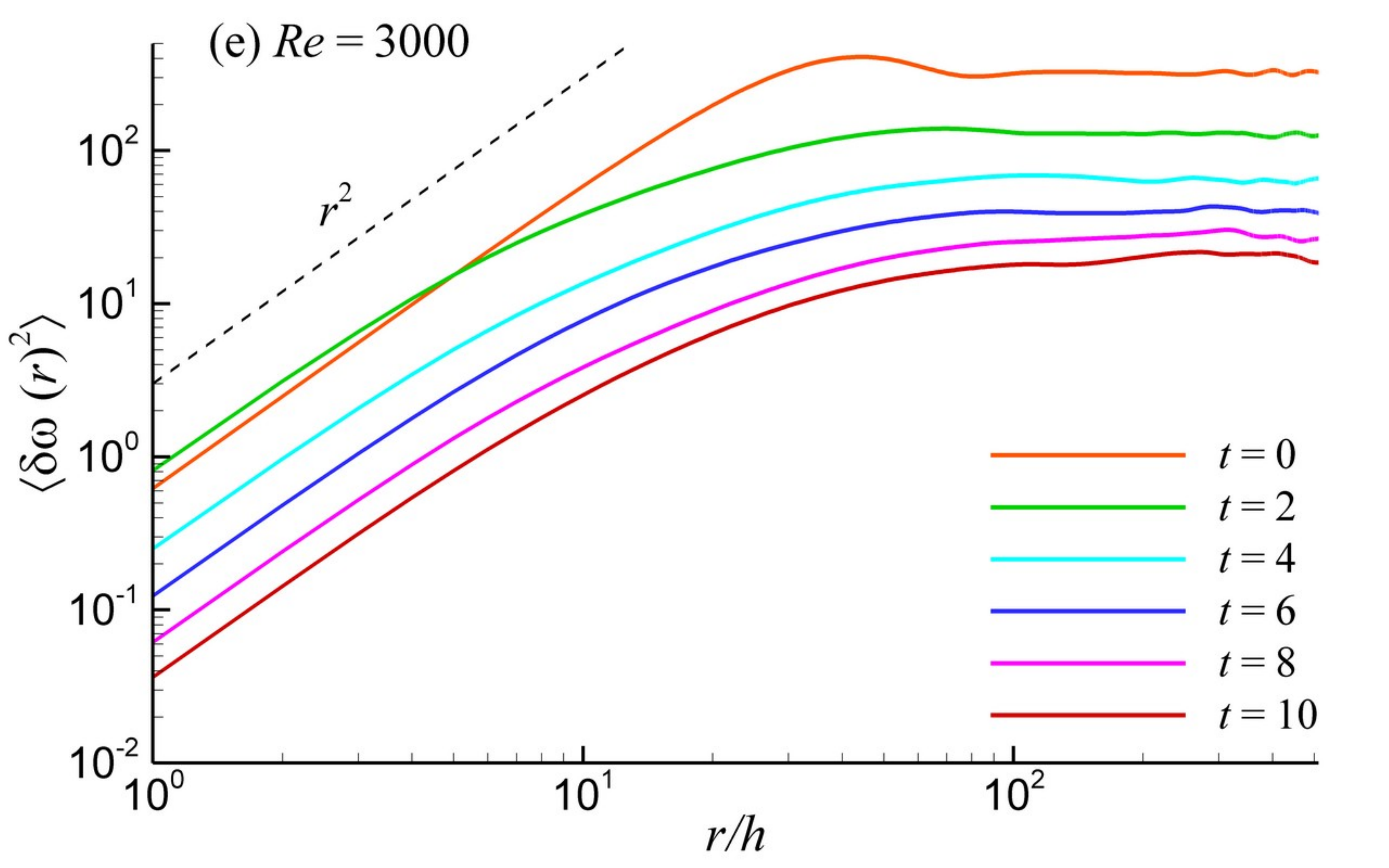}}
\subfigure{\includegraphics[width=0.33\textwidth]{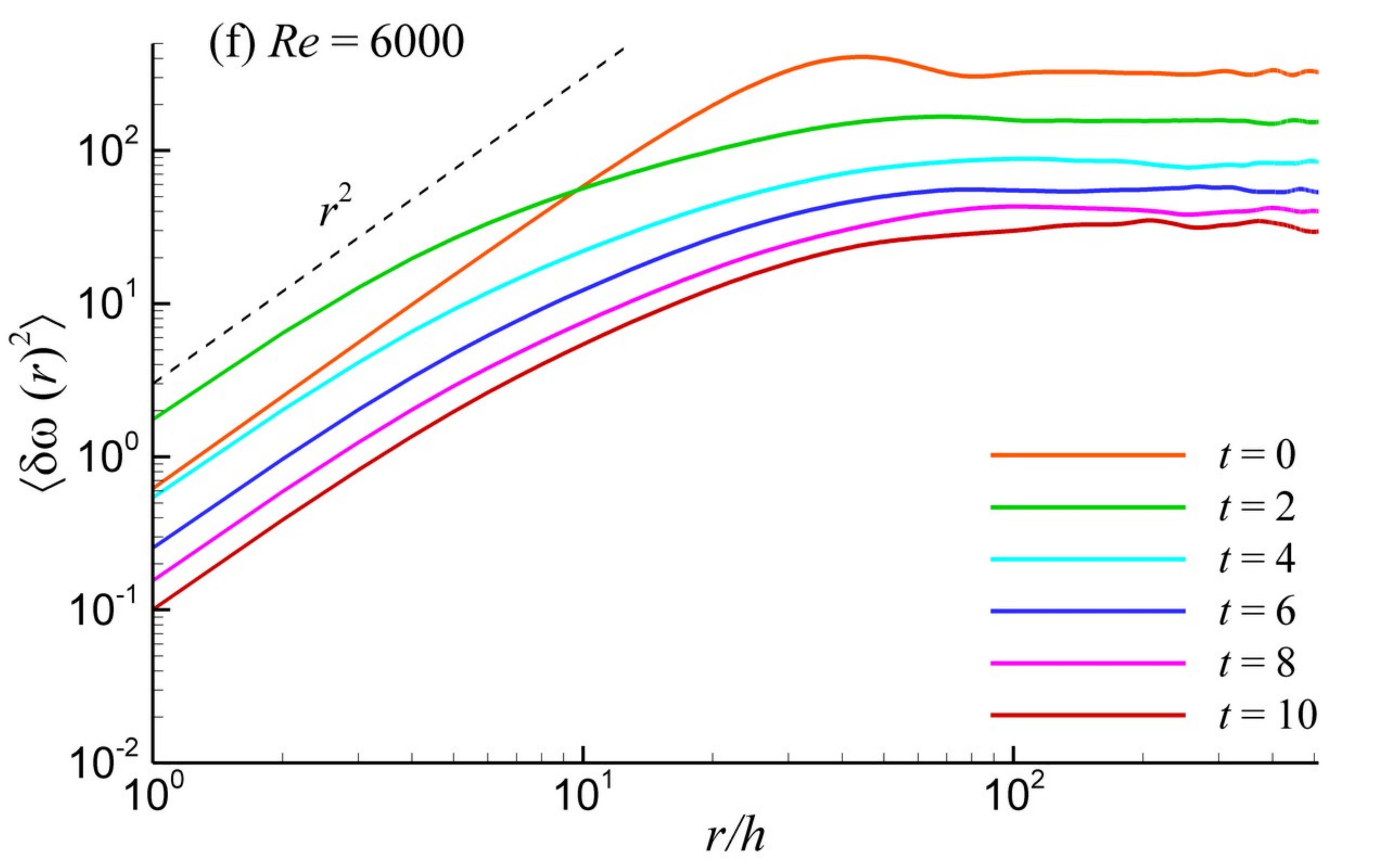}}
}
\caption{Evolution of the second-order vorticity structure functions in decaying turbulence for (a) $Re$=100, (b) $Re$=300, (c) $Re$=600, (d) $Re$=1000, (e) $Re$=3000,  and (f) $Re$=6000. The line $r^2$ is included in each subfigure.}
\label{fig:STR}
\end{figure*}

\begin{figure*}[t!]
\centering
\vspace{0.4in}
\mbox{
\subfigure{\includegraphics[width=0.33\textwidth]{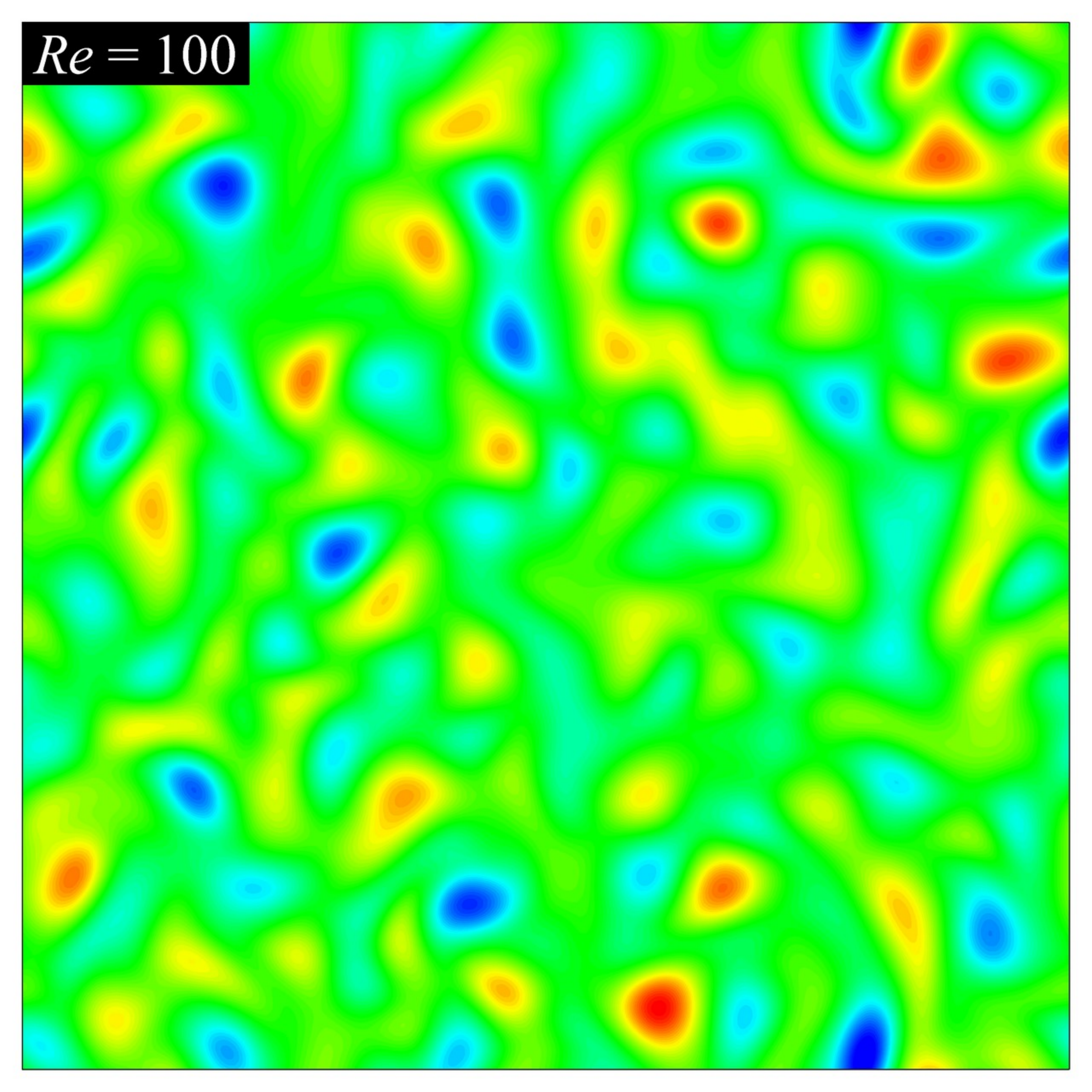}}
\subfigure{\includegraphics[width=0.33\textwidth]{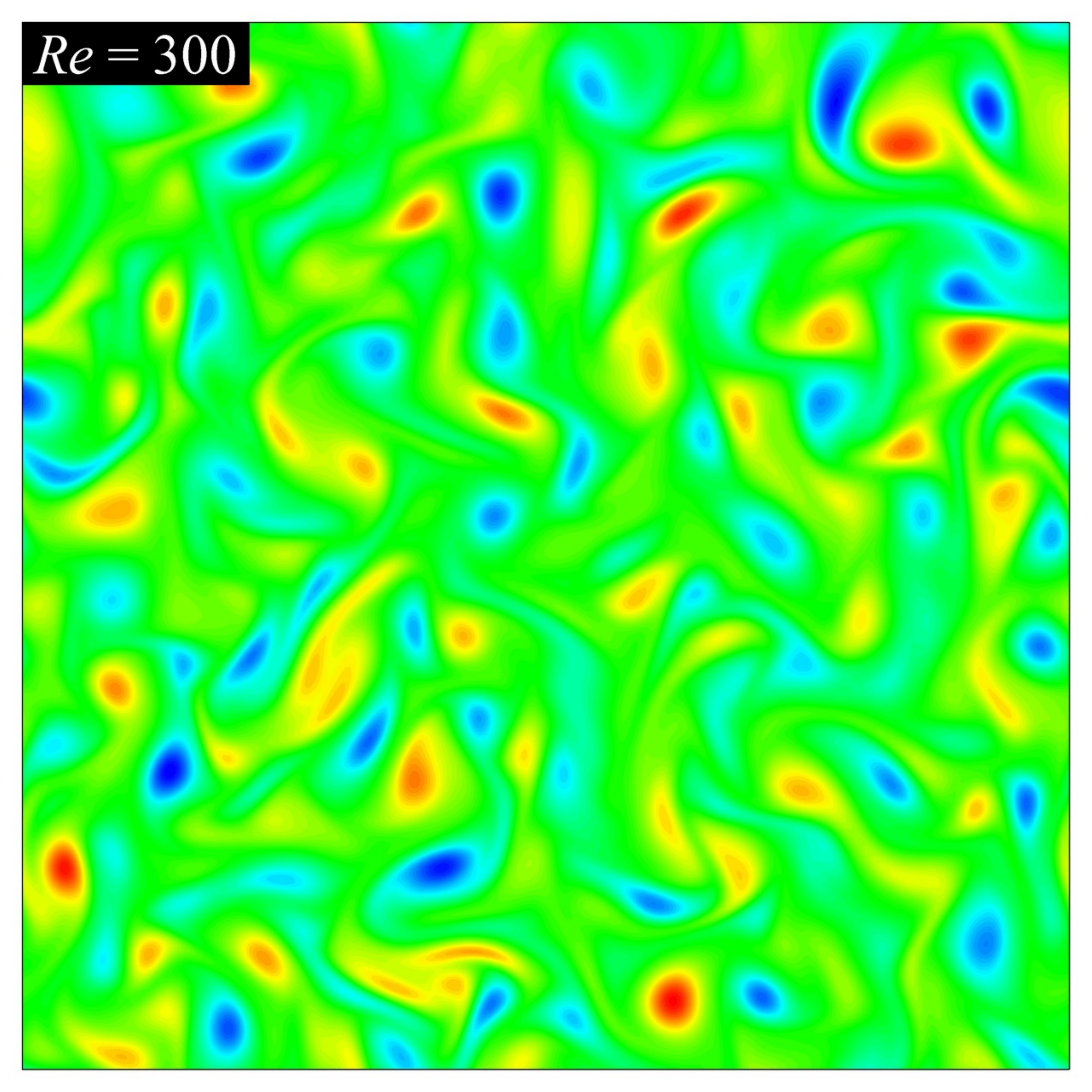}}
\subfigure{\includegraphics[width=0.33\textwidth]{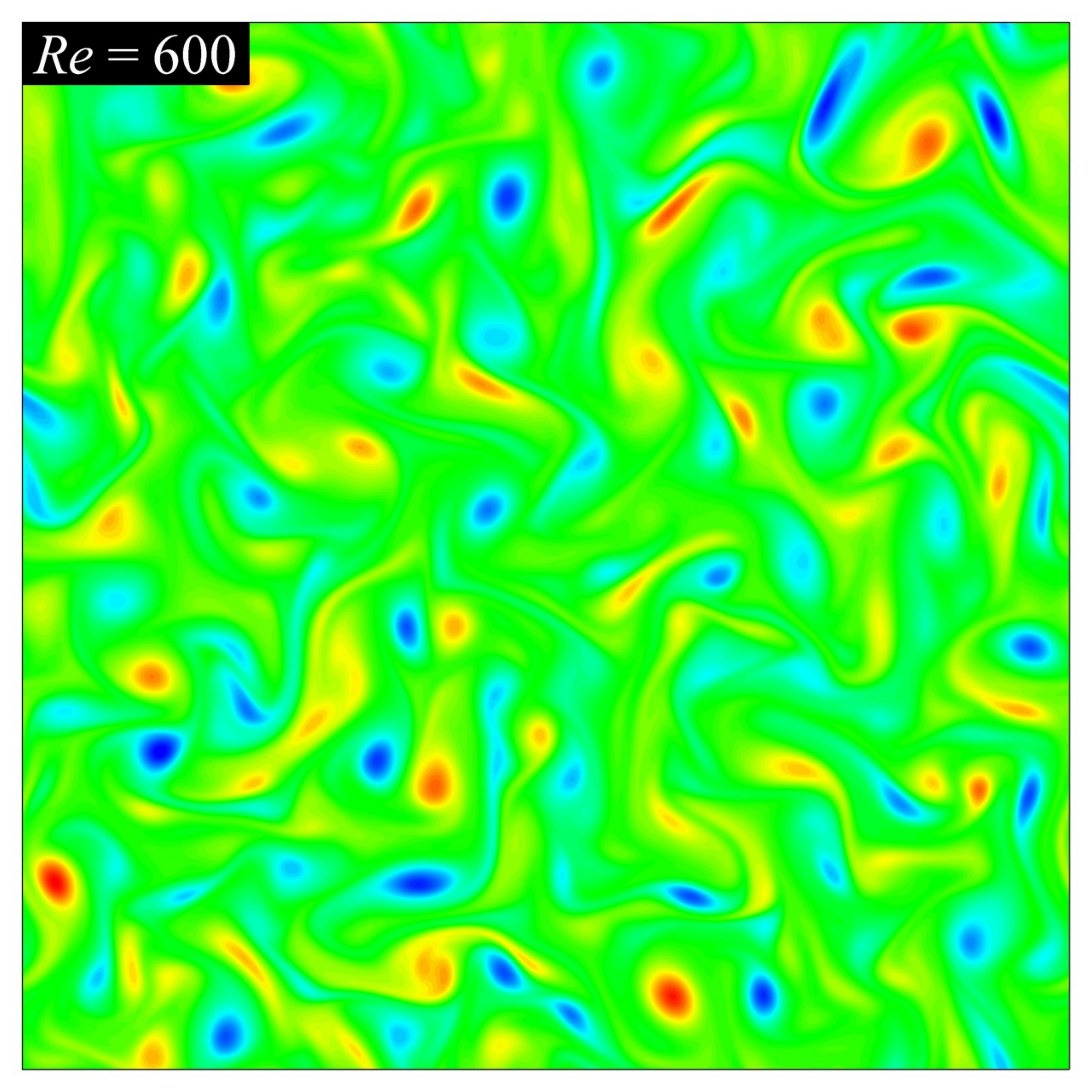}}
}
\\
\mbox{
\subfigure{\includegraphics[width=0.33\textwidth]{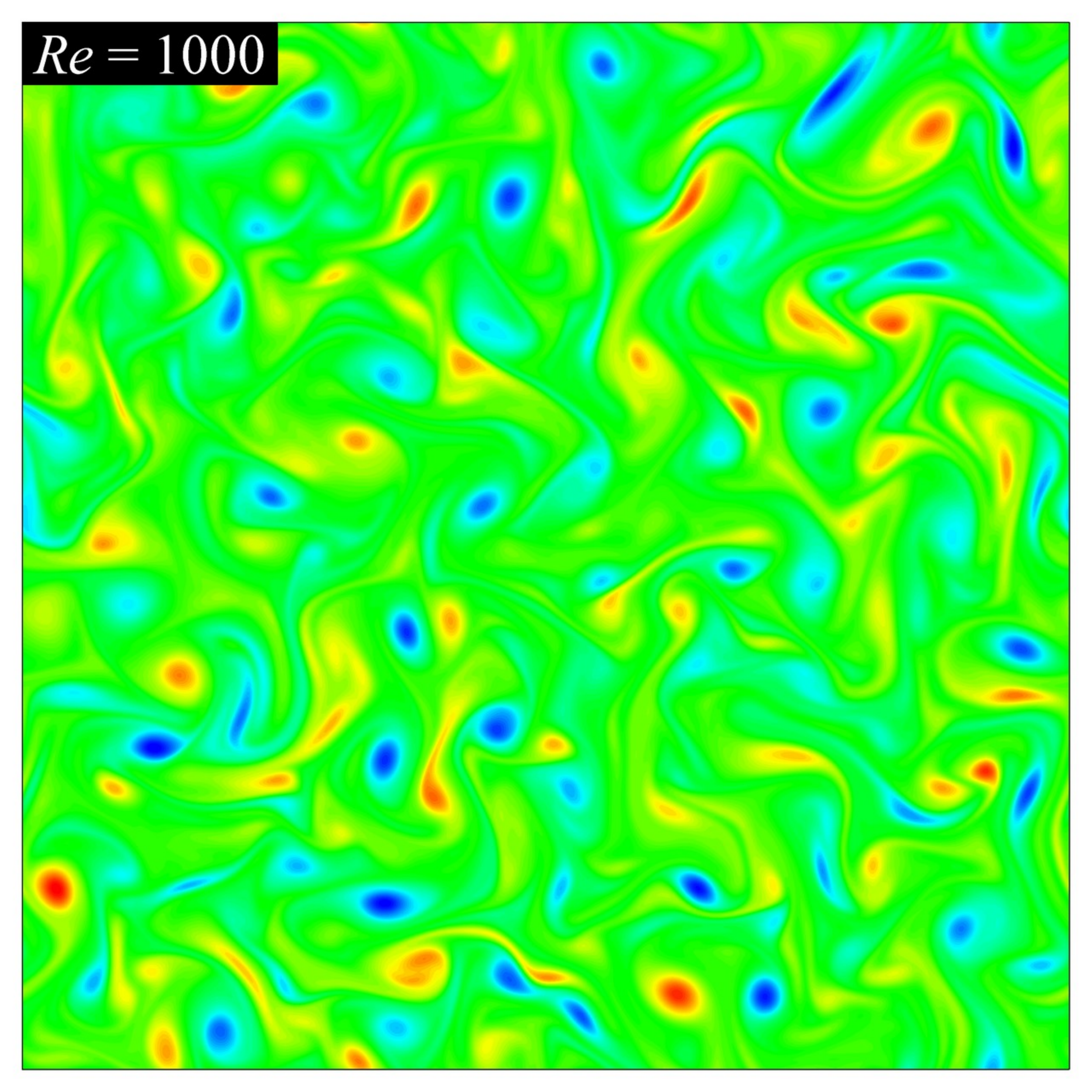}}
\subfigure{\includegraphics[width=0.33\textwidth]{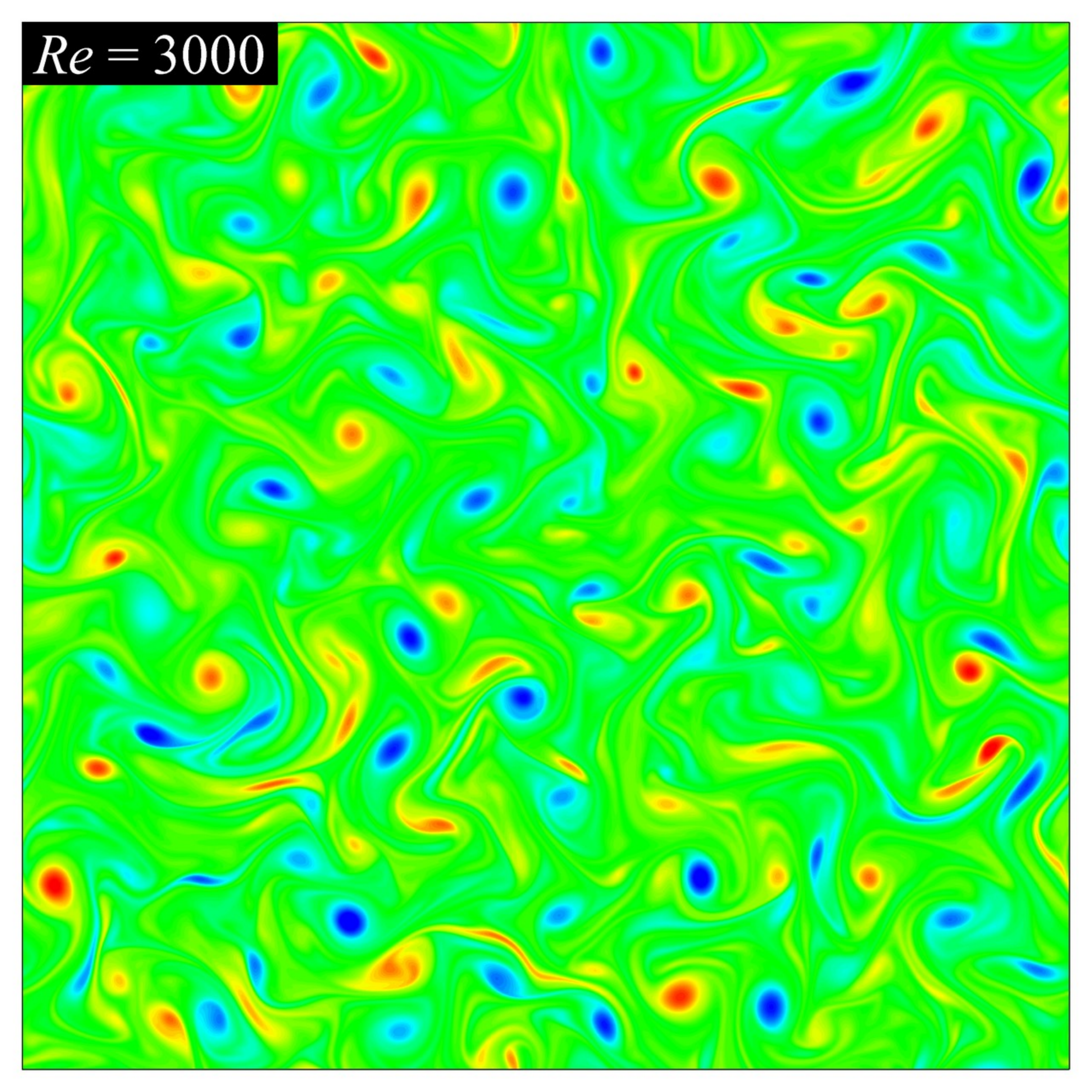}}
\subfigure{\includegraphics[width=0.33\textwidth]{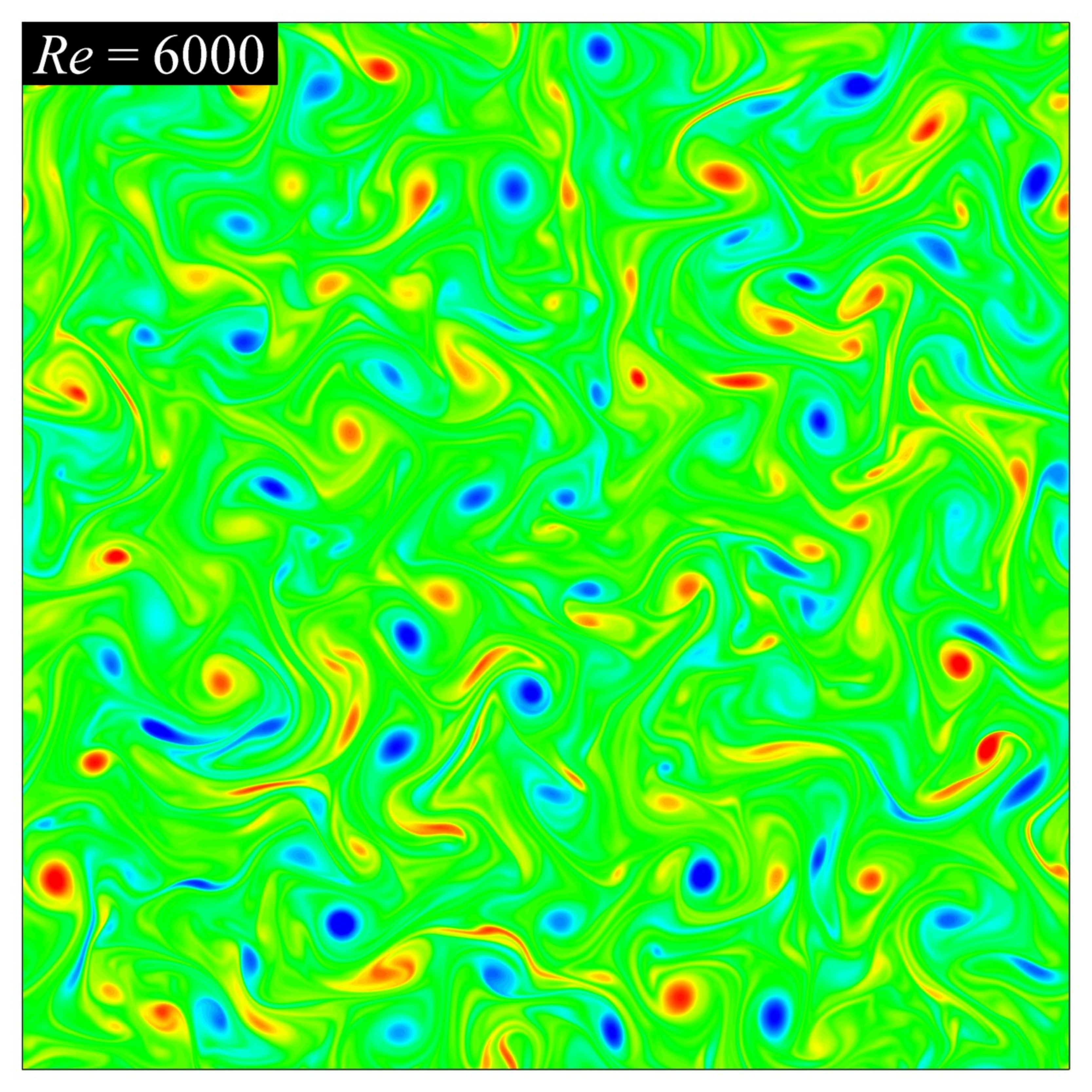}}
}
\caption{Vorticity contour plots at time $t=2$ for various Reynolds numbers. Stronger filamentation processes occur with increasing Reynolds number.}
\label{fig:Re-depend}
\end{figure*}

\begin{figure}[h!]
\centering
\includegraphics[width=0.5\textwidth]{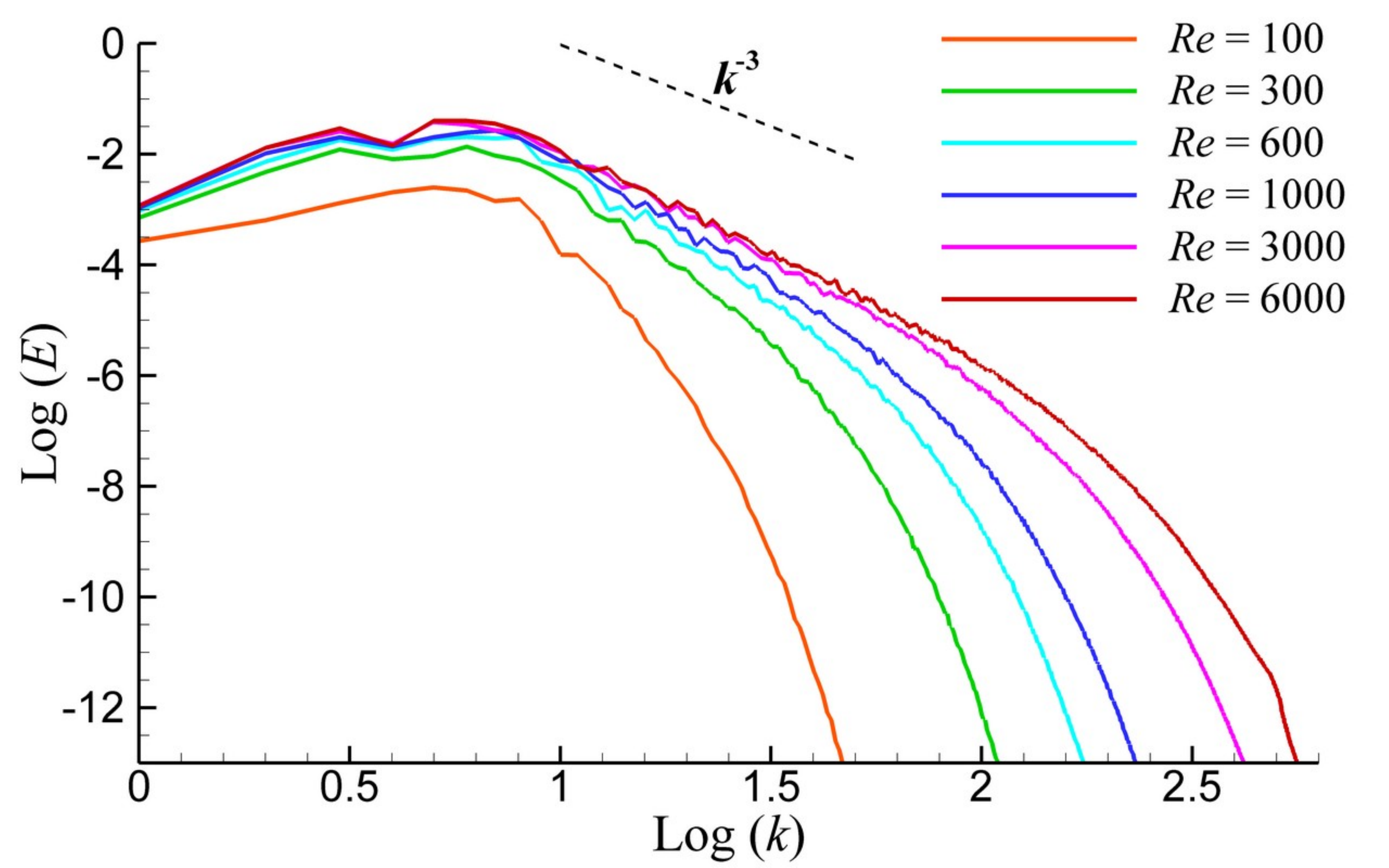}
\caption{Reynolds number dependence of the energy spectrum in wave space. In the inertial range, the angle averaged energy spectrum, defined by Eq.~(\ref{eq:Aesp}), asymptotically reaches $k^{-3}$ scaling with increasing Reynolds number.}
\label{fig:Re-spec}
\end{figure}

\begin{figure}[h!]
\centering
\includegraphics[width=0.5\textwidth]{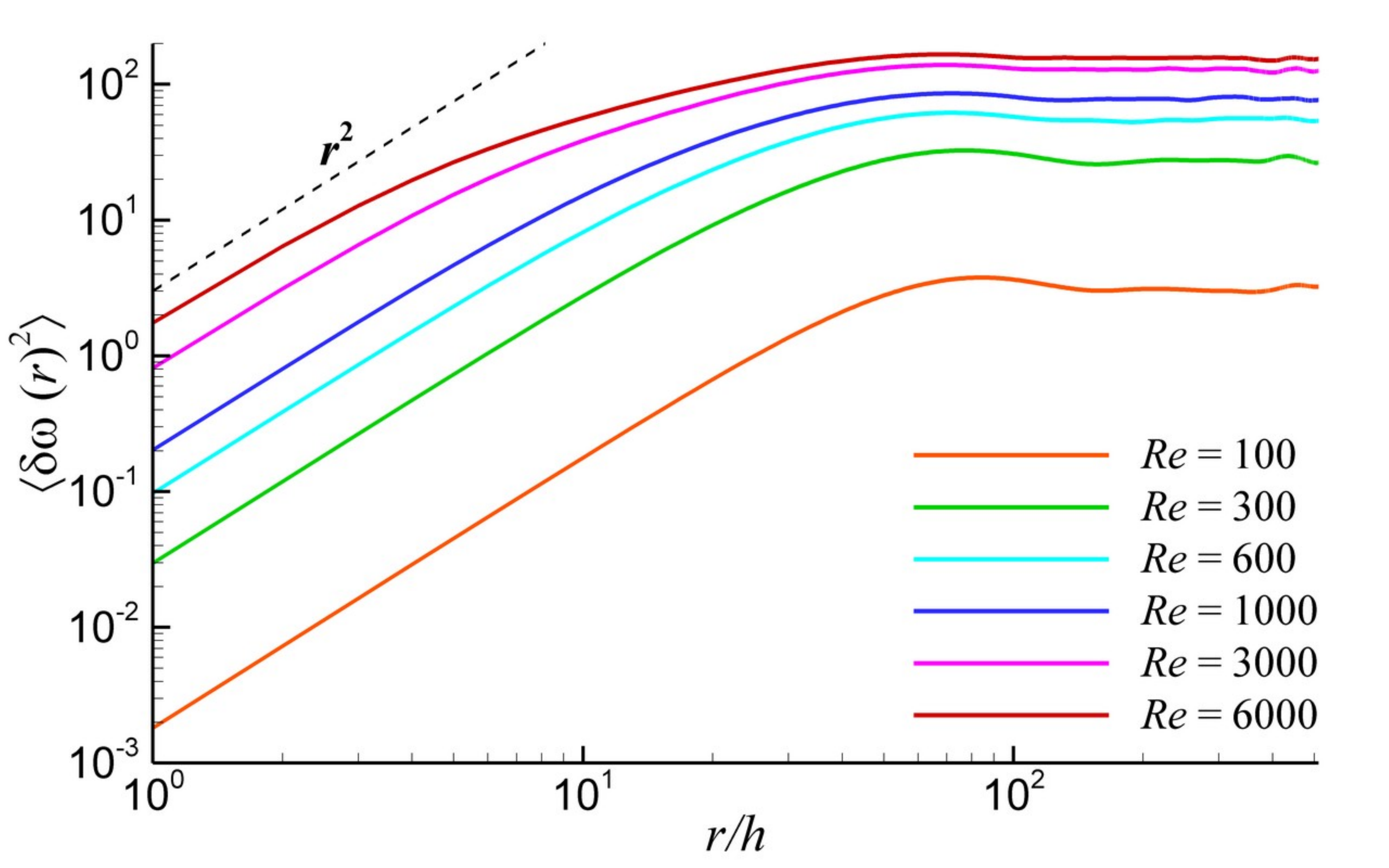}
\caption{Reynolds number dependence of the second-order vorticity structure functions in physical space showing the second-order vorticity structure function, as defined by Eq.~(\ref{eq:str}).}
\label{fig:Re-strf}
\end{figure}

In this part of our analysis, in order to elucidate the flow field evolution for various $Re$ and how the energy spectra and structure functions depend on $Re$, we perform a set of numerical experiments for $Re=100$, $300$, $600$, $1000$, $3000$, and $6000$. All numerical simulations are performed using a resolution of $1024^2$ and start from the same initial condition. Fig.~\ref{fig:ES} demonstrates the evolution of the angle averaged energy spectrum for different Reynolds numbers. As we can see, the energy spectrum becomes steeper than $k^{-3}$ for $Re=100$ due to the smaller filamentation and less interaction among vortices. However, increasing $Re$, the energy spectrum in the inertial range approaches towards the classical $k^{-3}$ limit, in agreement with KBL theory of two-dimensional turbulence. Similarly, the evolution of the second-order structure functions for various Reynolds numbers is exhibited in Fig.~\ref{fig:STR} showing $r^2$ scaling for small separations in all the cases. They flatten for large separations as predicted by the KBL theory of two-dimensional turbulence in the inviscid limit. Comparisons of the angle averaged energy spectra and the second-order vorticity structure functions at time $t=2$ are also shown in Fig.~\ref{fig:Re-spec} and Fig.~\ref{fig:Re-strf}, respectively. As demonstrated by Fig.~\ref{fig:Re-spec}, the angle averaged energy spectrum defined by Eq.~(\ref{eq:Aesp}) asymptotically reaches the $k^{-3}$ spectrum in the inertial range as $Re$ increases. We find that the Reynolds number dependency is more stringent if we look at the turbulence statistics in wave space such the angle averaged energy spectrum. The corresponding instantaneous vorticity fields at time $t=2$ are compared in Fig.~\ref{fig:Re-depend} for the same set of Reynolds numbers. As we can see from Fig.~\ref{fig:Re-depend}, the amount of filamentation increases for higher Reynolds numbers. Due to the smaller convection in lower Reynolds numbers, the interaction between two vortices is not as strong as that of the higher Reynolds number cases, such that less amount of filamentation occurs and results in bigger vortices with less magnitude of vorticity during the evolution because of faster decay rate.

Noticeably, our results provide numerical proof of the KBL theory for decaying two-dimensional turbulence using the sixth-order compact finite difference scheme for spatial discretization along with the third-order TVD Runge-Kutta   scheme for the time advancement. However, we emphasize that our primary goal in this study is to investigate the behavior of different numerical schemes for long time integration of turbulence simulations. In the following, we first test the temporal schemes and their numerical stabilities, and then focus on the evaluation of the behavior of spatial discretization schemes for computations of decaying turbulence.

\subsection{Comparison of temporal schemes}
\label{sec:temp}

We devote this section to investigating the characteristics of the temporal integration schemes described in Section~\ref{sec:time}. In general, the errors associated with temporal discretization are smaller than the errors associated with the spatial discretization, especially if we use high-order methods. Although time step limitations in an explicit time integration method can be an impediment to efficient solutions, for turbulent flow computations small time steps are indeed desirable to resolve the small turbulent time scales. In fact, a balanced spatio-temporal discretization dictates a CFL number of $O(1)$ and large time steps can have the same undesirable effect on the solution as a coarse spatial resolution or a low-order dissipative spatial discretization.

In this section, we evaluate the behavior of different explicit Runge-Kutta schemes introduced earlier for time integration procedure by performing a time step refinement study. The domain of absolute stability for the explicit Runge-Kutta schemes for scalar equations can be found in many references including Canuto \emph{et al.} \cite{canuto1988spectral} and Buthcher \cite{butcher2003numerical}. What is true for scalar equations is not necessarily true for systems as advocated by Pruett \emph{et al.} \cite{pruett1995spatial}. Our method of estimating stability limits is somewhat heuristic here, but it becomes a useful guideline in practical applications for a wide variety of numerical methods.

For this purpose, we first compare the accuracy and stability of the temporal schemes by performing a time refinement study using the the sixth-order compact difference scheme as a spatial discretization method with $512^2$ resolution. Table~\ref{tab:L2-CD6} shows the discrete $L_2$ error norms at $t=10$ for $Re=10^{3}$ obtained by the explicit Runge-Kutta schemes introduced earlier varying the time steps. The corresponding maximum CFL numbers for each computation are also documented in the table. Deviations of $\omega$ are computed from the base solution which is obtained by the forth-order Runge-Kutta method with a small time step of $\Delta t =2\times10^{-4}$. In terms of numerical stability, it is demonstrated in the table that the fourth-order Runge-Kutta scheme has a greater stability region than the third-order ones as we expect. Among the third-order Runge-Kutta schemes for the explicit time integration, the optimal TVD Runge-Kutta scheme becomes more accurate than all the low-storage schemes. However, the difference between the Williamson scheme and the TVD Runge-Kutta scheme is negligible. On the other hand, we find that the predictor/corrector type Runge-Kutta scheme exhibit substantially lower accuracy than other third-order methods.

\begin{table*}[t!]
\caption{Discrete $L_2$ error norms for the decaying turbulence problem at $t=10$ for $Re=10^{3}$ using the sixth-order compact difference scheme as a spatial discretization method. The reference solution for computing the $L_2$ norm is obtained using the RK4 method with a time step of $\Delta t =2\times10^{-4}$. A dash indicates the numerical instability.}
\begin{center}
\label{tab:L2-CD6}
\begin{tabular}{lcccccc}
\hline
\smallskip
Scheme  &$\|\omega\|_{L_2}^{CFL\approx0.1}$ &  $\|\omega\|_{L_2}^{CFL\approx0.2}$& $\|\omega\|_{L_2}^{CFL\approx0.4}$&$\|\omega\|_{L_2}^{CFL\approx0.8}$ &$\|\omega\|_{L_2}^{CFL\approx1.0}$&$\|\omega\|_{L_2}^{CFL\approx1.25}$   \\
        &($\Delta t=4\times10^{-4}$) & ($\Delta t=8\times10^{-4}$)&($\Delta t=16\times10^{-4}$)& ($\Delta t=32\times10^{-4}$)& ($\Delta t=40\times10^{-4}$) & ($\Delta t=50\times10^{-4}$)\\
\hline
RK4     &7.50E-7& 4.01E-6 &5.31E-5 &  7.87E-4 & 1.84E-3 & -- \\
TVDRK3  &3.92E-5& 3.11E-4 &2.46E-3 &  1.94E-2 & --      & -- \\
SYMRK3  &4.65E-4& 1.94E-3 &8.46E-3 &  3.98E-2 & --      & -- \\
P/CRK3  &1.32E-3& 5.23E-3 &2.06E-2 &  8.16E-2 & --      & -- \\
INHRK3  &4.02E-5& 3.18E-4 &2.51E-3 &  1.96E-2 & --      & -- \\
WILRK3  &3.99E-5& 3.14E-4 &2.49E-3 &  1.95E-2 & --      & -- \\
\hline
\end{tabular}
\end{center}
\end{table*}

\begin{table*}[t!]
\caption{Discrete $L_2$ error norms for the decaying turbulence problem at $t=10$ for $Re=10^{3}$ using the second-order conservative Arakawa scheme as a spatial discretization method. The reference solution for computing the $L_2$ norm is obtained using the RK4 method with a time step of $\Delta t =2\times10^{-4}$. A dash indicates the numerical instability.}
\begin{center}
\label{tab:L2-A2}
\begin{tabular}{lcccccc}
\hline
\smallskip
Scheme  &$\|\omega\|_{L_2}^{CFL\approx0.2}$ &  $\|\omega\|_{L_2}^{CFL\approx0.4}$& $\|\omega\|_{L_2}^{CFL\approx1.25}$&$\|\omega\|_{L_2}^{CFL\approx2.5}$ &$\|\omega\|_{L_2}^{CFL\approx3.0}$&$\|\omega\|_{L_2}^{CFL\approx3.5}$   \\
        &($\Delta t=8\times10^{-4}$) & ($\Delta t=16\times10^{-4}$)&($\Delta t=50\times10^{-4}$)& ($\Delta t=100\times10^{-4}$)& ($\Delta t=125\times10^{-4}$) & ($\Delta t=140\times10^{-4}$)\\
\hline
RK4     &2.05E-6& 1.94E-5 &1.76E-3 &  3.31E-2 & 9.89E-2 & -- \\
TVDRK3  &1.65E-4& 1.33E-3 &4.34E-2 &  4.16E-1 & --      & -- \\
SYMRK3  &2.05E-3& 8.61E-3 &1.04E-1 &  5.70E-1 & --      & -- \\
P/CRK3  &5.78E-3& 2.28E-2 &2.09E-1 &  7.18E-1 & --      & -- \\
INHRK3  &1.77E-4& 1.41E-3 &4.52E-2 &  4.20E-1 & --      & -- \\
WILRK3  &1.75E-4& 1.40E-3 &4.47E-2 &  4.16E-1 & --      & -- \\
\hline
\end{tabular}
\end{center}
\end{table*}

Next, we perform a similar analysis by using the second-order conservative Arakawa schemes for the same parameters. Results are demonstrated in Table~\ref{tab:L2-A2} showing the similar trends as observed in the Table~\ref{tab:L2-CD6} for the compact scheme. The interesting outcome is that the second-order scheme can run using almost triple the effective time step if we compare with the sixth-order scheme due to a wider stability region. Although the results are not shown here, we also performed similar studies for the fourth-order Arakawa and the fourth-order compact schemes. Between the second-order scheme and the fourth-order ones, the second-order spatial operator runs at nearly double the effective time step. These results suggest that the second-order scheme becomes more competitive with other high-order accurate spatial schemes by allowing larger time steps. On the other hand, as we will show in the following section, the second-order scheme usually requires double the resolution in each direction to be able to obtain a similar accuracy. Even if using a low order scheme allows us to use larger time steps, still, the high-order schemes become more effective in terms of the tradeoff between the accuracy and efficiency. We also demonstrate that although we use an optimal fast Poisson solver, the CPU time increases approximately 8 times by doubling the resolution in each direction. In fact, it would be larger if we use other suboptimal Poisson solver. In other worlds, that fraction of CPU time is much bigger in other geometries in which the solution of the elliptic equation dominates the computational time. It should also be noted that the efficiency of the high-order schemes would be more pronounced for three dimensional computations.

\subsection{Comparison of spatial schemes}
\label{sec:spat}

Our main objective is to investigate and analyze the behaviors of the spatial numerical schemes presented in Section~\ref{sec:NumS} for two-dimensional decaying turbulence simulations. In this section, four different families of high-order accurate finite difference representations are compared with pseudospectral simulations for various Reynolds numbers. We also compare high-order accurate schemes with the second-order accurate schemes. From our previous time refinement study, we conclude that the truncation error produced by Runge-Kutta time integration schemes is much smaller than the discretization schemes when we use a time step within the stability region. The TVD-RK scheme with $\Delta t=2\times10^{-4}$ is used as a time integration algorithm for the following analysis in which the CFL number becomes $O(0.1)$. Therefore, we can compare different spatial schemes without contaminating the solution due to time stepping algorithm. We also highlight that we use the sixth-order compact difference formulation for discretization of viscous terms in all schemes tested here.

\begin{figure*}
\centering
\mbox{
\subfigure{\includegraphics[width=0.33\textwidth]{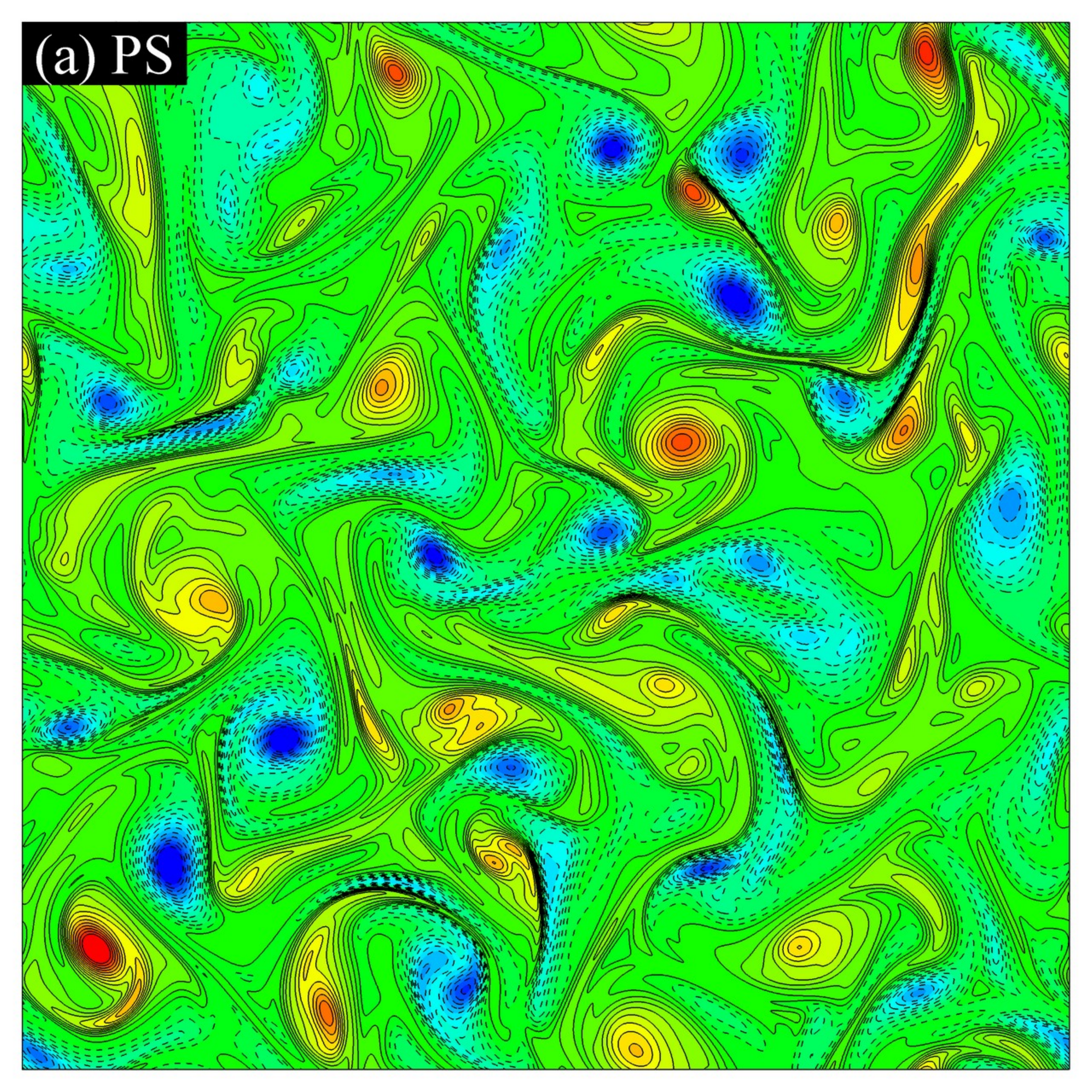}}
\subfigure{\includegraphics[width=0.33\textwidth]{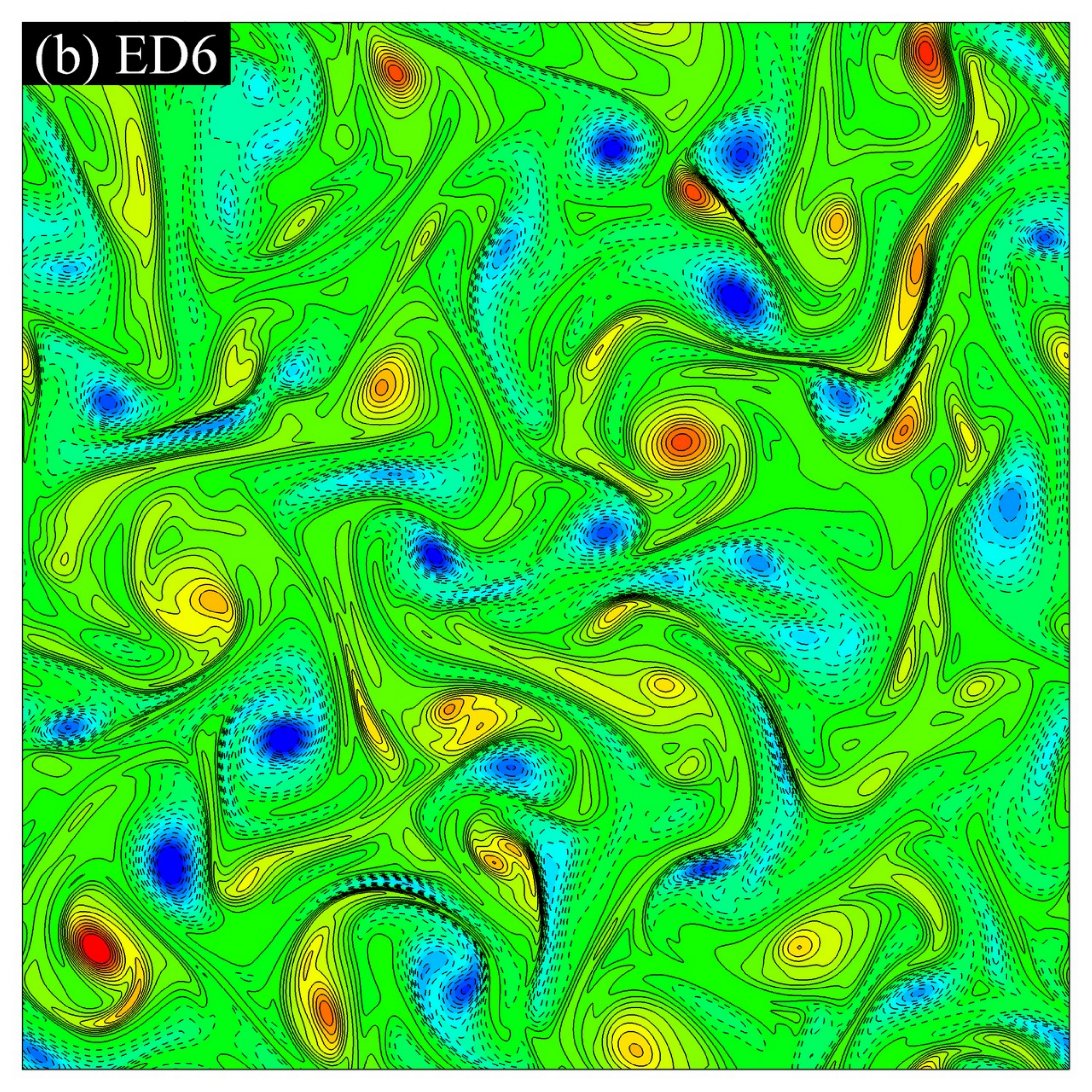}}
\subfigure{\includegraphics[width=0.33\textwidth]{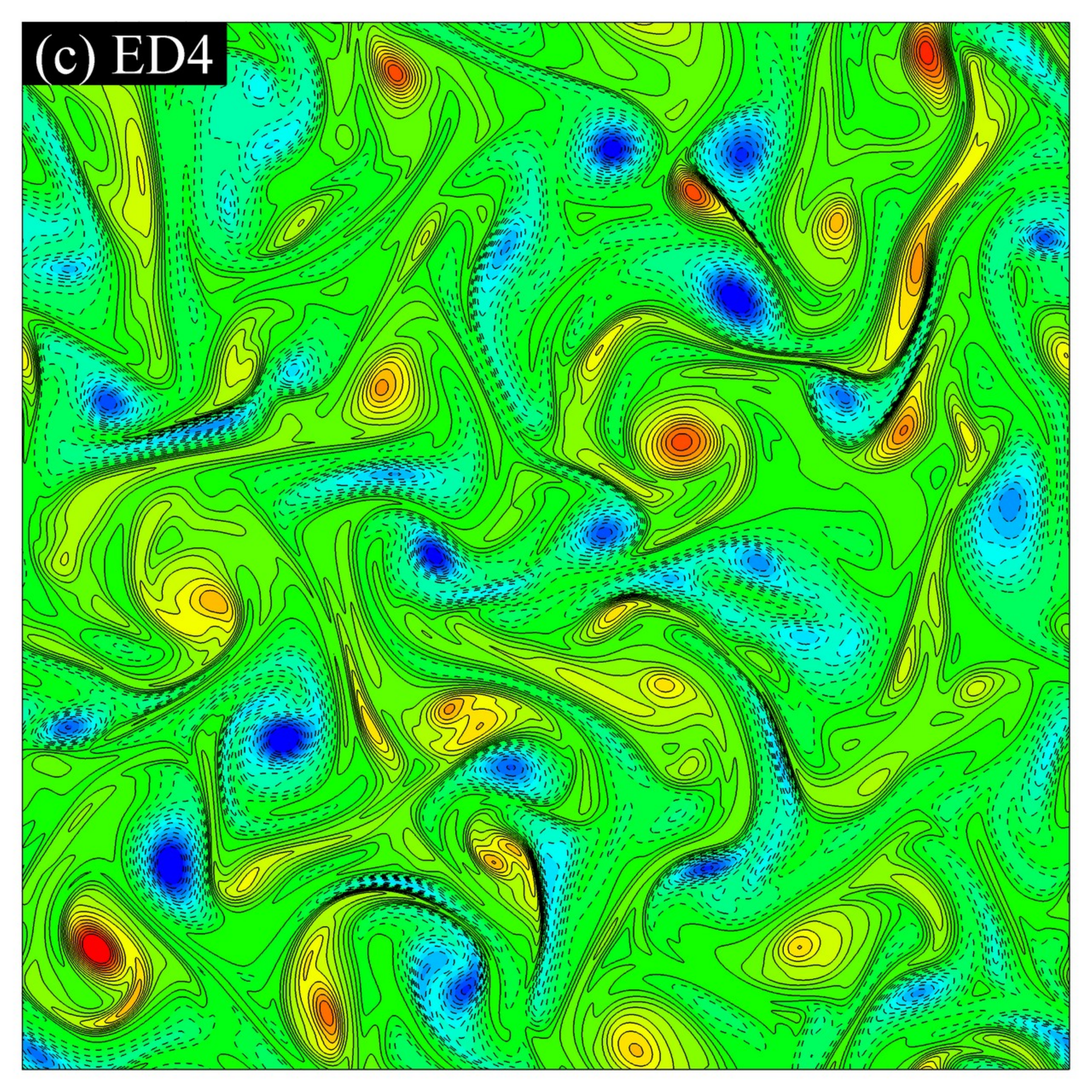}}
}
\\
\mbox{
\subfigure{\includegraphics[width=0.33\textwidth]{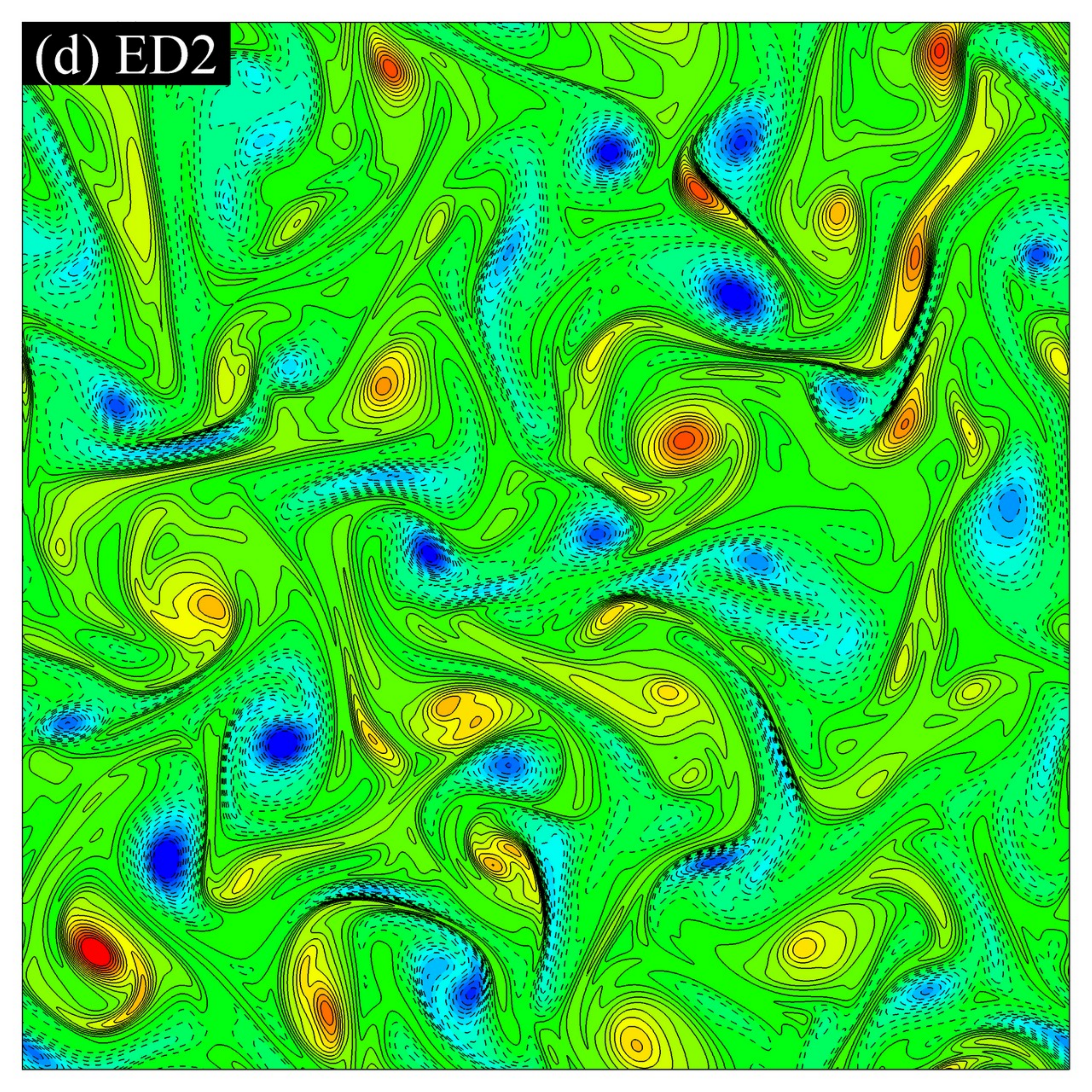}}
\subfigure{\includegraphics[width=0.33\textwidth]{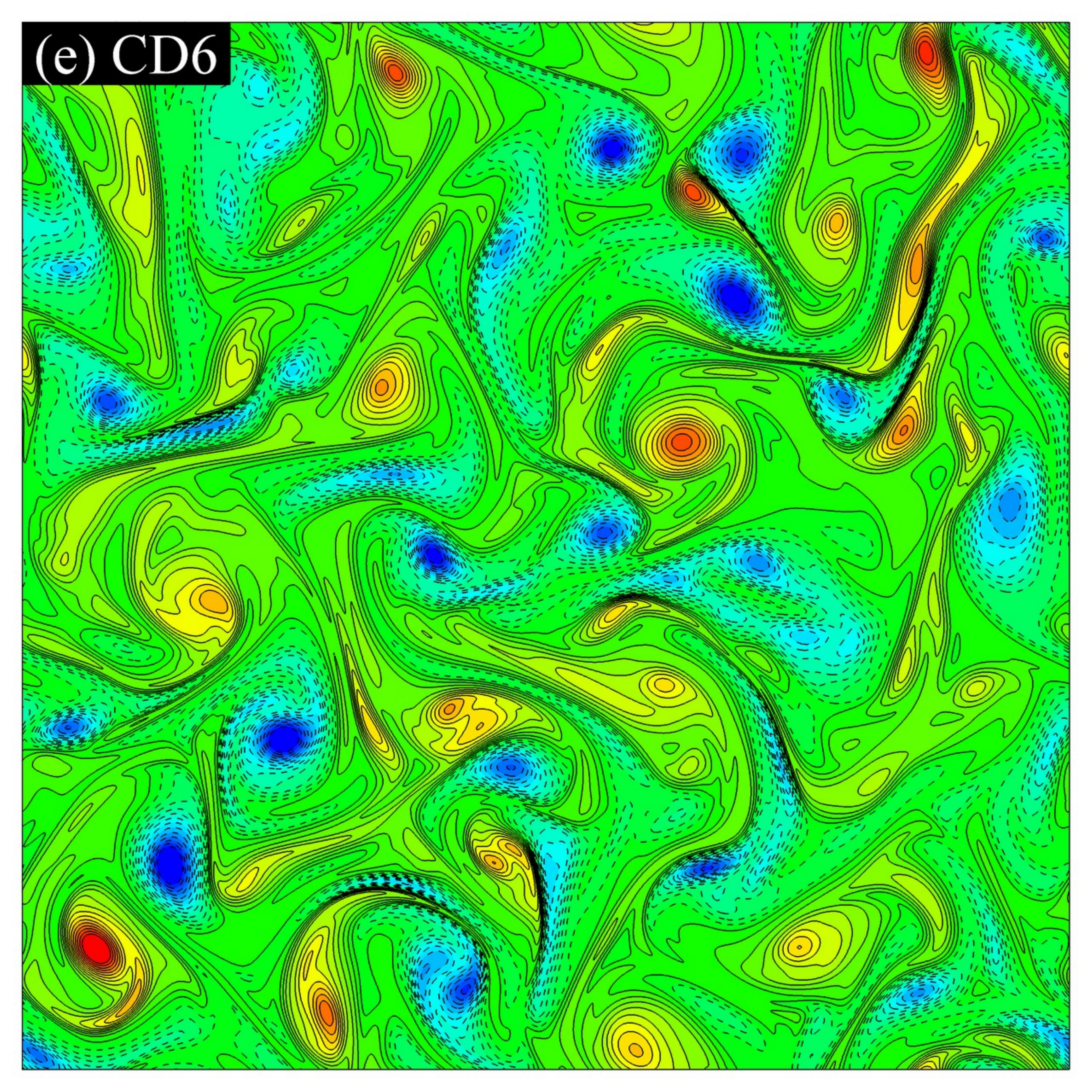}}
\subfigure{\includegraphics[width=0.33\textwidth]{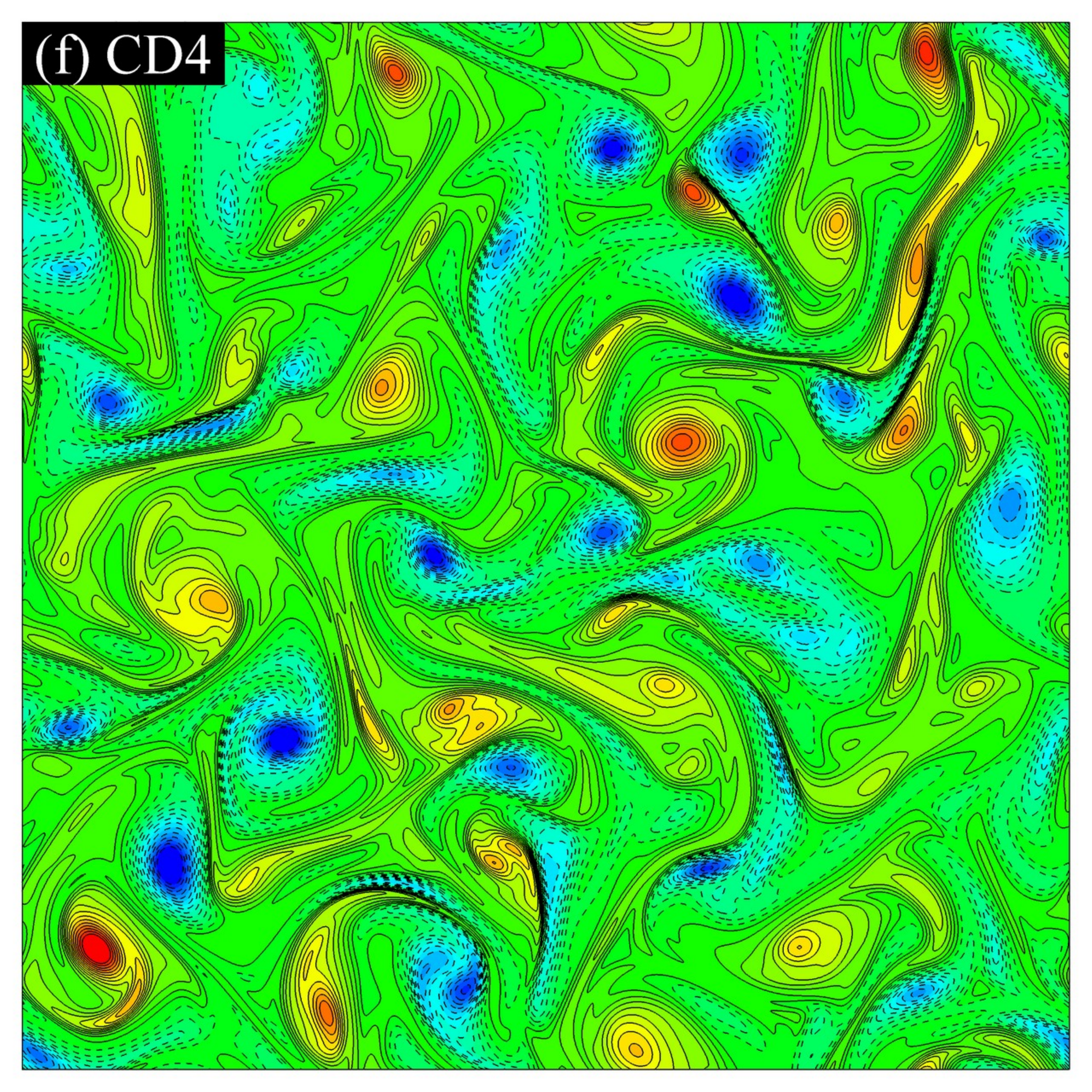}}
}
\\
\mbox{
\subfigure{\includegraphics[width=0.33\textwidth]{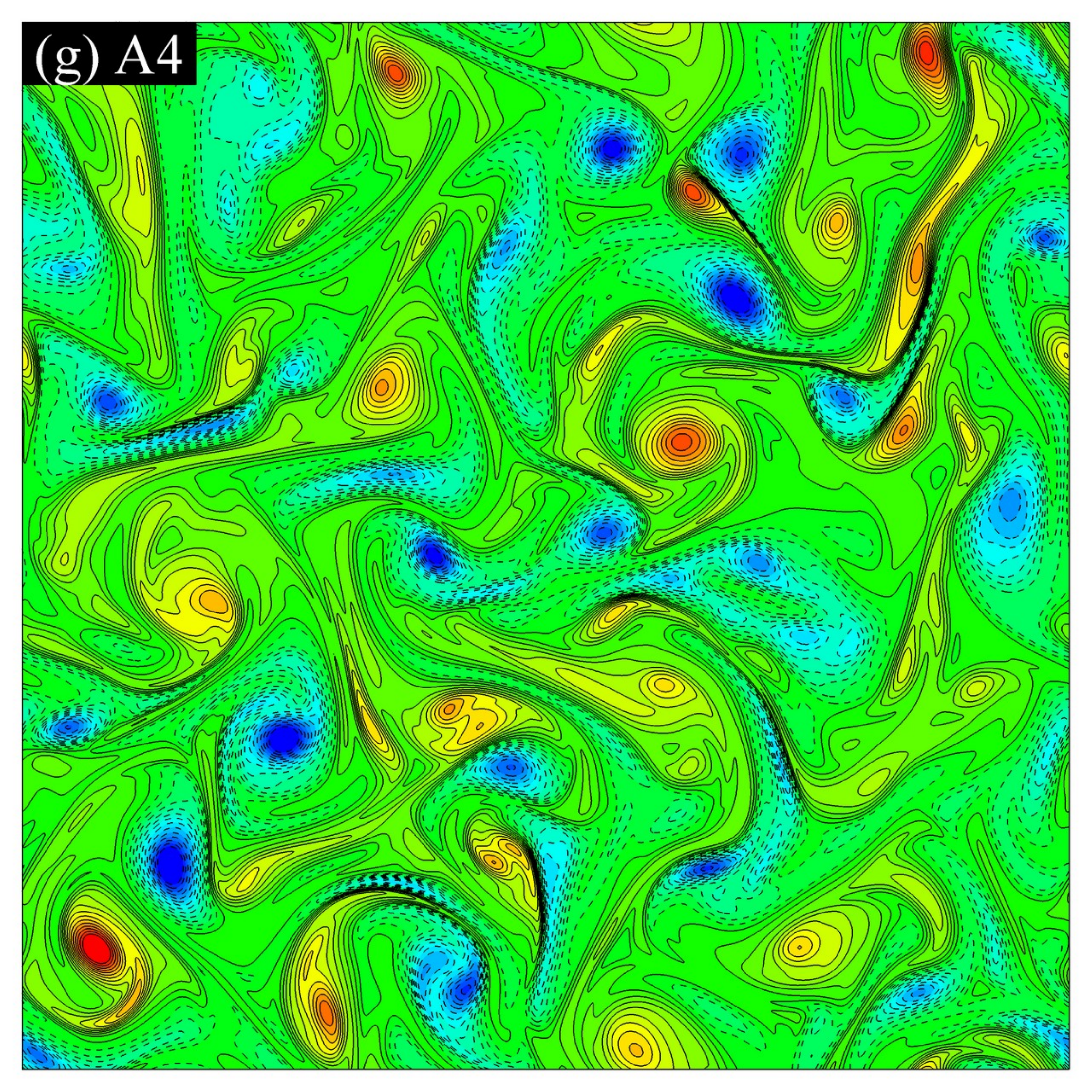}}
\subfigure{\includegraphics[width=0.33\textwidth]{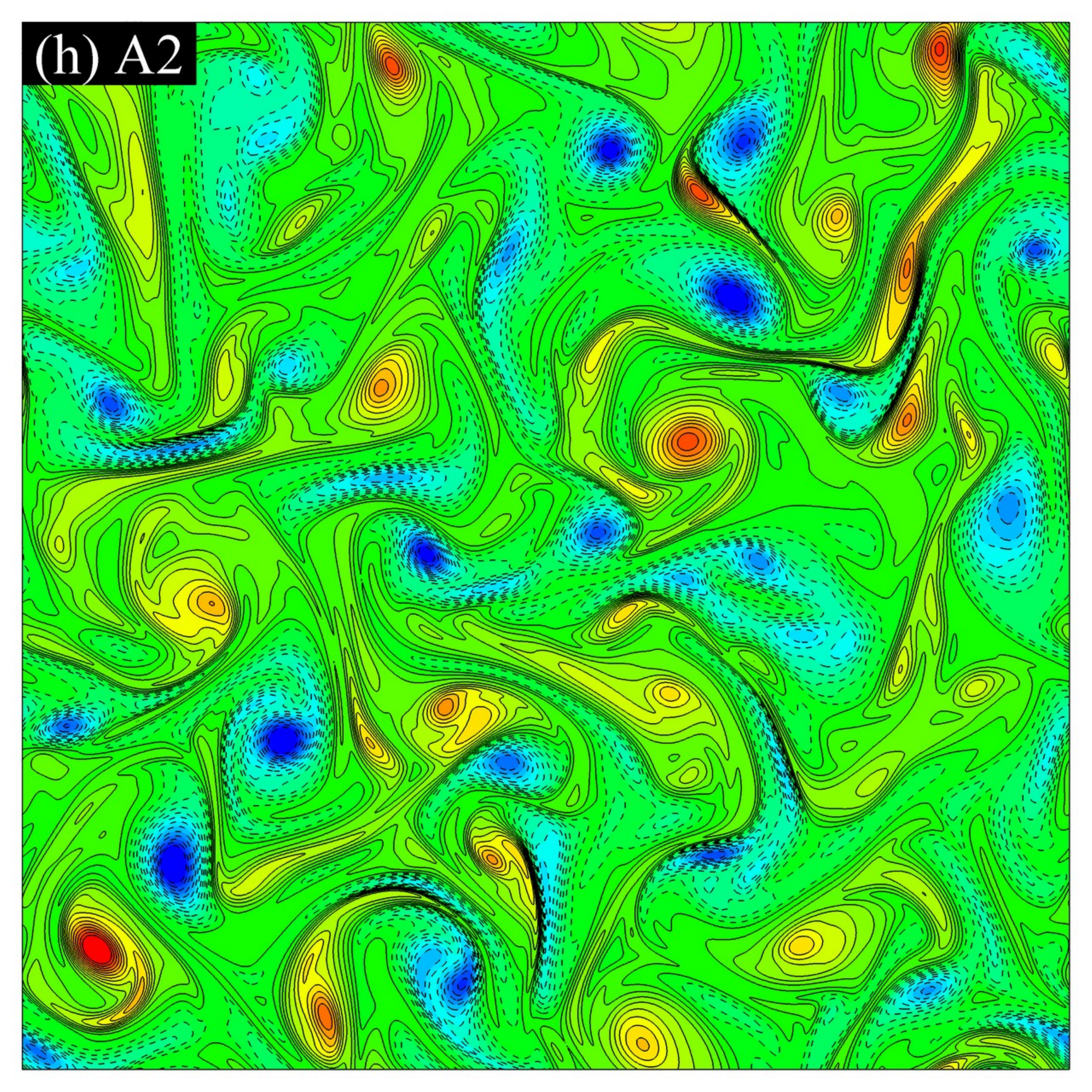}}
\subfigure{\includegraphics[width=0.33\textwidth]{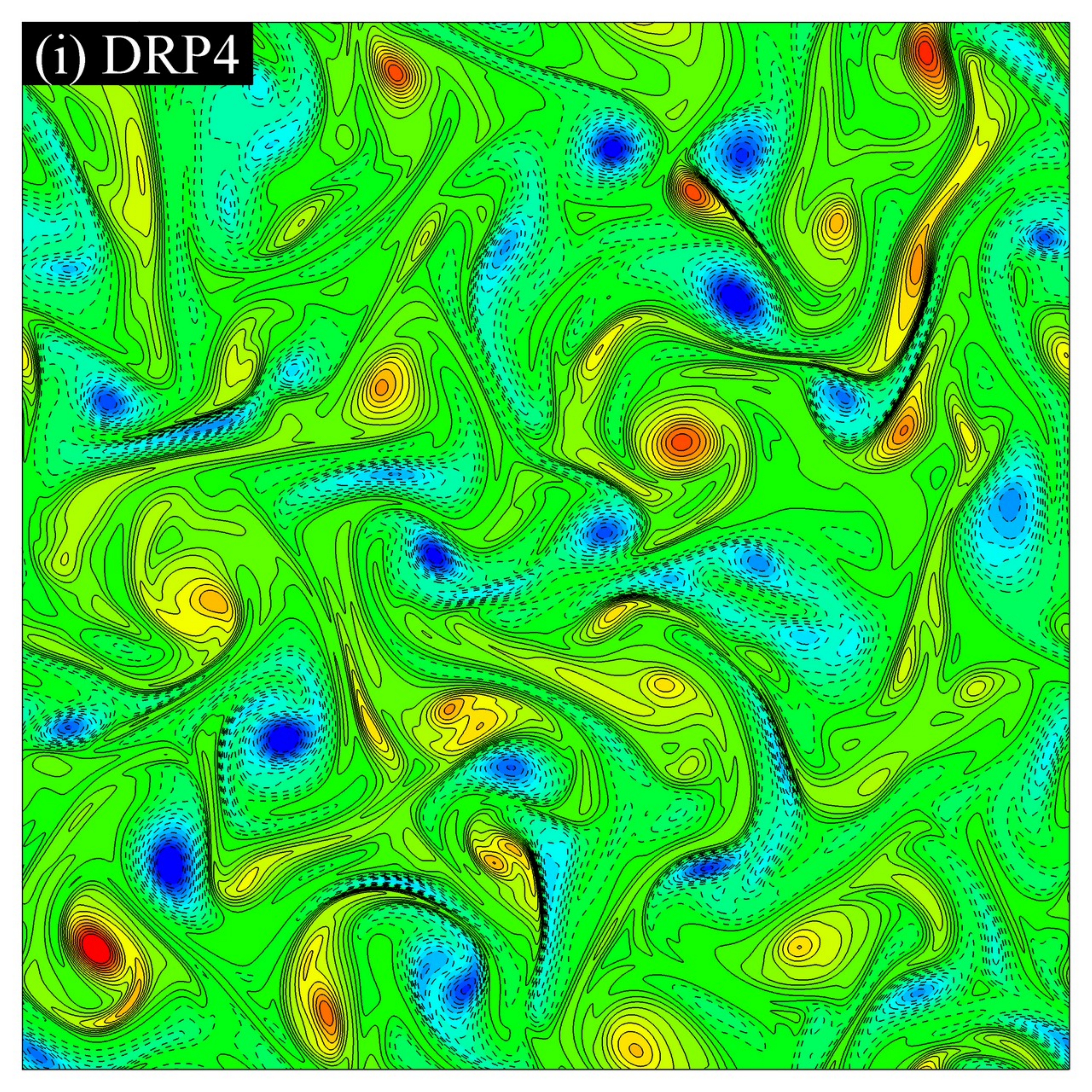}}
}
\caption{Comparison of the numerical schemes at time $t=5$ for $Re=1000$ with a resolution of $1024^2$ ($Re_c = 6.13$). (a) pseudospectral(PS) method, (b) sixth-order explicit difference (ED6) method, (c) fourth-order explicit difference (ED4) method, (d) second-order explicit difference (ED2) method, (e) sixth-order compact difference (CD6) method, (f) fourth-order compact difference (CD4) method, (g) fourth-order Arakawa (A4) method, (h) second-order Arakawa (A2) method, and (i) fourth-order dispersion-relation-preserving (DRP4) method. The vorticity contour layouts are identical in all nine cases illustrating 27 equidistant levels in the interval [-13, 13].}
\label{fig:Re1000-1024-f}
\end{figure*}

\begin{figure*}
\centering
\mbox{
\subfigure{\includegraphics[width=0.33\textwidth]{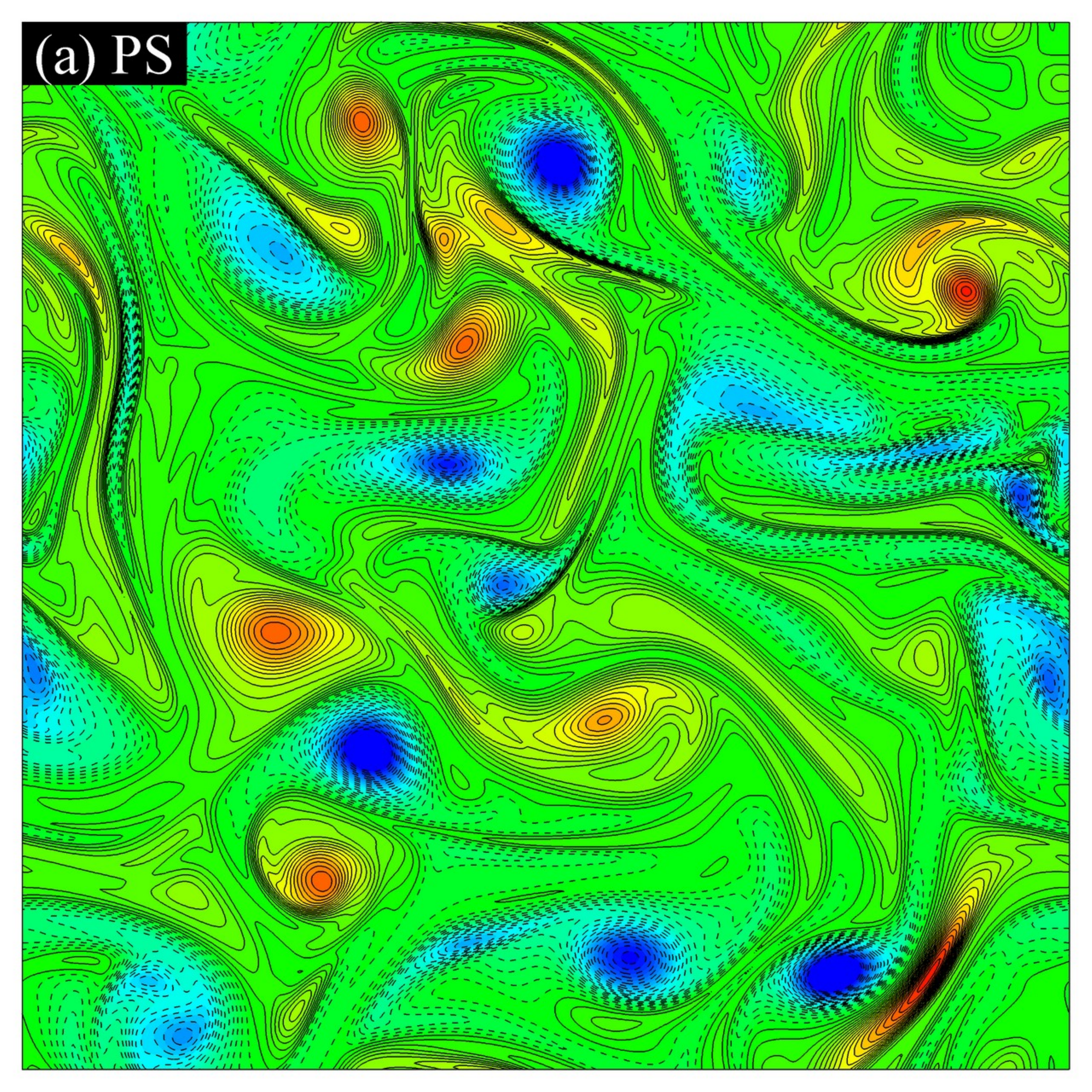}}
\subfigure{\includegraphics[width=0.33\textwidth]{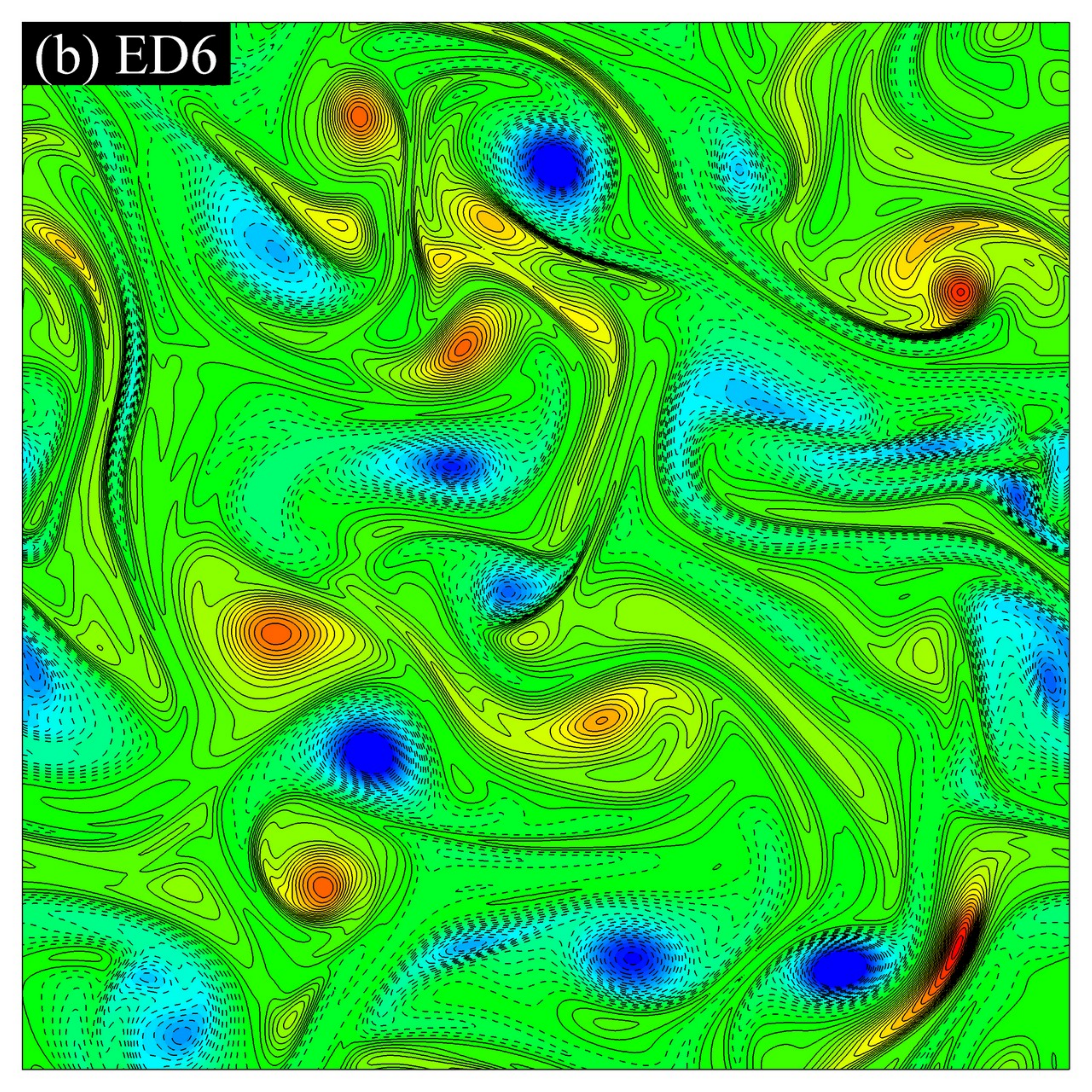}}
\subfigure{\includegraphics[width=0.33\textwidth]{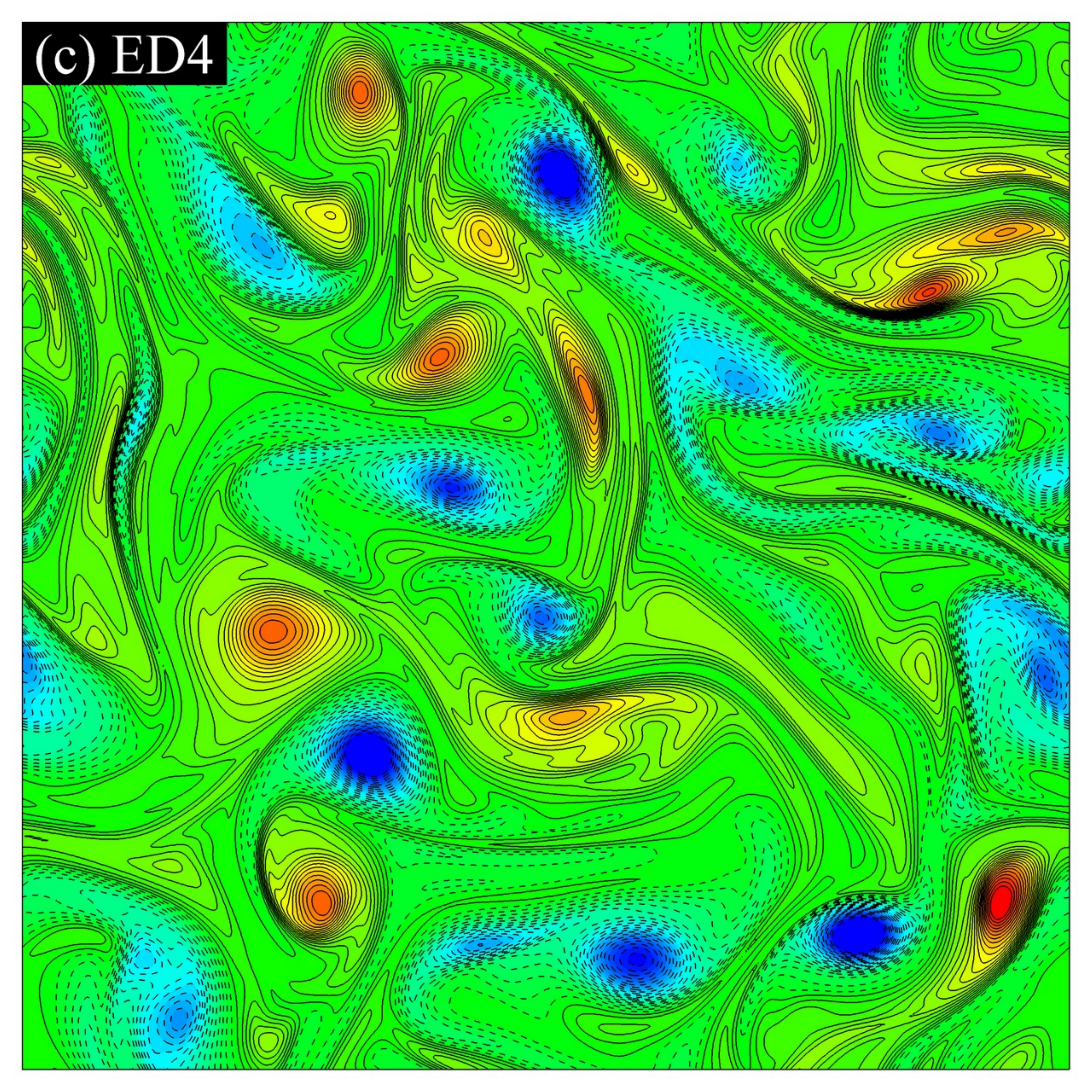}}
}
\\
\mbox{
\subfigure{\includegraphics[width=0.33\textwidth]{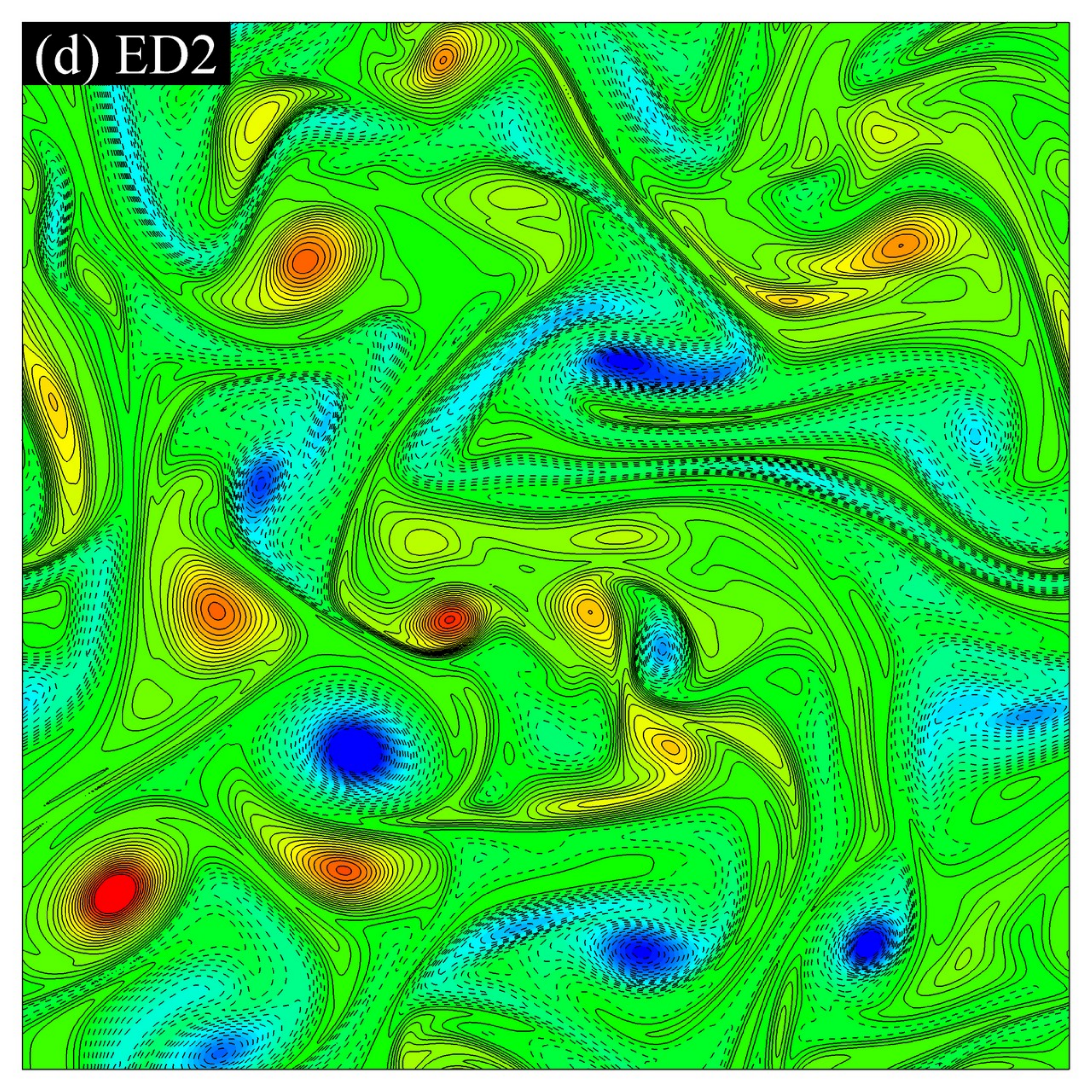}}
\subfigure{\includegraphics[width=0.33\textwidth]{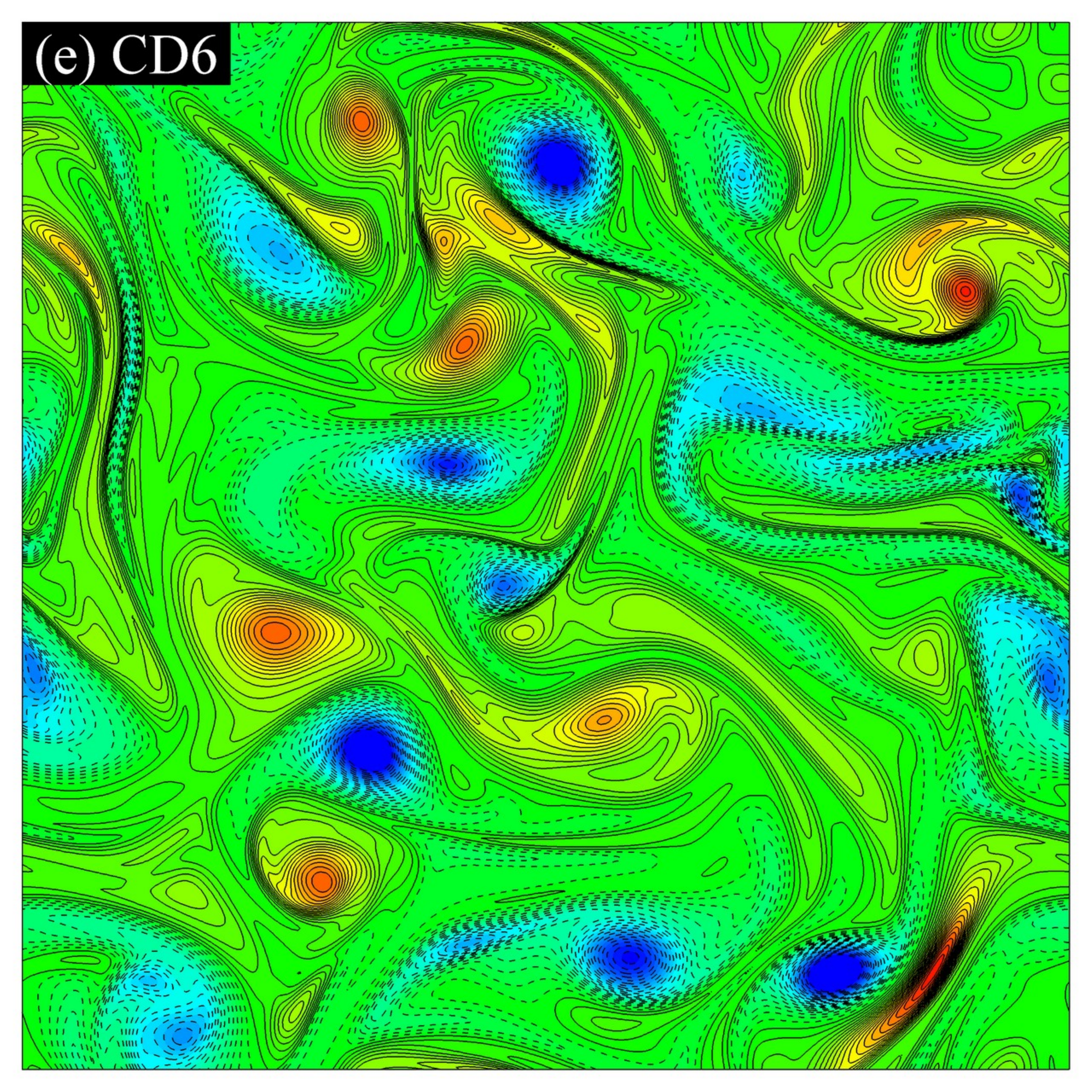}}
\subfigure{\includegraphics[width=0.33\textwidth]{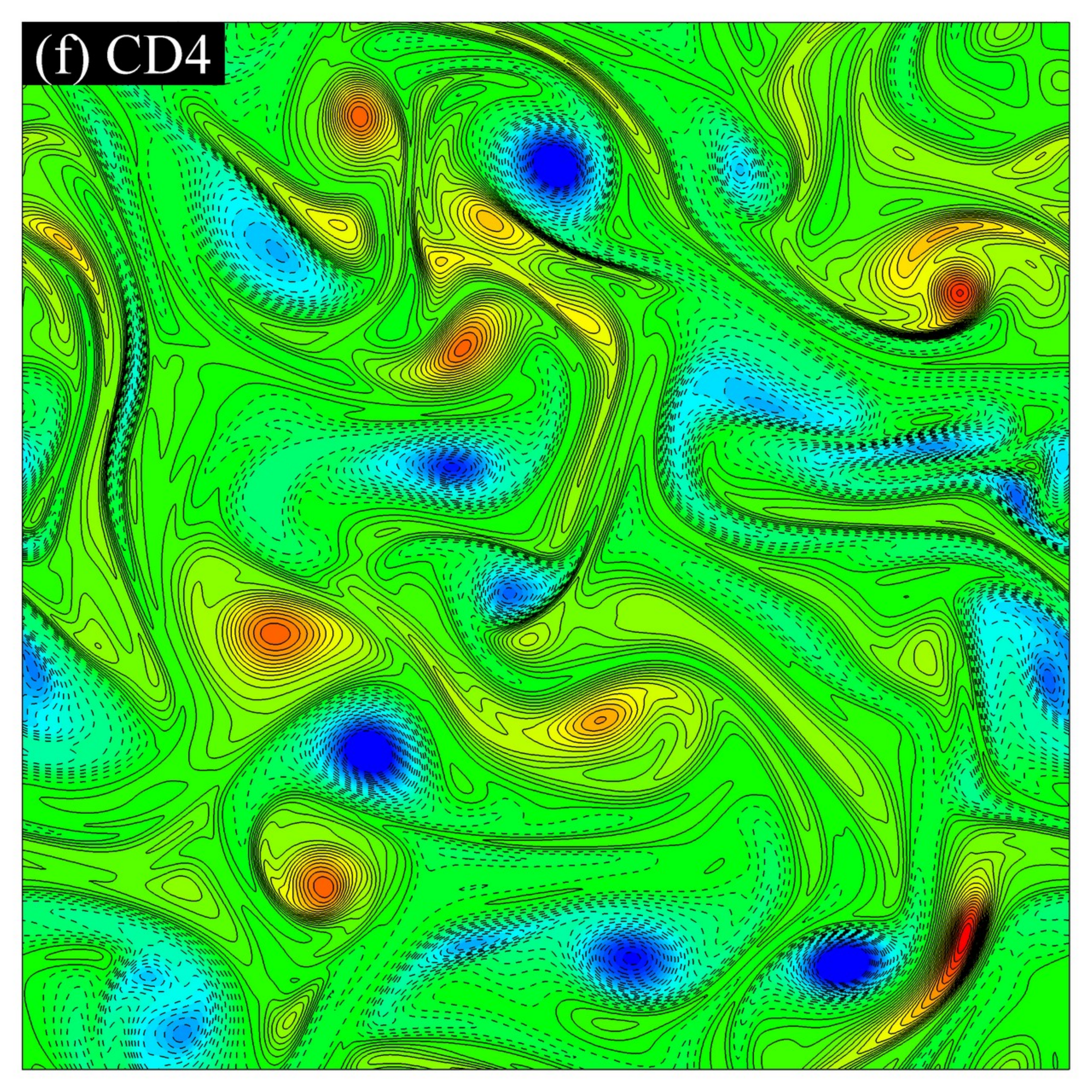}}
}
\\
\mbox{
\subfigure{\includegraphics[width=0.33\textwidth]{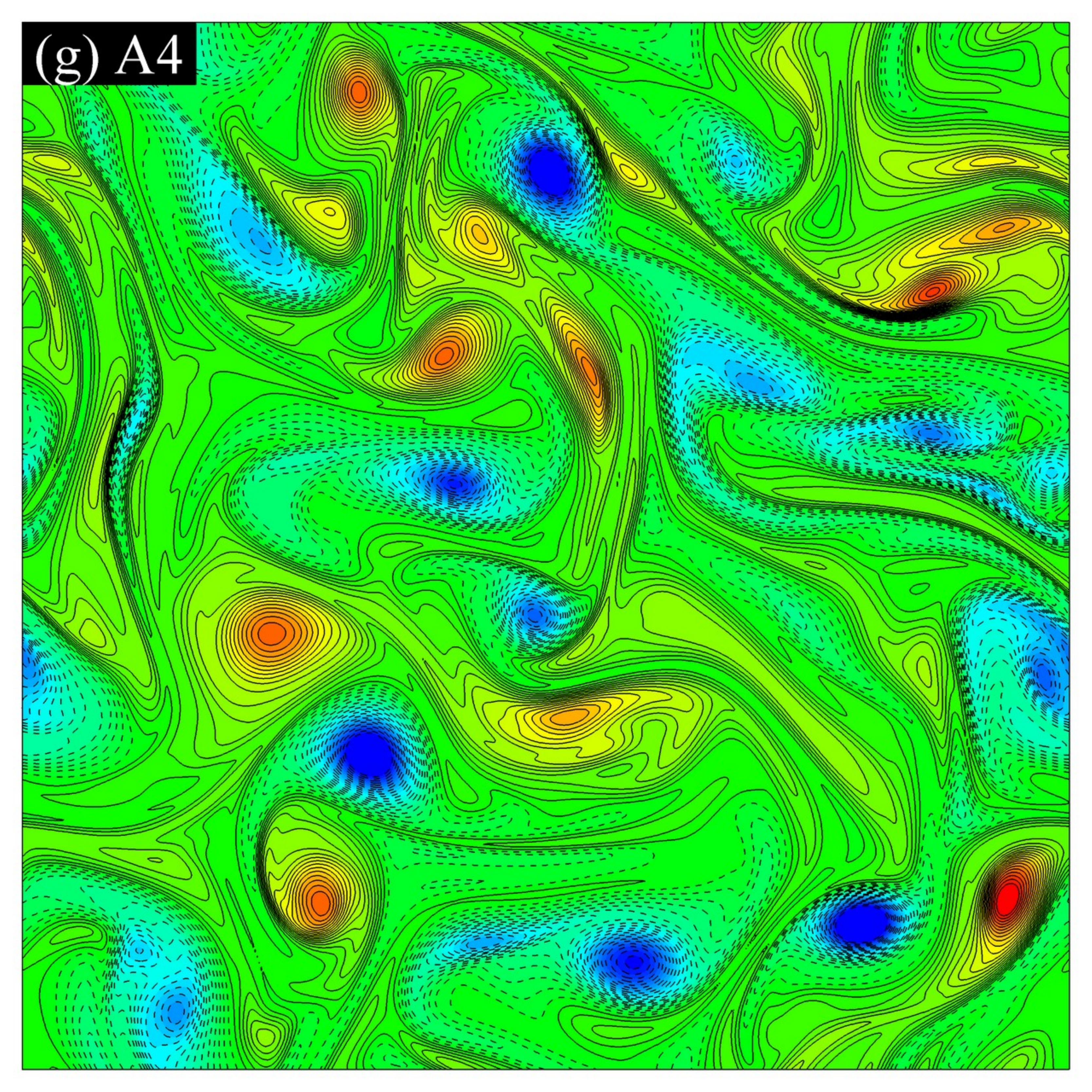}}
\subfigure{\includegraphics[width=0.33\textwidth]{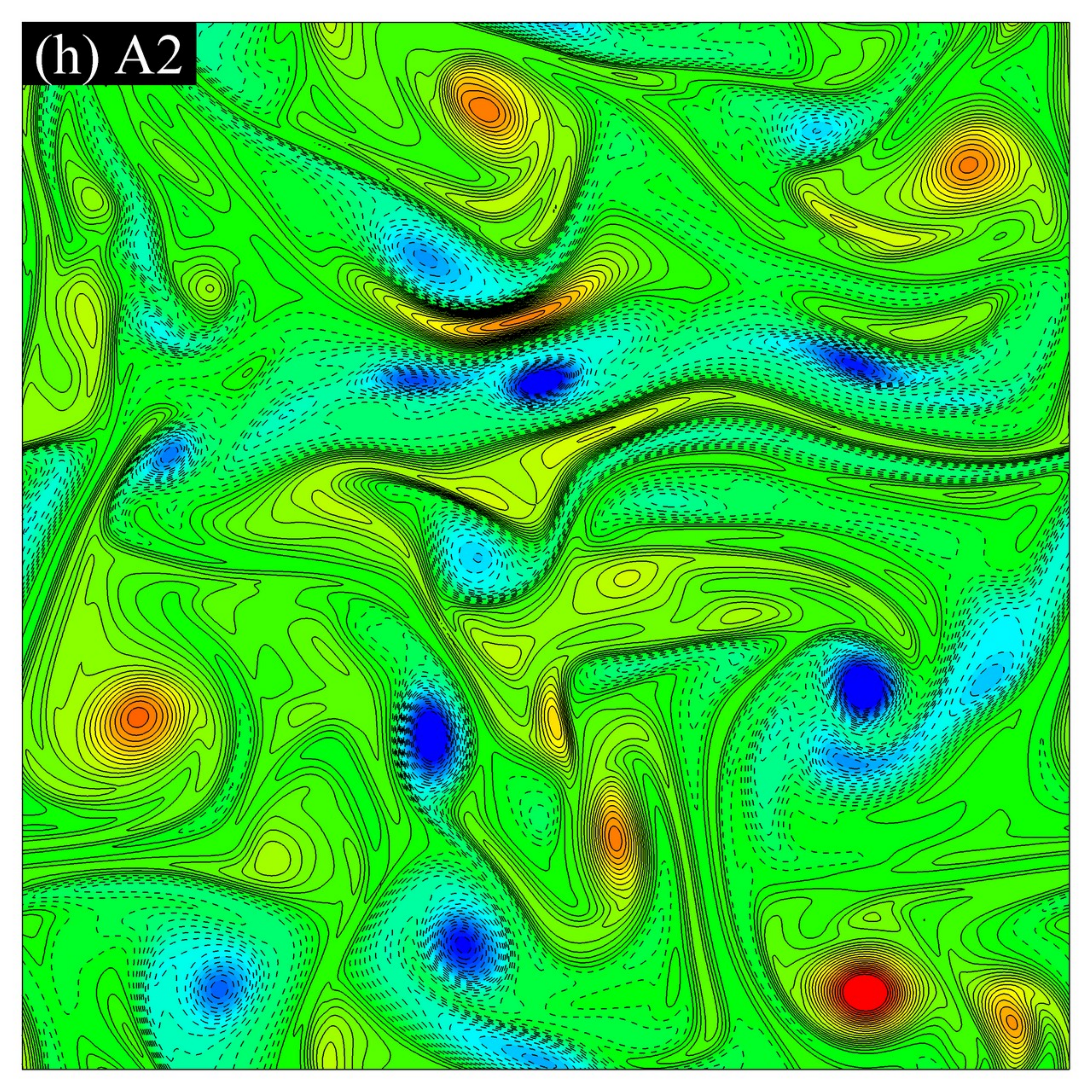}}
\subfigure{\includegraphics[width=0.33\textwidth]{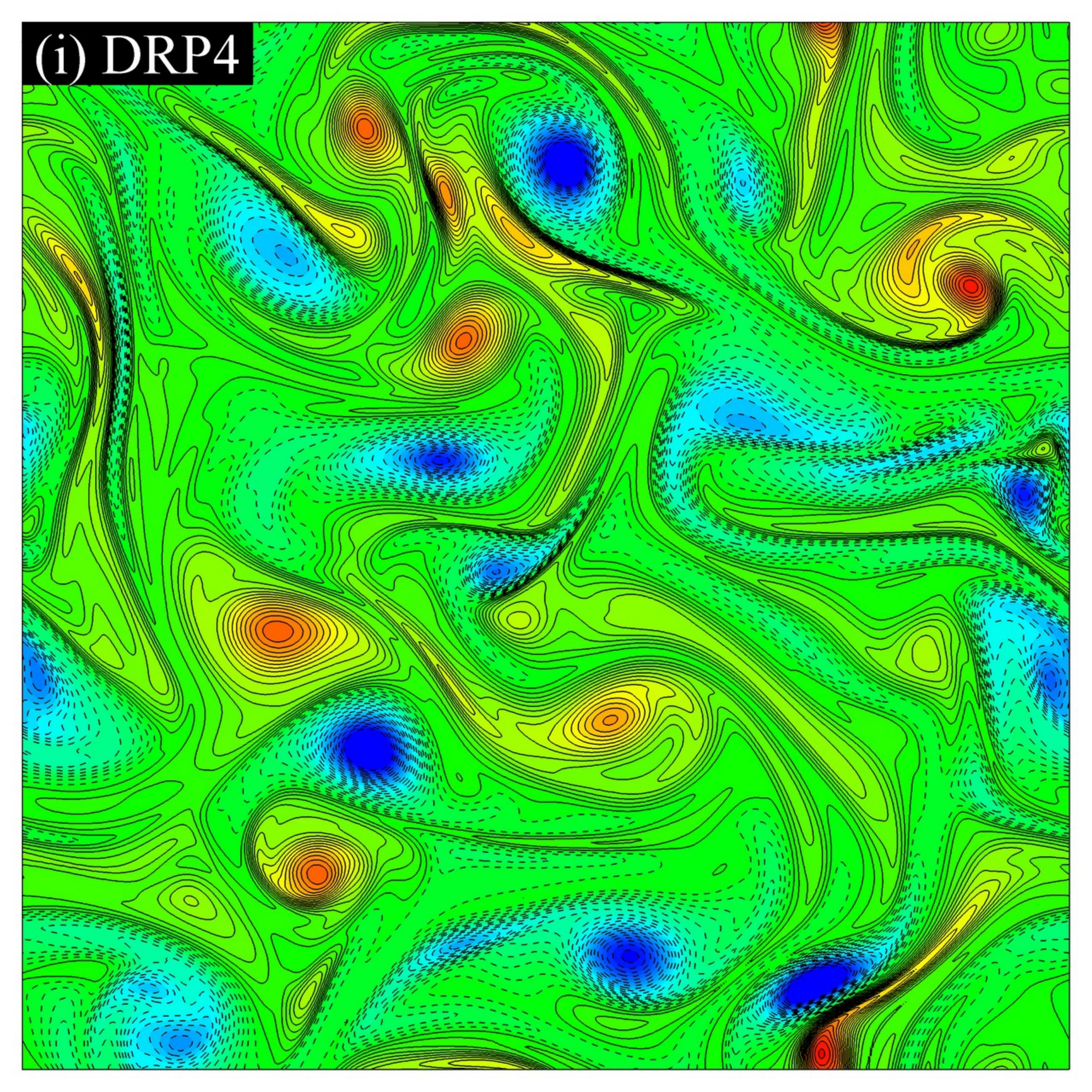}}
}
\caption{Comparison of the numerical schemes at time $t=10$ for $Re=1000$ with a resolution of $512^2$ ($Re_c = 12.27$). (a) pseudospectral(PS) method, (b) sixth-order explicit difference (ED6) method, (c) fourth-order explicit difference (ED4) method, (d) second-order explicit difference (ED2) method, (e) sixth-order compact difference (CD6) method, (f) fourth-order compact difference (CD4) method, (g) fourth-order Arakawa (A4) method, (h) second-order Arakawa (A2) method, and (i) fourth-order dispersion-relation-preserving (DRP4) method. The vorticity contour layouts are identical in all nine cases illustrating 41 equidistant levels in the interval [-8, 8].}
\label{fig:Re1000-512-f}
\end{figure*}

\begin{figure*}
\centering
\mbox{
\subfigure{\includegraphics[width=0.33\textwidth]{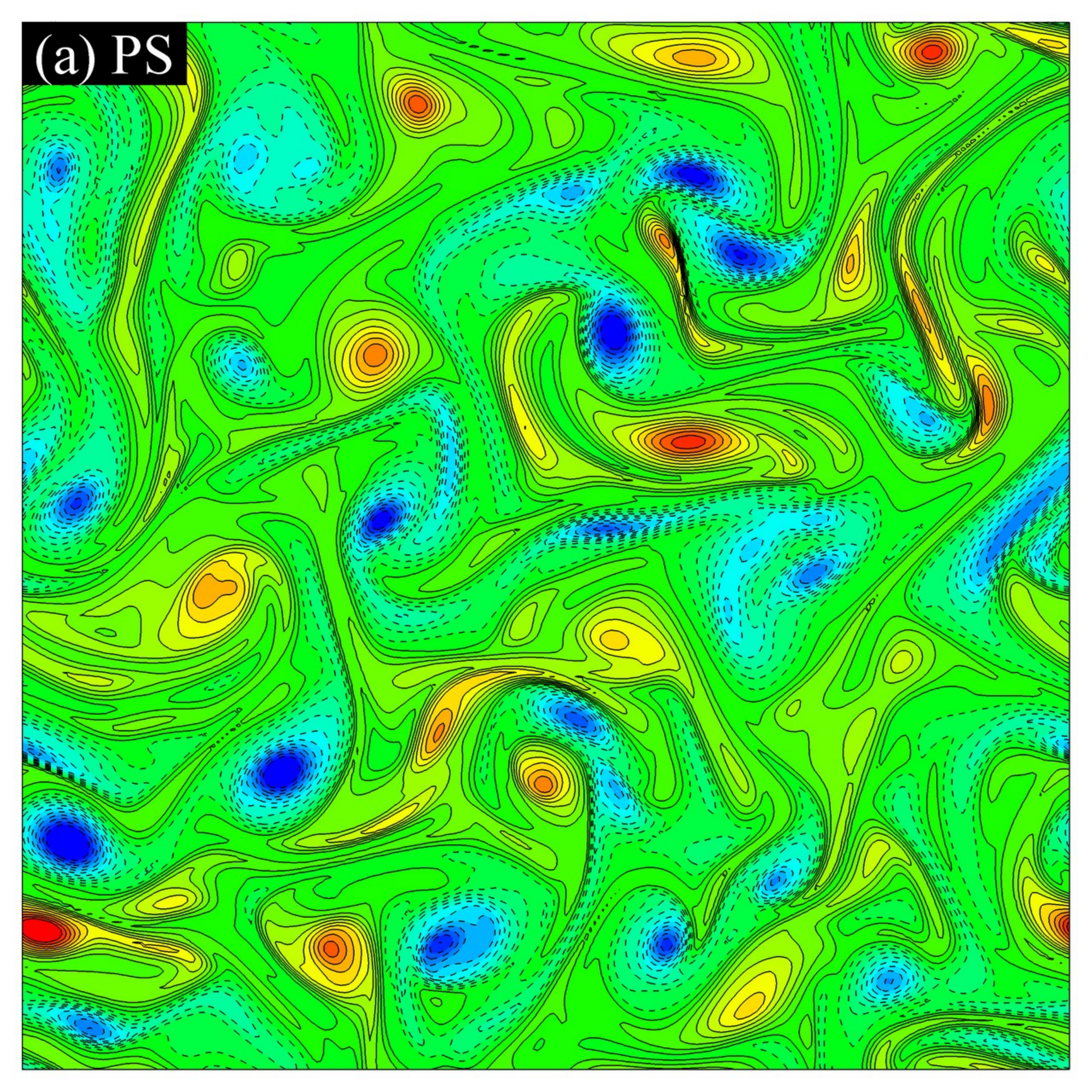}}
\subfigure{\includegraphics[width=0.33\textwidth]{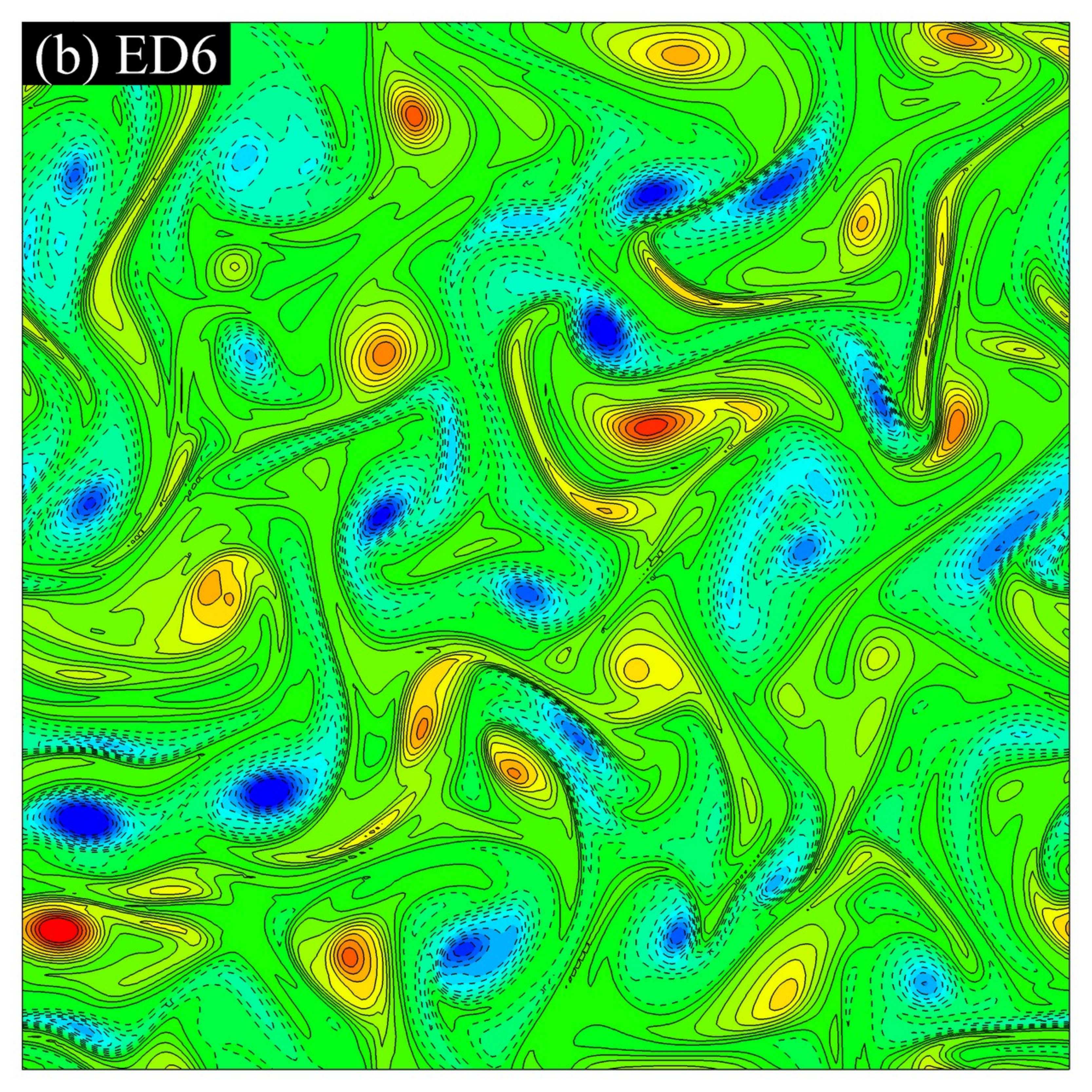}}
\subfigure{\includegraphics[width=0.33\textwidth]{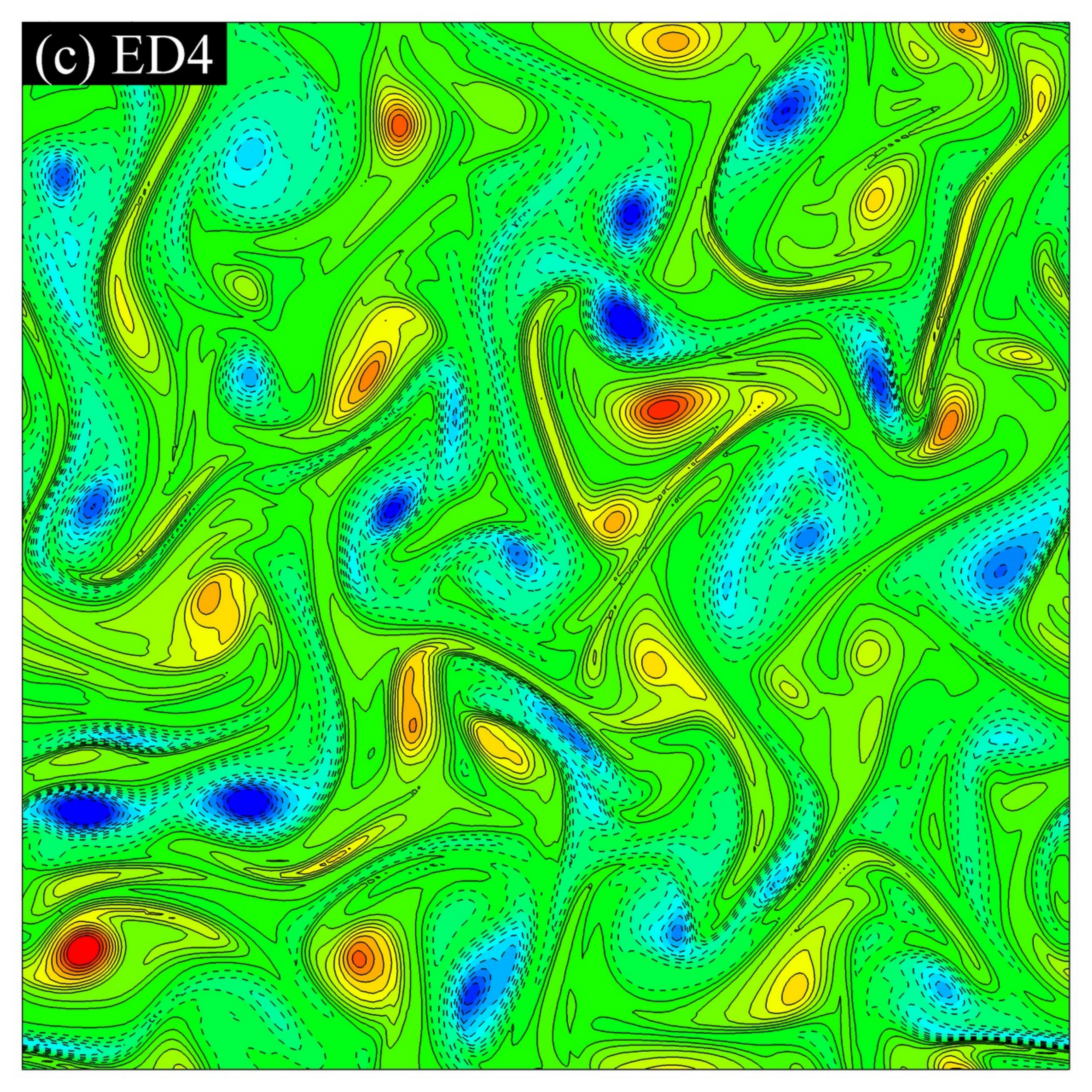}}
}
\\
\mbox{
\subfigure{\includegraphics[width=0.33\textwidth]{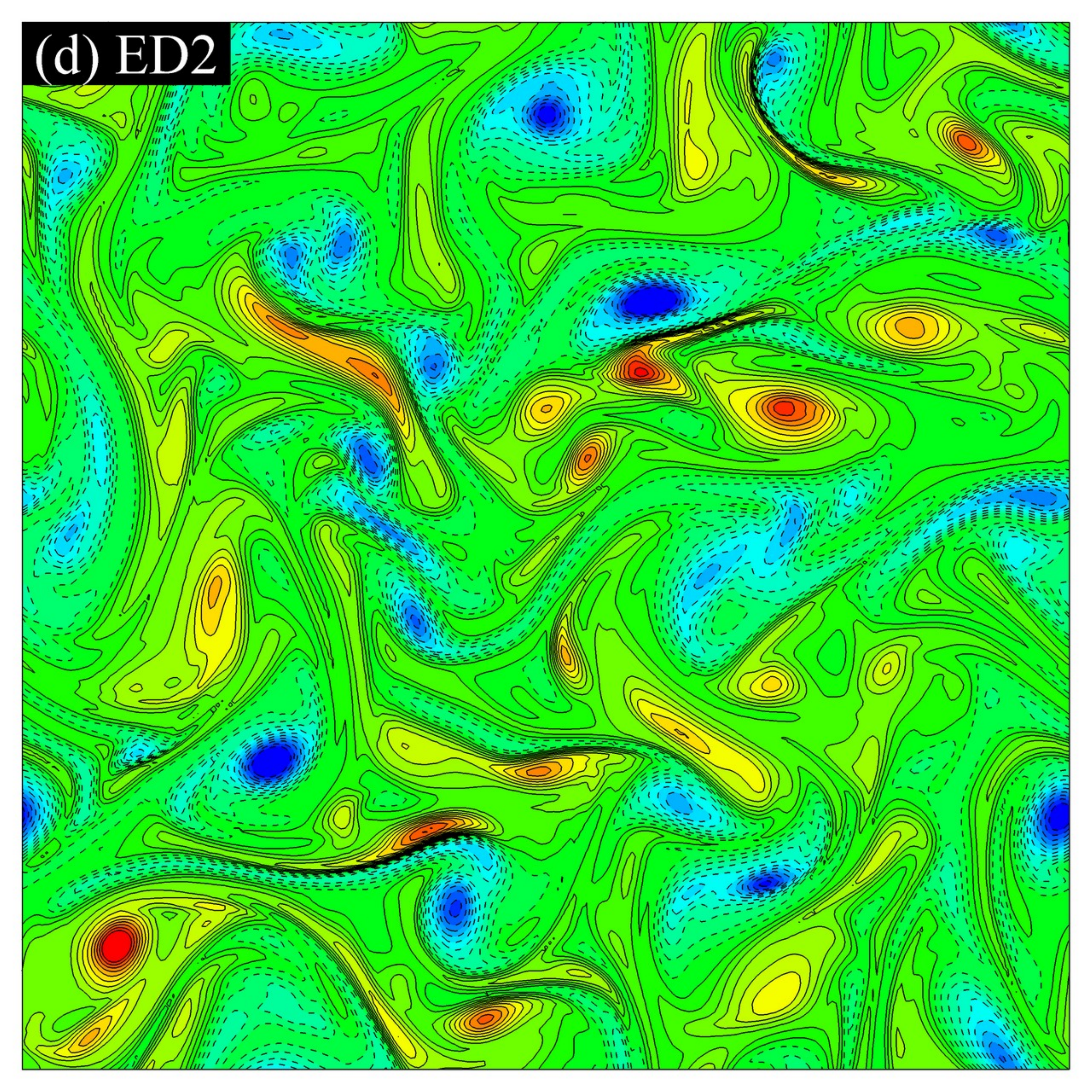}}
\subfigure{\includegraphics[width=0.33\textwidth]{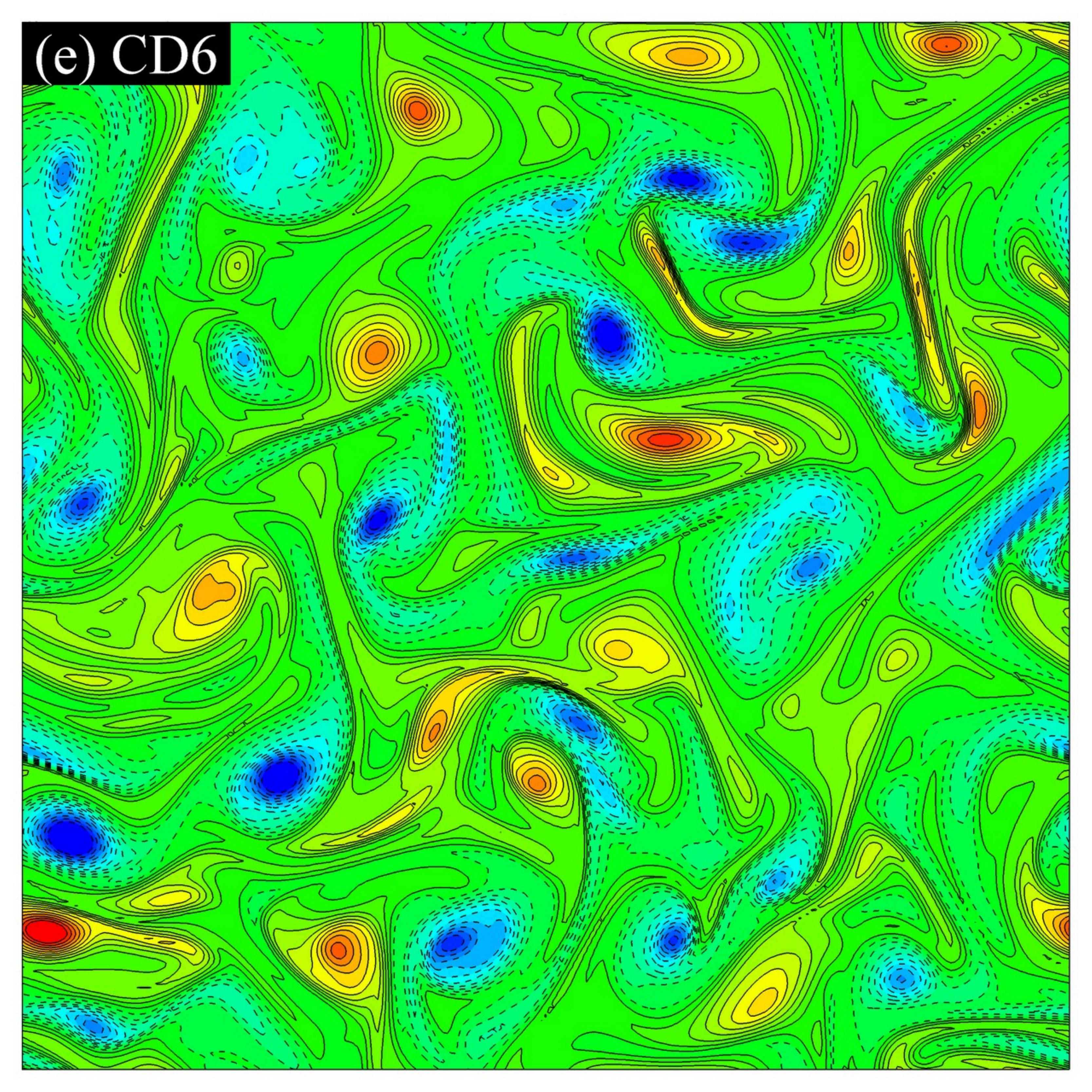}}
\subfigure{\includegraphics[width=0.33\textwidth]{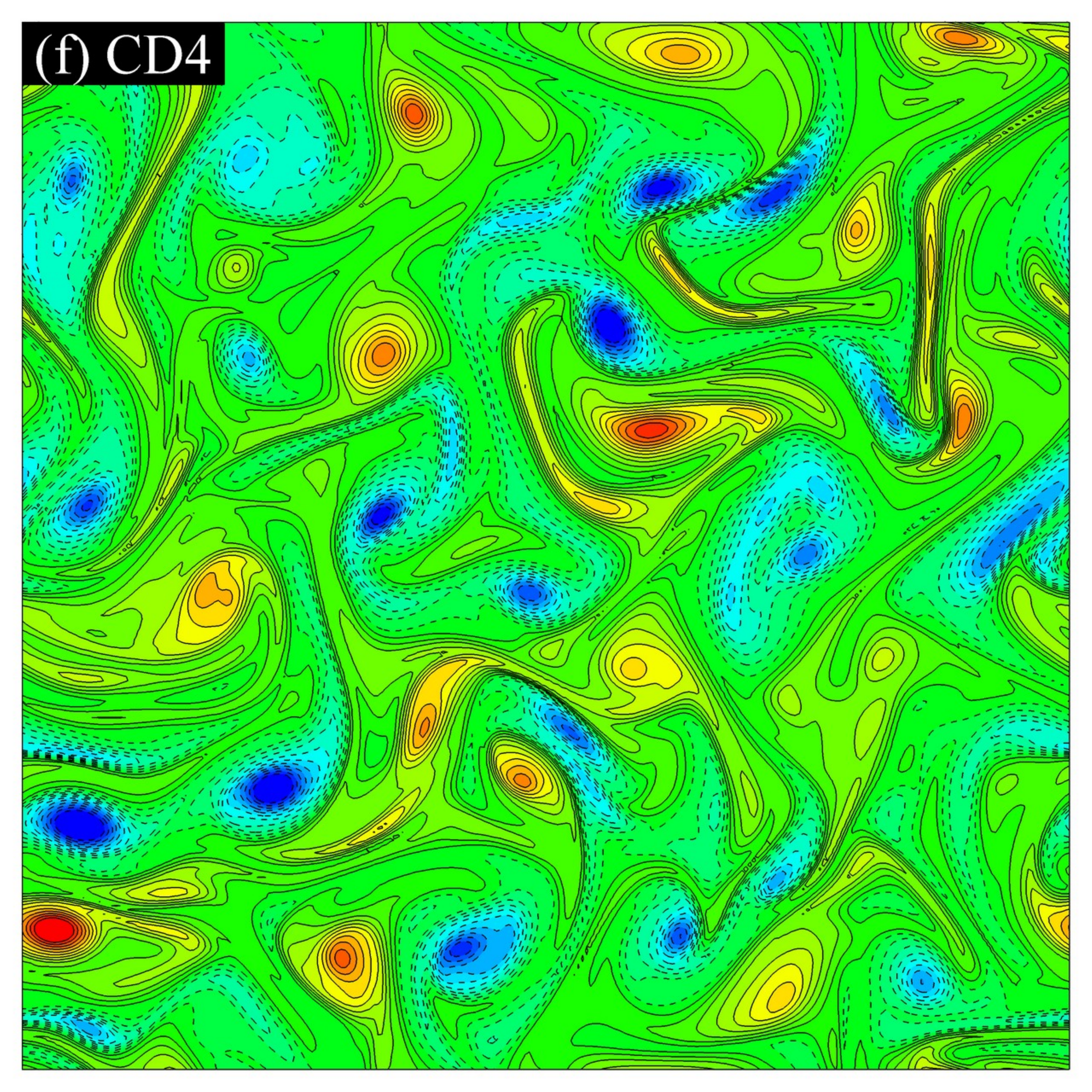}}
}
\\
\mbox{
\subfigure{\includegraphics[width=0.33\textwidth]{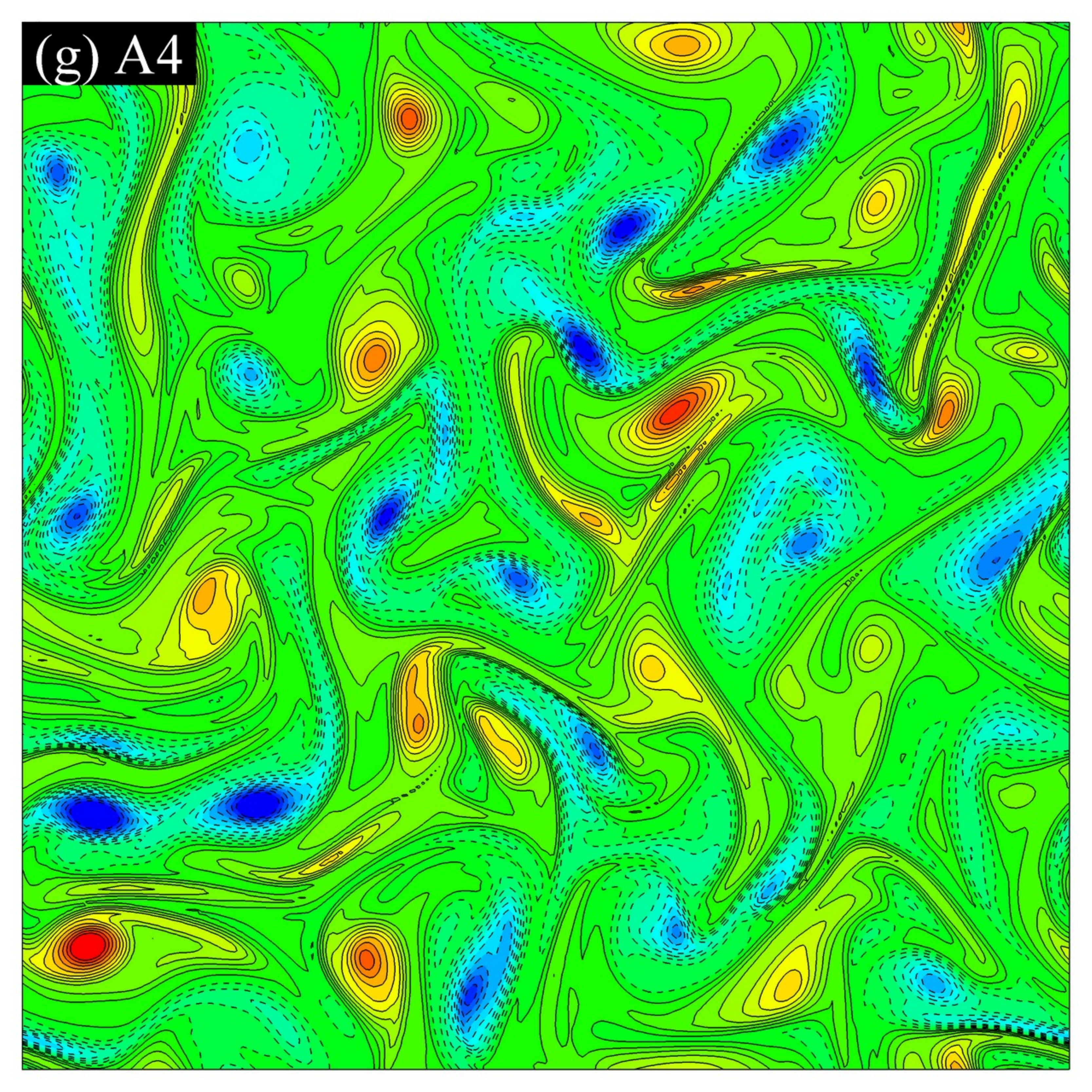}}
\subfigure{\includegraphics[width=0.33\textwidth]{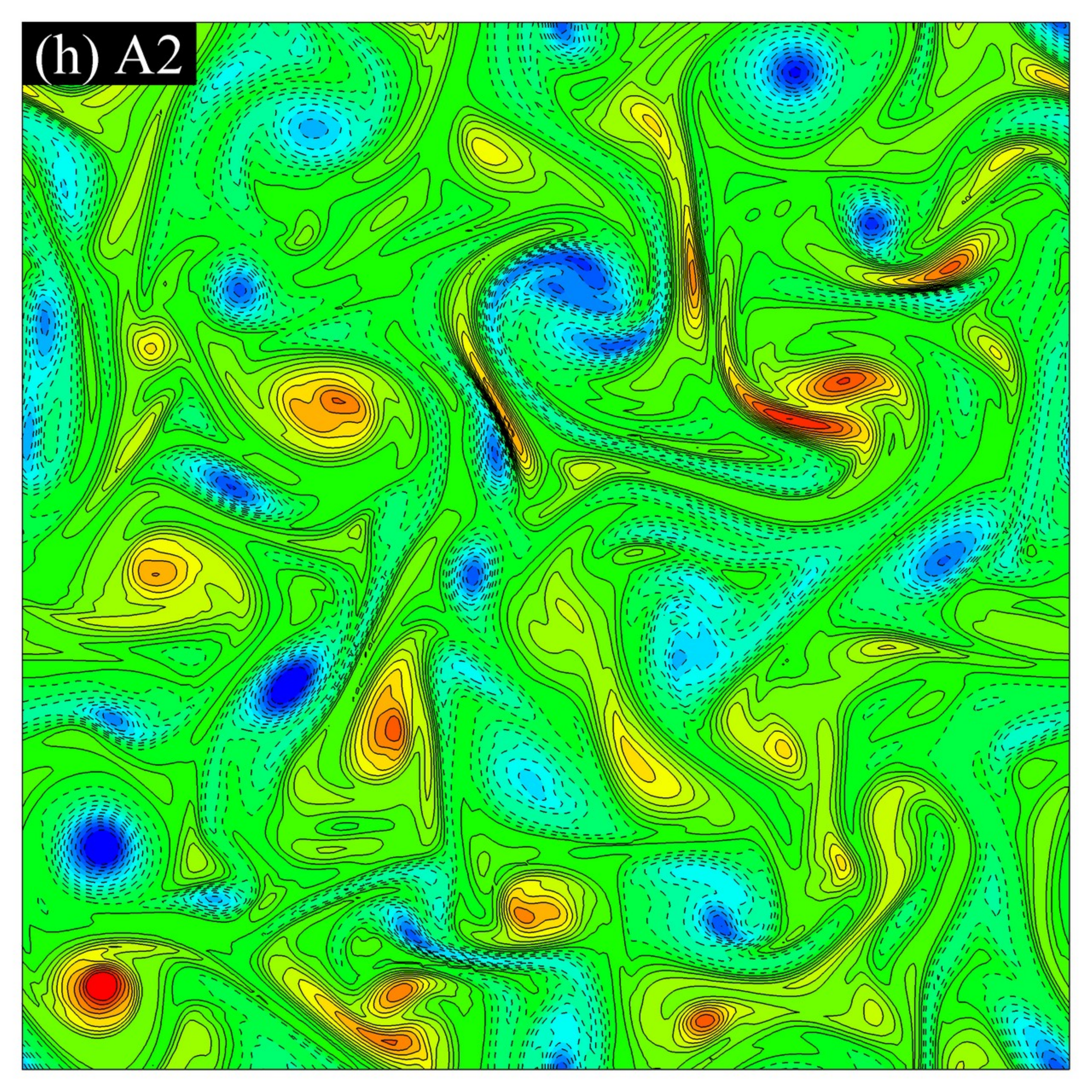}}
\subfigure{\includegraphics[width=0.33\textwidth]{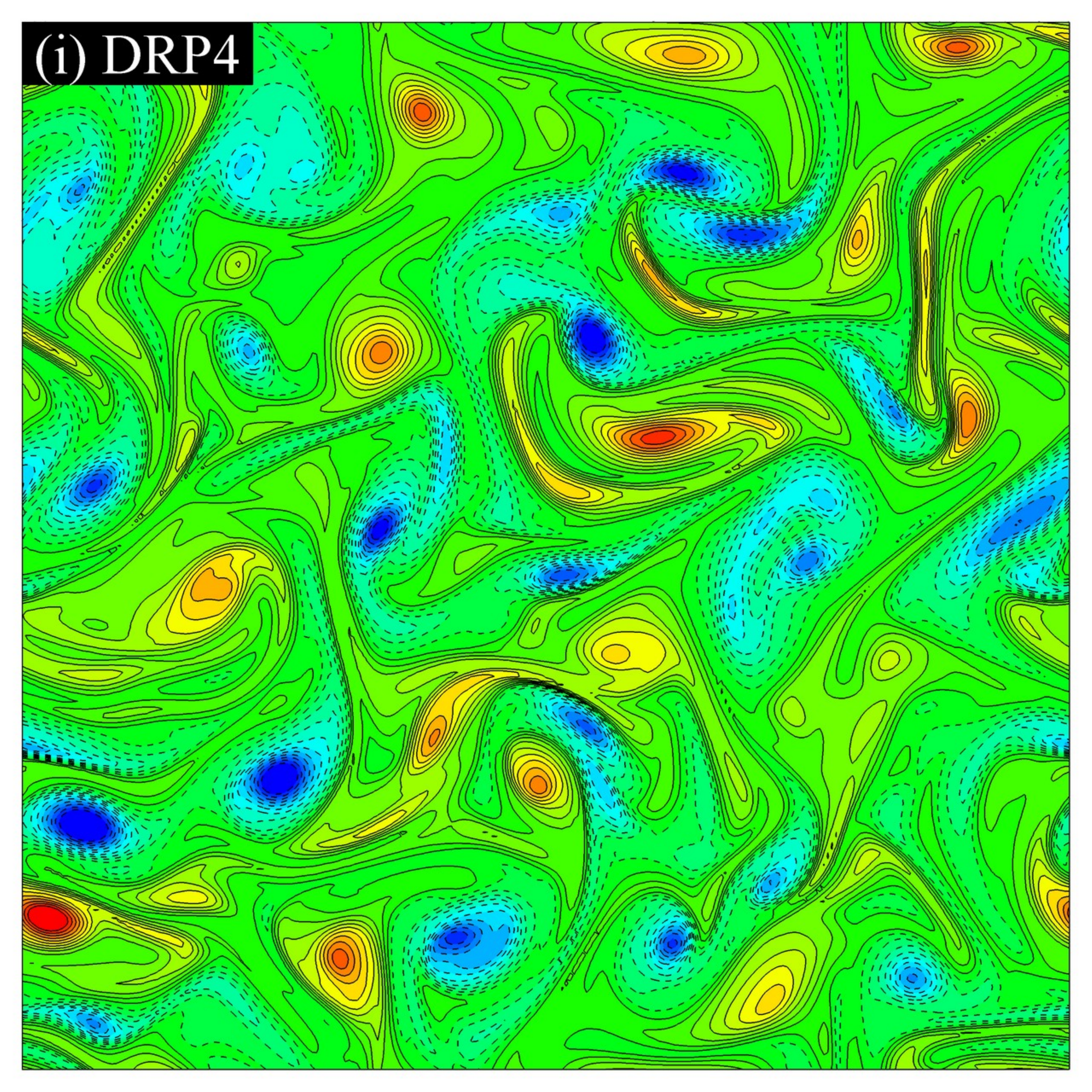}}
}
\caption{Comparison of the numerical schemes at time $t=6$ for $Re=1000$ with a resolution of $256^2$ ($Re_c = 24.54$). (a) pseudospectral(PS) method, (b) sixth-order explicit difference (ED6) method, (c) fourth-order explicit difference (ED4) method, (d) second-order explicit difference (ED2) method, (e) sixth-order compact difference (CD6) method, (f) fourth-order compact difference (CD4) method, (g) fourth-order Arakawa (A4) method, (h) second-order Arakawa (A2) method, and (i) fourth-order dispersion-relation-preserving (DRP4) method. The vorticity contour layouts are identical in all nine cases illustrating 23 equidistant levels in the interval [-11, 11].}
\label{fig:Re1000-256-f}
\end{figure*}

\begin{figure*}
\centering
\mbox{
\subfigure{\includegraphics[width=0.33\textwidth]{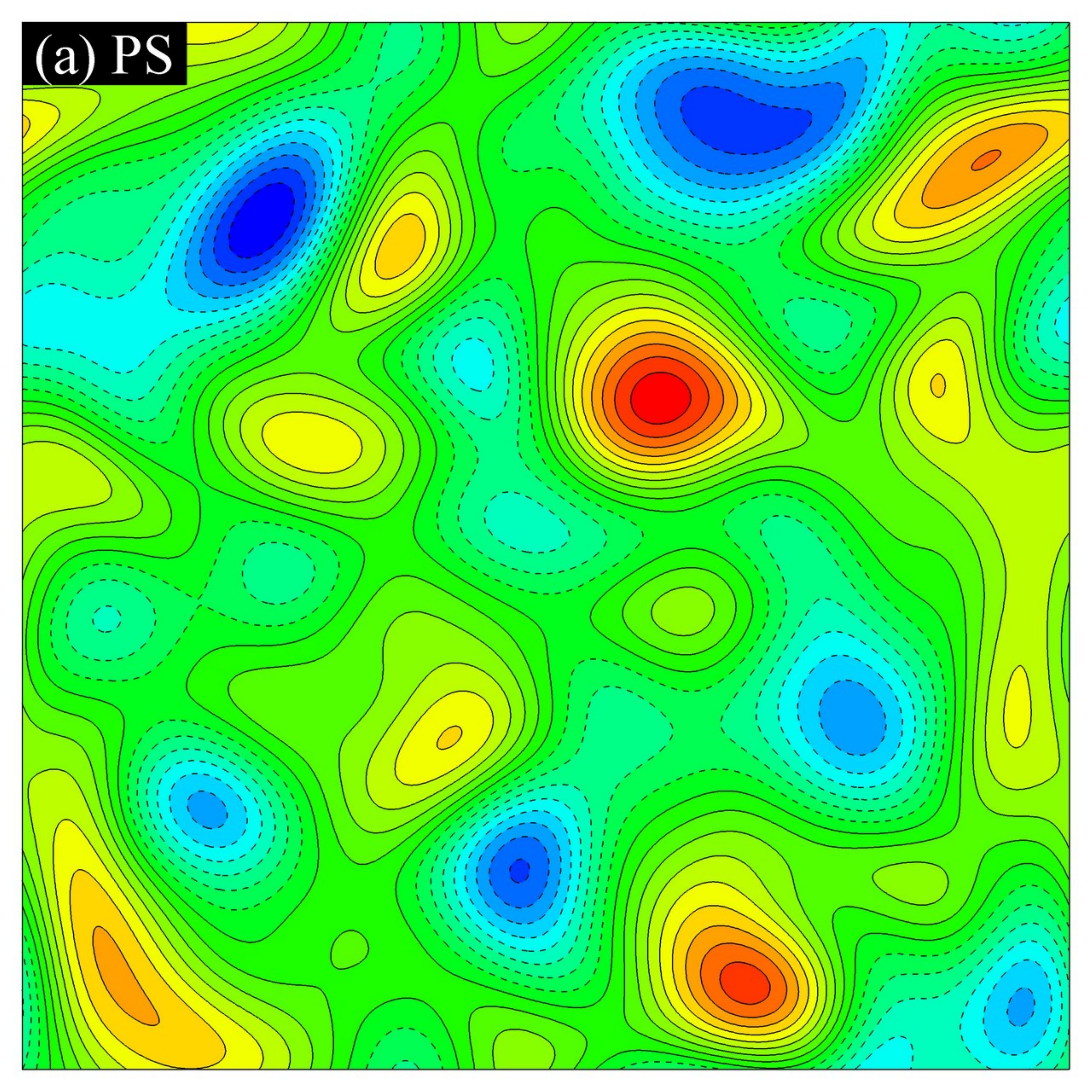}}
\subfigure{\includegraphics[width=0.33\textwidth]{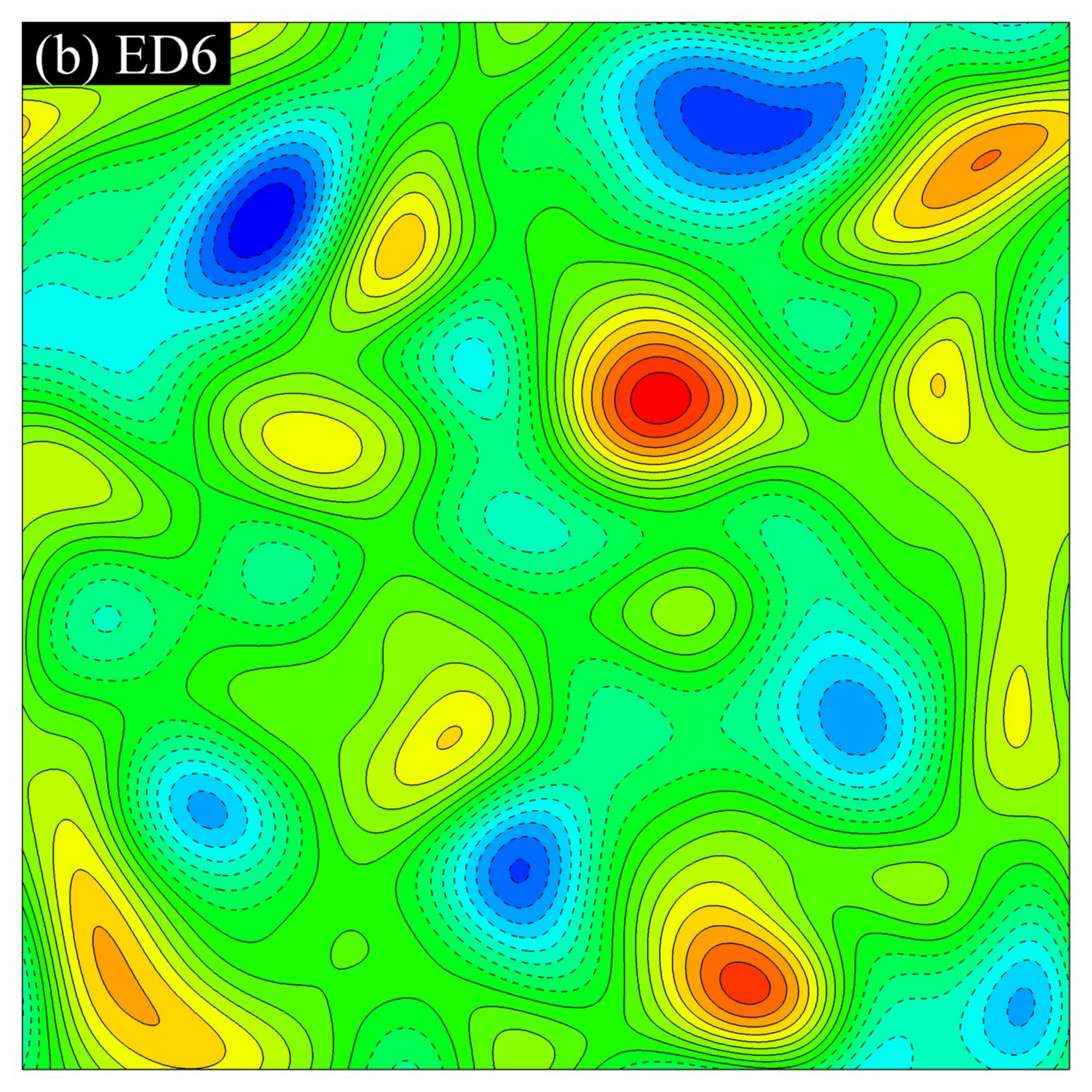}}
\subfigure{\includegraphics[width=0.33\textwidth]{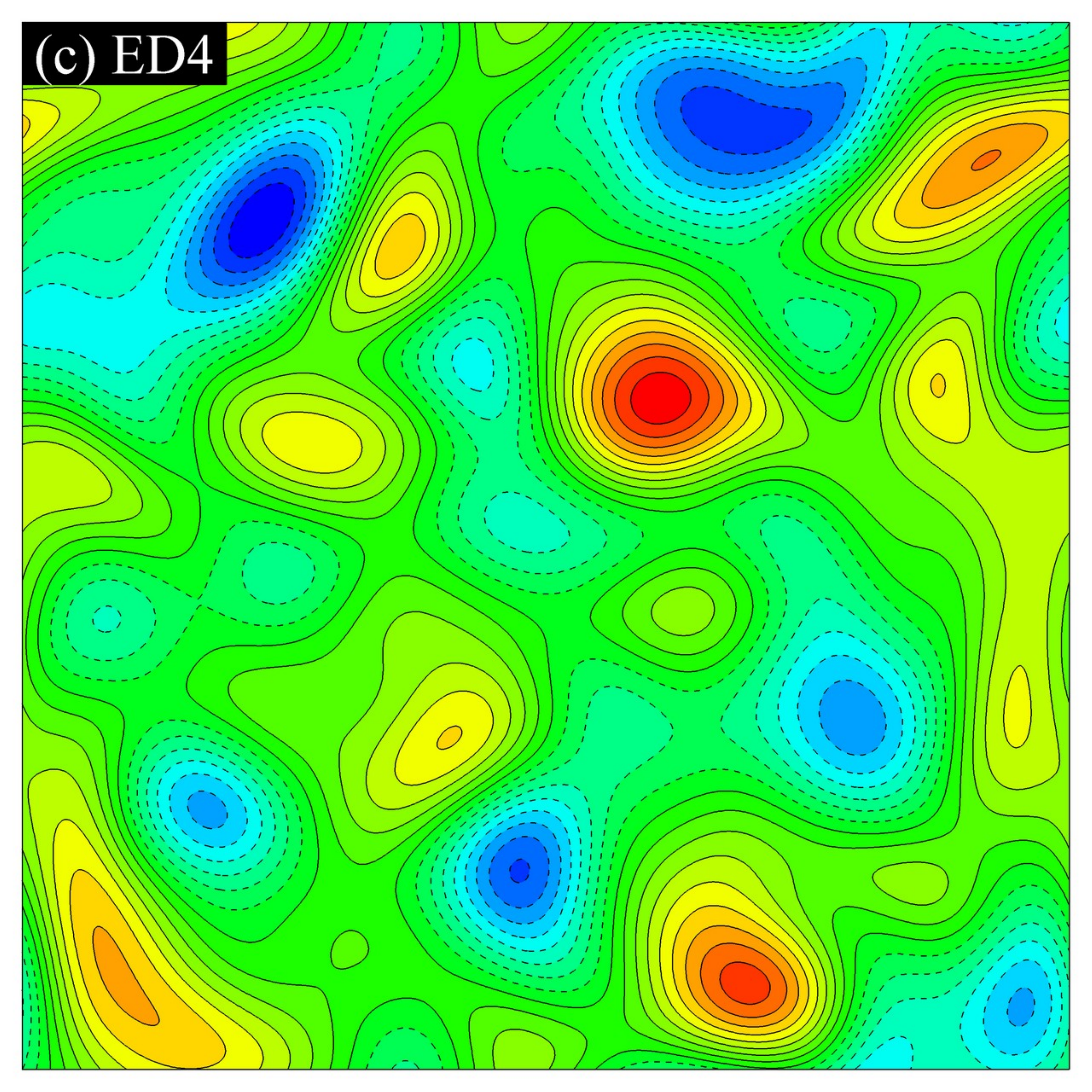}}
}
\\
\mbox{
\subfigure{\includegraphics[width=0.33\textwidth]{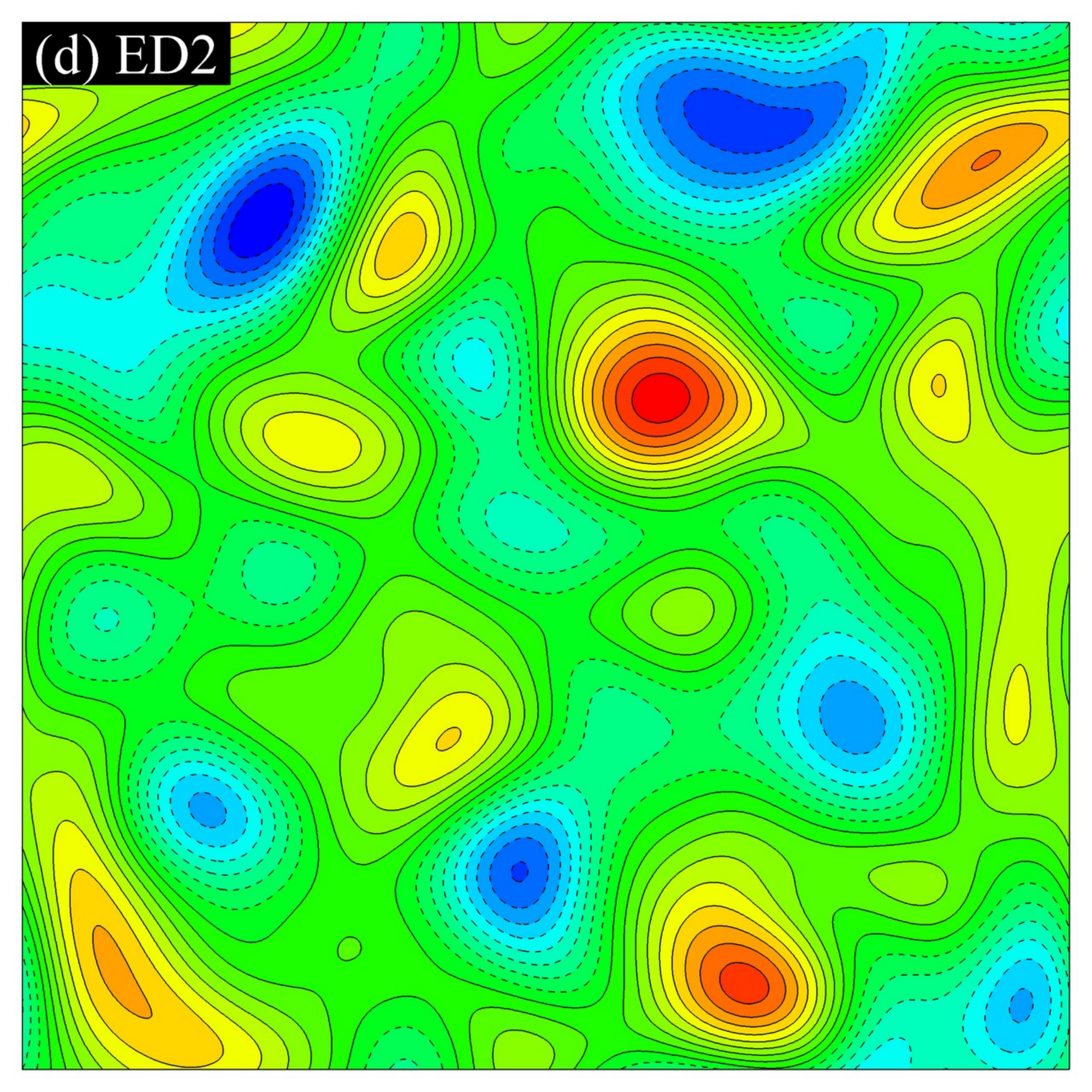}}
\subfigure{\includegraphics[width=0.33\textwidth]{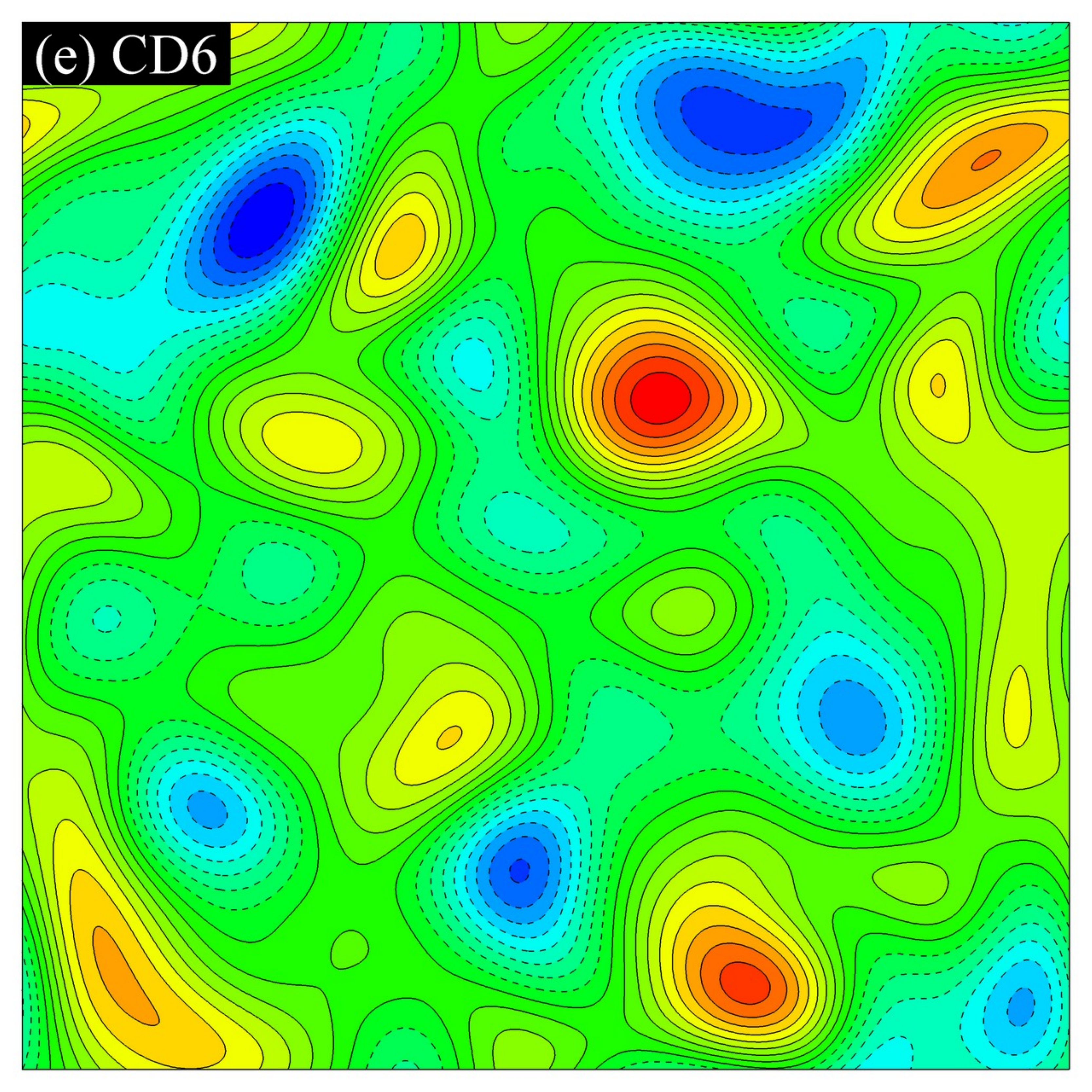}}
\subfigure{\includegraphics[width=0.33\textwidth]{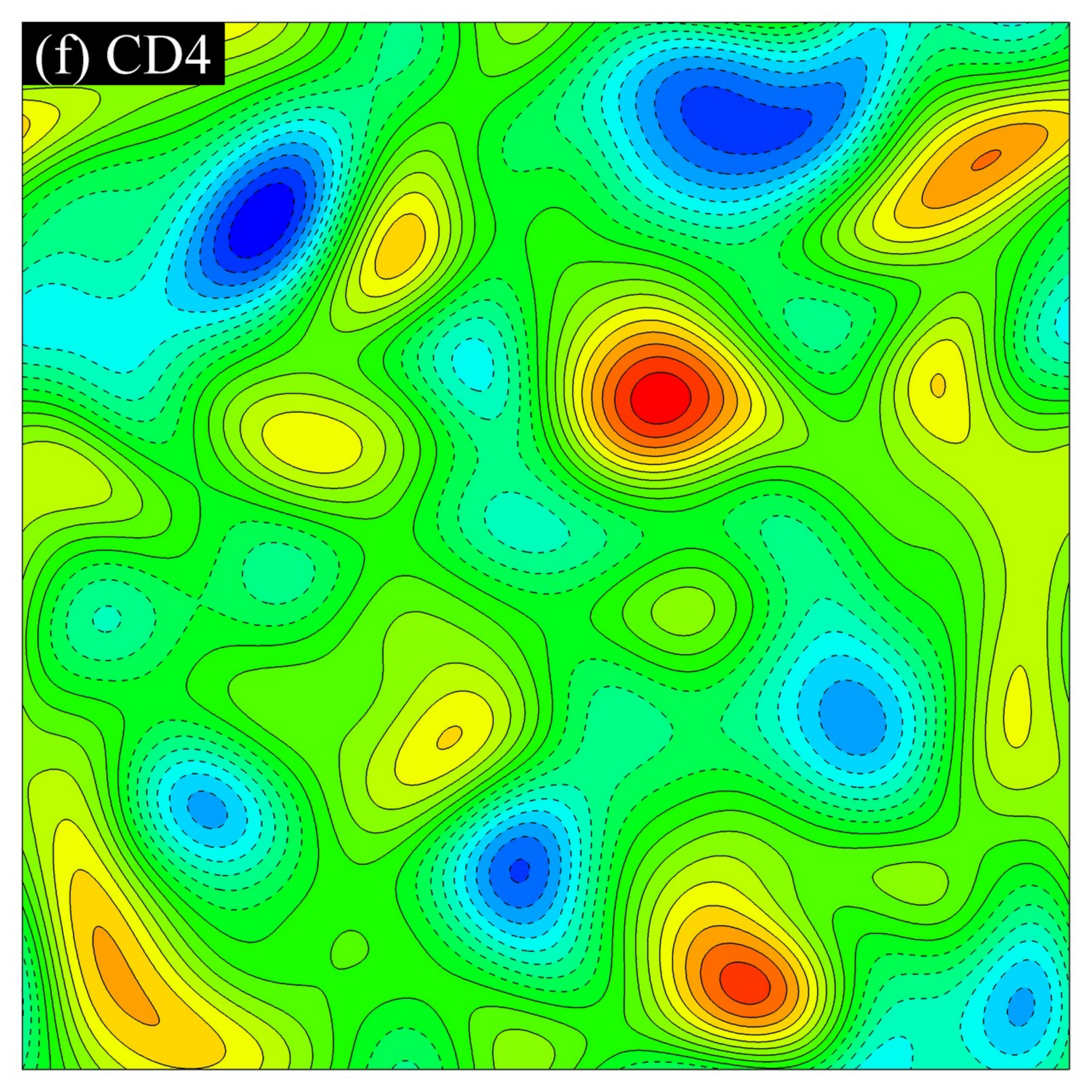}}
}
\\
\mbox{
\subfigure{\includegraphics[width=0.33\textwidth]{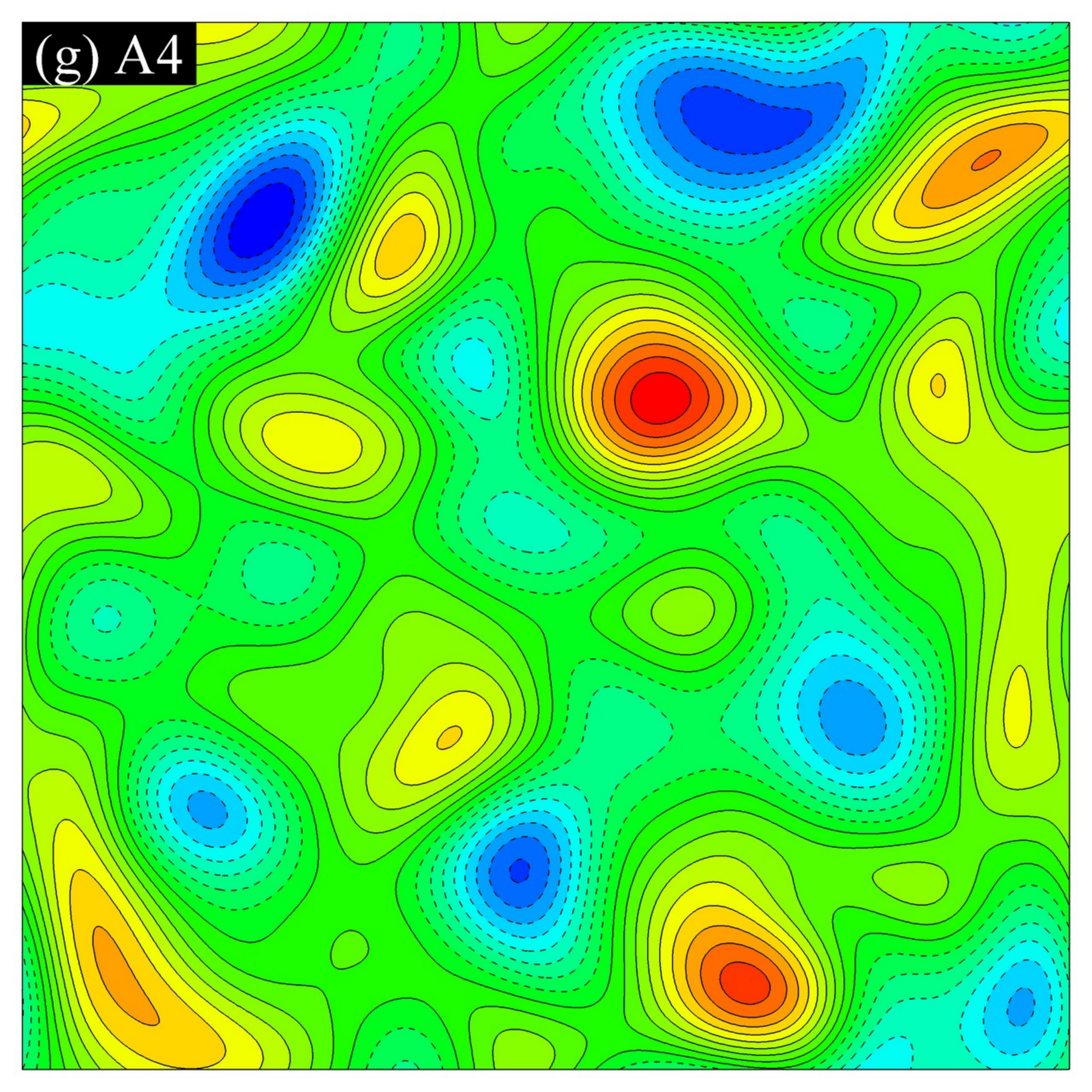}}
\subfigure{\includegraphics[width=0.33\textwidth]{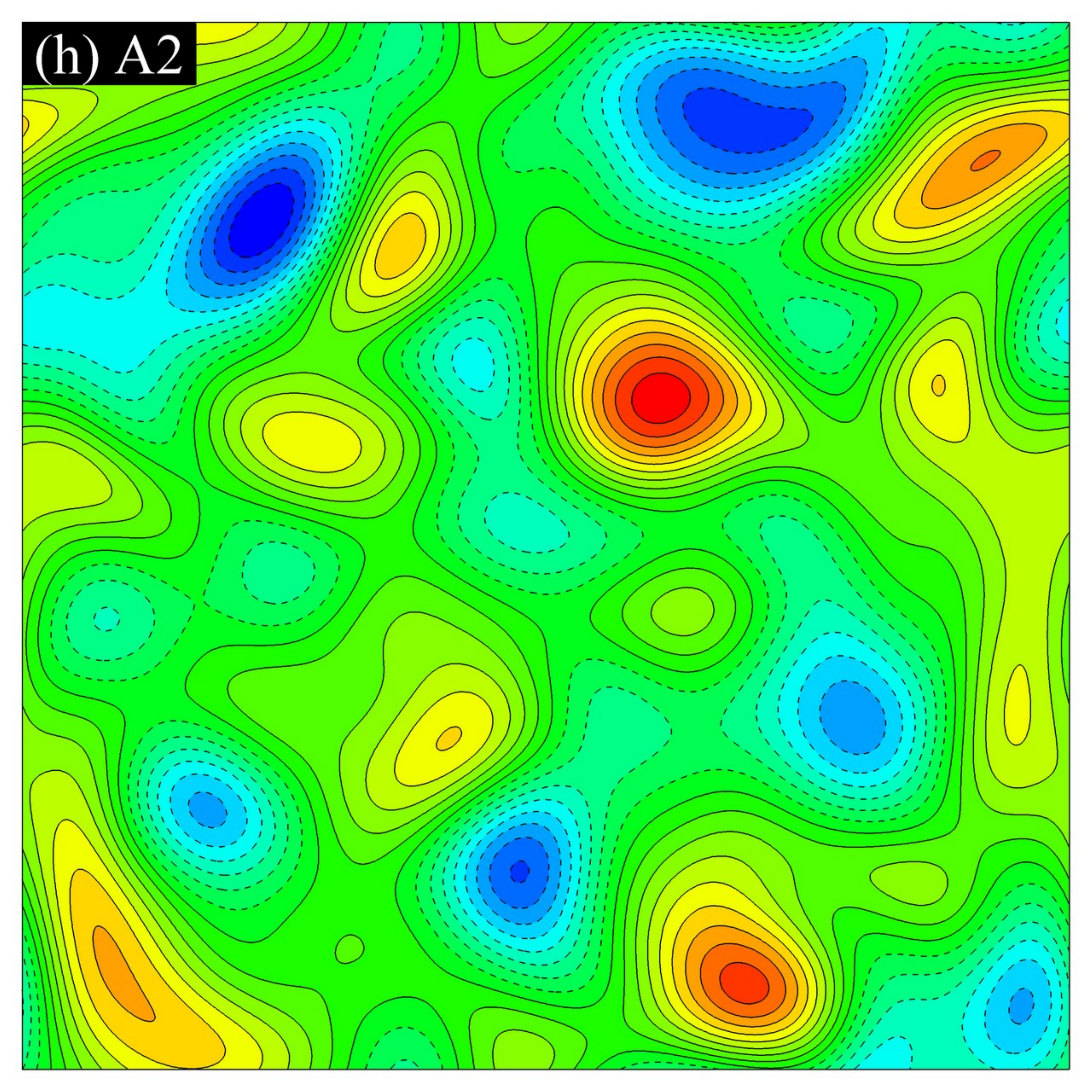}}
\subfigure{\includegraphics[width=0.33\textwidth]{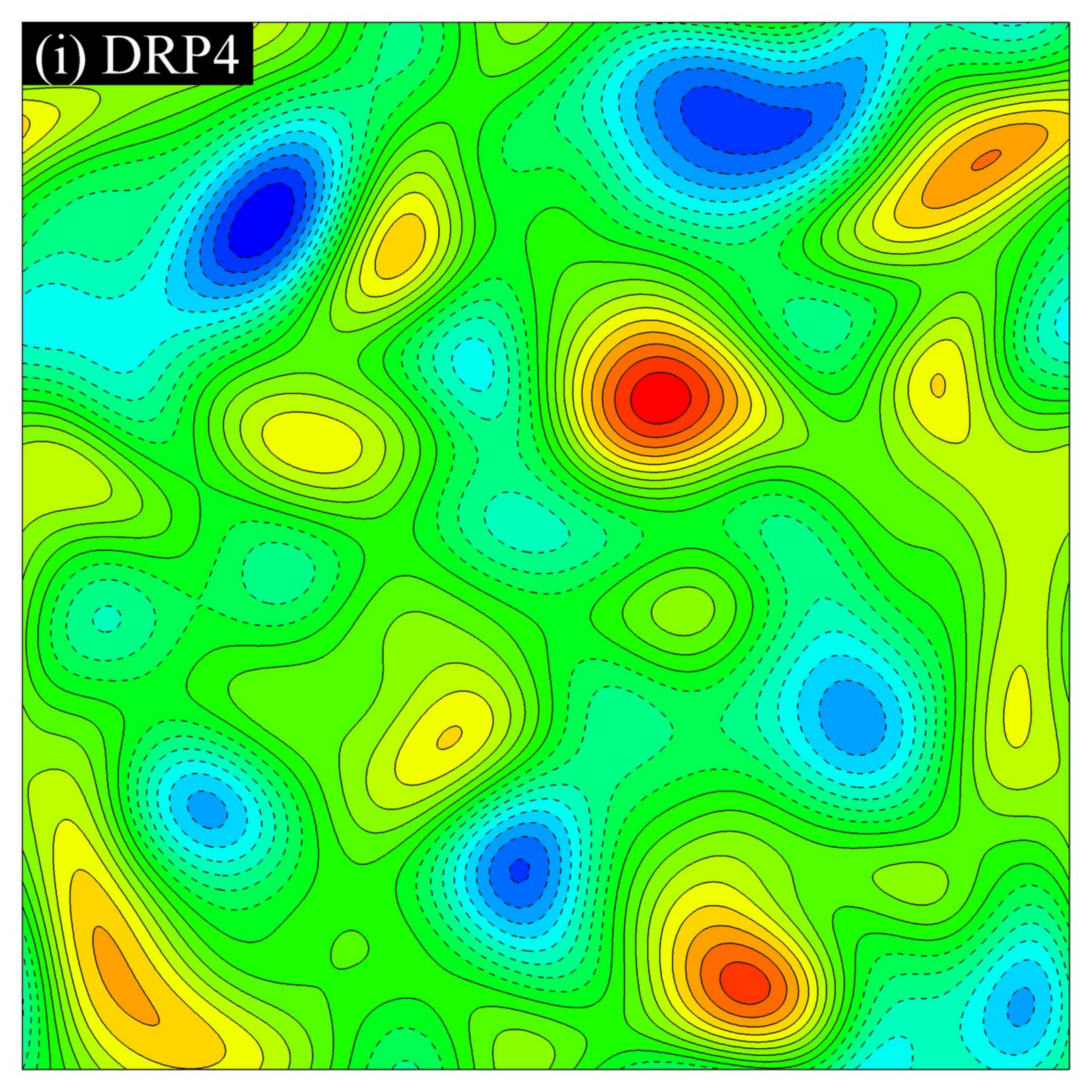}}
}
\caption{Comparison of the numerical schemes at time $t=10$ for $Re=100$ with a resolution of $512^2$ ($Re_c = 1.23$). (a) pseudospectral(PS) method, (b) sixth-order explicit difference (ED6) method, (c) fourth-order explicit difference (ED4) method, (d) second-order explicit difference (ED2) method, (e) sixth-order compact difference (CD6) method, (f) fourth-order compact difference (CD4) method, (g) fourth-order Arakawa (A4) method, (h) second-order Arakawa (A2) method, and (i) fourth-order dispersion-relation-preserving (DRP4) method. The vorticity contour layouts are identical in all nine cases illustrating 19 equidistant levels in the interval [-0.45, 0.45]. }
\label{fig:Re100-512-f}
\end{figure*}
\begin{figure}[h!]
\centering
\includegraphics[width=0.5\textwidth]{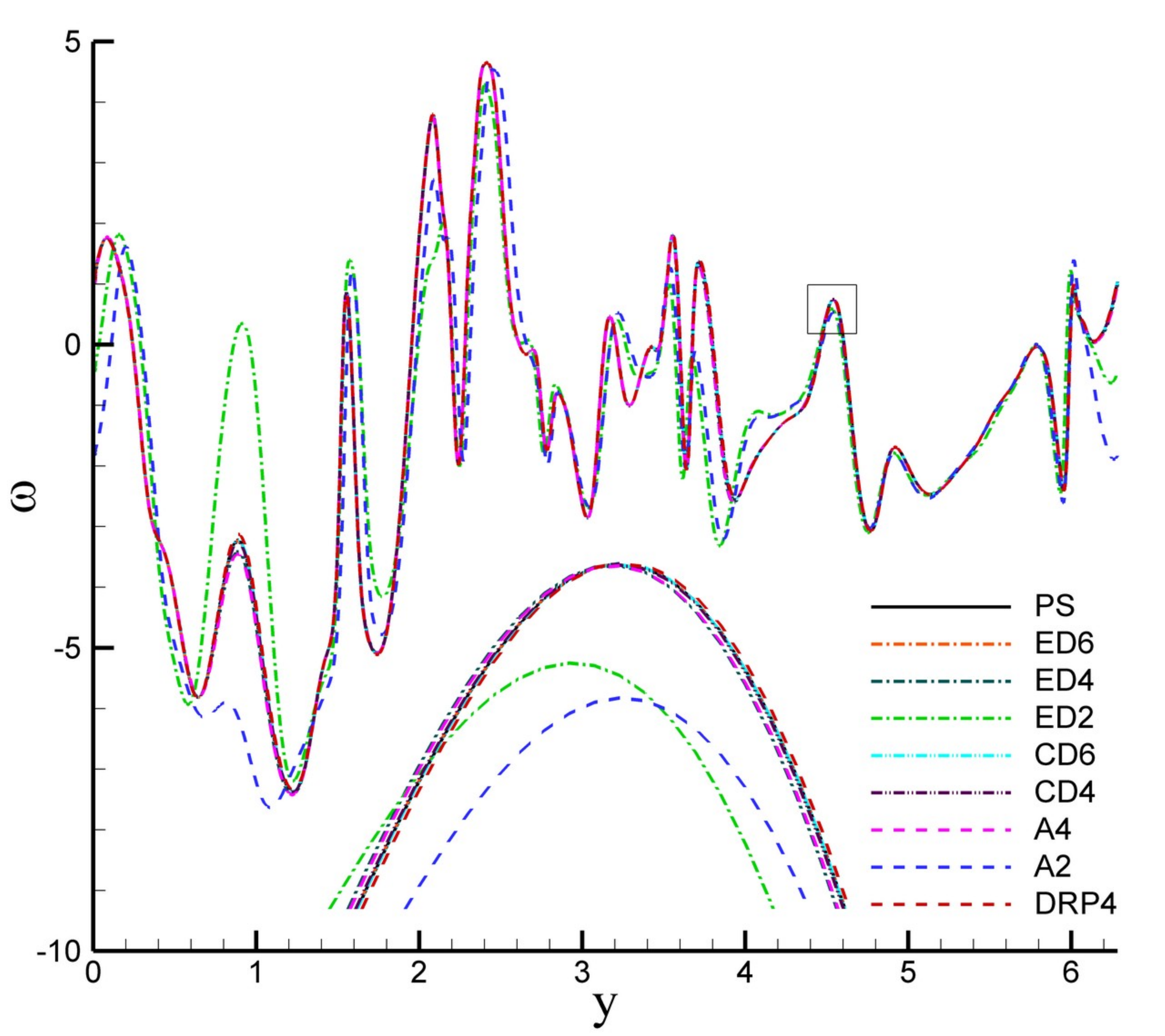}
\caption{Comparison of the numerical schemes for $Re=1000$ with a resolution of $1024^2$ ($Re_c = 6.13$) at time $t=5$. The centerline vorticity distributions at $x=\pi$ are plotted. Inset: close-up of boxed area.}
\label{fig:Re1000-1024-l}
\end{figure}

\begin{figure}[h!]
\centering
\includegraphics[width=0.5\textwidth]{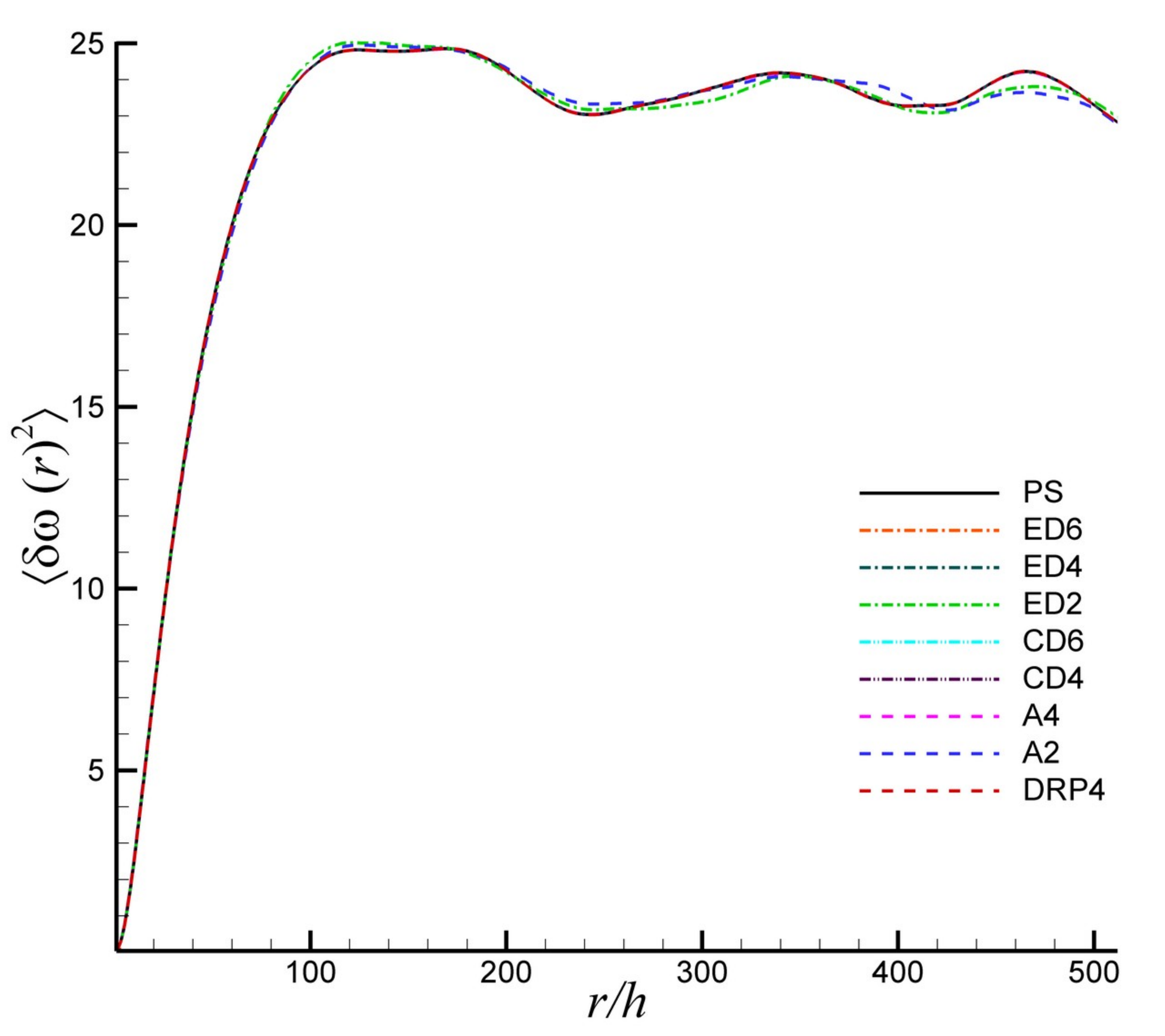}
\caption{Comparison of the second-order vorticity structure functions for $Re=1000$ with a resolution of $1024^2$ ($Re_c = 6.13$) at time $t=5$.}
\label{fig:Re1000-1024-s}
\end{figure}

First, we present a set of numerical experiments using a resolution of $1024^2$ for $Re=1000$. The vorticity fields obtained by nine different spatial schemes at time $t=5$ are presented and compared in Fig.~\ref{fig:Re1000-1024-f}. A comparison of the vorticity distributions on the vertical centerline is also illustrated in Fig.~\ref{fig:Re1000-1024-l} by showing also a close-up figure to make the comparison more clear. Furthermore, a comparison of statistical behavior in terms of the second-order vorticity structure functions is also demonstrated in Fig.~\ref{fig:Re1000-1024-s}. As we can see, the high-order schemes and the pseudospectral schemes yield similar flow fields. The second-order accurate schemes, both the explicit difference and the Arakawa schemes, predict slightly less accurate results. The sixth-order compact scheme exhibits the best prediction capability among the different formulations tested. In evaluating the performance of the different finite difference approximations, the deviation of vorticity field from the spectral solution is calculated by using the definition of discrete $L_2$ norm given by Eq.~(\ref{eq:err}). Table~\ref{tab:turb1} shows the $L_2$ norms, CPU times, and corresponding speed-up ratios for the finite difference formulations. The speed-up ratio is defined here as the ratio of CPU times for the computations performed using the pseudospectral method and those performed using finite difference schemes. First, all the finite difference methods are considerably more efficient than the pseudospectral method. The data in the table exhibits that the high-order schemes provide the same accuracy level as the Fourier-Galerkin pseudospectral method, while significantly decreasing its computational cost with a speed-up factor of 5 over the pseudospectral method. Among the fourth-order schemes, both the compact and dispersion-relation-preserving schemes show a better performance than the Arakawa and explicit difference schemes. This clearly shows that designing a scheme with reduced truncation error or optimizing the scheme with respect to the dispersion relation could result in a better prediction capability than the designing a scheme with its discrete conservation properties.

\begin{table}[t!]
\caption{Accuracy and efficiency of the finite difference approximations for the decaying turbulence problem at $Re=10^{3}$ using a resolution of $1024^2$. The reference solution for the $L_2$ norms is obtained by the pseudo-specral method which takes a CPU time of 621.32 hrs. The speed-up ratio is defined as the ratio of CPU times.}
\begin{center}
\label{tab:turb1}
\begin{tabular}{llclclc}
\hline
\smallskip
Scheme &\quad  &$\|\omega\|_{L_2}$ & \quad  & CPU (hrs.) &\quad  & Speed-up \\
\hline
ED6 & &1.96E-3& &128.39& &4.84  \\
ED4 & &3.51E-2& &124.12& &5.00  \\
ED2 & &1.07   & &121.63& &5.11  \\
CD6 & &1.63E-4& &149.52& &4.16  \\
CD4 & &6.23E-3& &142.79& &4.35  \\
A4  & &3.29E-2& &148.75& &4.17  \\
A2  & &1.11   & &127.29& &4.88  \\
DRP4& &1.80E-2& &126.59& &4.90  \\
\hline
\end{tabular}
\end{center}
\end{table}

\begin{figure}[h!]
\centering
\includegraphics[width=0.5\textwidth]{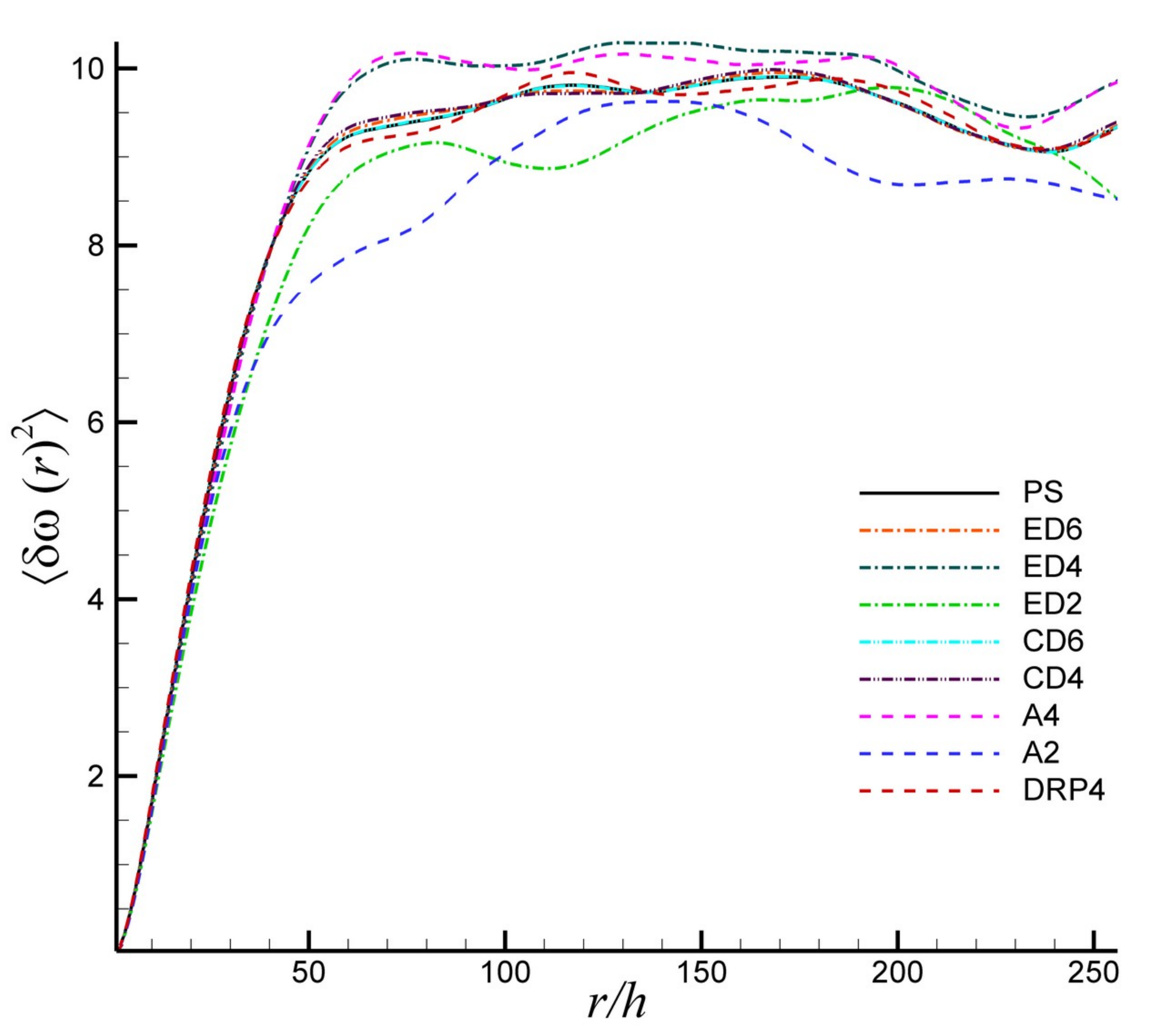}
\caption{Comparison of the second-order vorticity structure functions for $Re=1000$ with a resolution of $512^2$ ($Re_c = 12.27$) at time $t=10$.}
\label{fig:Re1000-512-s}
\end{figure}

\begin{figure}[h!]
\centering
\includegraphics[width=0.5\textwidth]{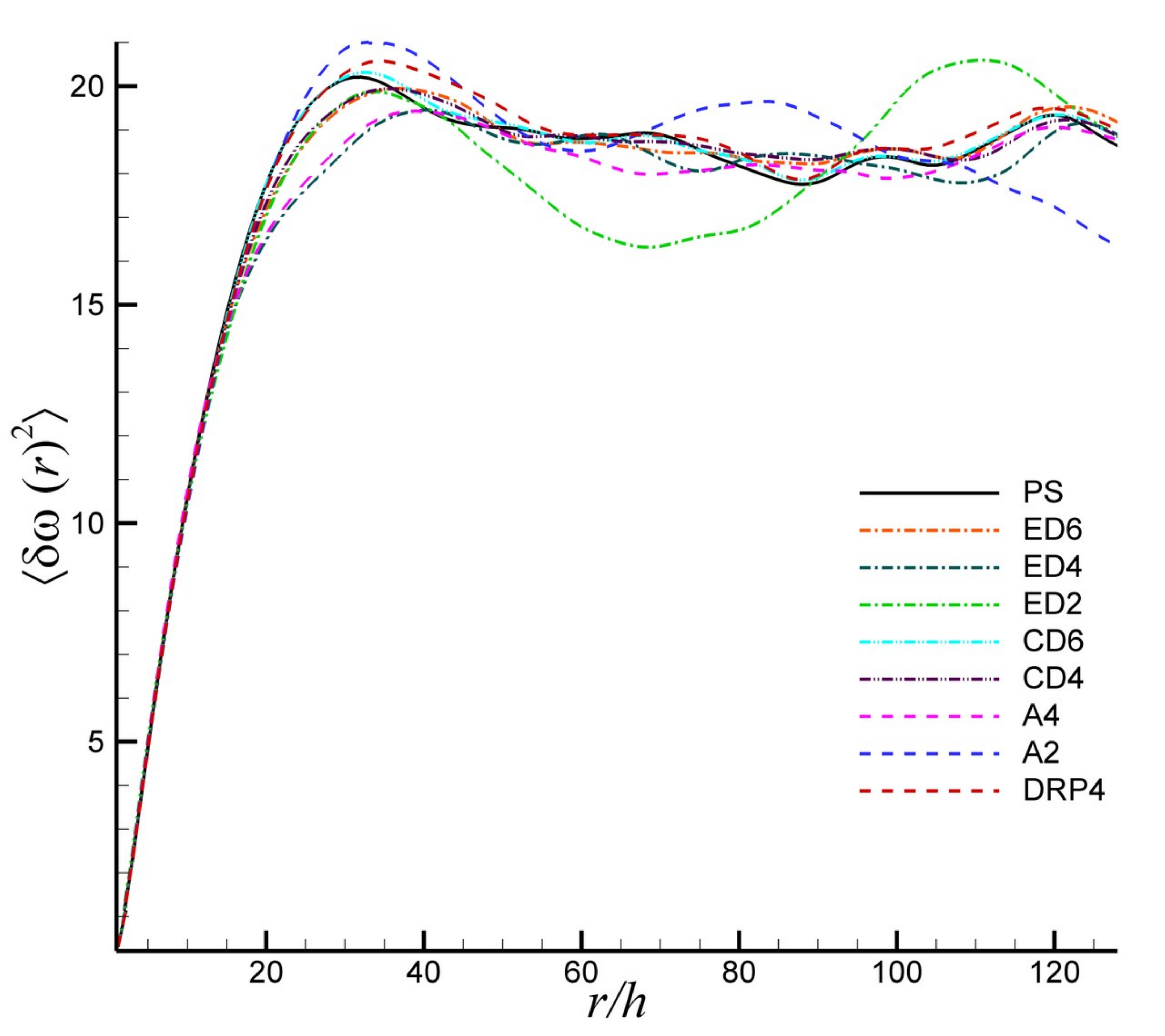}
\caption{Comparison of the second-order vorticity structure functions for $Re=1000$ with a resolution of $256^2$ ($Re_c = 24.54$) at time $t=6$.}
\label{fig:Re1000-256-s}
\end{figure}

In order to analyze the effects of the Reynolds number and required resolution to the performance of schemes, it is useful to define a cell Reynolds number in the following form
\begin{equation}
Re_c=Re\frac{2\pi}{N_x}
\label{eq:rec}
\end{equation}
Therefore, in our previous experiments having a resolution of $1024^2$ at $Re=1000$, the cell Reynolds number is $Re_c = 6.13$. Next, we perform similar experiments for $Re=1000$ with $512^2$ resolutions with an increasing cell Reynolds number of $Re_c = 12.27$. The vorticity fields at time $t=10$ obtained by nine different spatial schemes are plotted in Fig.~\ref{fig:Re1000-512-f}. The corresponding second-order vorticity structure functions are also illustrated in Fig.~\ref{fig:Re1000-512-s}. We can easily see that the difference between high-order and low-order schemes becomes amplified, and there is a substantial difference in the corresponding flow fields. If we decrease the resolution to the $256^2$ resulting in a higher cell Reynolds number, $Re_c = 24.54$, as shown in Fig.~\ref{fig:Re1000-256-f} and Fig.~\ref{fig:Re1000-256-s}, low-order accurate numerical schemes produce totally different flow fields. As the result of the series of calculations performed in the second and third sets of experiments, we can see that the sixth-order accurate schemes always predict the same accuracy results as the pseudospectral methods. In terms of fourth-order accurate schemes, we can see that the compact difference scheme produces a slightly better accuracy than the dispersion-relation-preserving scheme due to its lower truncation error properties. Between the compact scheme and Arakawa scheme, the compact scheme exhibits better accuracy than the Arakawa scheme. The computational efficiencies in terms of speed-up ratios for these sets of experiments are very similar to the previous high resolution computations given by Table~\ref{tab:turb1}. Making a comparison between low-order and high-order schemes, we can see that a slight increase in the computational time for a higher-order accurate scheme can result in significantly more accurate results than those obtained by the conventional second-order accurate schemes, and provide the same level of accuracy with the pseudospectral method in much less computational time.

\begin{figure}[h!]
\centering
\includegraphics[width=0.5\textwidth]{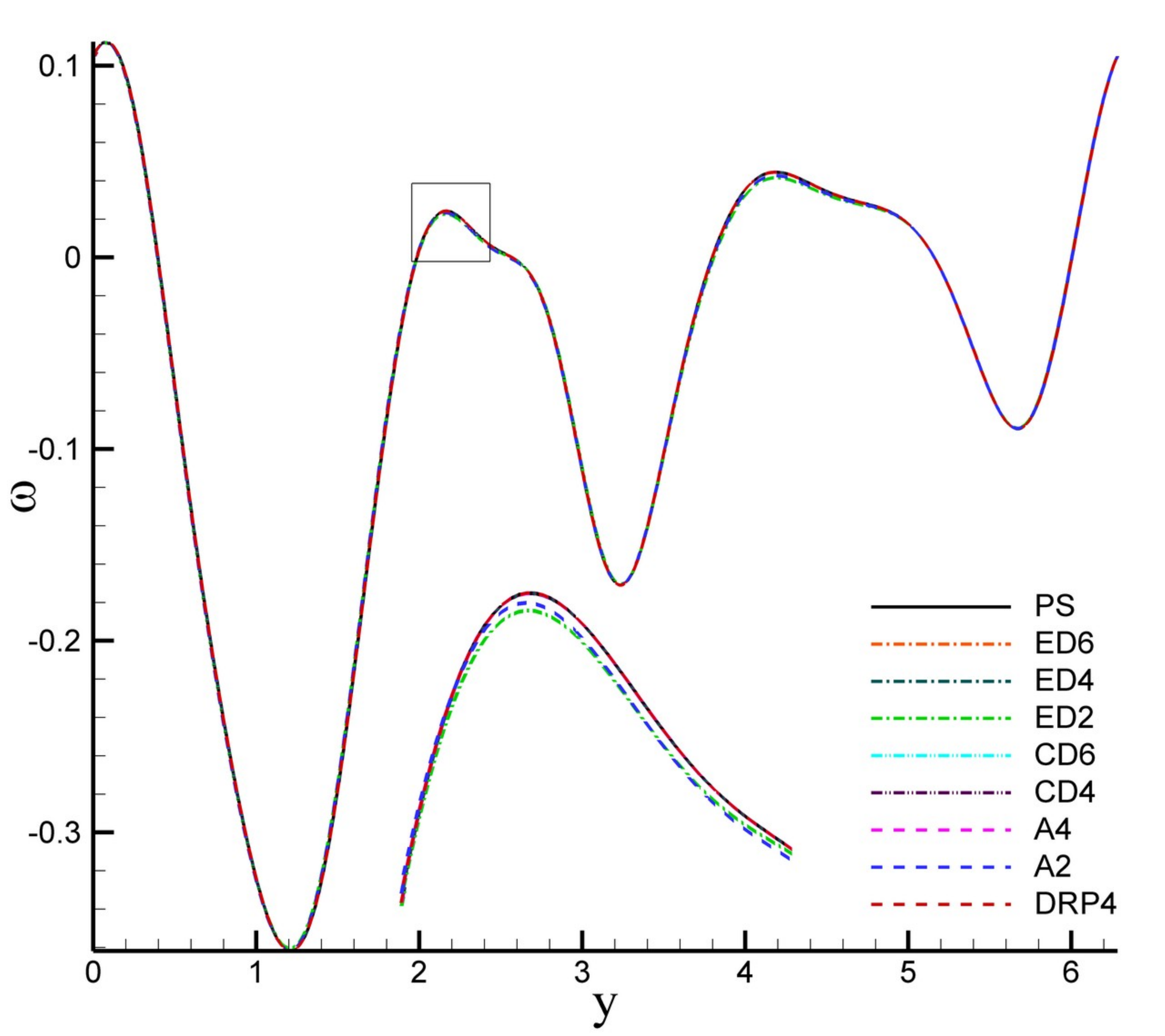}
\caption{Comparison of the numerical schemes for $Re=100$ with a resolution of $512^2$ ($Re_c = 1.23$) at time $t=10$. The centerline vorticity distributions at $x=\pi$ are plotted. Inset: close-up of boxed area. }
\label{fig:Re100-512-l}
\end{figure}

We perform one more set of numerical experiments in order to demonstrate that the results of the various spatial schemes converge the same flow field under well resolved conditions using a resolution of $512^2$ at a reduced $Re=100$ for which $Re_c = 1.23$. The vorticity contour plots obtained by these nine schemes are demonstrated in Fig.~\ref{fig:Re100-512-f} which shows that there is negligible difference among them. Both low-order and high-order difference schemes predict the same flow field with the pseudospectral method. This can also be seen clearly from Fig.~\ref{fig:Re100-512-l} showing the centerline vorticity distributions. Since the cell Reynolds number is smaller than 2, both second-order accurate and high-order accurate schemes produce similar results (i.e., well resolved direct numerical simulations). However, we can see that for larger cell Reynolds numbers, higher-order accurate schemes are required.

\subsection{Pile-up phenomenon for high cell Reynolds number}
\label{sec:pile}
\begin{figure*}
\centering
\mbox{
\subfigure[ED4 for $Re=3000$]{\includegraphics[width=0.43\textwidth]{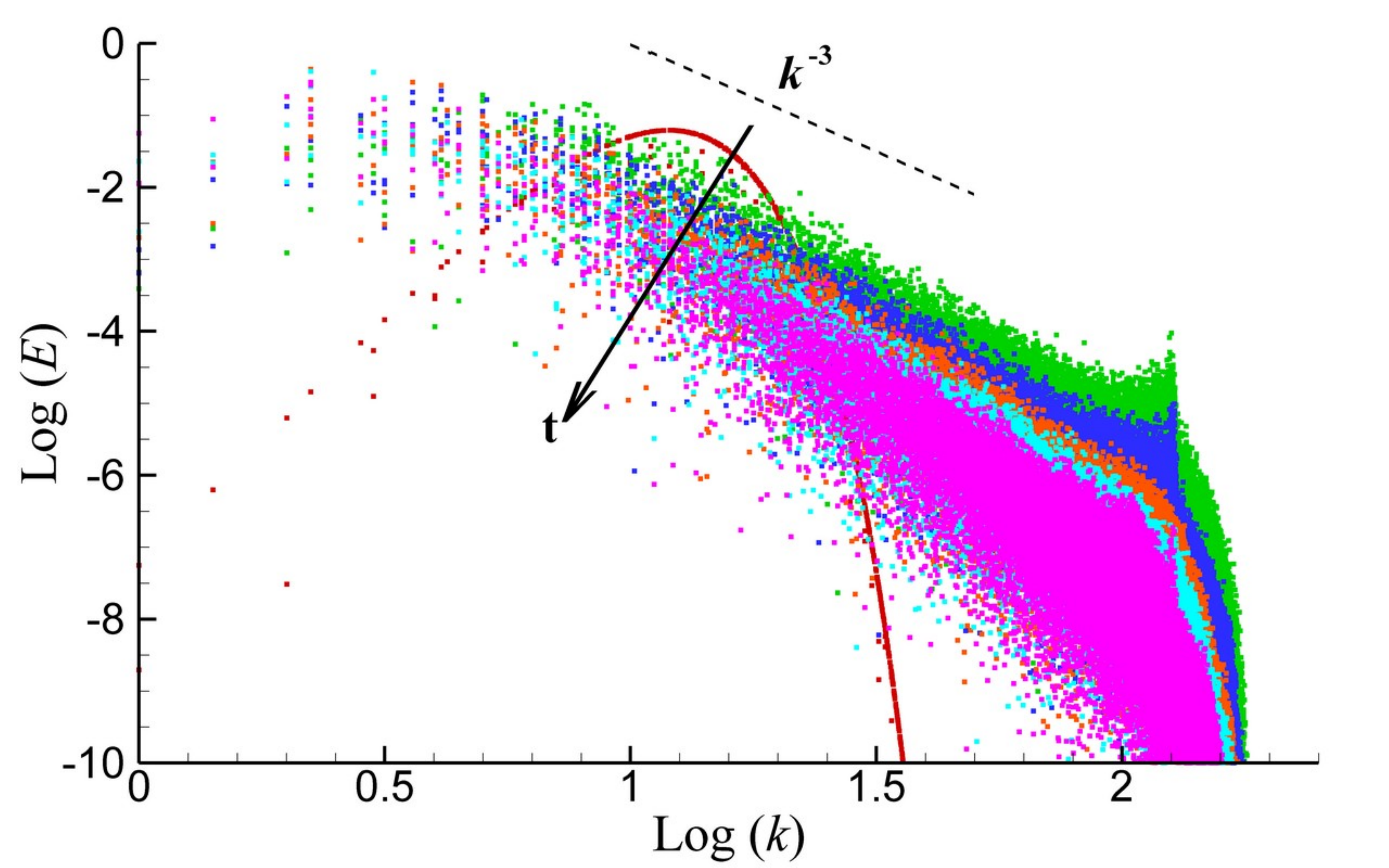}}
\subfigure[ED4 for $Re=6000$]{\includegraphics[width=0.43\textwidth]{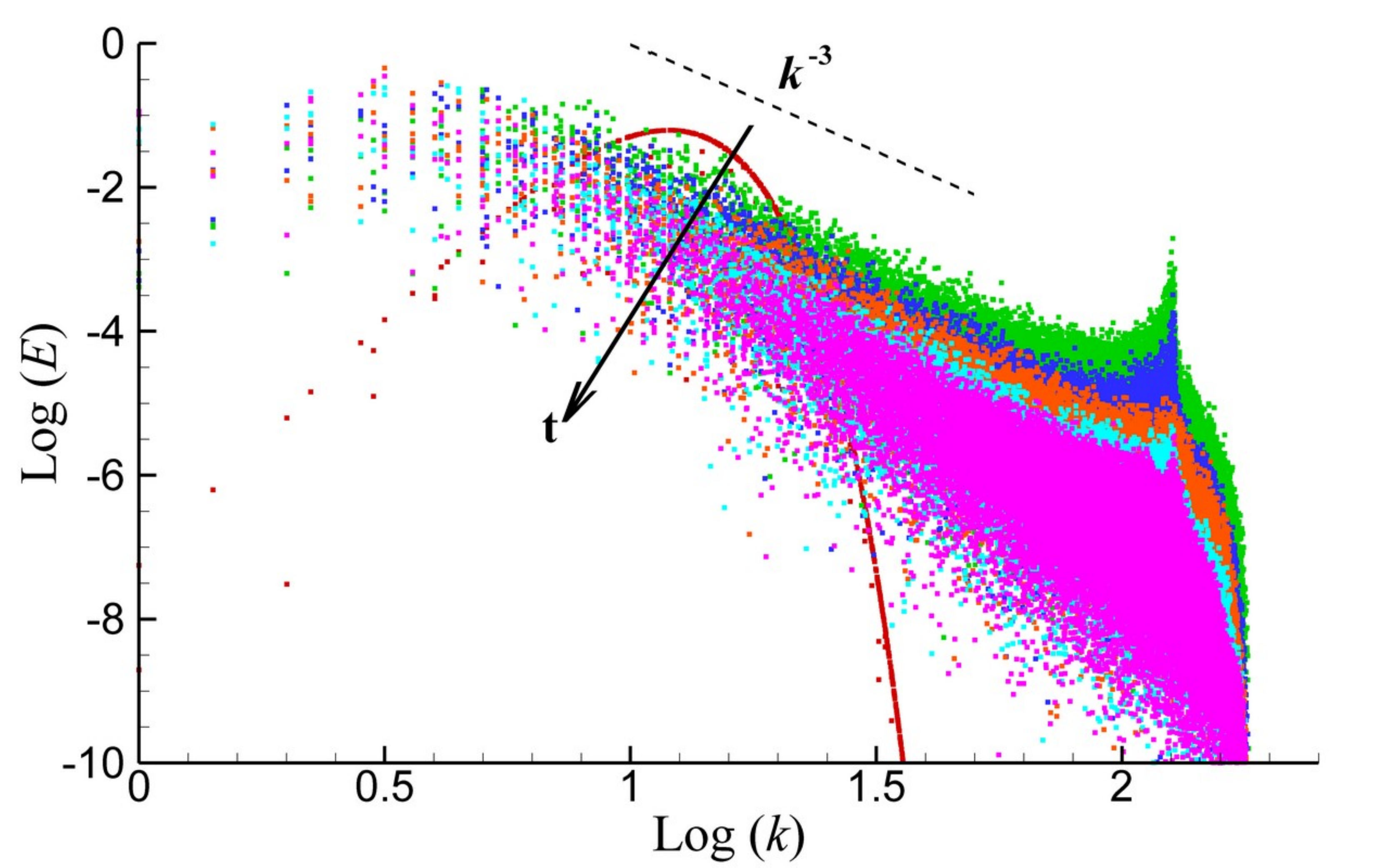}}
} \\
\mbox{
\subfigure[CD4 for $Re=3000$]{\includegraphics[width=0.43\textwidth]{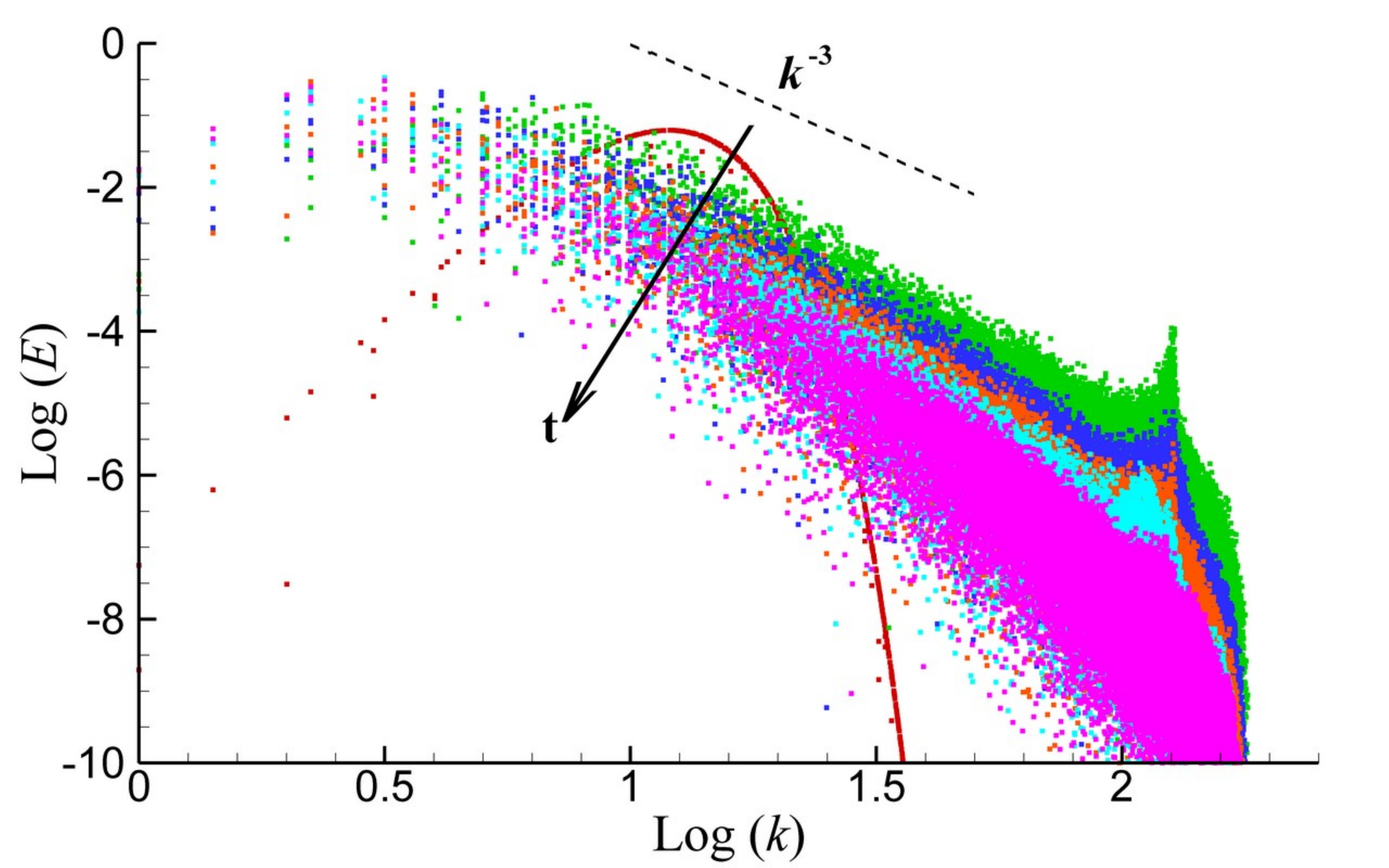}}
\subfigure[CD4 for $Re=6000$]{\includegraphics[width=0.43\textwidth]{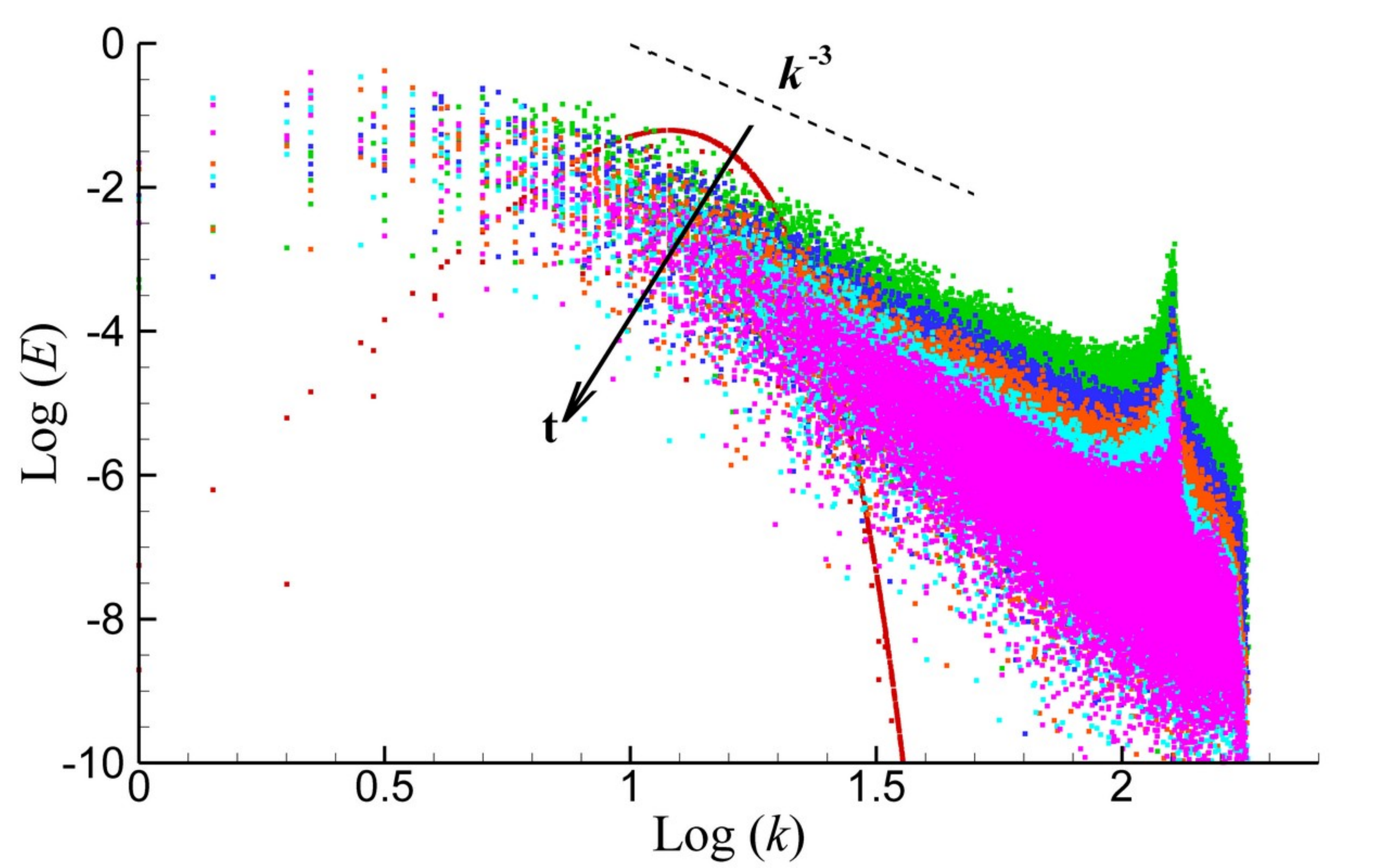}}
} \\
\mbox{
\subfigure[A4 for $Re=3000$]{\includegraphics[width=0.43\textwidth]{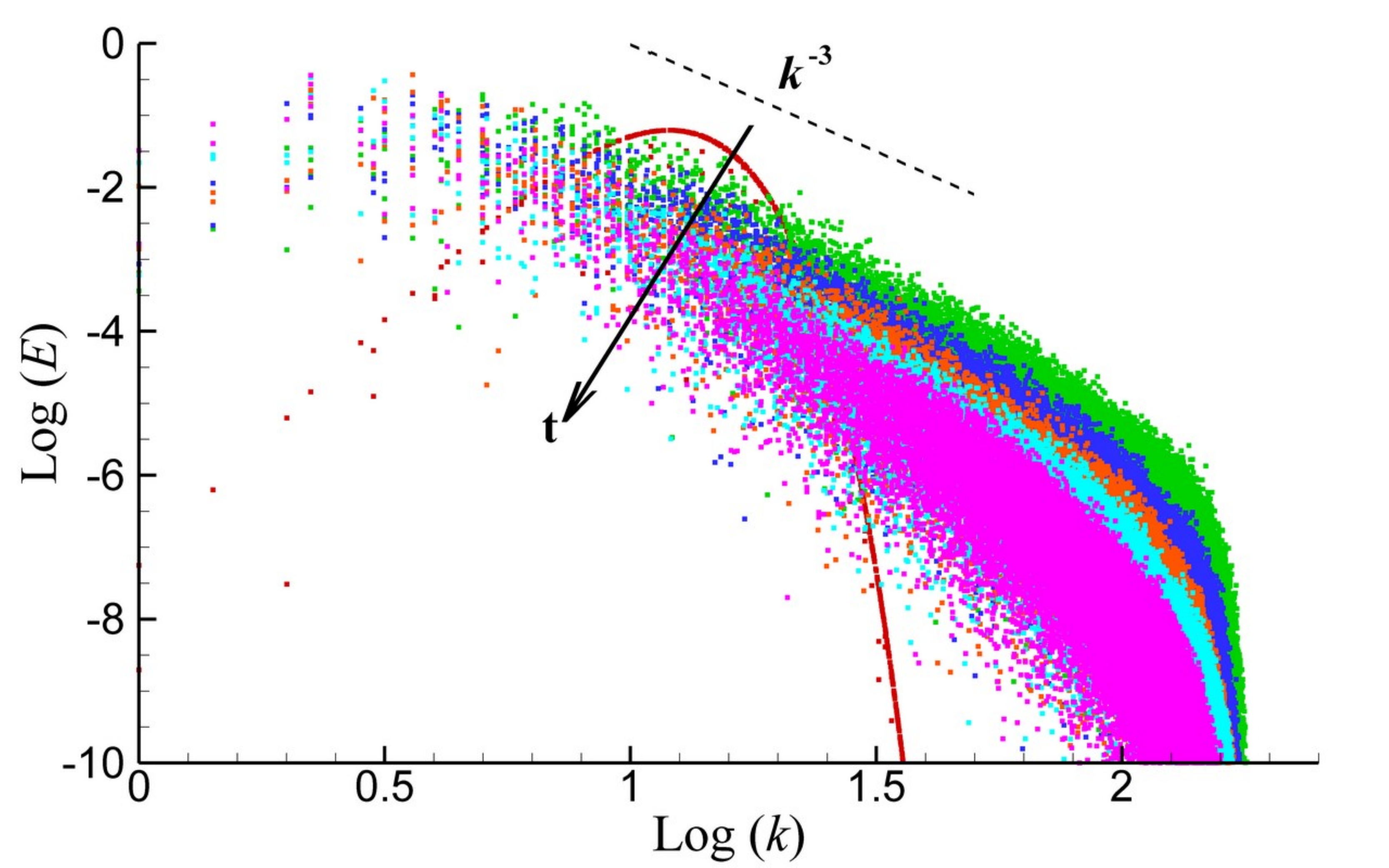}}
\subfigure[A4 for $Re=6000$]{\includegraphics[width=0.43\textwidth]{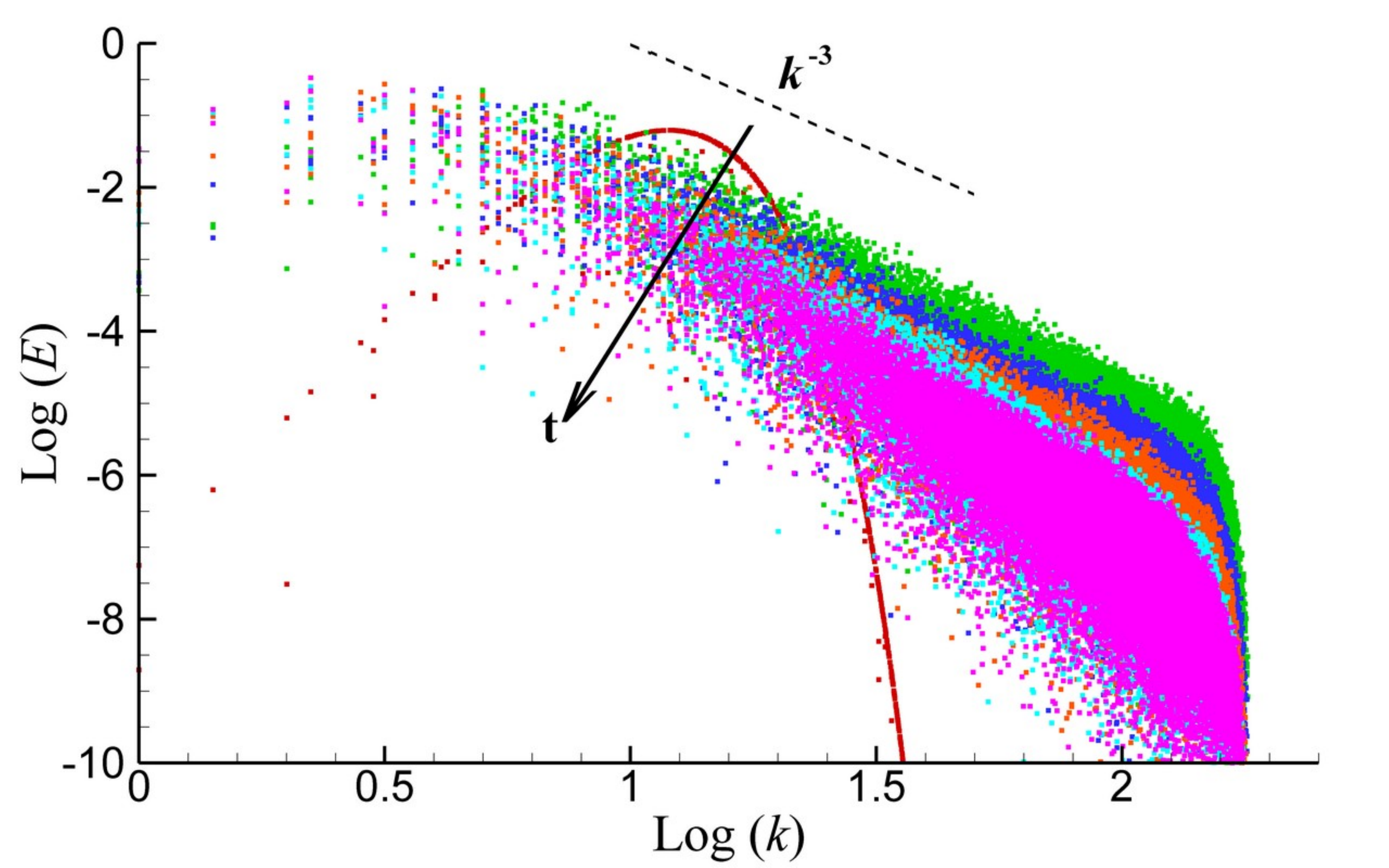}}
} \\
\mbox{
\subfigure[DRP4 for $Re=3000$]{\includegraphics[width=0.43\textwidth]{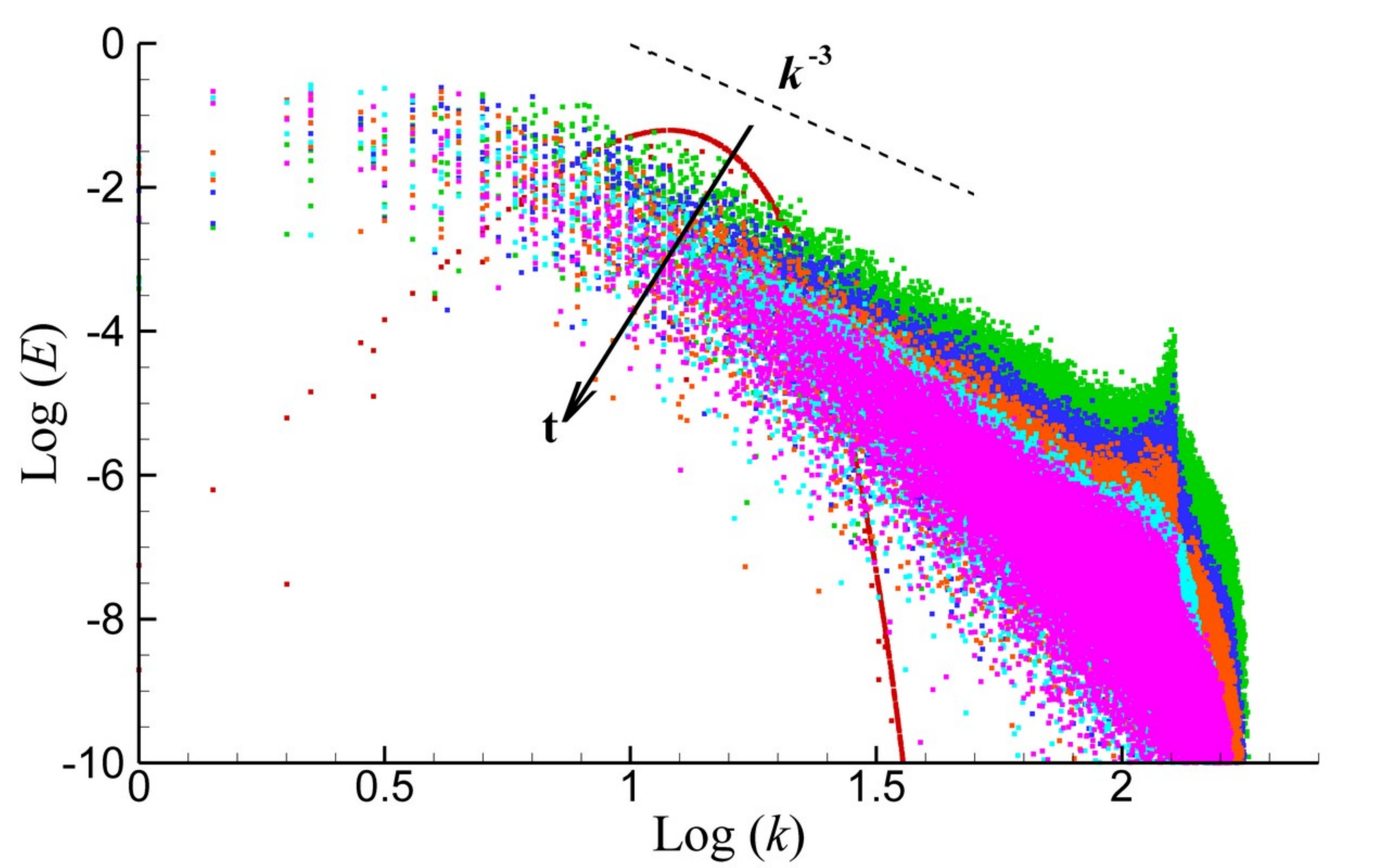}}
\subfigure[DRP4 for $Re=6000$]{\includegraphics[width=0.43\textwidth]{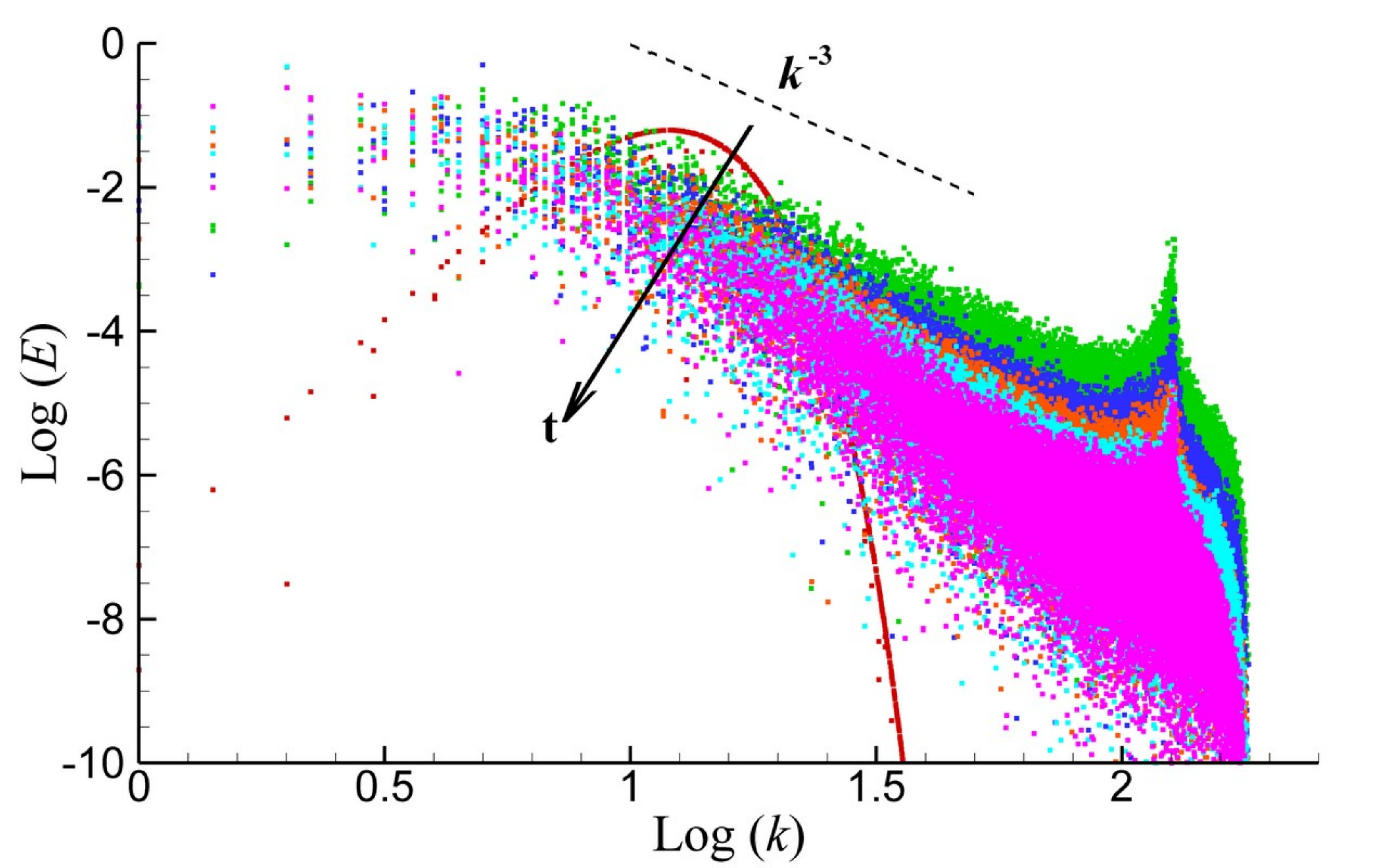}}
}
\caption{Evolution of energy spectra in decaying turbulence for two sets of Reynolds numbers: $Re=3000$ ($Re_c = 73.63$) in the left column and $Re=6000$ ($Re_c = 147.26$) in the right column. Comparison of the fourth-order finite difference schemes using a resolution of $256^2$: (a-b) explicit difference (ED4), (c-d) compact difference (CD4), (e-f) Arakawa (A4), and (g-h) dispersion-relation-preserving (DRP4) schemes. Energy spectra, defined by Eq.~(\ref{eq:esp}), are shown for times $t=0$, 2, 4, 6, 8, and 10.}
\label{fig:com-1}
\end{figure*}
\begin{figure*}[t!]
\centering
\mbox{
\subfigure[CD6 for $Re=1000$]{\includegraphics[width=0.5\textwidth]{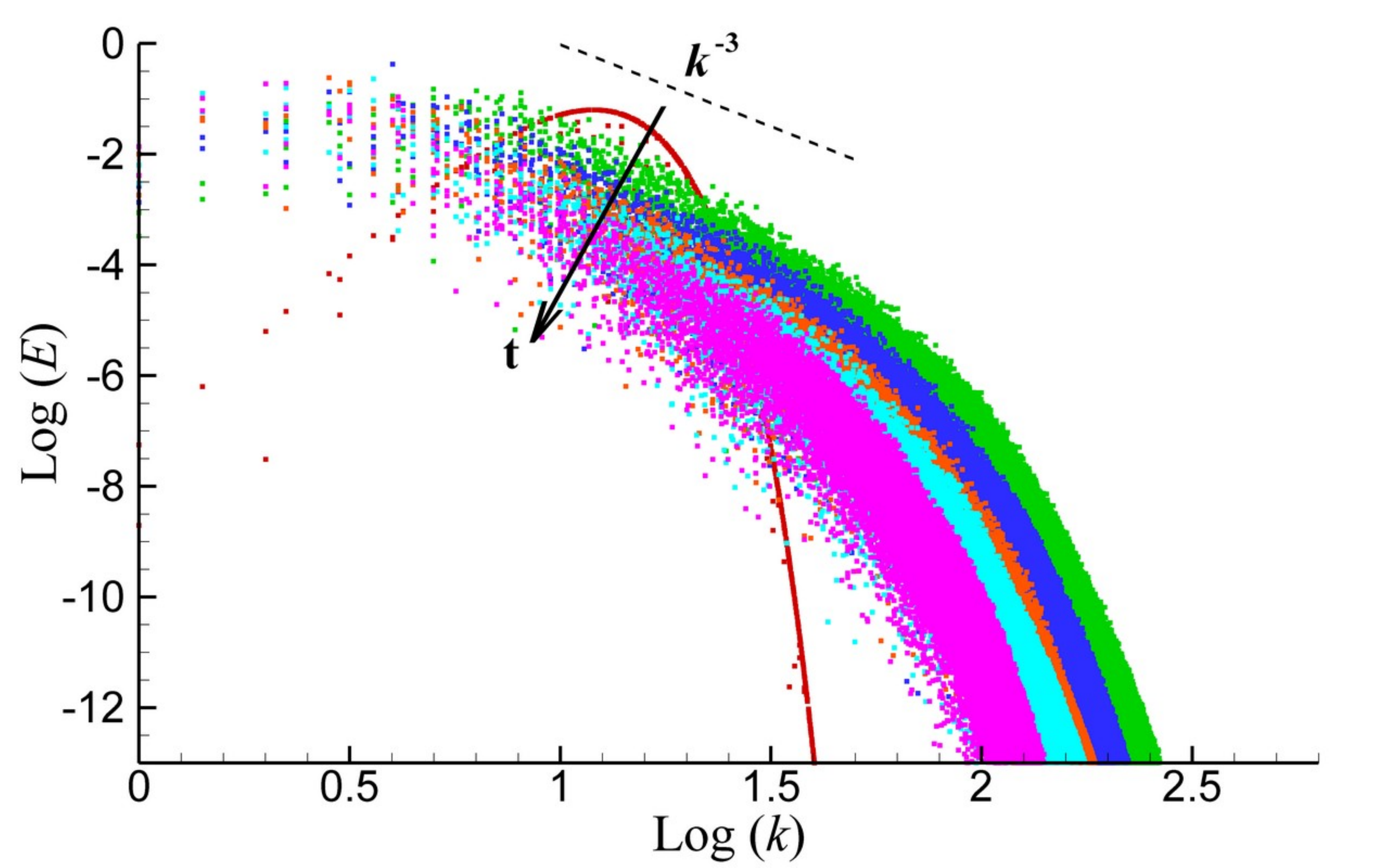}}
\subfigure[CD6 for $Re=6000$]{\includegraphics[width=0.5\textwidth]{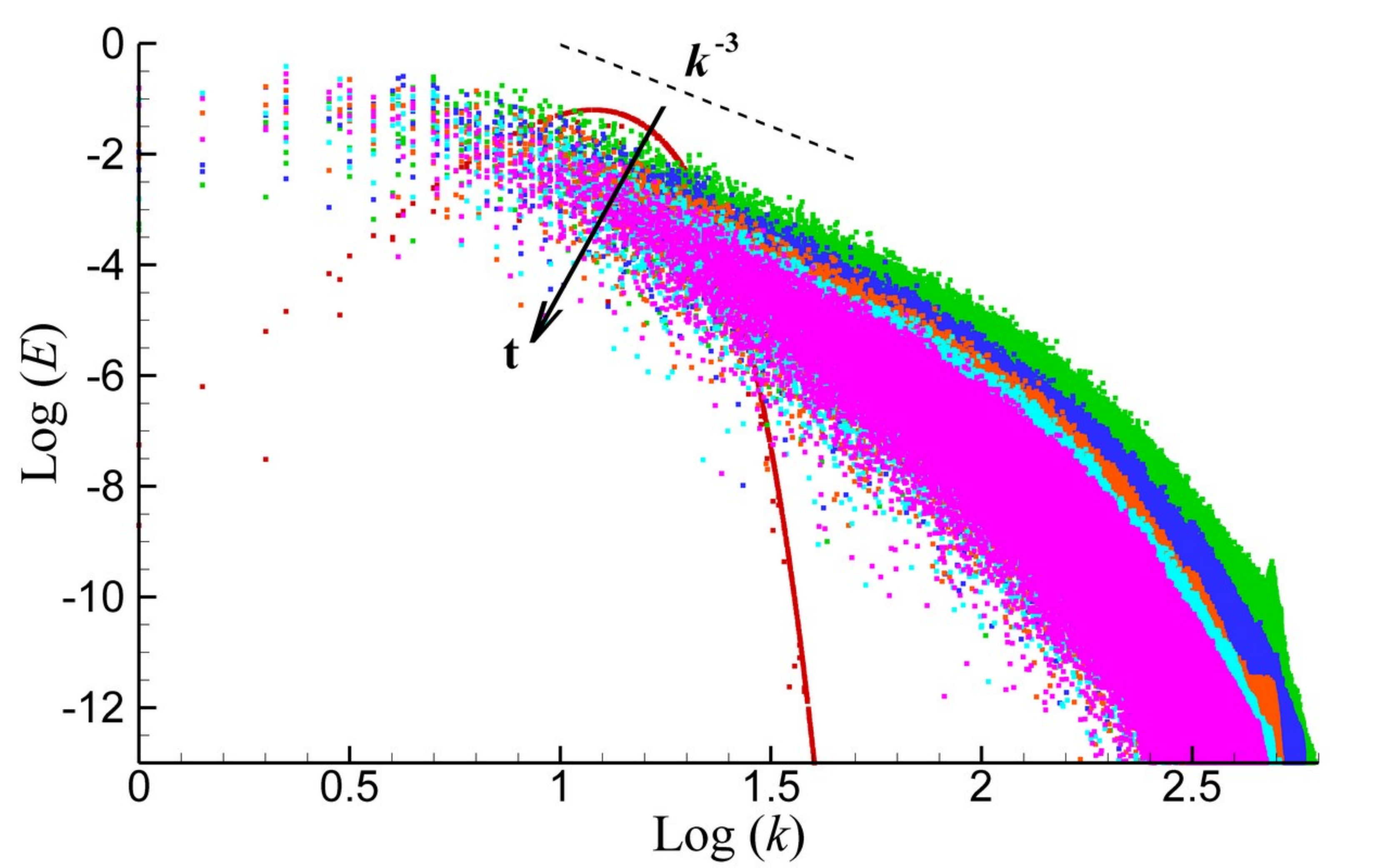}}
} \\
\mbox{
\subfigure[A4 for $Re=1000$]{\includegraphics[width=0.5\textwidth]{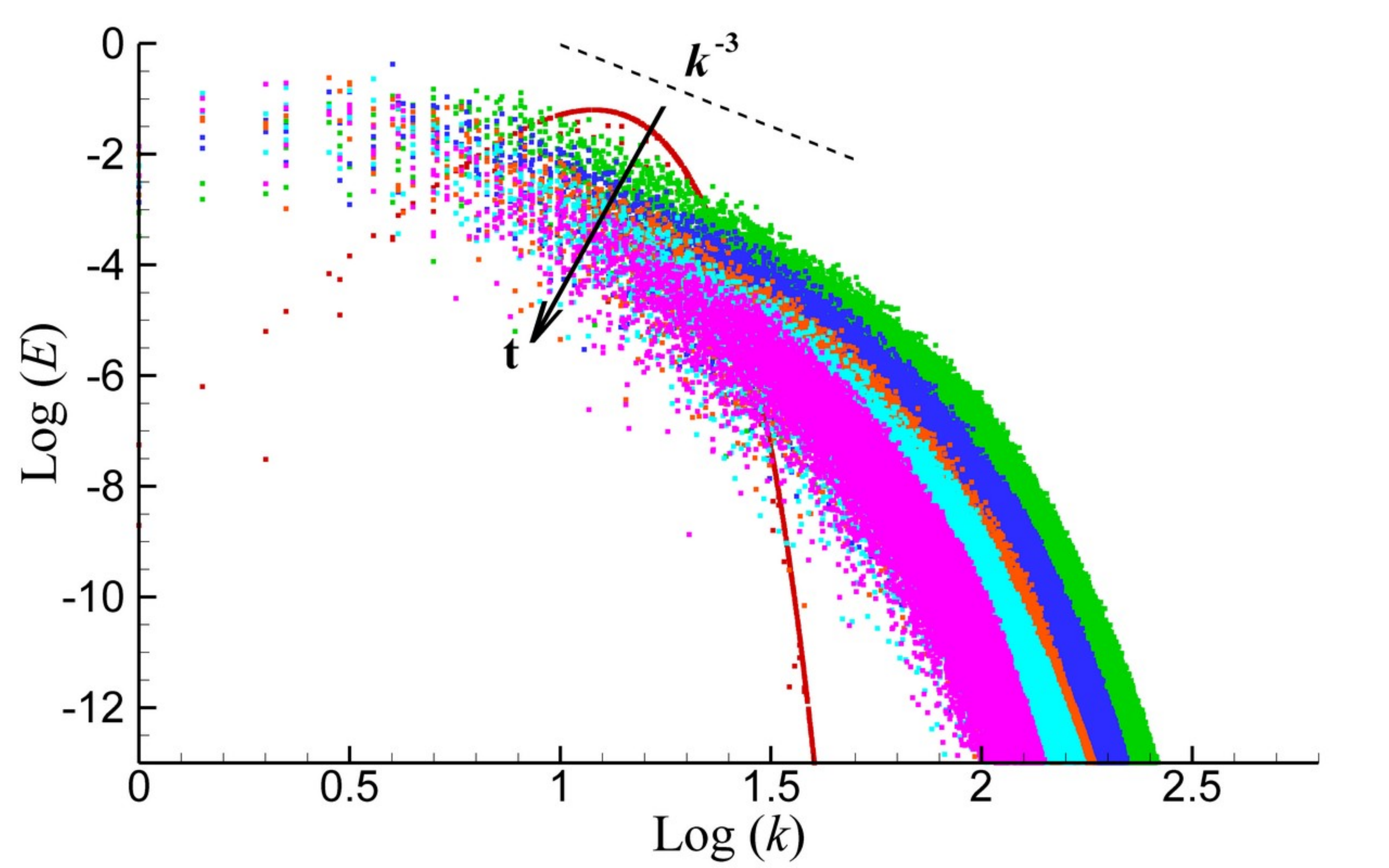}}
\subfigure[A4 for $Re=6000$]{\includegraphics[width=0.5\textwidth]{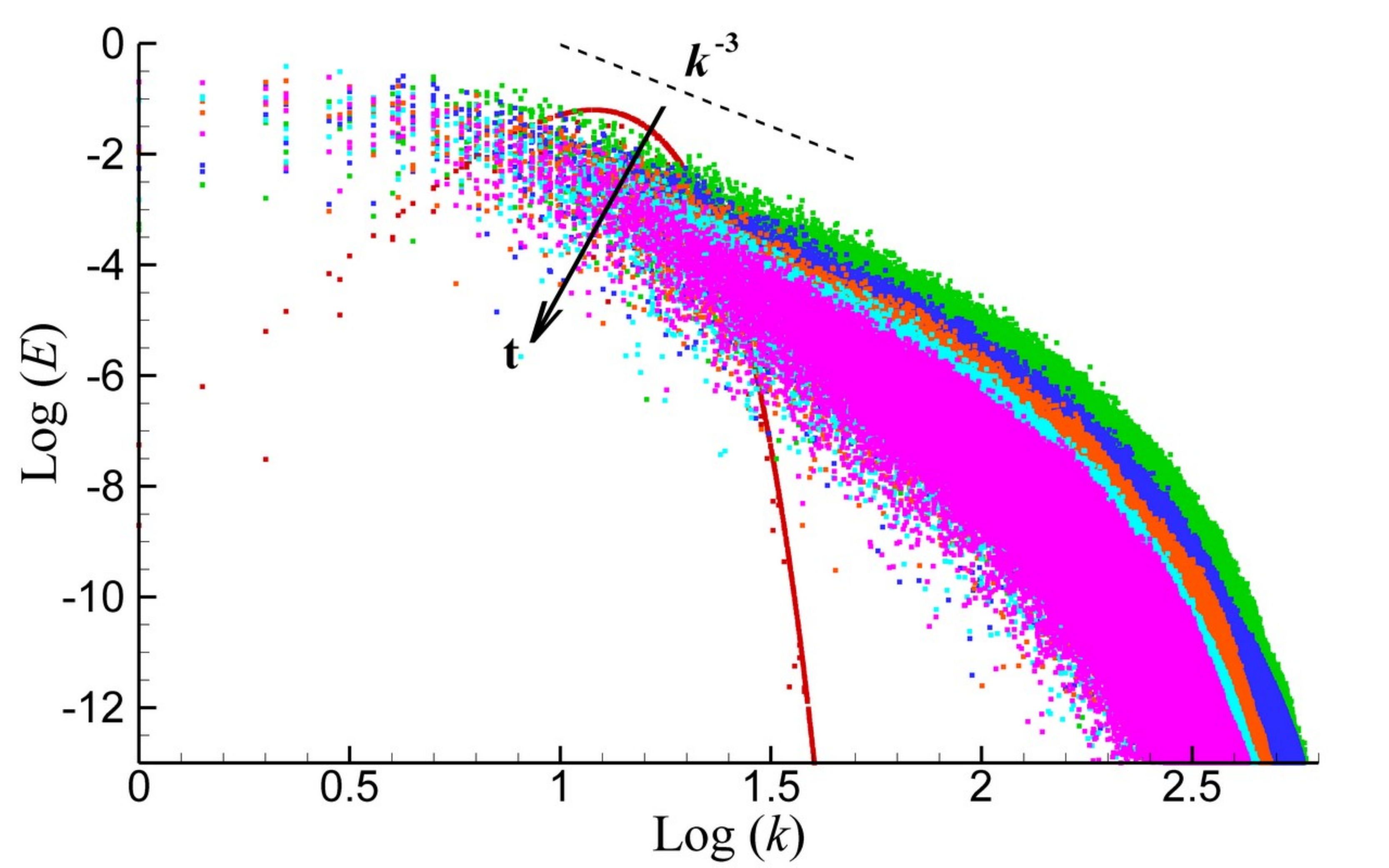}}
}
\caption{Evolution of energy spectra in decaying turbulence using a resolution of $1024^2$ obtained by (a) the sixth-order compact difference (CD6) scheme at $Re=1000$ ($Re_c = 6.13$), (b) the sixth-order compact difference (CD6) scheme at $Re=6000$ ($Re_c = 36.78$), (c) the forth-order Arakawa (A4) scheme at $Re=1000$ ($Re_c = 6.13$), and (d) the forth-order Arakawa (A4) scheme at $Re=6000$ ($Re_c = 36.78$). Energy spectra, defined by Eq.~(\ref{eq:esp}), are shown for times $t=0$, 2, 4, 6, 8, and 10.}
\label{fig:com-2}
\end{figure*}

Finally, we would like to test the performance of the finite difference formulations for extremely high cell Reynolds numbers. Two sets of numerical experiments are performed for a resolution of $256^2$ at $Re=3000$ and $Re=6000$. The corresponding cell Reynolds numbers in these sets are $Re_c = 73.63$ and $Re_c = 147.26$, respectively. A comparison of the fourth-order difference schemes is shown in Fig.~\ref{fig:com-1} illustrating the evolution of the energy spectra for these Reynolds number. Instead of presenting the angle averaged energy spectra, we prefer to show the full energy spectra, as defined by Eq.~(\ref{eq:esp}), for each pair $\textbf{k}=(k_x, k_y)$ in wave space to make the comparison more visible. This figure clearly illustrates that a higher degree of pile-up phenomenon occurs for all the finite difference schemes except the Arakawa scheme. Some sort of aliasing errors produce the growth of unphysical small scales. We demonsrate that these peaks of energy at small scales can be eliminated using the conservative Arakawa scheme. Although there is no big difference between the Arakawa scheme and the explicit difference scheme for smaller cell Reynolds numbers as shown in the previous sections, the discrete global conservation properties of the Arakawa schemes prevent the pile-up phenomena for higher cell Reynolds numbers.

Another sets of experiments using a higher resolution of $1024^2$ are also performed to demonstrate the limits of the difference schemes by considering the pile-up phenomenon. Evolution of the energy spectra for both sixth-order compact and Arakawa schemes are shown in Fig.~\ref{fig:com-2}. Based on our numerical experiments, it is concluded that the pile-up phenomenon occurs with non fully conservative schemes for higher cell Reynolds numbers ($Re_c > 30$). Therefore, the Arakawa scheme is a better candidate for underresolved simulations. It is mandatory to conserve energy to have a realistic simulations for higher cell Reynolds number. High-order centered schemes often show spurious wiggles for convective dominated problems when the numerical resolution is insufficient. The cell Reynolds number is restricted in some degree for these schemes. As the cell Reynolds number becomes larger, the nature of the finite difference equation is changed because of the cell size used. The spurious Nyquist signal would be generated almost instantaneously that propagates into the whole domain resulting an instability in the solution. In order to prevent these numerical instabilities for a centered difference scheme without enough resolution, the filtering methodology can be also used for underresolved flows.


\section{Summary and conclusions}
\label{sec:decayC}
A systematic comparison of a variety of high-order accurate finite difference schemes has been performed for two-dimensional decaying turbulence simulations by solving the vorticity-streamfunction formulation of two-dimensional incompressible Navier-Stokes equations. The following schemes were considered: the 6th-, 4th-, and 2nd-order explicit difference schemes, the 6th- and 4th-order compact difference schemes, the 4th- and 2nd-order Arakawa schemes, and the 4th-order dispersion-relation-preserving scheme. The objective of this study was to determine the accuracy and efficiency of these schemes for long-time integration of decaying turbulence simulations, which serves as an instance of a complex flow. We compared the schemes with the spectrally accurate Fourier-Galerkin pseudospectral method. A brief description of these algorithms is given in Section~\ref{sec:NumS}. In order to eliminate the errors coming from boundary conditions, we performed our experiments in a doubly periodic domain. This setting provides a suitable framework for testing the spatial discretization schemes.

We first validated the methods by solving the Taylor-Green vortex problem which is one of the available exact solutions for two-dimensional unsteady incompressible flows. We showed that the theoretical orders of accuracies of the schemes were obtained in practice. Then we solved a more challenging double shear layer benchmark problem which has been used to test the grid-independence of many incompressible algorithms, since the presence of thinner and thinner shear layers evolving with time is not captured by low grid resolution representations. We found that the low-order difference schemes require about twice the resolution in each direction to achieve the same results as the high-order difference schemes. We also demonstrated that numerical oscillations near the sharp vorticity gradients can be eliminated using the high-order difference formulations.

We tested the efficiency and accuracy of several variants of third- and fourth-order Runge-Kutta time stepping algorithms for decaying two-dimensional turbulence. The numerical stabilities of the spatio-temporal schemes were addressed by performing a time refinement study. We first showed that the fourth-order Runge-Kutta schemes have greater accuracy for a given time step and have larger allowable stability regions than the third-order Runge-Kutta schemes. We found that the spatially second-order accurate schemes can run using almost triple the effective time step compared with the sixth-order schemes, and can also run using almost double the effective time step compared with the fourth-order schemes, thereby making them more competitive with the high-order finite difference schemes. On the other hand, we showed that the second-order schemes usually require double the resolution in each direction to be able to obtain a similar accuracy. Therefore, we found that higher order schemes become more effective in terms of the tradeoff between the accuracy and efficiency.

We studied the Reynolds number dependence of freely evolving decaying turbulence. We showed that the predicted energy spectrum asymptotically converged to the theoretical $k^{-3}$ scaling as the Reynolds number increased, which is predicted by the KBL theory for forward cascading two-dimensional turbulence in the inviscid limit. We also demonstrated that the analysis was also relatively free from temporal discretization errors. We then tested the spatial finite difference schemes for various Reynolds numbers. We defined a cell Reynolds number $Re_c=Re 2\pi/N_x$. We demonstrated that for well resolved simulations, accurate results are obtained for all the schemes if $Re_c <2$. If we increase the Reynolds number, however, the difference between the high-order accurate and second-order accurate schemes increases dramatically. Compact difference schemes provide more accurate results if $Re_c < 30$. Spurious Nyquist signals have been observed for larger cell Reynolds numbers resulting in a pile-up phenomenon. In this case, the fully conservative Arakawa schemes gives more accurate results.

Our results demonstrate the importance of high-order accurate representations in convection-dominated complex flow problems, which, in general, require long time integrations. Although the order of accuracy is less important for well-resolved direct numerical simulations, it becomes significant for $Re_c > 2$. We show that, for fully resolved simulations, high-order accurate schemes do not require spectral-like simulation times. In fact, spectral-like accuracy is obtained with a speed-up factor of 5 over the pseudospectral method using the 6th-order compact scheme. We also demonstrate that the fourth-order Arakawa scheme is a better choice for higher cell Reynolds number computations.

\bibliographystyle{elsarticle-num}
\bibliography{references}

\end{document}